\documentclass[aps,amsmath,amssymb,reprint,superscriptaddress,longbibliography,UTF8]{revtex4-2}

\usepackage{float}
\usepackage{graphicx}
\usepackage{dcolumn}
\usepackage{bm}
\usepackage{mathrsfs,amssymb,amsmath}
\usepackage{ulem}
\usepackage[dvipsnames]{xcolor}
\usepackage{url}
\usepackage[colorlinks=true,linkcolor=blue,citecolor=blue,anchorcolor=blue,urlcolor=blue]{hyperref}
\usepackage{lineno}
\renewcommand*{\today}{October 29, 2022}

\expandafter\ifx\csname package@font\endcsname\relax\else
 \expandafter\expandafter
 \expandafter\usepackage
 \expandafter\expandafter
 \expandafter{\csname package@font\endcsname}%
\fi
\begin{document}

\title{Emerging versatile two-dimensional MoSi$_2$N$_4$ family}

\author{Yan Yin}
\affiliation{Institute for Frontier Science \& State Key Lab of Mechanics and Control of Mechanical Structures \&  Key Lab for Intelligent Nano Materials and Devices of Ministry of Education \& College of Aerospace Engineering, Nanjing University of Aeronautics and Astronautics (NUAA), Nanjing 210016, China}

\author{Qihua Gong\textcolor{blue}{$^*$}}
\affiliation{Institute for Frontier Science \& State Key Lab of Mechanics and Control of Mechanical Structures \&  Key Lab for Intelligent Nano Materials and Devices of Ministry of Education \& College of Aerospace Engineering, Nanjing University of Aeronautics and Astronautics (NUAA), Nanjing 210016, China}
\affiliation{MIIT Key Lab of Aerospace Information Materials and Physics \& College of Physics, Nanjing University of Aeronautics and Astronautics (NUAA), Nanjing 210016, China}

\author{Min Yi}
\email{yimin@nuaa.edu.cn}
\affiliation{Institute for Frontier Science \& State Key Lab of Mechanics and Control of Mechanical Structures \&  Key Lab for Intelligent Nano Materials and Devices of Ministry of Education \& College of Aerospace Engineering, Nanjing University of Aeronautics and Astronautics (NUAA), Nanjing 210016, China}

\author{Wanlin Guo\textcolor{blue}{$^*$}}
\affiliation{Institute for Frontier Science \& State Key Lab of Mechanics and Control of Mechanical Structures \&  Key Lab for Intelligent Nano Materials and Devices of Ministry of Education \& College of Aerospace Engineering, Nanjing University of Aeronautics and Astronautics (NUAA), Nanjing 210016, China}

\date{\today}

\begin{abstract}
The discovery of two-dimensional (2D) layered MoSi$_2$N$_4$ and WSi$_2$N$_4$ without knowing their 3D parents by chemical vapor deposition in 2020 has stimulated extensive studies of 2D MA$_2$Z$_4$ system due to its structural complexity and diversity as well as versatile and intriguing properties. Here, a comprehensive overview on the state-of-the-art progress of this 2D MA$_2$Z$_4$ family is presented. Starting by describing the unique sandwich structural characteristics of the emerging monolayer MA$_2$Z$_4$, we summarize and anatomize their versatile properties including mechanics, piezoelectricity, thermal transport, electronics, optics/optoelectronics, and magnetism. The property tunability via strain engineering, surface functionalization and layered strategy is also elaborated. Theoretical and experimental attempts or advances in applying 2D MA$_2$Z$_4$ to transistors, photocatalysts, batteries and gas sensors are then reviewed to show its prospective applications over a vast territory. We further discuss new opportunities and suggest prospects for this emerging 2D family. The overview is anticipated to guide the further understanding and exploration on 2D MA$_2$Z$_4$.
\end{abstract}

\maketitle

\section{introduction}
Two-dimensional (2D) materials have intrigued great attentions over the years since their fantastic characteristics are close to and even superior than their bulk counterparts, for instance, excellent mechanical properties~\cite{Akinwande2017-EML-Mechanics, Lee2008-science-Graphene-mechanics, Bertolazzi2011-ACSNano-MoS2-mechanics}, ultrahigh heat conduction~\cite{Balandin2011-NatMater-thermal, Li2020-AFM-boronthermal, Qian2021-NatMater-thermal}, unique quantum effects in low dimensions (e.g., superconductivity and quantum hall effect, etc.)~\cite{Wang2019-PRL-Superconductivity, Bekaert2019-PRL-H-MoB2-superconductivity, Li2021-MTP-review-superconductivity, Liu2019-NatPhys-BiGraphene-QHE, Shi2020-NatNanotech-WSe2-QHE}. 
Since graphene has been successfully prepared by mechanical exfoliation~\cite{Geim2004-Science-graphene, Geim2009-Science-graphene, Yi2015-JMCA-Graphene-mechanicalExfoliation}, most 2D members have been prepared by the top-down exfoliation of their naturally existed bulk parent materials, e.g., mechanical stripping materials (MoS$_2$, h-BH, NbSe$_2$, MnBi$_2$Te$_4$)~\cite{Novoselov2005-2D-mechanicalexfoliation, Geim2013-Nature-vdWs, Otrokov2019-Nature-MnBi2Te4}, interface-assisted exfoliation materials (black phosphorene, FeSe, Fe$_3$GeTe$_2$, RuCl$_3$, PtSe$_2$, PtTe$_2$, PdTe$_2$, and CrSiTe$_3$)~\cite{Dong2018-ChemRev-interface-assisted, Huang2020-NatCommun-2D-interfaceExfoliation}, and fluid dynamics assisted exfoliation~\cite{Yi2013-BN-hydrodynamics, Yi2014-kitchenBlender, Yi2016-fluiddynamics}. However, their structures are essentially limited by the parent materials.  
Bottom-up growth method, an another fabrication strategy, has been applied to synthesize dozens of novel 2D materials, e.g., monolayer borophene by direct evaporation~\cite{Mannix2015-Science-Borophene, Feng2016-NatChem-Borophene}, multilayer TMDs by chemical vapor deposition (CVD)~\cite{Zhan2012-Small-MoS2-SiO2, Zhang2019-AM-CVD-2DTMDs}, 2D van der Waals (vdW) heterostructures designed by mechanically assembled stacks~\cite{Novoselov2016-Science-vdWHetero}, and other materials prepared via layer-by-layer stacking in a specific sequence. However, 2D materials synthesized by this method face the challenge of discontinuous growth due to the surface energy constraints. 

\begin{figure*}[!t]
  \includegraphics[width=14cm]{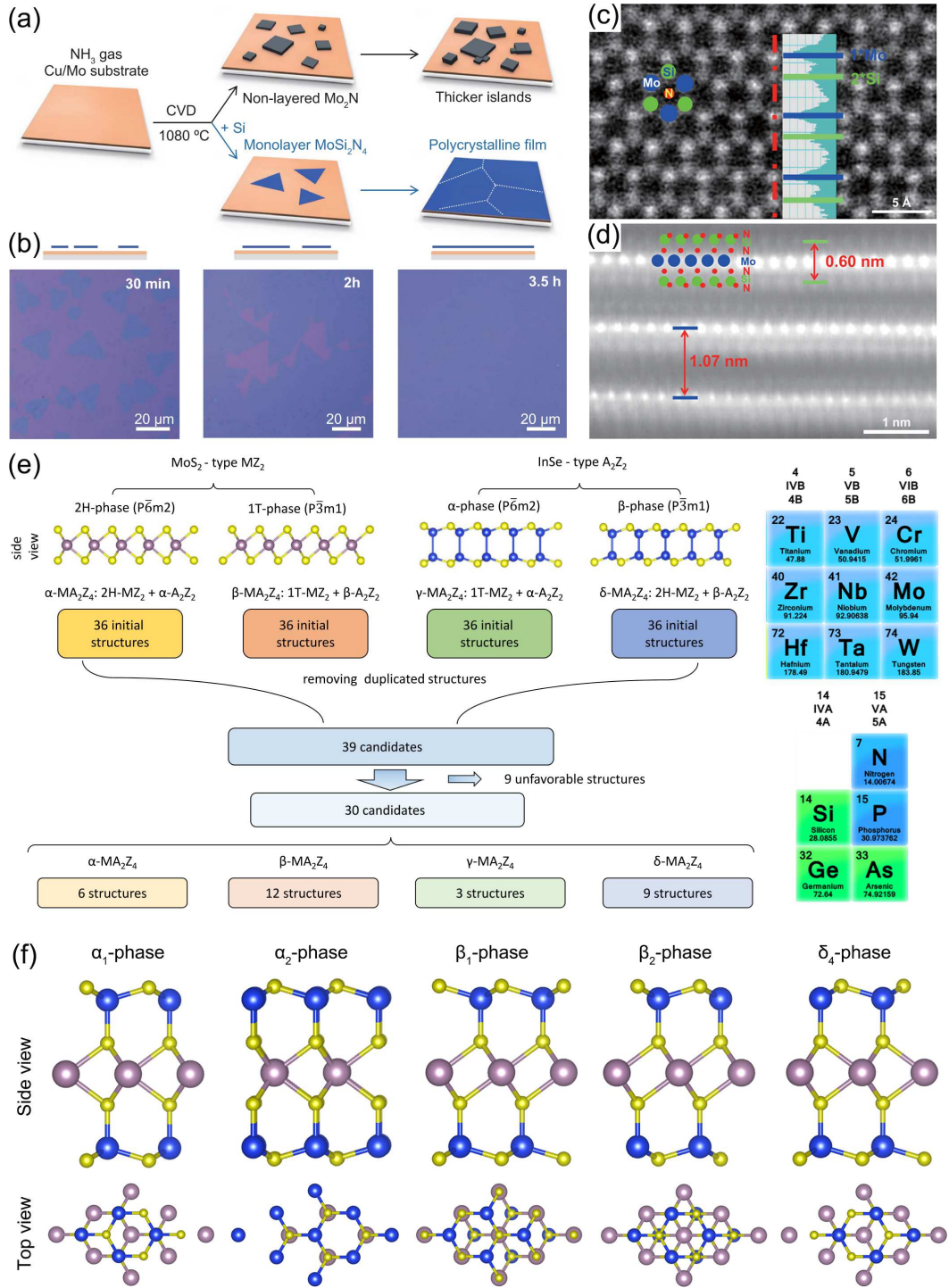}
  \caption{(a) Growth diagrams of MoN$_2$ and MoSi$_2$N$_4$ by CVD. (b) Optical images of MoSi$_2$N$_4$ grown by CVD for 30 min, 2 hours, and 3.5 hours. High-angle annular dark field scanning TEM (HAADF-STEM) imaging: (c) top view and (d) side view. (e) Schematic illustration of the intercalation method that uses the structures of a MoS$_2$-like monolayer (2H- and 1T phases) and those of an InSe-like monolayer ($\alpha$ and $\beta$ phases) to construct the structures of monolayer MA$_2$Z$_4$ family. (f) Unit cells for stable 2D MA$_2$Z$_4$ with different structures.~\cite{Hong2020-Science, Wang2021-NC}}
  \label{synthesis-component}
\end{figure*}

Recently, Ren and the coworkers successfully prepared novel 2D layer materials (MoSi$_2$N$_4$ and WSi$_2$N$_4$) without knowing their 3D parents, and broke through the obstacle of island growth~\cite{Hong2020-Science}. In their growing process, a Cu/Mo bilayer was used as the substrate and NH$_3$ gas as the nitrogen source. The crucial point of layer growth by CVD is that the appropriate atomic passivation of surface dangling bonds favors the decrement of surface energy. The growth diagrams of 2D molybdenum nitride without (MoN$_2$) or with Si (MoSi$_2$N$_4$) are shown in Fig.~\ref{synthesis-component}(a). The growing progress of MoN$_2$ without Si shows the obvious island domains and finally uneven micrometer-scale domains (appropriately 10~nm thick) form. On the contrary, MoN$_2$ with Si (MoSi$_2$N$_4$) firstly forms as triangular domains with uniform thickness, then expands to a centimeter-scale uniform polycrystalline film, and eventually maintains great ambient stabilization, as shown in Fig.~\ref{synthesis-component}(b). 
The high-angle annular dark field scanning TEM (HAADF-STEM) observation (Fig.~\ref{synthesis-component}(c) and (d)) indicates that MoSi$_2$N$_4$ is a MoN$_2$-derived septuple-atomic-layer compound built up in the order of N-Si-N-Mo-N-Si-N.  The vdW form of MoSi$_2$N$_4$ can be grown layer by layer due to the free of dangling bonds, indicating the possibility of large-scale preparation of this 2D compound.


Thanks to the breakthrough achievement in experimental synthesis, series of theoretical studies on MoSi$_2$N$_4$ and its derived materials have been further carried out. Density functional theory (DFT) calculations define this septuple-atomic-layer compounds as 2D MA$_2$Z$_4$ family~\cite{Wang2021-NC}. This 2D ternary material includes group \uppercase\expandafter{\romannumeral4}B, \uppercase\expandafter{\romannumeral5}B and \uppercase\expandafter{\romannumeral6}B elements for M-site atom, group \uppercase\expandafter{\romannumeral4}A elements for A-site atom and group \uppercase\expandafter{\romannumeral5}A elements for Z-site atom. The general approach to design MA$_2$Z$_4$ family layered van der Waals materials is proposed by intercalating MoS$_2$-type MZ$_2$ layer into an InSe-type A$_2$Z$_2$ monolayer (Fig.~\ref{synthesis-component}(e))~\cite{Wang2021-NC}.
If 2H and 1T phases of MZ$_2$ and $\alpha$ and $\beta$ phases of A$_2$Z$_2$ are considered, 4 types of monolayer MA$_2$Z$_4$ nanosheets can be obtained, i.e., $\alpha_i$-MA$_2$Z$_4$ ($i$~=~1--6), $\beta_i$-MA$_2$Z$_4$ ($i$~=~1--12), $\gamma_i$-MA$_2$Z$_4$ ($i$~=~1--3) and $\delta_i$-MA$_2$Z$_4$ ($i$~=~1--9). In total, there exist 30 structures of this family by removing duplicate symmetry and abandoning the structures that are energetically unstable. 
Five stable structures of 2D MA$_2$Z$_4$ are shown in Fig.~\ref{synthesis-component}(f). 
The experimentally synthesized and the most widely studied structure is $\alpha_1$-MA$_2$Z$_4$.
Due to the rich compositions and diverse structures, MA$_2$Z$_4$ family has exhibited intriguing physical and chemical characteristics, such as non-linear optics and second harmonic response~\cite{Yang2021-Nanoscale-Z-MA2Z4, Kang2021-PRB-2ndGeneration}, quantum behavior of strong exciton-phonon coupling in $\alpha_1$-MoSi$_2$N$_4$~\cite{Liang2022-APS-Exciton, Huang2022-AOM-Exciton-Phonon}, spin polarization and plasmon properties in $\alpha_1$-MoSi$_2$N$_4$~\cite{Wang2021-PRB-Ele-QuasiparticleModel, Zhao2021-APL-crcl3-hetero-spinvalley}, superconductivity in $\alpha_1$-TaSi$_2$N$_4$ and NbSi$_2$N$_4$~\cite{Yan2021-nanoscale-Ta-Nb-superconductivity}, topological insulating property in $\beta_2$-SrGa$_2$Se$_4$ and SrGa$_2$Te$_4$~\cite{Wang2021-NC}, ferromagnetic nature in $\delta_4$-VSi$_2$P$_4$~\cite{Wang2021-NC}, valley-half-semiconducting property in $\alpha_1$-VSi$_2$N$_4$~\cite{Zhou2021-npj-VSi2N4-framework}, Mott transition in XSi$_2$N$_4$~\cite{Wang2022-PRB-XSi2N4-mott}, etc. These intriguing properties could enable promising applications of MA$_2$Z$_4$ family in nanoelectronic devices such as magnetic tunnel junction, field effect transistors, highly sensitive and reusable gas sensors, etc.~\cite{Cui2021-PhysE-MoSi2N4-moleculardoped, Zhou2021-JPCL-monoWSi2N4-stacking, Islam2021-PRB-dimensionaleffect-spin, Wu2022-APL-MTJ, Ma2022-ASS-MoSi2N4defect, Zhao2021-ACSAEM-WGe2N4, Xiao2022-ACSomega-MoSi2N4-moleculardope, Gao2022-PRAppl-MoSi2P4}

In this paper, we aim to review the recent progress of the novel 2D layered MA$_2$Z$_4$ family, in terms of its structures, versatile properties and perspective applications.
After introducing the diverse structures in Fig.~\ref{synthesis-component}, we provide an extensive overview on the versatile properties regarding to mechanics, piezoelectricity, thermal transport, electronics, optics/optoelectronics, and magnetism. The tunability of each property via strain engineering, surface functionalization and layered strategy (e.g., multilayer or heterostructure) is also expounded.
Then, we introduce the perspective applications derived from the excellent properties of MA$_2$Z$_4$, including transistors, photocatalysts, batteries and sensors. Finally, we summarize the outstanding advantages of this family and suggest the conceivable outlook in the future.

\begin{table*}[htbp]
\centering
\caption{Structures and properties of semiconductors in MA$_2$Z$_4$ family: lattice constants ($a$), bandgap ($E_\text{g}^\text{PBE}$ for PBE and $E_\text{g}^\text{HSE}$ for HSE), carrier mobility ($\mu_{\text{e}}$ for electron and $\mu_{\text{h}}$ for holes), elastic modulus ($Y$), tensile strength ($E$), Poisson's ratio ($\nu$), thermal conductivity ($\kappa$).~\cite{Hong2020-Science, Wang2021-NC, Mortazavi2021-NanoEnergy, Liu2021-PLA-MoSi2P4, Yin2021-ACSAMI-MSi2Z4}} 
\renewcommand\arraystretch{1.5} 
\setlength{\tabcolsep}{0.4mm}
\begin{tabular}{ccccccccccc}
\hline
Struc. &Phase &$a$ &$E_\text{g}^\text{PBE}$ &$E_\text{g}^\text{HSE}$ &$\mu_{\text{e}}$ &$\mu_{\text{h}}$ &$Y$ &$E$ &$\nu$ &$\kappa$ \\ 
& &(\AA) &(eV) &(eV) &\multicolumn{2}{c}{(cm$^2$V$^{-1}$s$^{-1}$)} & &(GPa) & &(Wm$^{-1}$K$^{-1}$) \\ 
\hline
MoSi$_2$N$_4$ &$\alpha_1$ &2.91 &1.74($\Gamma$-K) &2.31($\Gamma$-K) &200 &1000 &491.4$\pm$139.1~GPa &65.8$\pm$18.3 &0.28 &417--439  \\ 
MoSi$_2$P$_4$ &$\alpha_1$  &3.47 &0.70(K-K) &0.99(K-K) &246--258 &1065--1429 &159~GPa &17.5-21.2 &* &116--122  \\
&$\alpha_2$ &3.46 &0.91(K-K) &1.19(K-K) &* &* &* &* &* &*  \\
MoSi$_2$As$_4$ &$\alpha_1$ &3.62 &0.56(K-K) &0.98(K-K) &* &* &* &* &* &46  \\
&$\alpha_2$ &3.61 &0.74(K-K) &1.02(K-K) &* &* &* &* &* &*  \\
MoGe$_2$N$_4$ &$\alpha_1$ &3.02--3.04 &0.91--0.99($\Gamma$K-K) &1.27--1.38($\Gamma$K-K) &490 &2190 &362~GPa &40.3--42.1 &* &286  \\
MoGe$_2$P$_4$ &$\alpha_1$ &3.55 &0.04($\Gamma$-K) &0.84($\Gamma$-K) &* &* &139~GPa &15.3--18.4 &* &63  \\
&$\alpha_2$ &3.53 &0.56(K-K) &0.95(K-K) &* &* &* &* &* &*  \\
MoGe$_2$As$_4$ &$\alpha_2$ &3.69 &0.47(K-K) &0.83($\Gamma$K-K) &* &* &* &* &* &*  \\

WSi$_2$N$_4$ &$\alpha_1$ &2.91 &2.08($\Gamma$-K) &2.57--2.66($\Gamma$-K) &320 &2026 &506~GPa &55.5--59.2 &0.27 &401--503  \\ 
WSi$_2$P$_4$ &$\alpha_1$ &3.48 &0.53(K-K) &0.81(K-K) &* &* &167~GPa &18.8--22.0 &* &129  \\
&$\alpha_2$ &3.46 &0.86(K-K) &1.11(K-K) &* &* &* &* &* &*  \\
WSi$_2$As$_4$ &$\alpha_2$ &3.61 &0.71(K-K) &0.95(K-K) &* &* &* &* &* &*  \\
WGe$_2$N$_4$ &$\alpha_1$ &3.02 &1.15--1.29($\Gamma$K-K) &1.51--1.69($\Gamma$K-K) &690 &2490 &384~GPa &42.6--44.5 &* &322  \\
WGe$_2$P$_4$ &$\alpha_1$ &3.55 &0.48(K-K) &0.73(K-K) &* &* &145~GPa &16.5--19.3 &* &64  \\
&$\alpha_2$ &3.54 &0.63($\Gamma$K-K) &0.89(K-K) &* &* &* &* &* &*  \\ 
WGe$_2$As$_4$ &$\alpha_2$ &3.69 &0.50($\Gamma$K-$\Gamma$K) &0.78(K-K) &* &* &* &* &* &*  \\

CrSi$_2$N$_4$ &$\alpha_1$ &2.84 &0.49($\Gamma$-K) &0.94(K-K) &* &* &468~GPa &55.4--57.8 &* &332--348  \\ 
CrSi$_2$P$_4$ &$\alpha_1$ &3.42 &0.28(K-K) &0.64(K-K) &* &* &154~GPa &18.5--21.2 &* &120  \\
&$\alpha_2$ &3.41 &0.34($\Gamma$K-K) &0.65(K-K) &* &* &* &* &* &*   \\
CrGe$_2$N$_4$ &$\alpha_1$ &2.98 &0.49($\Gamma$-K) &0.31($\Gamma$-K) &* &* &340~GPa &38.1--38.7 &* &198  \\ 
CrGe$_2$P$_4$ &$\alpha_2$ &3.49 &0.04($\Gamma$K-K) &0.36($\Gamma$K-K) &* &* &* &* &* &*  \\

TiSi$_2$N$_4$ &$\alpha_1$ &2.93 &1.57($\Gamma$M-M) &2.50($\Gamma$M-M) &* &* &* &* &* &107  \\

ZrSi$_2$N$_4$ &$\alpha_1$ &3.04 &1.55($\Gamma$M-M) &2.41($\Gamma$M-M) &* &* &382~N/m &* &0.32 &82   \\
&$\beta_2$ &3.05 &1.00($\Gamma$-M) &0.36($\Gamma$-M) &* &* &400~N/m &* &0.26 &*  \\
ZrGe$_2$N$_4$ &$\beta_2$ &3.19 &1.04($\Gamma$-$\Gamma$) &2.34($\Gamma$-$\Gamma$) &* &* &* &* &* &*  \\

HfSi$_2$N$_4$ &$\alpha_1$ &3.02 &1.80($\Gamma$M-M) &2.70($\Gamma$M-M) &* &* &406~N/m &* &0.32 &124  \\
&$\beta_2$ &3.04 &1.21($\Gamma$-M) &2.21($\Gamma$-M) &* &* &420~N/m &* &0.25 &*  \\
HfGe$_2$N$_4$ &$\beta_2$ &3.18 &1.15($\Gamma$-$\Gamma$) &2.45($\Gamma$-$\Gamma$) &* &* &* &* &* &*  \\

PdSi$_2$N$_4$ &$\beta_2$ &2.99 &2.50($\Gamma$-M) &3.80($\Gamma$-M) &* &* &356~N/m &* &0.29 &*  \\

PtSi$_2$N$_4$ &$\beta_2$ &3.02 &2.50($\Gamma$-M) &3.80($\Gamma$-M) &* &* &349~N/m &* &0.30 &*  \\

\hline
\end{tabular}
\label{parameters}
\end{table*}

\section{Versatile properties}

\subsection{Mechanical properties}

\begin{figure}[!t]
  \includegraphics[width=7cm]{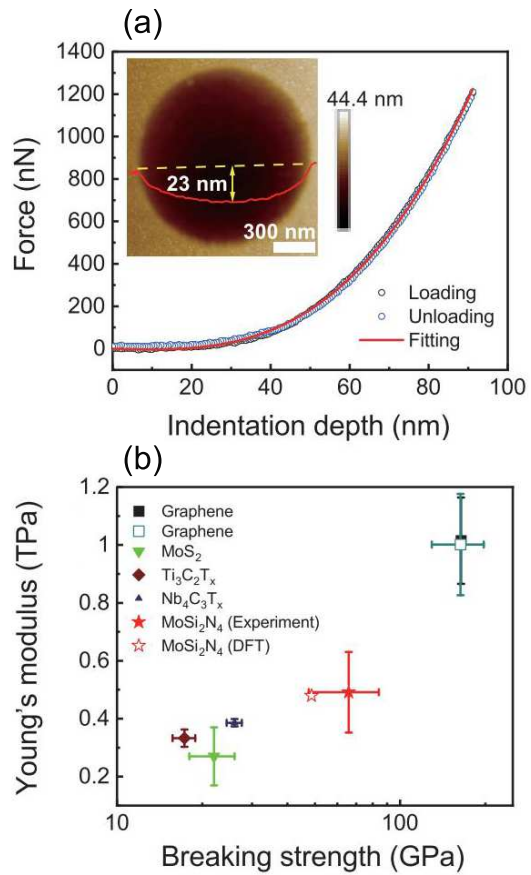}
  \caption{(a) Force-displacement curve of monolayer MoSi$_2$N$_4$ by atomic force microscopy nanoindentation, inset: the nanoindentation profile of suspended MoSi$_2$N$_4$. (b) Mechanical properties of MoSi$_2$N$_4$ and other 2D materials.~\cite{Hong2020-Science}}
  \label{mechanics}
\end{figure}

Since the extraordinary mechanical properties of graphene (103~GPa of the in-plane stiffness and nearly 1~TPa of the elastic modulus~\cite{Lee2013-Science-graphene-mechanical}), 2D materials with excellent mechanical properties have attracted great attentions. 
As for MoSi$_2$N$_4$, its experimentally measured tensile strength ($E$) and elastic modulus ($Y$) are 65.8$\pm$18.3~GPa and 491.4$\pm$139.1~GPa~\cite{Hong2020-Science}, respectively, which are nearly half of those in graphene and higher than those in most 2D TMDs (e.g., 22~GPa and 270$\pm$100~GPa of MoS$_2$)~\cite{Bertolazzi2011-ACSNano-MoS2-mechanics, Liu2014-NanoLett-MoS2-WS2-mechanics, Zhang2016-APL-WSe2-mechanics},
MXene (e.g., 17~GPa and 333~GPa of Ti$_3$C$_2$T$_x$~\cite{lipatov2018-Ti3C2Tx}, 26~GPa and 386~GPa of Nb$_4$C$_3$T$_x$~\cite{Lipatov2020-AEM-Nb4C3Tx}), and black phosphorene~\cite{Wei2014-APL-BlackP-mechanical} (18~GPa and 166~GPa). 
In detail, Ren et al.~\cite{Hong2020-Science} measured the mechanical properties of monolayer MoSi$_2$N$_4$ via atomic force microscopy nanoindentation (Fig.~\ref{mechanics}(a)). With a diamond tip of 11.1~nm, the indentation of hole is about 23~nm. The elastic behavior of monolayer MoSi$_2$N$_4$ is demonstrated due to the well fitting force-displacement curves of loading and unloading states. Theoretical prediction of mechanical properties of 2D materials is mainly based on the linear elastic model. The theoretically calculated tensile strength and elastic modulus are 48.3--57.8~GPa and 479.1--487.0~GPa, respectively~\cite{Mortazavi2021-NanoEnergy, Li2021-PhysE-mechanism, Shojaei2021-JPDAPP-MoSi2N4}, in good agreement with experimental values. 

\begin{figure*}[!t]
\centering
\includegraphics[width=16cm]{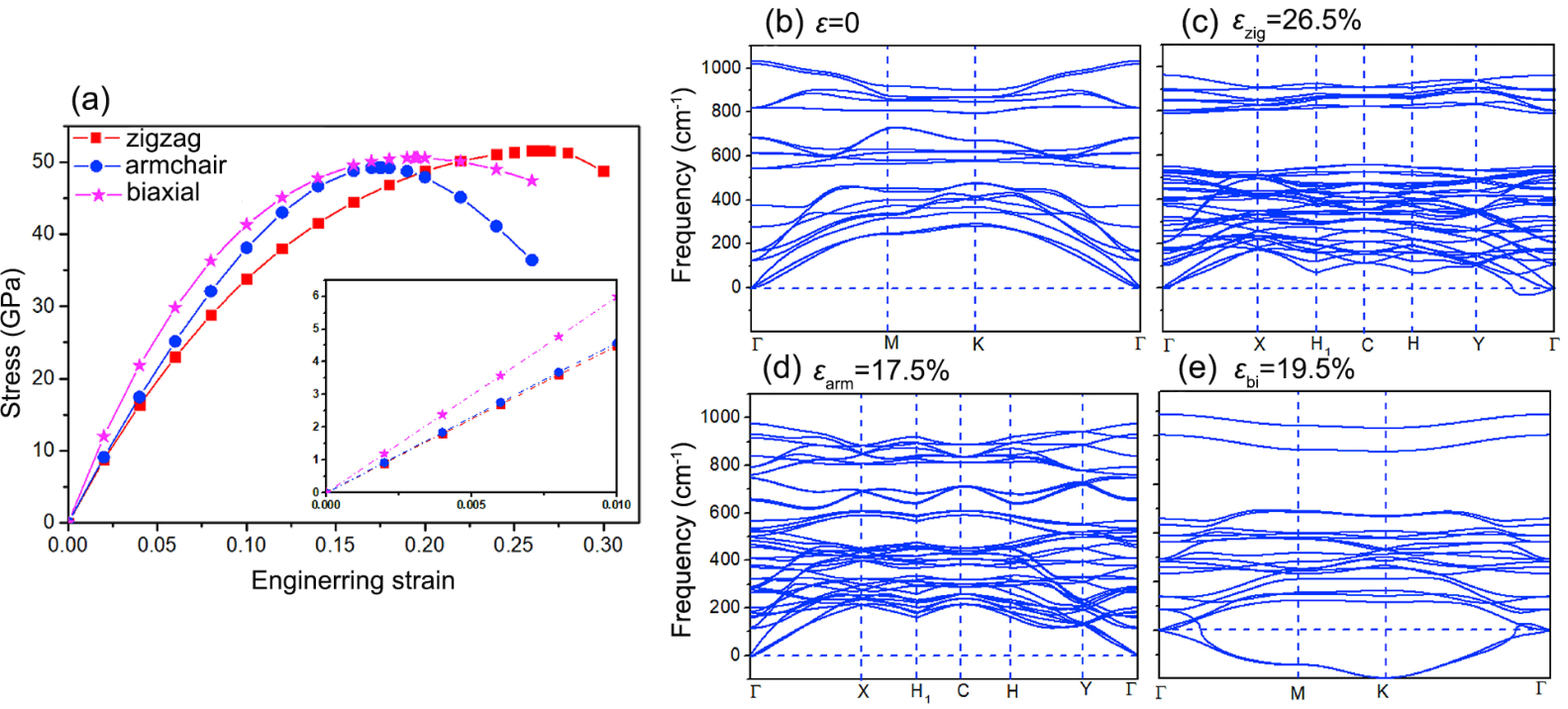}
\caption{(a) Strain-stress curve of monolayer MoSi$_2$N$_4$ and the corresponding phonon dispersion under different strains: (b) no strain, (c) zigzag uniaxial strain $\epsilon_{\text{zig}}$~=~26.5$\%$, (d) armchair uniaxial strain $\epsilon_{\text{arm}}$~=~17.5$\%$, (e) biaxial strain $\epsilon_{\text{bi}}$~=~19.5$\%$.~\cite{Li2021-PhysE-mechanism}}
\label{failure}
\end{figure*}

Actually, the critical strain and ideal tensile stress of 2D materials are vital indicators for practical applications, which depend on the elastic limit and lattice vibration~\cite{Marianetti2010-PRL-graphene-failure}. 
Li et al.~\cite{Li2021-PhysE-mechanism} focused on the elastic limit and failure mechanism of monolayer MoSi$_2$N$_4$. Under biaxial and uniaxial (zigzag or armchair) strains, the ideal strengths of monolayer MoSi$_2$N$_4$ are similar, about 50~GPa, while the corresponding critical strain is 19.5$\%$ (biaxial strain), 26.5$\%$ (zigzag strain) and 17.5$\%$ (armchair strain).
In Fig.~\ref{failure}, when the strains are below 20$\%$ ($\epsilon$~$<$~20$\%$), the tensile stresses ($\sigma$) are in the order of $\sigma_{\text{bi}}$~$>$~$\sigma_{\text{arm}}$~$>$~$\sigma_{\text{zig}}$. There exists an obvious yield phenomenon under biaxial and armchair strains when $\epsilon$~$\geq$~20$\%$, but the yield limit under a zigzag strain is about 25$\%$.
By fitting the initial strain-stress curve based on the linear regression up to 1$\%$ strain (the inset of Fig.~\ref{failure}(a)), the elastic moduli are calculated as $E_{\text{zig}}$~=~448.3~$\pm$~5.1~GPa and $E_{\text{arm}}$~=~457.8~$\pm$~3.9~GPa, which are more than twice those of MoS$_2$ ($E_{\text{zig}}$~=~197.9~$\pm$~4.3~GPa and $E_{\text{arm}}$~=~200~$\pm$~3.7~GPa)~\cite{Li2012-PRB-MoS2-mechanics}. The degeneracy of elastic moduli indicate a nearly elastic isotropy in monolayer MoSi$_2$N$_4$. 
On the other hand, the failure mechanism has been investigated on the aspect of lattice stability by phonon dispersion. When the tensile strength limit is reached under an armchair strain, the phonon dispersion has no imaginary frequency. This indicates lattice stability and further reveals that the failure phenomenon of monolayer MoSi$_2$N$_4$ is ascribed to the elastic failure of the SiN layer before reaching the critical strain. The obvious imaginary frequency of out-of-plane acoustic branch (ZA) before reaching tensile strength limit demonstrates that the failure mechanism of monolayer MoSi$_2$N$_4$ is attributed to phonon instability under the zigzag or biaxial strains. 

The mechanical parameters of other members in MA$_2$Z$_4$ family are listed in Table~\ref{mechanics}. Mortazavi et al.~\cite{Mortazavi2021-NanoEnergy} found in MA$_2$Z$_4$ family the mechanical properties are mainly affected by the terminating atom (Z-site) rather than the core atom (M-site). Three possible factors that influence the mechanical properties are proposed: Z-site atomic mass, structure and chemical bonds. Firstly, increasing Z-site atomic weight deteriorates the mechanical properties, but the elastic modulus has little change when M-site atomic mass is increased. For examples, MA$_2$N$_4$ has higher $Y$ and $E$ than MA$_2$P$_4$, while $Y$ and $E$ of MoSi$_2$N$_4$ and WSi$_2$N$_4$ are close. Secondly, since the M-A bonds are absolutely vertical, only M-Z and A-Z bonds participate in the deformation when applying an in-plane loading, indicating that the two bonds related to Z-site atoms determine the mechanical properties. Besides, chemical bonds formed with N atoms are always stronger than those with P/As atoms. Bonds formed with Si are sturdier than those with Ge, resulting in the highest elastic modulus of monolayer MSi$_2$N$_4$. 

The mechanical properties of some other MA$_2$Z$_4$-derived materials have also been investigated, e.g., CrC$_2$N$_4$, SnSi$_2$N$_4$, SnGe$_2$N$_4$ and XMoSiN$_2$ (X = S/Se/Te)~\cite{Mortazavi2021-MaterTodayEnergy-CrC2N4, Tian2021-PRB-SnSi2N4, Dat2022-RSCadv-SnGe2N4, Sibatov2022-ASS-XMoSiN2}. The predicted elastic modulus and tensile strength of CrC$_2$N$_4$ are as high as 676 and 54.8~GPa, respectively, while those of SnSi$_2$N$_4$ (478 and 47~GPa) are close to those of MoSi$_2$N$_4$. This indicates that the MA$_2$Z$_4$ structure offers excellent mechanical features, but the intrinsic mechanisms related to the elements and structures require further studies.

\subsection{Piezoelectricity, ferroelectricity and flexoelectricity}

\begin{figure*}[!t]
\centering
\includegraphics[width=17.5cm]{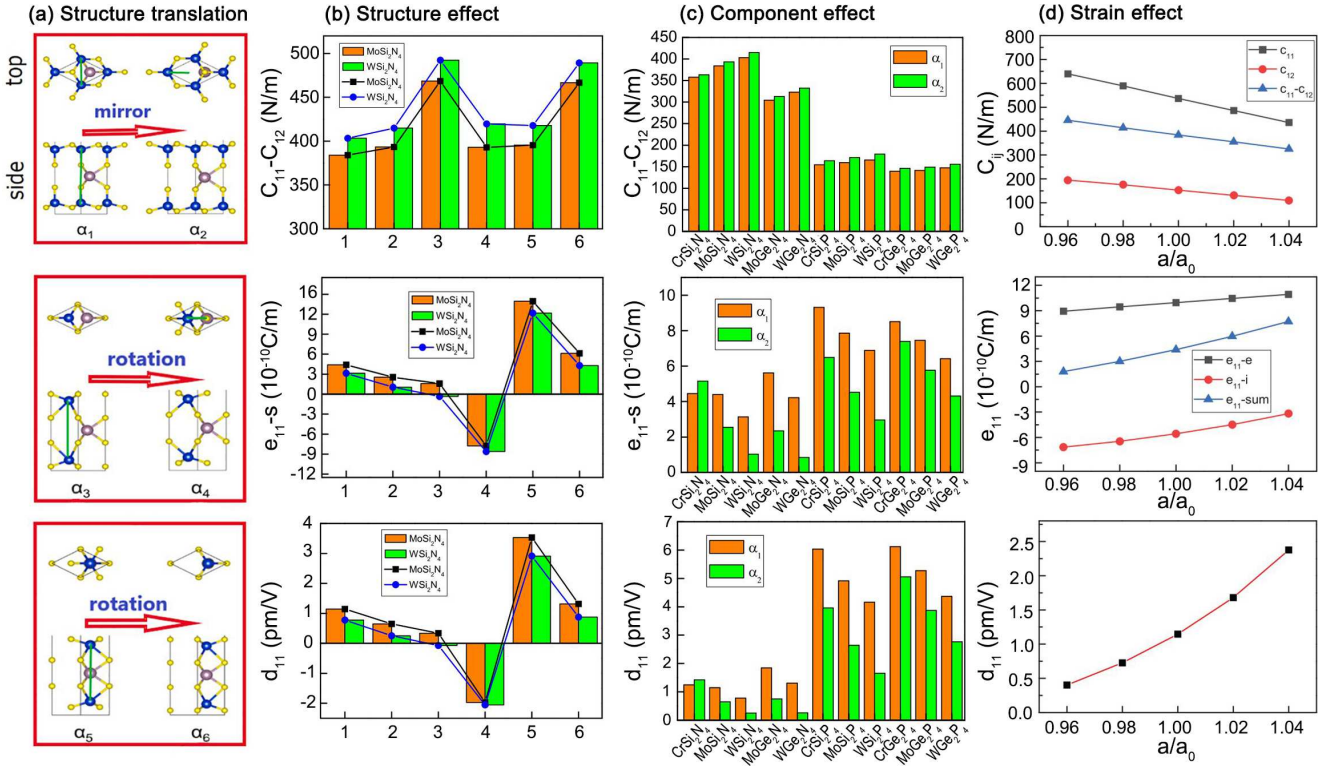}
\caption{(a) Structure translation of $\alpha_i$-MA$_2$Z$_4$ ($i$~=~1--6). (b--d) Structure, component and strain effects on piezoelectric parameters.~\cite{Guo2021-ComputMaterSci-structureEffect, Guo2020-EPL-MSi2N4-piezoelectricity}}
\label{piezoelectric}
\end{figure*}

Piezoelectricity of 2D materials can convert mechanical energy into electrical energy and $vice$ $versa$, which always occurs in semiconductors or insulators with broken inversion symmetry. The relax-ion piezoelectric coefficients ($e_{ijk}$ and $d_{ijk}$) are expressed as 
\begin{equation}
e_{ijk}~=~\frac{\partial P_{i}}{\partial \epsilon_{jk}}~=~e^{elc}_{ijk}+e^{ion}_{ijk}
\end{equation}
\noindent
and 
\begin{equation}
d_{ijk}~=~\frac{\partial P_{i}}{\partial \sigma_{jk}}~=~d^{elc}_{ijk}+d^{ion}_{ijk}
\end{equation}
\noindent
where $P_i$, $\epsilon_{jk}$, $\sigma_{jk}$ are polarization vector, strain and stress. The subscript $elc$ or $ion$ represent the electronic or ionic contributions. The relationship between $d_{ijk}$ and $e_{ijk}$ is built via elastic tensor ($C_{ij}$, with the Voigt notation). 
The independent parameters of tensor are reduced due to the symmetry of crystal structure.
In monolayer MA$_2$Z$_4$, the counterpart with $P\overline{6}m2$ space group prohibits the out-of-plane piezoelectric effect, so that only the in-plane piezoelectric strain and stress coefficients (i.e., $e_{11}$, $d_{11}$) and elastic coefficients (i.e., $C_{11}$, $C_{12}$) are considered~\cite{Guo2020-EPL-MSi2N4-piezoelectricity, Guo2021-ComputMaterSci-structureEffect}.

Structure effect on piezoelectricity of MA$_2$Z$_4$ has been investigated~\cite{Guo2021-ComputMaterSci-structureEffect}. The structures are divided into six configurations by different operations (translation, mirror and rotation) of A$_2$Z$_2$ layers, which are nominated as $\alpha_i$ ($i$~=~1--6), as shown in Fig.~\ref{piezoelectric}(a). It is observed that the values of $C_{11}-C_{12}$ of MSi$_2$N$_4$ (M~=~Mo, W) are close when $i$~=~1,~2,~4,~5, while those of $C_{11}-C_{12}$ are higher when $i$~=~3,~6, indicating $\alpha_3$ and $\alpha_6$ MSi$_2$N$_4$ (M~=~Mo/W) with high resistance to deformation. The structural sensibility of $d_{11}$ is homologous with that of $e_{11}$ from $\alpha_1$ to $\alpha_6$. $e_{11}$ of $\alpha_5$ exhibits the largest value among six structures, which is 13.95~$\times$~$10^{-10}$~C/m for MoSi$_2$N$_4$ and 12.17~$\times$~$10^{-10}$~C/m for WSi$_2$N$_4$. $d_{11}$ of $\alpha_5$-MoSi$_2$N$_4$ and $\alpha_5$-WSi$_2$N$_4$ is 3.53 and 2.91~pm/V, respectively. $d_{11}$ of $\alpha_1$-MoSi$_2$N$_4$ and $\alpha_1$-WSi$_2$N$_4$ (experimentally synthesized phases) is 1.15 and 0.78~pm/V, respectively.

Studies on the influence of different M/A/Z atoms reveal that $C_{11}-C_{12}$ of both $\alpha_1$ and $\alpha_2$ improves with the element periodicity increasing~(Fig.~\ref{piezoelectric}(c)). For instance, $C_{11}-C_{12}$ of MSi$_2$N$_4$ is in the order of CrSi$_2$N$_4$<MoSi$_2$N$_4$<WSi$_2$N$_4$. The larger elastic constants of MA$_2$N$_4$, compared with those of MA$_2$P$_4$ and other 2D materials (e.g., TMDs, metal oxides, and \uppercase\expandafter{\romannumeral3}--\uppercase\expandafter{\romannumeral5} semiconductors~\cite{Duerloo2012-2Dmater-inplane-piezo, Blonsky2015-2Dmater-inplane-piezo}), indicate that MA$_2$N$_4$ system is more rigid.
The change trend of $e_{11}$ and $d_{11}$ follows the opposite regularity, compared with elastic coefficients. $d_{11}$ of MA$_2$P$_4$ with the same M and A atoms is larger than that of MA$_2$N$_4$ for both $\alpha_1$ and $\alpha_2$ phases. 
$d_{11}$ of $\alpha_1$- and $\alpha_2$-MA$_2$Z$_4$ is in the range of 0.78--6.12~pm/V and 0.25--5.06~pm/V, respectively.
Monolayer MA$_2$P$_4$ nanosheets (e.g., $\alpha_1$-CrSi$_2$P$_4$, $\alpha_1$-MoSi$_2$P$_4$, $\alpha_1$-CrGe$_2$P$_4$, $\alpha_1$-MoGe$_2$P$_4$ and $\alpha_2$-CrGe$_2$P$_4$) show excellent piezoelectric response. In addition, the effect of A-site atoms on piezoelectric performance can be ignored so that $d_{11}$ of MoGe$_2$N$_4$ is close to that of MoSi$_2$N$_4$. $d_{11}$ of most monolayer MA$_2$P$_4$ nanosheets is even larger than that of 2D TMDs (e.g., $d_{11}$~=~3.65~pm/V of MoS$_2$, $d_{11}$~=~2.12~pm/V of WS$_2$, $d_{11}$~=~4.55~pm/V of MoS$_2$, and $d_{11}$~=~2.64~pm/V of WSe$_2$)~\cite{Cui2018-npj2D-piezo} and $d_{33}$ (3.1~pm/V) of bulk piezoelectric wurtzite GaN~\cite{Lueng1999-JAP-BulkAlN-GaN-piezo}. 

The compressive and tensile biaxial strain are shown to obviously improve and deteriorate the piezoelectric performance of MoSi$_2$N$_4$, respectively (Fig.~\ref{piezoelectric}(d))~\cite{Guo2020-EPL-MSi2N4-piezoelectricity}. 
Piezoelectric stress and strain responses are improved with the in-plane strain from -4$\%$ to 4$\%$. $d_{11}$ can be enhanced by 107$\%$  if a rensile biaxial strain of 4$\%$ is applied.  
VSi$_2$P$_4$, a spin-gapless semiconductor (SGS), possesses a wide range of properties due to its strain sensitivity. With the increasing strain, it presents as ferromagnetic metal (FMM), SGS, ferromagnetic semiconductor (FMS), or ferromagnetic half-metal (FMHM)~\cite{Guo2020-PCCP-VSi2P4}. In the strain range of 1$\%$--4$\%$, the coexistence of ferromagnetism and piezoelectricity can be achieved in FMS VSi$_2$P$_4$. Its $d_{11}$ under 1$\%$, 2$\%$ and 3$\%$ strain is 4.61, 4.94 and 5.27~pm/V, respectively.

There exist both in-plane and out-of-plane piezoelectric polarizations in Janus MSiGeN$_4$ (M~=~Mo/W) owing to the broken reflection symmetry along the out-of-plane direction~\cite{Guo2021-JMCC-Janus-MSiGeN4, Guo2021-JSemicond-Janus-MSiGeN4}. The in-plane piezoelectric coefficients of Janus MSiGeN$_4$ ($d_{11}$~=~1.494~pm/V for MoSiGeN$_4$, $d_{11}$~=~1.050~pm/V for WSiGeN$_4$) are between those of MSi$_2$N$_4$ and MGe$_2$N$_4$, while the out-of-plane stress piezoelectric coefficients ($d_{31}$) are --0.014 and 0.011~pm/V for MoSiGeN$_4$ and WSiGeN$_4$, respectively. 
Under an in-plane biaxial strain, $d_{11}$ of MSiGeN$_4$ is improved with the increasing $e_{11}$, which is similar with MA$_2$N$_4$. 
A tensile strain of 10$\%$ can increase $d_{11}$ of MoSiGeN$_4$ and WSiGeN$_4$ by several times, with the values up to 8.081 and 7.282~pm/V, respectively.
On the contrary, a compressive biaxial strain can effectively enhance the out-of-plane piezoelectric response. The MA$_2$Z$_4$-derived SrAlGaSe$_4$ has both in-plane ($d_{11}$~=~--1.865~pm/V/) and out-of-plane ($d_{31}$~=~--0.068~pm/V) piezoelectricity under uniaxial a tensile strain of 6$\%$~\cite{Guo2021-JMCC-SrAlGaSe4}.

In addition to piezoelectricity, the intrinsic ferroelectricity and its electrical switching in MA$_2$Z$_4$ are still open issues. For example, the sliding ferroelectricity is found in vdW MoA$_2$N$_4$ bilayer and multilayer~\cite{Zhong2021-JMCA-SlidingFerroelectricity}. 
The interlayer inequivalence caused by the stacking order directly generates the out-of-plane polarization.
Then, the induced vertical polarization can be switched via the interlayer sliding. The calculated vertical polarization is 3.36, 3.05, 2.49 and 3.44~pC/m for AB stacking bilayer MoSi$_2$N$_4$, MoGe$_2$N$_4$, CrSi$_2$N$_4$ and WSi$_2$N$_4$, respectively, which are higher than that of bilayer WTe$_2$ and BN~\cite{Fei2018-Nature-2D-ferroelectric, Li2017-ACSNano-2D-Verticalpolarization}.

The flexoelectricity of monolayer MA$_2$Z$_4$ is also of interest, which is often induced by bending deformation of 2D materials with a strain gradient~\cite{Mortazavi2021-NanoEnergy}. The out-of-plane bending flexoelectric coefficients of 12 kinds of MA$_2$Z$_4$ (M~=~Mo/Cr/W, A~=~Si/Ge, Z~=~N/P) are found in the range of 0.001--0.047~nC/m under a bending strain gradient of 0.3/Å. The highest flexoelectric coefficient of WGe$_2$N$_4$ is about 1.5 times higher than that of MoS$_2$~\cite{Zhuang2019-PRB-MoS2-flexo}. It is proposed that the enhancement of flexoelectricity is insufficient with the bending deformation, but is broadened by the construction of asymmetry structures (e.g., Janus counterparts)~\cite{Dong2017-ACSNano-JanusMXY-piezo, Javvaji2019-PRM-JanusTMDs-flexo}.

\subsection{Thermal conductivity}
Owing to the high tensile strength and thus strong bond interactions, the thermal conductivity of monolayer MA$_2$Z$_4$ family has attracted attentions. Based on thermal Boltzmann transport equations, the theoretical lattice thermal conductivity ($\kappa$) of MoSi$_2$N$_4$ is predicted as high as 400~Wm$^{-1}$K$^{-1}$ at room temperature~\cite{Mortazavi2021-NanoEnergy, Yin2021-ACSAMI-MSi2Z4, Yu2021-NewJPhys-Thermal, Shen2022-PCCP-Thermal},
which is larger than that of most other 2D materials, such as hydrogenated borophene (368~Wm$^{-1}$K$^{-1}$)~\cite{Li2020-NanoRes-BHBFBCl-thermal}, TMDs (23--142~Wm$^{-1}$K$^{-1}$)~\cite{Cai2014-PRB-MoS2-thermal, Gu2014-APL-TMD, Torres2019-2DMater-TMD}, group \uppercase\expandafter{\romannumeral4}A and \uppercase\expandafter{\romannumeral6}A compounds (0.26--9.8~Wm$^{-1}$K$^{-1}$)~\cite{Liu2018-PRB-GroupIVSe-thermal, Sun2019-JPCC-SnSe-thermal}.
But it is lower than that of graphene (3000--5000~Wm$^{-1}$K$^{-1}$)~\cite{Balandin2008-Nanolett-Graphene-thermal, Ghosh2008-APL-Graphene3000}, and h-BNs (600~Wm$^{-1}$K$^{-1}$)~\cite{Lindsay2011-PRB-BN600}. Such high thermal conductivity of monolayer MoSi$_2$N$_4$ is promising for heat conductors and thermal management in semiconductor devices.

\begin{figure*}[!t]
\centering
\includegraphics[width=16cm]{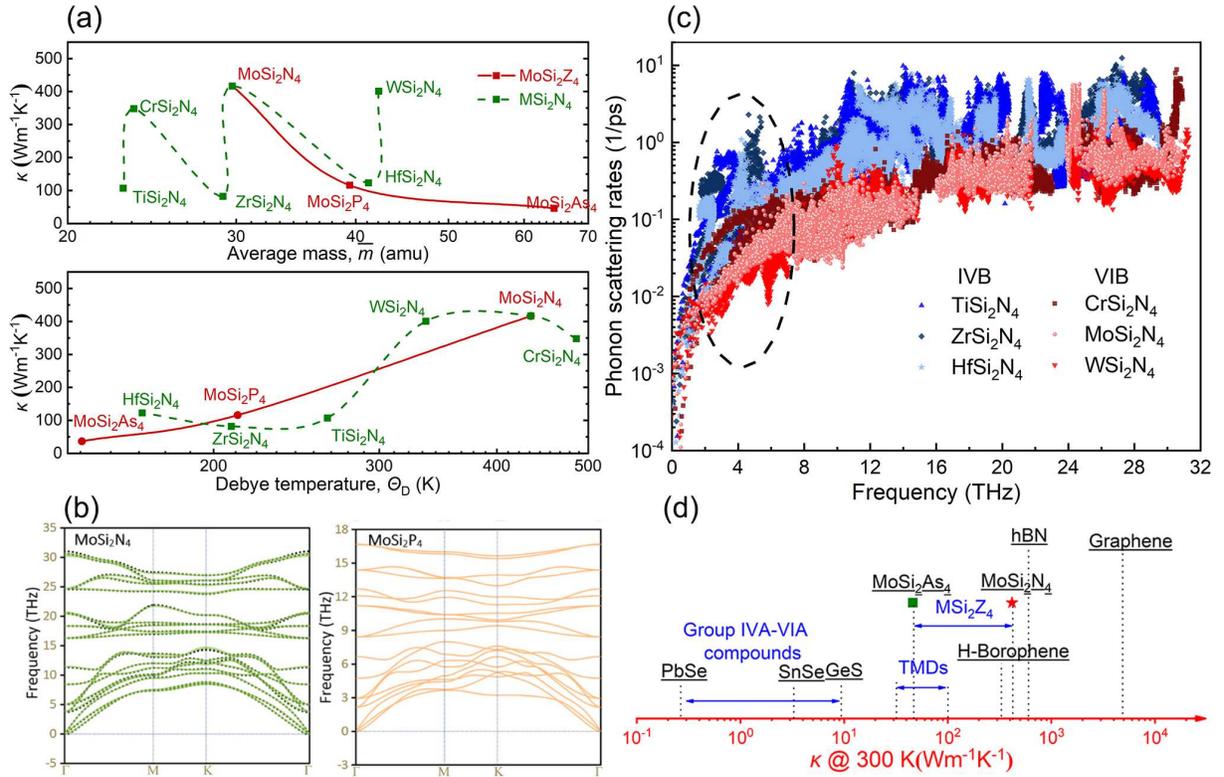}
\caption{(a) Thermal conductivity ($\kappa$) as a function of average mass (upper) and Debye temperature (down). (b) Phonon dispersion of MoSi$_2$N$_4$ and MoSi$_2$P$_4$, where the frequency range of MoSi$_2$N$_4$ is wider than that of MoSi$_2$P$_4$. (c) Phonon scattering of MA$_2$Z$_4$ with different M-site atoms, indicating that MA$_2$Z$_4$ with group \uppercase\expandafter{\romannumeral6}A atom has lower scattering rates. (d) Comparison of $\kappa$ at 300~K between MA$_2$Z$_4$ family and other 2D materials.~\cite{Mortazavi2021-NanoEnergy, Yin2021-ACSAMI-MSi2Z4}}
\label{kappa}
\end{figure*}

In Slack’s classic rules~\cite{Slack1962-PhysRev, Slack1973-JPCS}, crystals with high thermal conductivity follows four rules: simple crystal structure, light atomic masses, strong bonding and low anharmonicity. In the monolayer MA$_2$Z$_4$ family, by replacing M/A/Z atoms at different sites, the different contributions to $\kappa$ have been further examined~\cite{Mortazavi2021-NanoEnergy, Yin2021-ACSAMI-MSi2Z4}. It is found that when Z- site atoms are replaced, $\kappa$ of monolayer MA$_2$Z$_4$ obeys the Slack's rules (red lines in Fig.~\ref{kappa}(a)). For instance, $\kappa$ of MoSi$_2$Z$_4$ decreases by one order of magnitude from Z~=~N to As, indicating that Z atom plays a critical role in controlling the thermal conductivity of MoSi$_2$Z$_4$. The similar variation phenomena are found in MoGe$_2$Z$_4$ and WGe$_2$Z$_4$ as well. Meanwhile, in A-site-replaced MA$_2$Z$_4$, $\kappa$ is decreased by about 40.3$\%$--50.4$\%$ from A~=~Si to Ge. $\kappa$ of MoGe$_2$N$_4$ (286~Wm$^{-1}$K$^{-1}$) is about 40.3$\%$ lower than that of MoSi$_2$N$_4$ (439~Wm$^{-1}$K$^{-1}$). $\kappa$ of WGe$_2$P$_4$ (64~Wm$^{-1}$K$^{-1}$) is only half that of WSi$_2$P$_4$ (129~Wm$^{-1}$K$^{-1}$). 
The decreasing phenomena are intrinsically attributed to the phonon properties. 
The Z- or A-site atoms of MA$_2$Z$_4$ determine the phonon frequency range. Monolayer MA$_2$N$_4$ shows wider frequency range than MA$_2$P$_4$ and MA$_2$As$_4$ (Fig.~\ref{kappa}(b)). MSi$_2$Z$_4$ exhibits wider phonon dispersion than MGe$_2$Z$_4$. Thermal conductivity is related to the phonon group velocity that directly depends on phonon branches. Thus, wider phonon bands lead to higher phonon group velocity and result in higher thermal conduction.

When M atoms are from different group \uppercase\expandafter{\romannumeral4}B or \uppercase\expandafter{\romannumeral6}B atoms, the variation of $\kappa$ is abnormal and the conventional guideline for searching high $\kappa$ does not work. $\kappa$ exhibits an irregular oscillation with the increasing atomic mass and decreasing Debye temperature (green dash lines in Fig.~\ref{kappa}(a)). 
Mortazavi et al. proposed that the weight of core atoms is the dominant factor on thermal conduction of MA$_2$Z$_4$ and $\kappa$ increases with the weight of core atoms. This violates the Slack's rules but is consistent with the classical theory that stiffer systems favor a higher $\kappa$. However, in our recent work~\cite{Yin2021-ACSAMI-MSi2Z4}, we found that the influential factors on the variation of $\kappa$ is not limited to the atomic mass. For instance, the average mass of WSi$_2$N$_4$ is 1.5 times that of CrSi$_2$N$_4$, but $\kappa$ shows only 13$\%$ difference.  
We proposed that the abnormal thermal conductivity of M-site replaced MA$_2$Z$_4$ is relative to the group that M atoms belong to. $\kappa$ of MA$_2$Z$_4$ with group  \uppercase\expandafter{\romannumeral6}B M atom is about 3--4 times of that with group 
\uppercase\expandafter{\romannumeral4}B M atom. These abnormal phenomena with respect to M atoms are attributed to the fundamental vibrational properties and phonon scattering behavior. The acoustic branches of MA$_2$Z$_4$ with group  \uppercase\expandafter{\romannumeral6}B M  are more bunched and less flattened. Consequently, the phonon scattering rates (Fig.~\ref{kappa}(c)) of group \uppercase\expandafter{\romannumeral6}B M-site MA$_2$Z$_4$ are lower than that of group \uppercase\expandafter{\romannumeral4}B M-site MA$_2$Z$_4$, and higher $\kappa$ presents in the former. As shown in Fig.~\ref{kappa}(d), the thermal conductivity of monolayer MA$_2$Z$_4$ family is in a very wide range (10$^1$--10$^3$~Wm$^{-1}$K$^{-1}$), locating between that of 2D TMDs and hBN.

\begin{figure*}[!t]
\centering
\includegraphics[width=16cm]{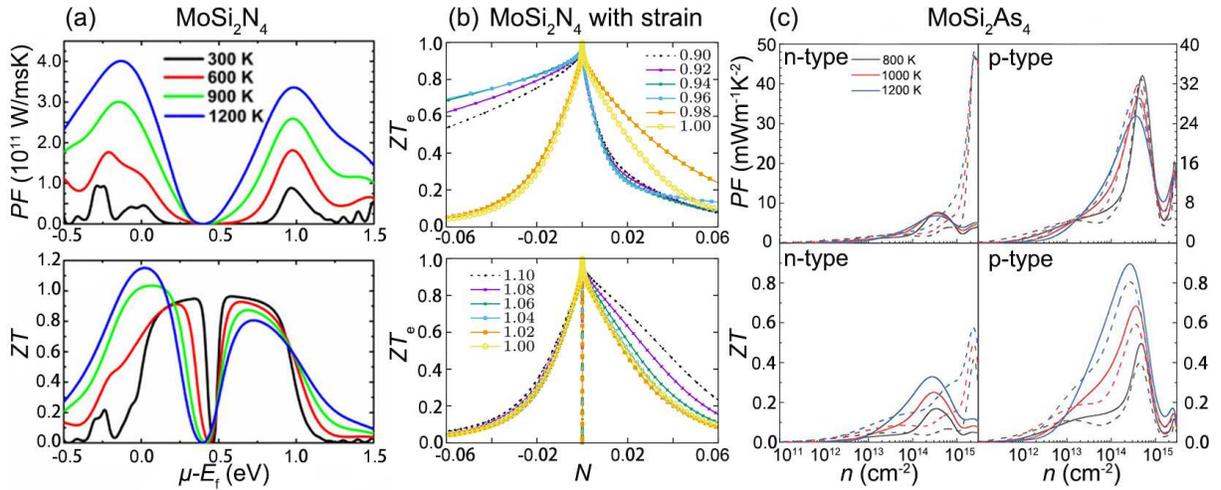}
\caption{(a) Thermoelectric performance of monolayer MoSi$_2$N$_4$ at different temperatures: Power factor ($PT$) (top) and figure of merit ($ZT$) (bottom). (b) Effect of compressive (top) and tensile (bottom) strain on $ZT$. (c) Thermoelectric performance of monolayer MoSi$_2$As$_4$.~\cite{Shojaei2021-JPDAPP-MoSi2N4, Guo2021-CPB-TransportCoefficients, Huang2022-EurophysLett-MoSi2As4}} 
\label{thermoelectric}
\end{figure*}

\subsection{Thermoelectric properties}

In addition to the thermal conductivity, the thermoelectric performance of monolayer MA$_2$Z$_4$ has been also investigated. 
The dimensionless figure of merit ($ZT$) is always used to measure the efficiency of thermoelectric conversion of a thermoelectric material, which is expressed as: 
\begin{equation}
\begin{split}
\mathcal
{\text{$ZT$}=\frac{\text{$S$}^\text{$2$}\text{$T$}}{\rho({\kappa_\text{$e$}}+{\kappa_\text{$l$}})}}
\\
\end{split}
\end{equation}
where $S$, $1/\rho$, $T$, $\kappa_e$, and $\kappa_l$ are the Seebeck coefficient, electrical conductivity, working temperature, electronic thermal conductivity and lattice thermal conductivity, respectively. 
The higher $ZT$ indicates the better heat-to-electricity conversion efficiency. 
Thus, 2D materials for the thermoelectric field are prone to semiconductors with high Seebeck coefficient, electrical conductivity and low thermal conductivities.

The power factor ($PF$~=~$\text{$S$}^\text{$2$}\text{$T$}/\rho$) of monolayer MoSi$_2$N$_4$ is studied to evaluate the capability of producing electricity~\cite{Shojaei2021-JPDAPP-MoSi2N4}. The heat conversion into electricity is favored by the temperature since $PF$ increases considerably with temperature. The maximum $PF$ is located at a chemical potential range of 0.12--0.98~eV (upper in Fig.~\ref{thermoelectric}(a)). $PF$ in the negative chemical potential region is more sensitive to temperature than in the positive region. Meanwhile the maximum $ZT$ increases with temperature in the chemical potential range of -0.45--0.5~eV. The maximum $ZT$ of monolayer MoSi$_2$N$_4$ is up to 1.2 at 1200~K~\cite{Shojaei2021-JPDAPP-MoSi2N4}, while $ZT$ of MoGe$_2$N$_4$ is up to 1.0 at 900~K~\cite{Zhang2022-JSSC-MoSiN-MoGeN-thermoelec}.
Guo et al.~\cite{Guo2021-CPB-TransportCoefficients} studied the effect of strain on the thermoelectric performance of monolayer MoSi$_2$N$_4$, as shown in Fig.~\ref{thermoelectric}(b). They found that the compressive strain rather than the tensile strain has significant effect on $S$.
Besides, $S$ with n-type doping has observable improvement under a compressive strain, which is ascribed to the strain-driven conduction band degeneracies. 

As mentioned above, the high lattice thermal conductivity of monolayer MoSi$_2$N$_4$ brings negative effect on $ZT$ and further restrains the thermoelectric performance. It is imperative for seeking other MA$_2$Z$_4$ semiconductors with low thermal conductivity to satisfy the requirement of high $ZT$. Monolayer MoSi$_2$As$_4$ semiconductor has no doubt to be a good candidate due to its low lattice thermal conductivity. 
However, Huang et al.~\cite{Huang2022-EurophysLett-MoSi2As4} found that $ZT$ of MoSi$_2$As$_4$ is 0.33 for n-type and 0.90 for p-type at 1200~K, which are even lower than that of MoSi$_2$N$_4$ (Fig.~\ref{thermoelectric}(c)). 
The underlying reason still remains unknown. Hence, whether there is any member with ultra-high thermoelectric properties needs to be further explored.

\begin{figure*}[!t]
\centering
\includegraphics[width=14cm]{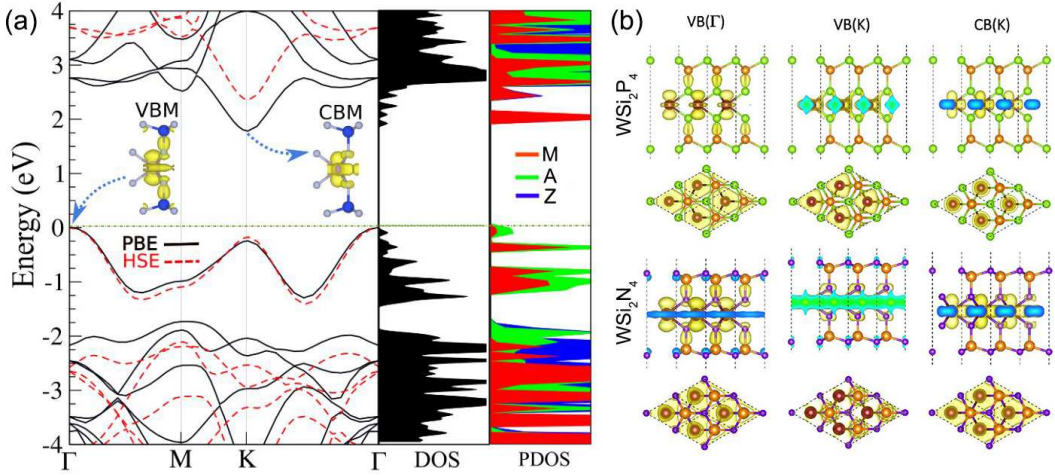}
\caption{(a) Band structure of monolayer MoSi$_2$N$_4$. 
(b) Charge density distribution of WSi$_2$P$_4$ and WSi$_2$N$_4$ at VB($\Gamma$), VB(K) and CB(K) with the isosurface value as 0.01~e/\AA$^3$.~\cite{Shojaei2021-JPDAPP-MoSi2N4, Mortazavi2021-NanoEnergy}}
\label{project-band}
\end{figure*}

\begin{figure}[!t]
\centering
\includegraphics[width=8.6cm]{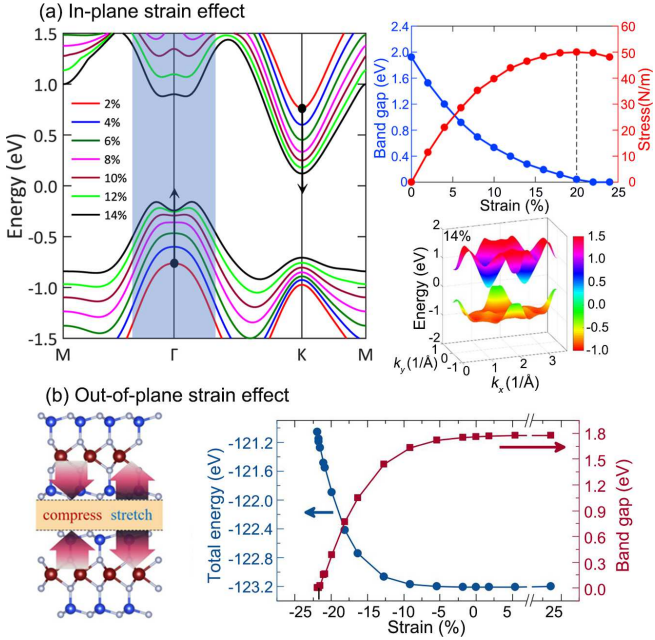}
\caption{(a) In-plane strain effect on monolayer MoSi$_2$N$_4$: variation of band structure (left), bandgap and stress versus strain (top right), and 3D `Mexican hat' band under a tensile strain of 14$\%$ (bottom right). (b) Out-of-plane strain effect on bilayer MoSi$_2$N$_4$.~\cite{Alavi-rad2022-SST-monoMoSi2N4-strain, Zhong2021-PRB-bilayer-verticalstrain}}
\label{strain-band}
\end{figure}

\subsection{Electrical properties}
Till now, nearly hundred members in monolayer MA$_2$Z$_4$ family have been predicted. 
Wang et al.~\cite{Wang2021-NC} systematically investigated 72 thermodynamically and dynamically stable MA$_2$Z$_4$ compounds with a septuple-atomic-layer which possess the same intercalated architecture as MoSi$_2$N$_4$. The electric properties can be classified according to the total numbers of valence electrons. Most of monolayer MA$_2$Z$_4$ nanosheets with 32 or 34 valence electrons are semiconductors, while those with 33 valence electrons are non-magnetic metals or ferromagnetic semiconductors.  
Ding et al.~\cite{Ding2021-JPCC-Y-Cd-Hf-Hg} identified 12 stable MSi$_2$N$_4$ with trigonal prismatic (H-phase) or octahedral (T-phase) structures, including six new members. The M-site atoms contain both the early transition metal element (group \uppercase\expandafter{\romannumeral3}B--\uppercase\expandafter{\romannumeral6}B) and the late transition metal (group \uppercase\expandafter{\romannumeral8}B, e.g., Pd and Pt). Group \uppercase\expandafter{\romannumeral3}B and \uppercase\expandafter{\romannumeral4}B MSi$_2$N$_4$ with H- and T-phase geometries are all stable, while group \uppercase\expandafter{\romannumeral5}B and \uppercase\expandafter{\romannumeral6}B ones are only stable with H-phase. 
Moreover, other MoSi$_2$N$_4$ derived structures have emerging in an overwhelming trend, such as, bilayer or multilayer MA$_2$Z$_4$, heterostructures, Janus MA$_2$Z$_4$, etc. 
These MA$_2$Z$_4$ nanosheets exhibit versatile electronic properties depending on the number of valence electrons and structural phases.
In this subsection, we emphatically discuss the intrinsic electrical properties of these monolayers and their tunability.

H-phase monolayer MA$_2$Z$_4$ nanosheets, where M~=~Cr/Mo/W/Ti/Zr/Hf, A~=~Si/Ge, Z~=~N/P/As,
are semiconductors with a bandgap around 0.04--1.79~eV (PBE) and 0.31--2.57~eV (HSE) calculated by DFT~\cite{Mortazavi2021-NanoEnergy, Yin2021-ACSAMI-MSi2Z4}. Monolayer MA$_2$N$_4$ nanosheets are indirect bandgap materials ($\Gamma$-K), while monolayer MA$_2$P$_4$ nanosheets are direct gap semiconductors (K-K). The electronic contribution to CB and VB derived primarily from interior atomic orbitals leads to the robust band edge states~\cite{Wu2022-npjComputMater-robustbandedge}. From the projected band structure and charge density distribution (Fig.~\ref{project-band}), valence band maximum (VBM) at $\Gamma$ point is mainly contributed by M-$d_{z^2}$ orbital and marginally contributed by Z-$p_z$ and A-$p_z$, while at K point it is contributed by M-($d_{x^2-y^2}$,~$d_{xy}$) and Z-($p_x$,~$p_y$,~$p_z$) orbitals which locate at the centre area of the structure.
Conduction band minimum (CBM) is solely occupied by M-($s,~d_{z^2}$) orbitals. 
The strong interaction of $d$ orbitals of M atom renders to the highly dispersion between VB and CB, indicating high charge carrier mobilities and small effective masses. The calculated electron and hole mobilities are 200 and 1100~cm$^2$V$^{-1}$s$^{-1}$ for MoSi$_2$N$_4$, 490 and 2190~cm$^2$V$^{-1}$s$^{-1}$ for MoGe$_2$N$_4$, 320 and 2026~cm$^2$V$^{-1}$s$^{-1}$ for WSi$_2$N$_4$, 690 and 2490~cm$^2$V$^{-1}$s$^{-1}$ for WGe$_2$N$_4$~\cite{Mortazavi2021-NanoEnergy}. 
The carrier mobility ($\mu$) of monolayer MoSi$_2$P$_4$ is comparable with that of MoSi$_2$N$_4$, where $\mu$ in zigzag (armchair) direction is 245.992 (257.985)~cm$^2$V$^{-1}$s$^{-1}$ for electrons and 1065.023 (1428.885)~cm$^2$V$^{-1}$s$^{-1}$ for holes~\cite{Liu2021-PLA-MoSi2P4}, which are much higher than those of MoS$_2$ (72.16 and 200.52~cm$^2$V$^{-1}$s$^{-1}$~\cite{Cai2014-JACS-MoS2-elec}). 
Besides, SnGe$_2$N$_4$~\cite{Dat2022-RSCadv-SnGe2N4}, a MA$_2$Z$_4$-derived structure without transition metals, possesses a high electron mobility of 1061.66~cm$^2$V$^{-1}$s$^{-1}$, but a low hole mobility of 28.35~cm$^2$V$^{-1}$s$^{-1}$, owing to a strongly concave downward conduction band and a flat valence band. Recent DFT calculations also indicate 2D MoSi$_2$N$_4$ as an ideal platform for the exploration of exciton-involved physics~\cite{KONG2022100814Acomprehensive}.

The bandgap of bilayer MoSi$_2$Z$_4$ (Z~=~P/As) is similar with that of monolayers due to the weak vdW interaction between layers. In terms of carrier mobilities, there exist several interesting phenomena~\cite{Yao2021-Nanomaterials-Bilayer}. (1) In monolayer and bilayer MoSi$_2$Z$_4$, the hole carrier mobilities are about 3--4 times larger than those of electrons. The difference of carrier mobilities effectively facilitates the spatial separation of electrons and holes, restraining the recombination probability of photo-excited carriers. (2) Carrier mobility of bilayer MoSi$_2$Z$_4$, especially hole carrier mobility, exhibits anisotropic behaviors. 
(3) the carrier mobilities (electron and hole) of bilayer MoSi$_2$Z$_4$ are about 2 times that of monolayer MoSi$_2$Z$_4$. 
Monolayer and bilayer MoSi$_2$Z$_4$ with high carrier mobilities exhibit pronounced carrier polarization and are promising materials for high-performance nanoscale electronic and optoelectronic devices.

The electronic transition of monolayer MA$_2$Z$_4$ from semiconductor to metal can be triggered by applying in-plane strains~\cite{Alavi-rad2022-SST-monoMoSi2N4-strain, Liu2021-PLA-MoSi2P4}.
The bandgap of monolayer MoSi$_2$N$_4$ decreases with the increasing in-plane biaxial strain (Fig.~\ref{strain-band}(a)). The relation between bandgap ($E_g$) and strain ($\epsilon$) is fitted as~\cite{Alavi-rad2022-SST-monoMoSi2N4-strain} $E_g~=~0.004~\epsilon^2-0.17~\epsilon+1.87$.
Under an in-plane biaxial strain of 4$\%$ (6$\%$), the effective mass of holes is --2.33~$m_e$ (--3.84~$m_e$)  and the effective mass of electrons is 0.48~$m_e$ (0.43~$m_e$). This suggests that the in-plane biaxial strain simultaneously enhances the localization of holes and free electrons, which would achieve fascinating features like ferromagnetism and superconductivity. 
When the in-plane strain is over 10$\%$, there exists an inversion at the edge of VBM,  known as a `Mexican hat' in which two VBMs are induced and the transport properties would be improved. As the strain increases, the variation of `Mexican hat' dispersion is more noticeable. 
When $\epsilon$~=~20$\%$, MoSi$_2$N$_4$ becomes semimetal.
For MoSi$_2$P$_4$ under an in-plane armchair uniaxial strain, the bandgap transfers to be indirect at $\epsilon$~=~2$\%$, 3$\%$ and returns to be direct at $\epsilon$~=~--10$\%$~\cite{Liu2021-PLA-MoSi2P4}. Semiconductor-metal transition is predicted at $\epsilon_{\text{arm}}$~=~--12$\%$ and $\epsilon_{\text{zig}}$~=~--12$\%$ or 12$\%$.
Under an in-plane compressive strain, the bandgap of bilayer MoSi$_2$N$_4$ and WSi$_2$N$_4$ is transferred from the indirect to direct state and the bandgap value changes slightly, while the bandgap rapidly decreases with the increasing tensile strain~\cite{Wu2021-APL-biMoSi2N4andWSi2N4}.

The vertical (out-of-plane) compressive strain is also demonstrated to effectively tune the electronic properties of bilayer vdW MA$_2$Z$_4$~\cite{Zhong2021-PRB-bilayer-verticalstrain}.
With the increase of vertical compressive strain, the bandgap of bilayer MoSi$_2$N$_4$ monotonically decreases and reaches 0~eV at $\epsilon~=~22\%$ (Fig.~\ref{strain-band}(b)). This is attributed to the opposite energy shift of the states in different layers. This shift is driven by the asymmetric charge redistribution on the inner Z-Z sublayer at the interface. The similar transition is confirmed in other bilayer MA$_2$Z$_4$, and the pressure to realize such a transition ranges from 2.18 to 32.04~GPa.

\begin{figure*}[!t]
\centering
\includegraphics[width=13.5cm]{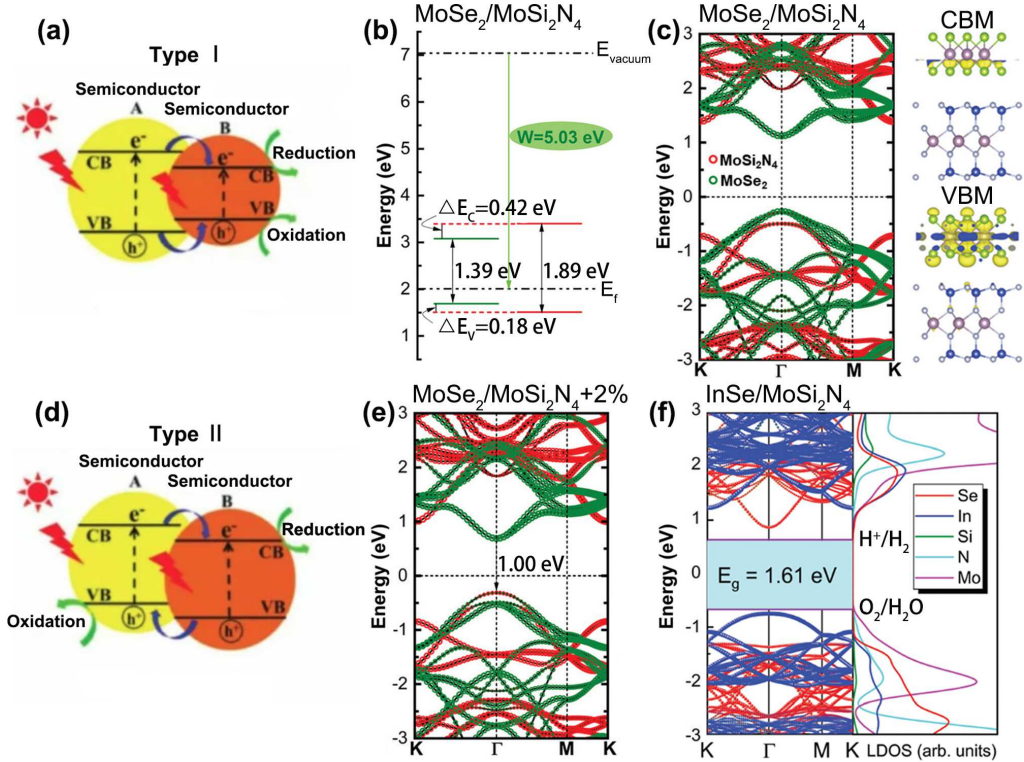}
\caption{Semiconducting heterostructures: (a) schematic diagrams of type~\uppercase\expandafter{\romannumeral1} (straddling gap) and (d) type~\uppercase\expandafter{\romannumeral2} (staggered gap) bands. (b) Type~\uppercase\expandafter{\romannumeral1} band alignment of MoSe$_2$/MoSi$_2$N$_4$ and (c) charge distribution at CBM and VBM. Type~\uppercase\expandafter{\romannumeral2} band structures of (e) MoSe$_2$/MoSi$_2$N$_4$ under 2$\%$ strain and (f) InSe/MoSi$_2$N$_4$.~\cite{Cai2021-JMCC-MoSe2-hetero, He2022-PCCP-InSe-MoSi2N4-elec-optic-HER}}
\label{semi-hetero}
\end{figure*}

The potential of 2D materials in industrial-grade low-dimensional nanodevices is further boosted by the enormous design flexibility offered by the vertical vdW heterostructures (vdWHs), in which physical properties can be customized by the vertically stacking of different 2D atomic layers.
Interlayer coupling of semiconducting vdWHs form 3 types of band alignment: type~\uppercase\expandafter{\romannumeral1} (straddling gap binds electrons/holes with the critical ratio of conduction band offset and valence band offset), 
type~\uppercase\expandafter{\romannumeral2} (staggered gap promotes the separation of electrons and holes and suppresses the recombination of electrons and holes), and type~\uppercase\expandafter{\romannumeral3} (broken gap). 
Several previous studies have revealed the electronic properties of vdWHs by stacking 2D semiconductors with MA$_2$Z$_4$.

TMDs/MA$_2$Z$_4$ vdWHs at the ground state belong to the type~\uppercase\expandafter{\romannumeral1} alignment, due to the mismatch induced by strain and layer interaction between TMDs and MoSi$_2$N$_4$ (Fig.~\ref{semi-hetero})~\cite{Cai2021-JMCC-MoSe2-hetero, Cai2021-SSRN-WSe-hetero}. TMDs provide the main contribution to CBM and VBM. The carrier mobilities of MoSe$_2$/MoSi$_2$N$_4$ and WSe$_2$/MoSi$_2$N$_4$ are up to 10$^4$~cm$^2$V$^{-1}$s$^{-1}$, which are higher than those of monolayer MoSe$_2$ and WSe$_2$. 
Under an external electric field or strain, the band structures of TMDs/MA$_2$Z$_4$ vdWHs are transferred from type~\uppercase\expandafter{\romannumeral1} to type~\uppercase\expandafter{\romannumeral2}. 
Actually, some other vdWHs by stacking 2D materials (e.g., C$_3$N$_4$, ZnO, InSe and Cs$_3$Bi$_2$I$_9$) with MoSi$_2$N$_4$ exhibit semiconducting characteristic with type~\uppercase\expandafter{\romannumeral2} band structure, indicating promising application in photocatalytic field~\cite{Nguyen2022-PRB-C3N4-hetero, Ng2022-APL-GaN-ZnO-hetero, He2022-PCCP-InSe-MoSi2N4-elec-optic-HER, Liu2022-JAP-Cs3Bi2I9-hetero}. The carrier mobility of type~\uppercase\expandafter{\romannumeral2} InSe/MoSi$_2$N$_4$ is up to 10$^4$~cm$^2$V$^{-1}$s$^{-1}$, indicating more suitable for photocatalytic nano devices when compared with type~\uppercase\expandafter{\romannumeral1} TMDs/MA$_2$Z$_4$ vdWHs. 
In addition, BP/MoSi$_2$P$_4$ and BP/MoGe$_2$N$_4$ vdWHs possess direct bandgap and belong to type~\uppercase\expandafter{\romannumeral2} alignment~\cite{Guo2022-BP-MoSi2P4-hetero, Nguyen2021-JPCL-BP-MoGe2N4-hetero}. Combined with high optical absorption properties, these MA$_2$Z$_4$-based vdWHs would be promising candidates for solar cell devices. 

\begin{figure*}[!t]
\centering
\includegraphics[width=12.5cm]{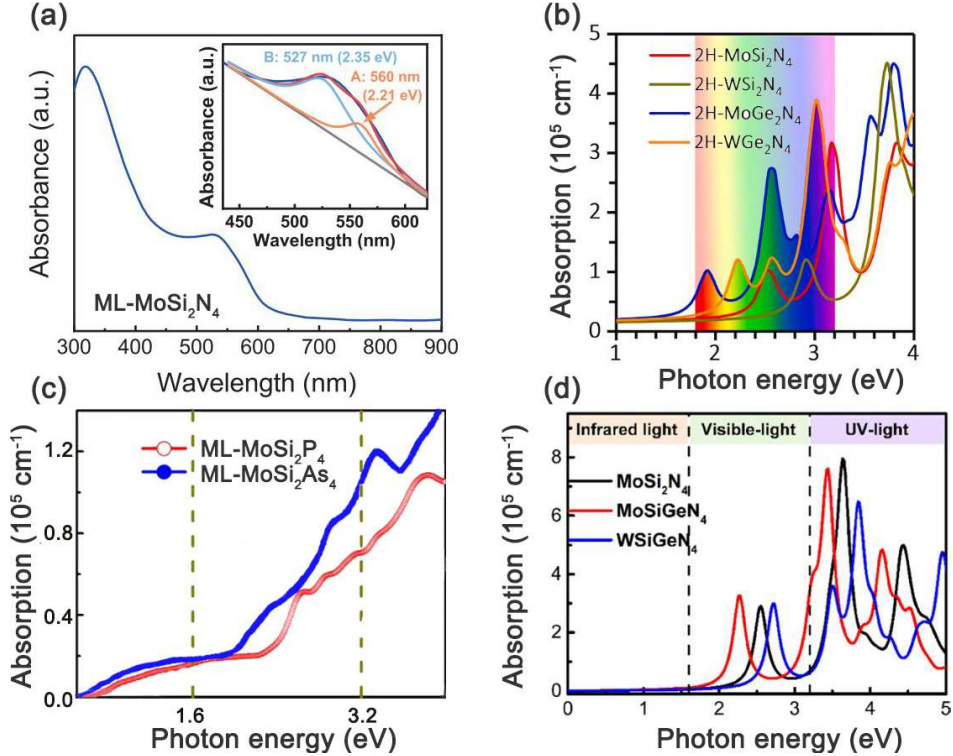}
\caption{Optical absorption spectra of monolayer MA$_2$Z$_4$ system: (a) MoSi$_2$N$_4$, (b) MA$_2$N$_4$ (M~=~Mo/W, A~=~Si/Ge), (c) MoSi$_2$Z$_4$ (Z~=~P/As), and (d) Janus MSiGeN$_4$.~\cite{Hong2020-Science, Mortazavi2021-NanoEnergy, Yao2021-Nanomaterials-Bilayer, Guo2021-JSemicond-Janus-MSiGeN4}} 
\label{optical}
\end{figure*}

In metal and semiconductor contacts, the Schottky phenomenon with rectification effect in vdWHs occurs. Reducing the Schottky barrier height (SBH) or tuning the Schottky contacts to Ohmic contacts are key challenges for achieving energy efficient and high-performance power devices.
MA$_2$Z$_4$-based vdWHs provide opportunities for reconfigurable and tunable nanoelectronic devices.
In graphene/MA$_2$Z$_4$ vdWHs, there exists a tiny bandgap~\cite{Liang2022-CPB-graphene-hetero}. P-type Schottky contacts are formed at the interfaces of two heterojunctions, where the electrons are transferred from graphene to MoSi$_2$N$_4$ (WSi$_2$N$_4$).
The n-type SBH ($\Phi_\text{n}$) and the p-type SBH ($\Phi_\text{p}$) are 0.922~eV and 0.797~eV, respectively, indicating the presence of a p-type Schottky contact~\cite{Cao2021-APL-GrapheneNbS2hetero}. In contrast, graphene/MoGe$_2$N$_4$ vdWH forms an n-type Schottky contact with a barrier of 0.63 eV~\cite{Pham2021-NJP-GR-schottky}.
Monolayer Janus MoSiGeN$_4$ maintains the semiconducting property with an indirect bandgap of 1.436 (2.124)~eV obtained by PBE (HSE06)~\cite{Binh2021-JPCL-GR-MoGeSiN4-hetero}. 
The vdWHs formed by graphene and Janus MoGeSiN$_4$ or MoSiGeN$_4$ have been investigated. The n-type SBH of graphene/MoGeSiN$_4$ is 0.63~eV, while the p-type SBH of graphene/MoSiGeN$_4$ is 0.74~eV. These SBHs are close to those of MA$_2$Z$_4$ (MoSi$_2$N$_4$ and MoGe$_2$N$_4$), but more adjustable, providing a useful guidance for the design of controllable Schottky nanodevices by using MA$_2$Z$_4$ family. 
Compared with graphene-based vdWHs, the vdWHs formed by MA$_2$Z$_4$ and other 2D materials possess better performance in Schottky contact~\cite{Cao2021-APL-GrapheneNbS2hetero, Nguyen2022-JPCL-MoSH-hetero}.
In NbS$_2$/MoSi$_2$N$_4$, $\Phi_n$ and $\Phi_p$ are 1.642 eV and 0.042 eV, respectively~\cite{Cao2021-APL-GrapheneNbS2hetero}. This ultralow p-type SBH of NbS$_2$/MoSi$_2$N$_4$ vdWH suggests the potential of NbS$_2$ as an efficient 2D electrical contact to MoSi$_2$N$_4$ with high charge injection efficiency, particularly at room-temperature.

In addition, the Schottky contact types of multilayer vdWHs vary with the order of stacking layers. 
For instance, the graphene/MoSi$_2$N$_4$/MoGe$_2$N$_4$ vdWH has an n-type Schottky contact with a SBH of 0.33~eV. While MoSi$_2$N$_4$/graphene/MoGe$_2$N$_4$ and MoSi$_2$N$_4$/MoGe$_2$N$_4$/graphene vdWHs are both p-type with a SBH of 0.41 and 0.46~eV, respectively~\cite{Pham2021-RSCAdv-MoGeN-GR-MoSiN-hetero}. The contact barriers in the multilayer vdWHs are smaller than those in the bilayer vdWHs, suggesting that the graphene/MoSi$_2$N$_4$/MoGe$_2$N$_4$ vdWHs provide an effective pathway to reduce the Schottky barrier. This is highly beneficial for improving the charge injection efficiency of contact heterostructures. 
The p-type Schottky contacts of MoSi$_2$N$_4$/graphene or WSi$_2$N$_4$/graphene at the interface can be transferred to n-type by a compressive strain. When a compressive strain of 10$\%$ is applied, the transition from Schottky to Ohmic contacts occurs in MoSi$_2$N$_4$/graphene and WSi$_2$N$_4$/graphene.
However, no transition is induced by a tensile strain~\cite{Liang2022-CPB-graphene-hetero, Pham2021-NJP-GR-schottky}. 

\begin{figure*}[!t]
\centering
\includegraphics[width=17cm]{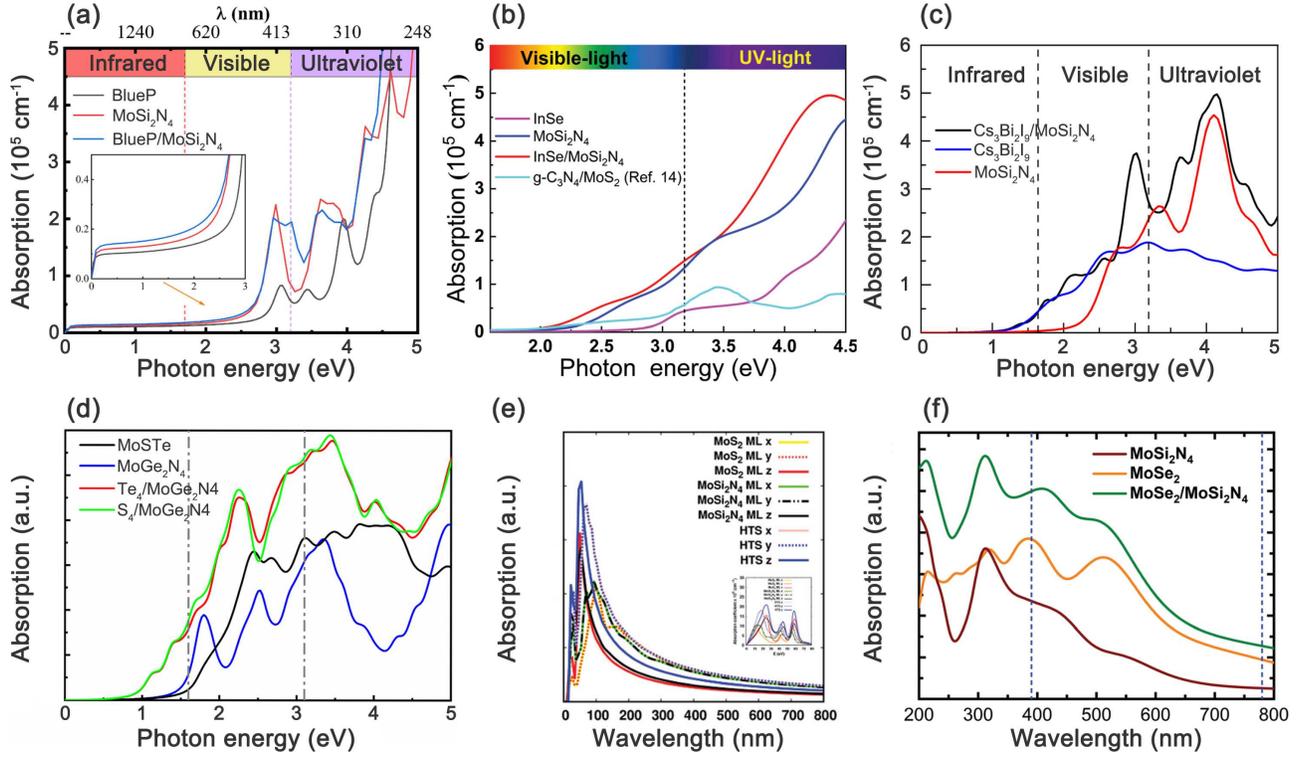}
\caption{Optical absorption spectra of MA$_2$Z$_4$ based heterostructures: (a) BlueP/MoSi$_2$N$_4$ (b) InSe/MoSi$_2$N$_4$, (c) Cs$_3$Bi$_2$I$_9$/MoSi$_2$N$_4$, (d) Janus MoSTe/MoGe$_2$N$_4$, (e) MoS$_2$/MoSi$_2$N$_4$, and (f) MoSe$_2$/MoSi$_2$N$_4$.~\cite{Xuefeng2022-JPD-BlueP-hetero, Cai2021-JMCC-MoSe2-hetero, Liu2022-JAP-Cs3Bi2I9-hetero, He2022-PCCP-InSe-MoSi2N4-elec-optic-HER, Bafekry2021-NJC-MoS2-Hetero, Wang2021-Nanomaterials-hetero-strain}}
\label{hetero-optical}
\end{figure*}

\subsection{Optical properties}


The optical absorption of monolayer MA$_2$Z$_4$ system is up to 10$^5$~cm$^{-1}$ in the visible range~\cite{Hong2020-Science, Shojaei2021-JPDAPP-MoSi2N4, Yadav2021-MA2N4-photovoltaic-photocatalytic, Norouzi2021-PhysScrip-WSi2N4-G0W0}, which is comparable with that of graphene, phosphorene and MoS$_2$. 
In monolayer MoSi$_2$N$_4$ (Fig.~\ref{optical}(a)), the experimental results show that a strong peak of optical absorption appears at about 320~nm and a broad peak in the range of 500--600~nm~\cite{Hong2020-Science}. The calculated results are in good agreement with the experimental data.  

Different atomic compositions bring the tunable optical properties to this system~\cite{Mortazavi2021-NanoEnergy, Yao2021-Nanomaterials-Bilayer, Guo2021-JSemicond-Janus-MSiGeN4, Yu2021-AMI-Janus-HER-OER}. The obvious redshift phenomena of optical absorption spectra are triggered by heavy atoms in M-site MA$_2$N$_4$ (Fig.~\ref{optical}(b)), while the absorption coefficient is insensitive to the M-site counterparts. The absorption of MoSi$_2$N$_4$ and WSi$_2$N$_4$ is both 10$^5$~cm$^{-1}$ in the visible region. 
The similar red-shift of absorption spectra exhibits in A- or Z-site MoA$_2$Z$_4$ as well (Fig.~\ref{optical}(c)). The absorption coefficient, however, is improved with the increasing mass of A or Z atoms. For instance, the absorptions of MoSi$_2$Z$_4$ (Z~=~N/P/As) are in the order of MoSi$_2$N$_4$<MoSi$_2$P$_4$<MoSi$_2$As$_4$.
In Janus MA$_2$Z$_4$ (Fig.~\ref{optical}(d)), the redshift of absorption spectra and the absorption enhancement exist in the lighter MoSiGeN$_4$, compared with WSiGeN$_4$. 
As for MA$_2$Z$_4$-derived materials (e.g., CrC$_2$N$_4$~\cite{Mortazavi2021-MaterTodayEnergy-CrC2N4}, SnGe$_2$N$_4$~\cite{Dat2022-RSCadv-SnGe2N4} and XMoSiN$_2$~\cite{Sibatov2022-ASS-XMoSiN2}), the absorption of visible light can be up to 10$^5$~cm$^{-1}$ as well. When Cr is replaced by Mo or W in CrC$_2$N$_4$, the first absorption peak appears in ultraviolet spectra and high frequencies.

Furthermore, optical properties of MA$_2$Z$_4$ are shown to be manipulated by strain, surface functionalization, and heterostructure. 
Firstly, the optical absorption spectrum of MoSi$_2$N$_4$ shows redshift (blueshift) under a tensile (compressive) strain~\cite{Alavi-rad2022-SST-monoMoSi2N4-strain, Jian2021-JPCC-monoMoSi2N4-strain, Lv2022-PhysE-monolayer}. The absorption edges 
decrease with the increasing tensile strain. A 4--10$\%$ tensile strain enhances the optical absorption capacity in the visible region up to 43--70$\%$.
The reflectance ability (average reflectance rate), meanwhile, is improved from 16$\%$ to 23$\%$ with the tensile strain from 0$\%$ to 10$\%$. Yang et al.~\cite{Lv2022-PhysE-monolayer} found monolayer MoSi$_2$N$_4$ exhibits more outstanding optical absorption capacity in the ultraviolet range than that in the visible range, especially  under biaxial compressive strain.
The optical bandgap is evaluated by the slope of absorption peak, in which the corresponding value is 2.47~eV without strain, 2.9 eV with --3$\%$ strain and 3.05~eV with --4$\%$ strain. 
Furthermore, compared with monolayer MA$_2$Z$_4$, the bilayer or multilayer MA$_2$Z$_4$ shows strong optical absorbance and broad absorption areas as well~\cite{Yao2021-Nanomaterials-Bilayer, Mwankemwa2022-ResPhys-MoSi2N4-MoSiGeN4, Bafekry2021-JAP-biMoSi2N4-ElecField-strain}. When the vertical (out-of-plane) compressive strain increases from 0 to 12$\%$, there exists strong blueshift but only slight decrease of absorption and reflectivity. This manifests that MA$_2$Z$_4$ possesses stable optical absorption capacity independent of vertical strain and the number of layers, which will be more convenient for experimental fabrication of 2D optoelectronic devices~\cite{Bafekry2021-JAP-biMoSi2N4-ElecField-strain}.

Secondly, it has been reported that surface functionalization plays an active role in regulating the optical properties of MoSi$_2$N$_4$, where the adatoms can be Au, F, and Alkali elements (Li, Na, K)~\cite{Xu2022-JPCS-AuAdsorption, Sun2022-CommuniTheoerticPhys-Li-Na-K, Chen2021-ResPhys-F-MoSi2N4}. In Au-MoSi$_2$N$_4$, the optical absorption capacity is significantly improved with the increment of Au concentration (6.25--56.3$\%$/unit cell). It can be increased by 1--2 times in the visible light region and by 52$\%$ in the ultraviolet region. The results show that Au absorption is beneficial for the photocatalytic activity, making Au-MoSi$_2$N$_4$ a potential candidate for photoelectrochemical applications and short-wavelength optoelectronic devices. 
Alkali-metal adsorption not only enhances the optical absorption coefficient but enlarges the absorption area~\cite{Sun2022-CommuniTheoerticPhys-Li-Na-K}. The absorption coefficient at a wavelength of 380~nm is 0.34$\times$10$^5$~cm$^{-1}$ for pristine MoSi$_2$N$_4$, 0.57$\times$10$^5$~cm$^{-1}$ for Li-MoSi$_2$N$_4$, 1.09$\times$10$^5$~cm$^{-1}$ for Na-MoSi$_2$N$_4$, and 0.89$\times$10$^5$~cm$^{-1}$ for K-MoSi$_2$N$_4$. 
The light at wavelength of 780~nm is almost unabsorbable by pristine MoSi$_2$N$_4$. However, the absorption at this wavelength is 0.92$\times$10$^5$~cm$^{-1}$ for Li-MoSi$_2$N$_4$, 1.34$\times$10$^5$~cm$^{-1}$ for Na-MoSi$_2$N$_4$, and 1.14$\times$10$^5$~cm$^{-1}$ for K-MoSi$_2$N$_4$. Na-MoSi$_2$N$_4$ exhibits the strongest optical response among these three alkali-metal-decorated MoSi$_2$N$_4$.

Thirdly, in Fig~\ref{hetero-optical}, the optical absorption capacity of monolayer MoSi$_2$N$_4$ is comparable with that of monolayer MoS$_2$ and MoSe$_2$, and better than that of other monolayer 2D materials, such as BlueP, InSe and Cs$_3$Bi$_2$I$_9$. It is interesting that MA$_2$Z$_4$-based vdWHs have more excellent optical response than the monolayer MA$_2$Z$_4$ and other monolayer 2D materials~\cite{Xuefeng2022-JPD-BlueP-hetero, Cai2021-JMCC-MoSe2-hetero, Liu2022-JAP-Cs3Bi2I9-hetero, He2022-PCCP-InSe-MoSi2N4-elec-optic-HER, Bafekry2021-NJC-MoS2-Hetero, Wang2021-Nanomaterials-hetero-strain, Nguyen2021-JPCL-BP-MoGe2N4-hetero, Zeng2021-PCCP-optical-C2N-herto}. These enhancements mainly appear in the visible and ultraviolet light range. The high performance of BlueP/MoSi$_2$N$_4$ vdWH is in the visible light range of 460--780~nm (the insert of Fig.~\ref{hetero-optical}(a)), which is attributed to the overlap of electronic states between valence bands caused by the interlayer coupling and charge transfer~\cite{Xuefeng2022-JPD-BlueP-hetero}. InSe/MoSi$_2$N$_4$ vdWH shows great improvement of absorption in the ultraviolet range~\cite{He2022-PCCP-InSe-MoSi2N4-elec-optic-HER}. 
The absorption of monolayer Cs$_3$Bi$_2$I$_9$ is superior than that of monolayer MoSi$_2$N$_4$ in the photon energy range of 1--2.6~eV, while the relationship reverses in the ultraviolet range (high photon energy), as shown in Fig.\ref{hetero-optical}(c). Surprisingly, the absorption is enhanced in the whole range by constructing the Cs$_3$Bi$_2$I$_9$/MoSi$_2$N$_4$ vdWH~\cite{Liu2022-JAP-Cs3Bi2I9-hetero}. Similarily, this increasing phenomena appear in Janus MoSTe/MoGe$_2$N$_4$, MoS$_2$/MoSi$_2$N$_4$ and MoSe$_2$/MoSi$_2$N$_4$ vdWHs (Fig.\ref{hetero-optical}(d--f))~\cite{Cai2021-JMCC-MoSe2-hetero, Bafekry2021-NJC-MoS2-Hetero, Wang2021-Nanomaterials-hetero-strain}, indicating the tunability of optical absorption by forming vdWHs.
As for strain effect on the absorption of MA$_2$Z$_4$-based heterostructures, the obvious redshift (blueshift) of optical spectrum appears under a tensile (compressive) strain. In BlueP/MoSi$_2$N$_4$ vdWH, it is found that a compressive strain restrains the absorption on coefficient in the visible region, while a tensile strain (0--8$\%$) promotes the optical absorption ability~\cite{Xuefeng2022-JPD-BlueP-hetero}. In Janus MoSTe/MoGe$_2$N$_4$ vdWH, a compressive strain increases the absorption coefficient in the visible and ultraviolet regions, while a tensile strain enhances it in the infrared region, regardless of which side (S or Te atoms) approaching to MoSi$_2$N$_4$~\cite{Wang2021-Nanomaterials-hetero-strain}.

\subsection{Magnetic properties}

It has been reported that group-\uppercase\expandafter{\romannumeral5}B (V/Nb/Ta) MA$_2$Z$_4$ nanosheets exhibit robust intrinsic magnetism, and group-\uppercase\expandafter{\romannumeral6}B (Cr/Mo/W) MA$_2$Z$_4$ ones show antiferromagnetic ground states~\cite{Wang2021-NC, Chen2021-CAEJ-MA2Z4-magnetic, Formed2022-JPDAP-MoSi2N4-NRs, Ding2022-ASS-MoN2X2Y2}.  
In Table~\ref{mag}, NbSi$_2$N$_4$, NbSi$_2$As$_4$, NbGe$_2$Z$_4$, NbGe$_2$P$_4$, TaSi$_2$N$_4$ and TaGe$_2$P$_4$ are ferromagnetic metals with the magnetic moment from 0.37 to 0.78~$\mu_\text{B}$ per metal atom, while VA$_2$Z$_4$ (except VGe$_2$N$_4$) are ferromagnetic semiconductors with a magnetic moment of 1~$\mu_\text{B}$ per V atom. 
On the other hand, NbSi$_2$P$_4$, TaSi$_2$P$_4$, TaSi$_2$As$_4$, TaGe$_2$N$_4$ and group-\uppercase\expandafter{\romannumeral6}B MA$_2$Z$_4$ (e.g., CrSi$_2$N$_4$, CrSi$_2$P$_4$, CrGe$_2$N$_4$, CrGe$_2$As$_4$, MoGe$_2$As$_4$, WGe2As$_4$ and WGe$_2$P$_4$) are antiferromagnetic.


\begin{table*}[htbp]
\centering
\caption{Structural properties of magnetic MA$_2$Z$_4$: lattice constants ($a$), magnetic moment (Mag$^\text{PBE}$ for PBE and Mag$^\text{HSE}$ for HSE), magnetic type (FM for ferromagnetic and AFM for antiferromagnetic), magnetic anisotropic energy (MAE), Curie temperature ($T_\text{C}$) with PBE, PEB+$U$ and HSE.~\cite{Hong2020-Science, Wang2021-NC, Ding2021-JPCC-Y-Cd-Hf-Hg}}
\renewcommand\arraystretch{1.5} 
\setlength{\tabcolsep}{1.5mm}
\begin{tabular}{ccccccccccc}
\hline
Struc. &Phase &$a$ &Mag$^\text{PBE}$ &Mag$^\text{HSE}$ &Mag$^\text{HSE}$ &Type &MAE &$T_\text{C}^\text{PBE}$ &$T_\text{C}^{\text{PBE}+U}$ &$T_\text{C}^\text{HSE}$ \\ 
& &(\AA) &($\mu_\text{B}$) &($\mu_\text{B}$) &($\mu_\text{B}$) & &($\mu$eV) &(K) &(K) &(K) \\  
\hline
VSi$_2$N$_4$ &$\alpha_1$ &2.88 &0.97 &1.05 &1.00--1.19 &FM &75.0--76.4 &230 &350 &506--687 \\ 
VSi$_2$P$_4$ &$\alpha_1$ &2.88 &0.96 &1.04 &1.00 &FM &68.5 &230 &350 &506--687 \\ 
             &$\delta_4$ &3.48 &1.00 &* &1.00 &FM &* &235 &452 &* \\
             
VSi$_2$As$_4$ &$\alpha_2$ &3.72 &1.00 &* &1.00 &FM &* &250 &* &*  \\
VGe$_2$P$_4$ &$\alpha_2$ &3.56 &1.00 &* &1.00 &FM &* &* &* &* \\
VGe$_2$As$_4$ &$\alpha_2$ &3.72 &1.00 &* &1.00 &FM &* &* &* &*  \\

NbSi$_2$N$_4$ &$\alpha_1$ &2.96 &0.32--0.57 &* &1.00 &FM &95.5 &* &* &*  \\
NbSi$_2$P$_4$ &$\alpha_1$ &3.53 &* &* & *&AFM &* &* &* &* \\
NbGe$_2$N$_4$ &$\alpha_1$ &3.09 &0.72 &* &1.00 &FM &* &* &* &197  \\

TaSi$_2$P$_4$ &$\alpha_1$ &3.54 &* &* &* &AFM &* &* &* &* \\
TaSi$_2$As$_4$ &$\alpha_2$ &3.68 &* &* &* &AFM &* &* &* &* \\
TaGe$_2$N$_4$ &$\alpha_1$ &3.08 &0.49 &* &1.00 &FM &* &* &* &*  \\

YSi$_2$N$_4$ &$\alpha_1$ &3.07 &1.00 &* &1.00 &FM &* &* &* &90   \\
YSi$_2$N$_4$ &$\beta_2$ &3.08 &1.00 &* &1.00 &FM &* &* &* &85  \\

\hline
\end{tabular}
\label{mag}
\end{table*}

\begin{figure}[!t]
\centering
\includegraphics[width=8.6cm]{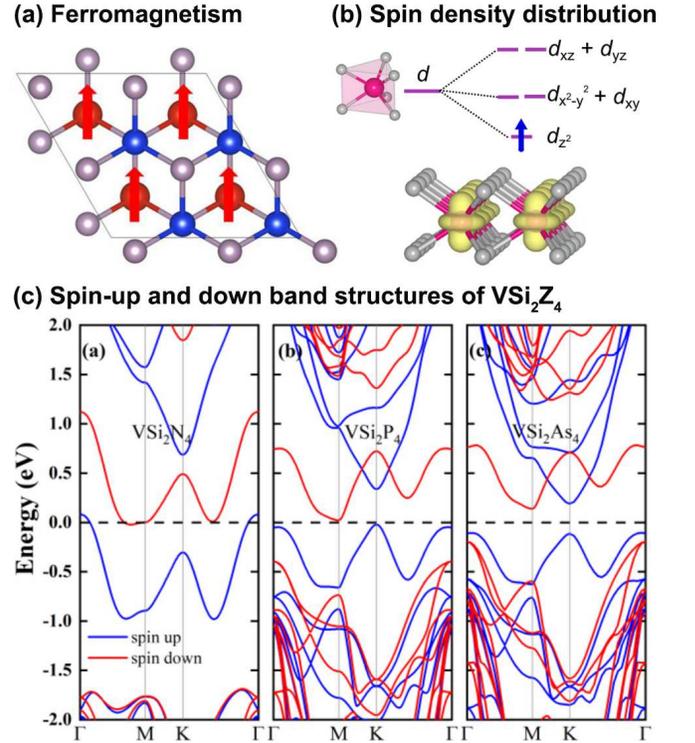}
\caption{(a) Magnetic ground states of monolayer VSi$_2$Z$_4$. (b) Spin density distribution. (c) Spin-up (blue) and spin-down (red) band structures of VSi$_2$Z$_4$ by PBE functional.~\cite{Feng2022-APL-VSi2X4-mag, Dey2022-arxiv-MAZ-VSiXN4-Janus-mag}}
\label{band-VSi2Z4}
\end{figure}

For ferromagnetic semiconductors, there exist three types: type-\uppercase\expandafter{\romannumeral1} spin gapless semiconductors (SGSs) (a zero bandgap in one spin channel but a bandgap in another spin channel), type-\uppercase\expandafter{\romannumeral2} SGSs (bandgaps of spin-up or spin-down channel but zero gap for the valence band and conduction band in opposite spin channels), and bipolar magnetic semiconductors (BMSs) (the valence band and conduction band in opposite spin channels approaching to Fermi level). 2D SGSs and BMSs with high Curie temperatures are highly desirable for advanced spintronic applications due to their unique electronic structure and high spin polarization~\cite{Wang2017-NSR-SpinGaplessSemiconductor, Wang2018-APL-SpinGaplessSemiconductor}. 
The intrinsic magnetic mechanisms of monolayer MA$_2$Z$_4$ nanosheets (e.g., VSi$_2$N$_4$, VSi$_2$P$_4$ and NbSi$_2$N$_4$) have been investigated by dissecting the spin-up and spin-down band structures, magnetic anisotropic energy (MAE) and Curie temperature ($T_\text{C}$). 

The magnetic ground states of monolayer VSi$_2$Z$_4$ are ferromagnetic (FM) by considering the spin polarization, as shown in Fig.~\ref{band-VSi2Z4}(a)~\cite{Feng2022-APL-VSi2X4-mag}. 
Based on GGA-PBE functional calculations, for the spin-up channel of monolayer VSi$_2$P$_4$ both VBM and CBM are located at K point and a direct bandgap of 0.36~eV presents, while VBM at $\Gamma$ point and CBM at M point indicate an indirect bandgap of 0.42~eV for the spin-down channel (Fig~\ref{band-VSi2Z4}(c)). 
The zero indirect bandgap between spin-up VBM and spin-down CBM implies that monolayer VSi$_2$P$_4$ and VSi$_2$N$_4$ are type-\uppercase\expandafter{\romannumeral2} SGSs via PBE calculations. In VSi$_2$As$_4$, the VBM and CBM for spin channels nearby the Fermi level are in the opposite direction, indicating that the spin-polarized currents would be easily generated with tunable spin polarization by applying a small gate voltage. 
The results show that the intrinsic ferromagnetism is attributed to the unpaired electron from the metal atom.
It can be noticed that in VSi$_2$Z$_4$ (Fig~\ref{band-VSi2Z4}(b)), the $d$ orbital electrons from transition metal occupy the bands approaching to the Fermi level, accompanying with the mixed orbital (e.g., $d_{z^2}$+ $d_{x^2+y^2}$) or transition of orbital composition (e.g., the transition between $d_{z^2}$ and $d_{x^2+y^2}$)~\cite{Dey2022-arxiv-MAZ-VSiXN4-Janus-mag, Yadav2021-MA2N4-photovoltaic-photocatalytic, Cui2021-PRB-VSi2N4-mag-spinvalley}. 

The increment of bandgap is 0.3~eV under the consideration of Coulomb interaction ($U$) effect~\cite{Wang2021-NC, Akanda2021-APL-Nb-VSi2N4-VSi2P4-mag}.
The bandgap of VSi$_2$Z$_4$ by HSE functional is larger than that by PBE and PBE+$U$. For instances, by HSE functional, VSi$_2$N$_4$ possesses a direct bandgap of 0.78~eV at K point, VSi$_2$P$_4$ has an indirect bandgap of 0.84~eV with spin-up CBM at $\Gamma$ point and spin-down VBM at M point, and NbSi$_2$P$_4$ exhibits an indirect bandgap of 0.54~eV with spin-up VBM at $\Gamma$ point and spin-down CBM at M point. 
The comparison with the three calculation method shows that the bandgap increases from PBE, PBE+$U$ to HSE.  
However, the spin polarizations under these methods are still high since there exists only one spin channels around the Fermi level. 

\begin{figure*}[!t]
\centering
\includegraphics[width=14cm]{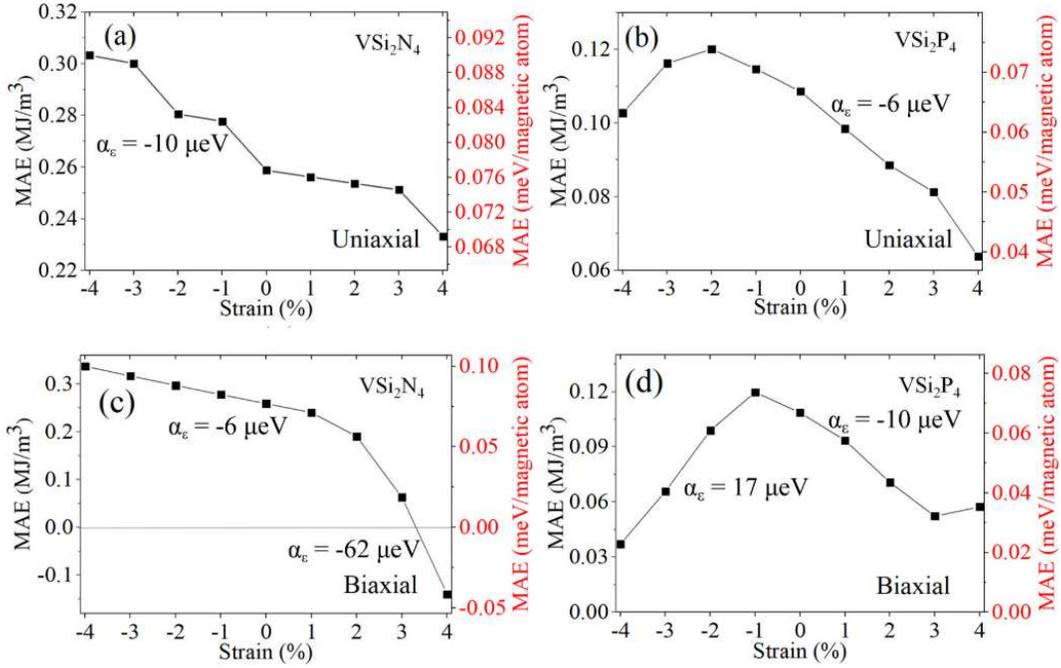}
\caption{Strain effect on MAE calculated with PBE(SOC)~+~$U$: (a) VSi$_2$N$_4$ and (b) VSi$_2$P$_4$ under a uniaxial strain, (c) VSi$_2$N$_4$ and (d) VSi$_2$P$_4$ under a biaxial strain.~\cite{Akanda2021-APL-Nb-VSi2N4-VSi2P4-mag}}
\label{strain-MAE}
\end{figure*}

Magnetic anisotropic energy (MAE) of NbSi$_2$N$_4$, VSi$_2$N$_4$ and VSi$_2$P$_4$ are summarized in Table~\ref{mechanics}. The positive values indicate the in-plane alignment of magnetic moments. Therefore, these ferromagnetic monolayer MA$_2$Z$_4$ nanosheets have easy magnetization plane, i.e., none energy consumes when magnetization rotates in the plane. In the finite energy resolution (1~eV), it is found that the total energy has no dependence on the angle of the magnetic moment within the plane, implying the weak in-plane anisotropy of MAE~\cite{Akanda2021-APL-Nb-VSi2N4-VSi2P4-mag}, similar to the recently reported 2D MnPS$_3$~\cite{XUE20221Nonlinear}.  
MAE of VSi$_2$Z$_4$ (Z~=~N/P) in the range of 0.11 to 0.25~meV per magnetic atom is lower than other 2D materials (e.g., 0.8~meV of CrI$_3$~\cite{Webster2018-PRB-CrX3-Mag}, 0.72~meV of Fe$_3$P~\cite{Zheng2019-JPCL-Fe3P-Mag} and 1~meV of Fe$_3$GeTe$_2$~\cite{Zhuang2016-PRB-FeGeTe-Mag}), and close to CrCl$_3$ (0.02~meV), CrBr$_3$ (0.16~meV), NiI$_2$~\cite{Han2020-PCCP-NiI2-magnetic} (0.11~meV) and FeCl$_2$~\cite{Torun2015-PRB-FeCl2-Mag} (0.07~meV). 

\begin{figure}[!t]
\centering
\includegraphics[width=7cm]{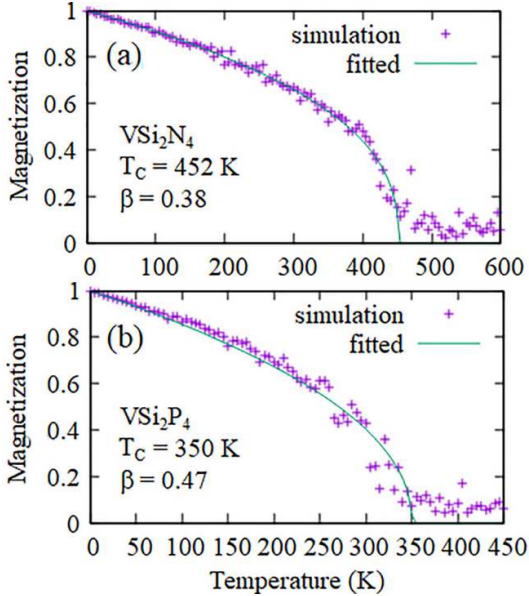}
\caption{Monte Carlo calculations of the normalized magnetization as a function of temperature for (a) VSi$_2$N$_4$ and (b) VSi$_2$P$_4$. The solid lines show the best fits to the analytical expression.~\cite{Akanda2021-APL-Nb-VSi2N4-VSi2P4-mag}}
\label{fitting-Tc}
\end{figure}

Curie temperature ($T_\text{C}$) of VSi$_2$N$_4$, VSi$_2$P$_4$ and VSi$_2$As$_4$ is 230, 235, and 250~K, respectively.~\cite{Feng2022-APL-VSi2X4-mag}, which is much higher than 45~K for monolayer CrI$_3$ ~\cite{Huang2017-Nature-CrI3-ferromagnetism} and 130~K for monolayer Fe$_3$GeTe$_2$~\cite{Fei2018-NatMater-Fe3GeTe2-ferromagnetism}. 
In Akanda's paper~\cite{Akanda2021-APL-Nb-VSi2N4-VSi2P4-mag}, the normalized magnetization of VSi$_2$N$_4$ and VSi$_2$P$_4$ based on Monte Carlo calculations follows to the analytic expression: $m(T)~=~(1-T/T_\text{C})^\beta$, as shown in Fig.~\ref{fitting-Tc}. The results from PBE+$U$ show that $T_\text{C}$ of VSi$_2$N$_4$ and VSi$_2$P$_4$ are 350 and 452~K, respectively, which are above room temperature. The difference of predicted $T_\text{C}$ is ascribed to the unequal calculation method using different DFT functionals. 
$T_\text{C}$ of VSi$_2$N$_4$ by HSE functional is up to 506--687~K. $T_\text{C}$ of Janus VSiGeN$_4$ and VSiSnN$_4$ is 507~K and 347~K, respectively~\cite{Dey2022-arxiv-MAZ-VSiXN4-Janus-mag}. With the same functional (e.g., PBE+$U$), the transition temperature of monolayer MA$_2$Z$_4$ is higher than that of other 2D magnets (e.g., TcGeSe$_3$, TcGeTe$_3$, ScCl, YCl, LaCl, LaBr$_2$, CrSBr, etc.) with high $T_\text{C}$ over room temperature~\cite{You2020-PRRes-2D-Tc, Jiang2018-ACSAMI-2D-Tc}. 


The magnetic properties of MA$_2$Z$_4$ family can be tuned by strain, adatoms and defects~\cite{Akanda2021-APL-Nb-VSi2N4-VSi2P4-mag, Dey2022-arxiv-MAZ-VSiXN4-Janus-mag, Cui2021-PRB-VSi2N4-mag-spinvalley, Li2021-AnnalenderPhysik-VSi2N4-CrSi2N4-straininducedMagnetic, Ray2021-ACSomega-MoSi2N4-halfmetal, Abdelati2022-PCCP-MetalDoping}.
Similar to the case of magnetic thin films \cite{Yi2019strain,Sander1999the,Gong2020Electric}, strain or stress also has a significant effect on magnetic behaviors of monolayer MA$_2$Z$_4$. The strain coefficient ($\alpha_\epsilon$) is defined as the sensitivity of MAE to strain, which is expressed as $\alpha_\epsilon$~=~$dE_\text{MAE}/d\epsilon$. In VSi$_2$N$_4$, $\alpha_\epsilon$ is --10~$\mu$eV/$\%$ under a compressive uniaxial strain and --6~$\mu$eV/$\%$ under a biaxial strain, while in VSi$_2$P$_4$, $\alpha_\epsilon$~=~--6~$\mu$eV/$\%$ under a tensile uniaxial strain, --10~$\mu$eV/$\%$ under a tensile biaxial strain, and 17~$\mu$eV/$\%$ under a compressive biaxial strain (Fig.~\ref{strain-MAE}). The strain sensitivity of both two materials is lower than that of CrSb (32~$\mu$eV/$\%$)~\cite{Park2020-PRB-CrSb-AFM}. When the biaxial strain is up to 3--4$\%$, the spin direction of VSi$_2$N$_4$ rotates from in-plane to out-of-plane state~\cite{Akanda2021-APL-Nb-VSi2N4-VSi2P4-mag, Cui2021-PRB-VSi2N4-mag-spinvalley}.

\begin{figure*}[!t]
\centering
\includegraphics[width=14.5cm]{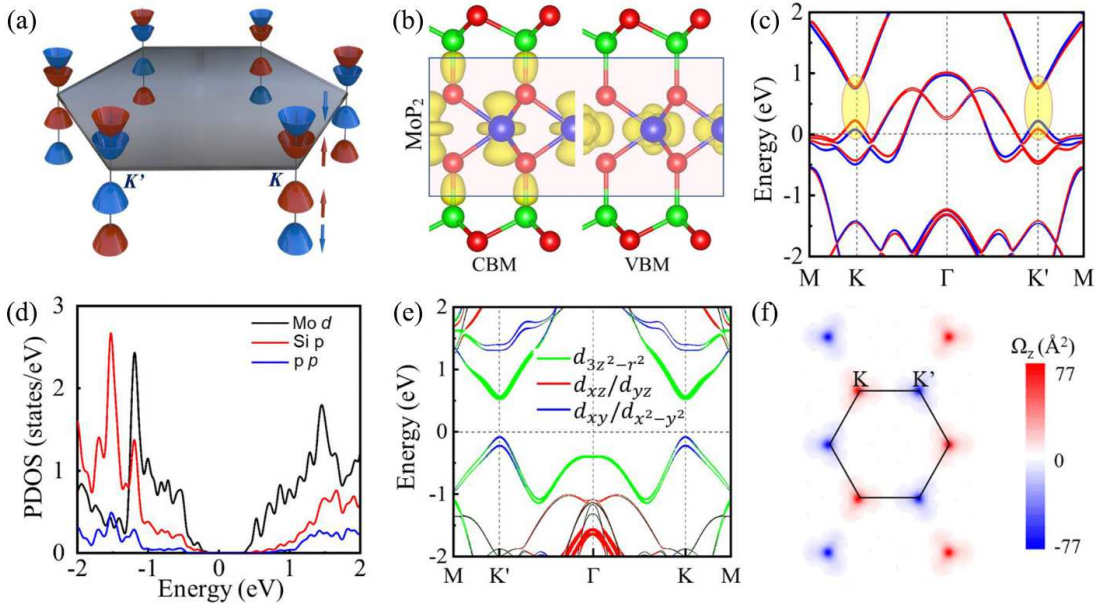}
\caption{(a) Schematic of the valleys near the $K$ and $K'$ points. (b) Charge densities at the VBM and CBM of MoSi$_2$P$_4$. (c) Band structure of MoP$_2$. (d) Partial densities of states (PDOSs) of MoSi$_2$P$_4$. (e) Orbital-projected band structure of MoSi$_2$P$_4$. (f) Berry curvature of MoSi$_2$P$_4$ with nonzero Berry curvature indicating the abnormal transverse velocity.~\cite{Yuan2022-PRB-MA2Z4-valley}}
\label{spin-valley-mechanism}
\end{figure*}

Additionally, substitutional doping and atomic vacancies induce the transition of nonmagnetic state in MoSi$_2$N$_4$ to magnetic state. Transition metal substitutional doping (e.g., Ag, Au, Bi, Fe, Mn, Pb and V) induces the asymmetric spin channels in MoSi$_2$N$_4$ with generating a magnetic moment of 1.00--5.87~$\mu_\text{B}$~\cite{Abdelati2022-PCCP-MetalDoping, Cui2022-metalAtomAdsorption}.
Schwingenschl$\ddot{\text{o}}$gl et al.~\cite{Ray2021-ACSomega-MoSi2N4-halfmetal} reported that N-site vacancy in MoSi$_2$N$_4$ leads to the spin-majority bands crossing the Fermi level. N- and Si-site vacancies generate a magnetic moment of 1.0 and 2.0~$\mu_\text{B}$, respectively. The nonmagnetic state in MA$_2$Z$_4$-derived MoN$_2$X$_2$Y$_2$ (XY~=~AlO, GaO, InO) nanosheets is transferred to ferromagnetic state by a hole doping~\cite{Ding2022-ASS-MoN2X2Y2}. Li et al.~\cite{Li2021-Nanoscale-MSi2CxN4-x} found the element substitution of nitrogen by carbon causes different magnetic moments in monolayer MSi$_2$C$_x$N$_{4-x}$ (M~=~Cr/Mo/W, $x$~=~1 or 2). The position of carbon atoms determines the ground state of magnetic moments. 
Monolayer CrSi$_2$CN$_3$ (C bridged Si and Mo) is predicted as ferromagnetic half-metal, where the magnetic moments is mainly originated from Cr atoms (0.72~$\mu_\text{B}$ per atom). There exist three AFM structures, i.e., CrSi$_2$N$_2$C$_2$ (two C atoms located at two outermost sites), MoSi$_2$N$_2$C$_2$ and CrSi$_2$N$_3$C (one C atoms located at outermost site). The magnetic moments are ascribed to the C and metal atoms. The FM states in half-metallic 2D systems are attributed to the hole-mediated double exchange, while the AFM states are originated from the super-exchange.

MA$_2$Z$_4$ family also has spin-valley properties.
As an emerging degree of freedom, valley refers to the presence of multiple energy extremal points in the Brillouin zone (BZ) for low-energy carriers in a semiconductor. Analogous to charge and spin, the valley degree of freedom can be exploited for information encoding and processing, leading to the concept of valleytronics~\cite{Li2020-PRB-MoSi2N4-WSi2N4-MoSi2As4-spinvalley}.
The spin-valley feature of semiconductors in MA$_2$Z$_4$ family has been systematically studied. Taking into account the spin-orbit coupling (SOC), valley spin splittings of MA$_2$Z$_4$ appear near both the CBM and VBM. The remarkable splitting near VBM is 139--500~meV, but the spin splitting near CBM is tiny~\cite{Ai2021-PCCP-MoSi2X4-spinvalley, Yuan2022-PRB-MA2Z4-valley, Liu2021-JPCL-CrSiN-CrSiP,HUSSAIN2022169897Emergence}.


Six kinds of monolayer MA$_2$Z$_4$ (M~=~Mo/W, A~=~C/Si/Ge, Z~=~N/P/Ge) nanosheets, i.e., MoSi$_2$P$_4$, MoSi$_2$As$_4$, WSi$_2$P$_4$, WSi$_2$As$_4$, WGe$_2$P$_4$ and WGe$_2$As$_4$, possess strong spin-valley coupling~\cite{Yuan2022-PRB-MA2Z4-valley, Sheoran2022-arxiv-WSi2N4-spinvally}. They are all direct bandgap semiconductors with band extrema located at the inequivalent K and K' point (Fig.~\ref{spin-valley-mechanism}(a)). 
A remarkable spin splitting near the VBM is shown in all the six monolayers with SOC, and the splitting in W-based compounds is larger than that in Mo-based ones. The spin valleys are attributed to the MZ$_2$ layer by analysing DOS and charge distribution (Fig.~\ref{spin-valley-mechanism}(b)).
Furthermore, the comparison of spin band structures between monolayer MoP$_2$ (Fig.~\ref{spin-valley-mechanism}(c)) and MoSi$_2$P$_4$ (Fig.~\ref{spin-valley-mechanism}(e)) shows that there exist other interfering states in the energy range of the valleys of MoP$_2$. This manifests that the AZ layers play the role of structural stabilizer and protect the valley states from the interference of the $p$ orbitals of Z atoms and $d$ orbitals of M atom. 
Time-reversal symmetry always constrains the valley-polarized current. 
Nonzero Berry curvature of MA$_2$Z$_4$ in Fig.~\ref{spin-valley-mechanism}(f) indicates the abnormal transverse velocity proportional to the Berry curvature would be generated, leading to a spatial separation of the carriers coupled to two different valleys. The carriers from different valleys are accumulated at opposite sides, resulting in an anomalous valley Hall effect.~\cite{Yuan2022-PRB-MA2Z4-valley, Yang2020-PRB-MoSi2N4-MoSi2As4-spinvalley}

\begin{figure}[!b]
\centering
\includegraphics[width=8.8cm]{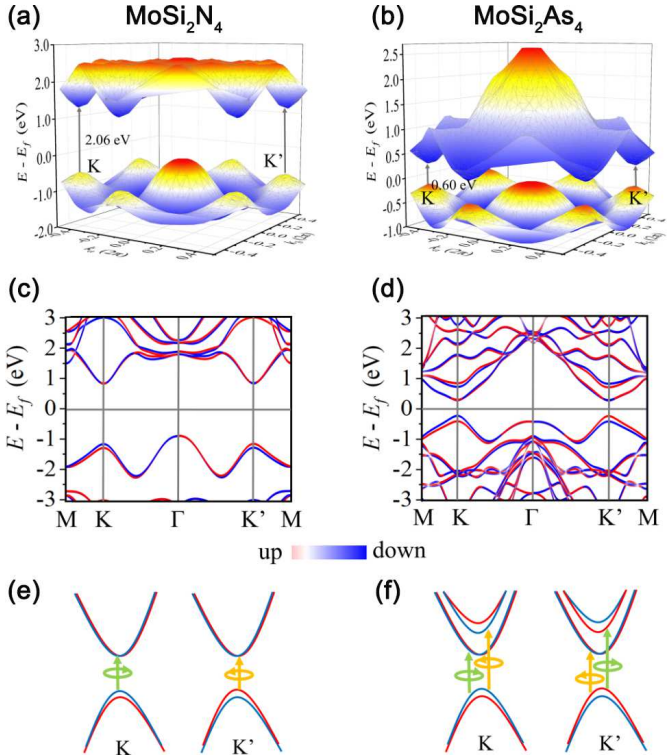}
\caption{Spin-valley band structures of monolayer MoSi$_2$N$_4$ and MoSi$_2$As$_4$: (a) (b) 3D CBM and VBM of MoSi$_2$N$_4$ and MoSi$_2$As$_4$, (c) (d) spin down and spin up band with SOC. Comparison between (e) traditional valleys and (f) multiple-folded valleys.~\cite{Yang2020-PRB-MoSi2N4-MoSi2As4-spinvalley}}
\label{Spin-Valley-MoSi2As4}
\end{figure}

\begin{figure}[!t]
\centering
\includegraphics[width=8.4cm]{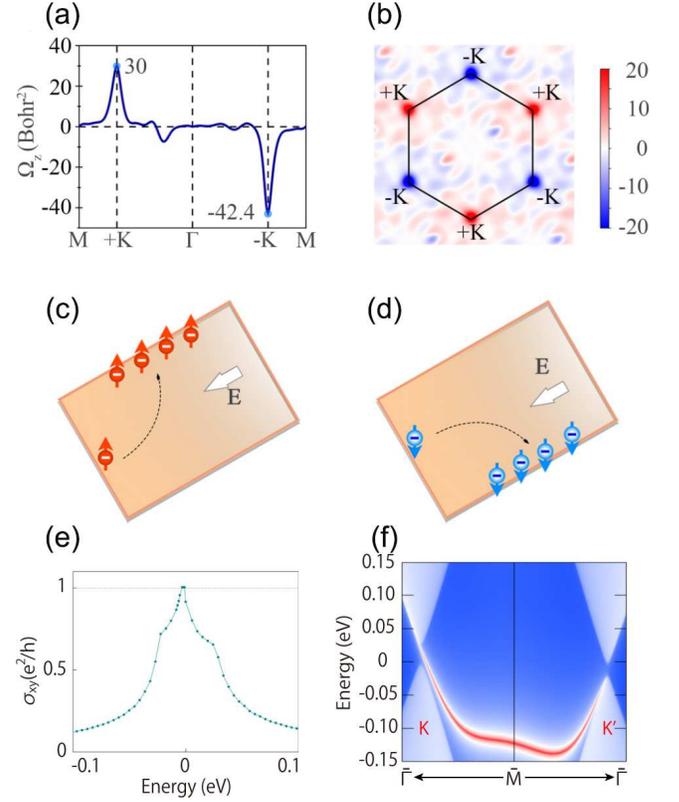}
\caption{Berry curvature of monolayer VSi$_2$P$_4$ (a) as a curve along the high-symmetry points and (b) as a counter map over the 2D Brillouin zone. (c) (d) Diagrams of the anomalous valley Hall effect under an electron doping and an in-plane electric field, but with opposite magnetization. (e) Anomalous Hall conductivity versus chemical potential with SOC. (f) The corresponding edge spectrum for the QAH state with SOC. Note that the plot is centred at the M point, so the left valley is K and the right valley is K'.~\cite{Feng2021-PRB-VSi2P4-spinvalley, Li2021-PRB-VSi2P4-QAH}}
\label{QAH-spinvalley}
\end{figure}


Monolayer MoSi$_2$As$_4$ is found to exhibit `perfect valleys' with no interference from other part of Brillouin zone, as well as multiple-folded valleys (Fig.~\ref{Spin-Valley-MoSi2As4})~\cite{Yang2020-PRB-MoSi2N4-MoSi2As4-spinvalley}.
The another valley is originated from the second unoccupied conduction band, which enlarges the degree of freedom in energy. The calculation results show that the electron can be selectively pumped from VBM to CBM with a low photon energy of 0.41~eV. With 1.00 eV energy input, the electron can be excited to the VBM and second unoccupied conduction band, indicating the potential application of MoSi$_2$As$_4$ for multiple-information operator and storage in valleytronic devices.

 
The experimentally synthesized MoSi$_2$N$_4$ and WSi$_2$N$_4$ are indirect bandgap semiconductors in which CBM is located at K (or K’) and VBM at $\Gamma$ point, limiting their potential utilization in valleytronic devices.
Nevertheless, the difference between the uppermost valence band at K and $\Gamma$ point is small (e.g., 144.4~meV in MoSi$_2$N$_4$). Strain engineering could eliminate this small difference.
The indirect bandgap of MoSi$_2$N$_4$ is transferred to the direct bandgap under a compressive strain of 2$\%$ and the direct-gap state maintains until a compressive strain up to 4$\%$~\cite{Ai2021-PCCP-MoSi2X4-spinvalley, Li2020-PRB-MoSi2N4-WSi2N4-MoSi2As4-spinvalley}. 
Monolayer MoSi$_2$N$_4$ with --4$\%$ strain possesses a large spin splitting and strong spin valley coupling (valleys K and K’) and thus would be an ideal valleytronic materials. 

For other magnetic members in MA$_2$Z$_4$ family, the coupling between magnetism and valley provokes the quantum anomalous Hall (QAH) effect.
The spin splitting of VSi$_2$N$_4$ is 102.3~meV (27.3~meV) at VBM (CBM)~\cite{Zhou2021-npj-VSi2N4-framework}. In contrast, the spin splitting of VSi$_2$P$_4$ is 49.4~meV at CBM but can be neglected at VBM~\cite{Feng2021-PRB-VSi2P4-spinvalley}. 
Ma et al.~\cite{Feng2021-PRB-VSi2P4-spinvalley} separately considered the effect of SOC or magnetic exchange interaction on spin splitting of VSi$_2$P$_4$, and proposed that the combined effect causes its spontaneous valley polarization (Fig.~\ref{QAH-spinvalley}). The spin splitting energy at K and K' points is expressed as $\Delta^{\text{K/K'}}$~=~$E^{\text{K/K'}}_\uparrow$-$E^{\text{K/K'}}_\downarrow$, $\uparrow$ and $\downarrow$ mean spin-up and spin-down channels. With the magnetic exchange interaction, the broken time-reversal symmetry leads to $\Delta^{\text{K}}_{\text{mag}}$~=~$\Delta^{\text{K'}}_{\text{mag}}$; with SOC, the time-reversal symmetry results in $\Delta^{\text{K}}_{\text{SOC}}$~=~$-\Delta^{\text{K'}}_{\text{SOC}}$; with the synergistic effect, the net spin splitting should be the $\Delta^{\text{K}}_{\text{mag}}$+$\Delta^{\text{K}}_{\text{SOC}}$ at K point and $\Delta^{\text{K'}}_{\text{mag}}$+$\Delta^{\text{K'}}_{\text{SOC}}$ at K' point.
Li et al.~\cite{Li2021-PRB-VSi2P4-QAH} found Hubbard-$U$ can effectively tune the phase diagram of MA$_2$Z$_4$ family, and result in intriguing magnetic, valley and topological features. When $U$~=~2.25 and 2.36~eV, the QAH phase and valley structure coexist, as shown in Fig.~\ref{QAH-spinvalley}(e) and (f). 
Besides, strain effect promotes the trivial topology in VSi$_2$P$_4$ and VA$_2$Z$_4$-derived materials (VN$_2$X$_2$Y$_2$)~\cite{Feng2021-PRB-VSi2P4-spinvalley, Li2021-PRB-VSi2P4-QAH, Wang2021-APL-VN2X2Y2-mag}. There exists an edge state connecting the valence and conduction bands under a compressive strain of 2.1~$\%$~\cite{Feng2021-PRB-VSi2P4-spinvalley} and a biaxial tensile strain~\cite{Li2021-PRB-VSi2P4-QAH}.
Recent first-principles calculations reveal 2D MSi$_2$Z$_4$ with 1T' structure as larg-bandgap and tunable quantum spin Hall insulator, which possesses a protected spin-polarized edge state and a large spin-Hall conductivity~\cite{Islam2022Switchable}.


\section{Prospective Application}

\begin{figure}[!t]
\centering
\includegraphics[width=8.6cm]{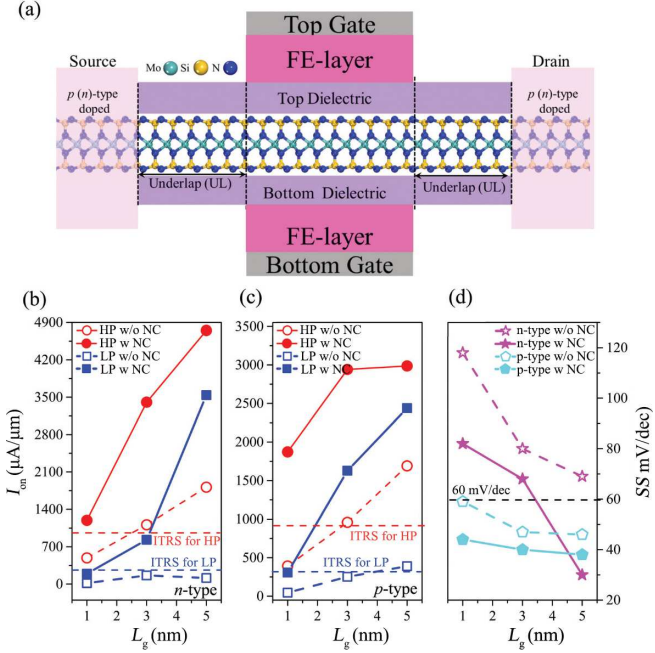}
\caption{(a) Nanodevice design of monolayer MA$_2$Z$_4$-based transistor. On-current ($I_\text{on}$) of (b) n-type and (c) p-type versus gate length ($L_\text{g}$). (d) Subthreshold swing ($SS$) versus $L_\text{g}$.~\cite{Sun2021-JMCC-MoSi2N4-transistor}}
\label{transistor}
\end{figure}

\subsection{Transistors}
Semiconductors in 2D MA$_2$Z$_4$ family have atomic-scale thickness, ambient stability, suitable bandgap and excellent electronic mobilities, indicating that this family could be promising candidates for the new-generation miniaturized field-effect transistors (FETs)~\cite{Sun2021-JMCC-MoSi2N4-transistor, Huang2021-PRAppli-MoSi2N4-transistor5nm, Ye2022-PCCP-MoSi2N4-transistor3nm, Ghobadi2022-IEEE-MoSi2N4-Transistor, Nandan2022-IEEE-MoSi2N4-Transistor}. Double-gate metal-oxide-semiconductor FET (DG MOSFET) based on monolayer MoSi$_2$N$_4$ has been designed and investigated~\cite{Sun2021-JMCC-MoSi2N4-transistor, Huang2021-PRAppli-MoSi2N4-transistor5nm}. This kind of FET has advantages of high operating frequency, good gain controllability and low feedback capacitance. There are three factors to evaluate the performance of DG MOSFET, i.e., on-current ($I_\text{on}$), gate control, and intrinsic delay time and power consumption. 

$I_\text{on}$ represents the device operating speed, while off-current ($I_{\text{off}}$) means the static power dissipation. According to the International Technology Roadmap for Semiconductors (ITRS) published in 2013, the standard of $I_\text{on}$ ($I_\text{on}/I_\text{off}$) is 900~$\mu$A$\mu$m$^{-1}$ (9.0$\times$10$^3$).
The highest $I_\text{on}$ of n-type (p-type) DG monolayer MoSi$_2$N$_4$ MOSFET is up to 1813~$\mu$A$\mu$m$^{-1}$ (1690~$\mu$A$\mu$m$^{-1}$), as shown in Fig.~\ref{transistor}(b) and (c)~\cite{Sun2021-JMCC-MoSi2N4-transistor}, which is 70$\%$ (80$\%$) higher than n-type (p-type) DG monolayer MoS$_2$ MOSFET~\cite{Zhang2021-ACSAEM-MoS2-DGMOSFET}. 
Subthreshold swing ($SS$) can reflect the switching rate of MOSFET between the on and off states, which always describes the gate-control ability in the subthreshold region.  
$SS$ of DG monolayer MA$_2$Z$_4$ MOSFET is predicted as 69~mV~dec$^{-1}$ for the n-type and 52~mV~dec$^{-1}$ for the p-type when the optimum $I_\text{on}$ is reached. As for sub-5~nm scale MOSFETs, the value of $SS$ should be below the Boltzmann limit at room temperature (60~mV~dec$^{-1}$) since the short channel effect would cause the coexist of tunneling current and thermoionic current, and further influence the overall performance of devices. When gate length ($L_\text{g}$) is below 5~nm, $SS$ of p-type and n-type MA$_2$Z$_4$ MOSFETs is 46--52~mV~dec$^{-1}$ (Fig.~\ref{transistor}(d))~\cite{Sun2021-JMCC-MoSi2N4-transistor, Huang2021-PRAppli-MoSi2N4-transistor5nm}.
Switching speed in MOSFETs is described by the intrinsic delay time and power consumption. The ITRS standard of delay time for the high performance and low power is 0.423 and 1.493~ps, respectively. Besides, ITRS requirements of power consumption for the high performance and low power are 0.24 and 0.28~fJ$\mu$m$^{-1}$, respectively. DG monolayer MA$_2$Z$_4$ MOSFET with an optimum assembly can meet the standards. 
In brief, the design of device determines whether DG monolayer MA$_2$Z$_4$ MOSFET satisfies the ITRS standard in terms of the structural parameters, like gate length, doping concentration to the source and drain, and underlap length between gate and electrodes.

\subsection{Photocatalyst}
Photocatalytic process can be described by four important steps: (1) light absorption to generate electron-hole pairs, (2) separation of excited charges to suppress the electron-hole  recombination, (3) transfer of electrons and holes to the surface of photocatalysts, and (4) utilization of charges on the surface for redox reactions~\cite{Zhu2017-AdvEnergyMater-Photocatlysis, Li2019-MaterTodayChem-2DPhotocatalysts}.  
The typical reactions are water splitting and CO$_2$ reduction, which are important for the environment and dual-carbon confinement strategy.
2D MA$_2$Z$_4$ systems with high optical absorption could be prospective candidates for photocatalytic applications. 

\begin{figure}[!t]
\centering
\includegraphics[width=8.6cm]{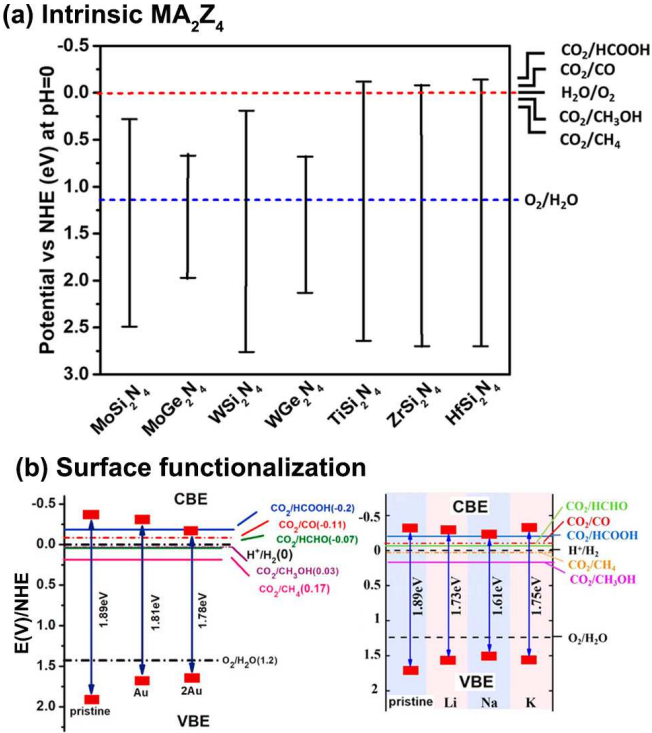}
\caption{Band edge positions with respect to water redox and CO$_2$ reduction levels of (a) monolayer MA$_2$Z$_4$ members and (b) MoSi$_2$N$_4$ after surface functionalization. The normal hydrogen electrode (NHE) potential is set as 0~eV.~\cite{Yadav2021-MA2N4-photovoltaic-photocatalytic, Xu2022-JPCS-AuAdsorption, Sun2022-CommuniTheoerticPhys-Li-Na-K}}
\label{band-alignment}
\end{figure}

The band edges of photocatalyst candidate in 2D MA$_2$Z$_4$ systems must straddle the standard redox potentials. The conduction band edge (CBE) of monolayer MoSi$_2$N$_4$ is about 0.71~eV, higher than the H$^+$/H$_2$ reduction level. The valence band edge (VBE) is about 0.19~eV, lower than the O$_2$/H$_2$O oxidation level. Band edges of monolayer MoSi$_2$N$_4$ straddle the water redox potentials in both highly acidic and neutral conditions (Fig.~\ref{band-alignment}(a)).
In contrast, the band edges of monolayer WSi$_2$N$_4$ (WGe$_2$N$_4$) only satisfy the standard in the highly acidic (neutral) condition~\cite{Mortazavi2021-NanoEnergy, Jian2021-JPCC-monoMoSi2N4-strain}. Group-\uppercase\expandafter{\romannumeral4}B MSi$_2$N$_4$ nanosheets with appropriate band edges are considered as the most suitable materials among 2D MA$_2$Z$_4$ family for both water splitting and CO$_2$ reduction application~\cite{Yadav2021-MA2N4-photovoltaic-photocatalytic}. 
It is revealed that adatoms reduce the bandgap of MoSi$_2$N$_4$ and further promote the possibility of photocatalytic process~\cite{Xu2022-JPCS-AuAdsorption, Sun2022-CommuniTheoerticPhys-Li-Na-K, Shi2022-JMS-MoSi2N4-HER-adatoms}. The CBE of Au-MoSi$_2$N$_4$ is closer to the level for CO$_2$/HCOOH reduction and thus can easily transfer charges to CO$_2$ and produce natural gas (HCOOH, HCHO, CO, CH$_4$, and CH$_3$OH). The VBE of Au-MoSi$_2$N$_4$ approaching to the O$_2$/H$_2$O redox potential is beneficial for the interaction between holes and H$_2$O, and restrains the electron-hole recombination (Fig.~\ref{band-alignment}(b))~\cite{Xu2022-JPCS-AuAdsorption}. Meanwhile, Li and Na absorption can improve the capacity of water splitting by using MoSi$_2$N$_4$~\cite{Sun2022-CommuniTheoerticPhys-Li-Na-K}.
The adjustable CBE and VBE are observed in MA$_2$Z$_4$-based heterostructures (e.g., BlueP/, InSe/ and MoSiGeN$_4$/MoSi$_2$N$_4$)~\cite{Xuefeng2022-JPD-BlueP-hetero, He2022-PCCP-InSe-MoSi2N4-elec-optic-HER, Mwankemwa2022-ResPhys-MoSi2N4-MoSiGeN4, Zhao2021-JPCL-stackingEngineering, Hussain2022-PhysE-MoSi2N4-XSi2N4-hetero, Ren2022-PRM-MSi2N4-bandAlignments}, providing more possibility of this family for photocatalytic devices.

\begin{figure}[!t]
\centering
\includegraphics[width=7.5cm]{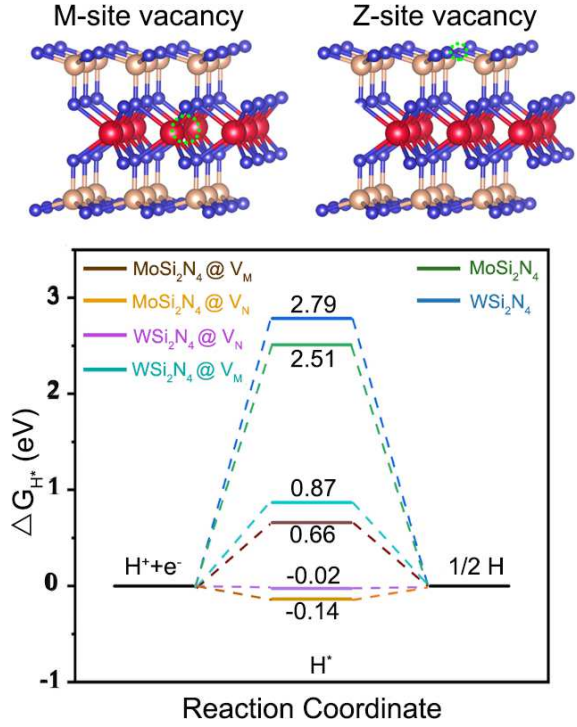}
\caption{Crystal structures of monolayer MSi$_2$N$_4$ with M- (V$_\text{M}$) and Z-site (V$_\text{N}$) vacancy, and the corresponding Gibbs free energy diagram of HER processing.~\cite{Zang2021-PRM-MoSi2N4-WSi2N4-HER}}
\label{GibbsFreeEnergy}
\end{figure}

The other important factor of photocatalysis is the absorption of reactive atoms or molecules, which determines the electron transition to the surface and the charge utilization for redox reactions. 
The optimal site for hydrogen or CO$_2$ absorption in monolayer MA$_2$Z$_4$ is reported as Z site due to the lowest spontaneous binding energy compared with other sites~\cite{Zang2021-PRM-MoSi2N4-WSi2N4-HER, Yadav2021-MA2N4-photovoltaic-photocatalytic}.
The Gibbs free energy for the hydrogen adsorption ($\Delta\!G_{\text{H}^\ast}$) on monolayer MoSi$_2$N$_4$ and WSi$_2$N$_4$ is calculated to be 2.51 and 2.79~eV, respectively, which is much larger than the ideal value ($\Delta\!G_{\text{H}^\ast}$~=~0~eV). This indicates the weak binding of hydrogen in pristine monolayer MSi$_2$N$_4$ (M~=~Mo/W) and inertial hydrogen evolution reaction (HER) activity~\cite{Zang2021-PRM-MoSi2N4-WSi2N4-HER}. The HER performance of MSi$_2$N$_4$, as well as N$_2$ reduction reaction (NRR), can be triggered by introducing N-site vacancy~\cite{Zang2021-PRM-MoSi2N4-WSi2N4-HER, Qian2022-JMST-MoSi2N4-x-HER, Xiao2021-ASS-defectiveMoSi2N4-HER, Luo2021-JMCA-MSi2N4-adsorption}. As shown in Fig.~\ref{GibbsFreeEnergy}, it is obvious that the vacancy significantly influence $\Delta\!G_{\text{H}^\ast}$. $\Delta\!G_{\text{H}^\ast}$ is decreased about 2--3 times by M-site vacancy and approaches to zero by N-site vacancy ($\Delta\!G_{\text{H}^\ast}$~=~--0.14~eV in MoSi$_2$N$_4$ and --0.02~eV in WSi$_2$N$_4$). This manifests that the HER performance of MSi$_2$N$_4$ with the outermost N-site vacancy is comparable with and even better than that of Pt. On the other hand, transition metallic atomic doping and strain engineering are verified as the effective strategies to tune $\Delta\!G_{\text{H}^\ast}$ and trigger HER, oxygen evolution reaction (OER) or oxygen reduction reaction (ORR) activity~\cite{Xiao2021-ASS-defectiveMoSi2N4-HER, Lu2022-MetalEmbeddedMoSi2N4ASS-OER-ORR, Sahoo2022-ACSomega-VGe2N4-NbGe2N4-strain}. Particularly, $|\Delta\!G_{\text{H}^\ast}|$ is calculated as only 0.05~eV in WSi$_2$N$_4$ with Fe-doping at Si-site. A 3$\%$ tensile strain results in $\Delta\!G_{\text{H}^\ast}$~=~0.015~eV in NbGe$_2$N$_4$.

In the MA$_2$Z$_4$ family, there exist other good photocatalysts with appropriate $\Delta\!G$~\cite{Chen2021-CAEJ-MA2Z4-magnetic, Liu2021-JPCC-MAZ-HER, Zheng2021-AMI-MoSi2N4-Catalysts, Yu2021-AMI-Janus-HER-OER}, for examples TiSi$_2$N$_4$, HfA$_2$Z$_4$, ZrA$_2$Z$_4$, Janus MSiGeN$_4$, etc.
Recently, a theoretical method in multilevel is proposed to screen appropriate candidates for hydrogen evolution in MA$_2$Z$_4$ family~\cite{Liu2021-JPCC-MAZ-HER}. There exist four screening criteria: (1) small structural deformation after hydrogen absorption to maintain the potential stability during the HER processes, (2) low absolute value of Gibbs free energy ($|\Delta G|$~$\rightarrow$~0) with hydrogen adsorption, (3) suitable bandgaps, (4) high environmental stability. Taking these factors in mind, the multilevel screening workflow discovers seven MA$_2$Z$_4$ with great stability and highly active HER among 144 MA$_2$Z$_4$ structures, i.e., $\alpha_1$-VGe$_2$N$_4$, $\alpha_1$-NbGe$_2$N$_4$, $\alpha_1$-TaGe$_2$N$_4$, $\alpha_1$-NbSi$_2$N$_4$, $\alpha_2$-VGe$_2$N$_4$, $\alpha_2$-NbGe$_2$N$_4$, and $\alpha_2$-TiGe$_2$P$_4$. Monolayer $\alpha_1$-NbSi$_2$N$_4$ with the lowest formation energy is considered as the most promising 2D MA$_2$Z$_4$ material for the HER application along the synthesis routine. At low H coverage ($\theta < 25 \%$), the optimal $|\Delta G|$ of these seven MA$_2$Z$_4$ is lower than 0.1~eV, which is comparable with or even superior to that of Pt (--0.09~eV at $\theta$~=~25$\%$). A descriptor ($E_{\text{LUS}}$) is proposed as energy level of the lowest unoccupied state to evaluate the capacity of H adsorption, since hydrogen adsorption is often accompanied with the electron transition to the CBM (for semiconductors) or the Fermi level (for metals). A higher $E_{\text{LUS}}$ is prone to restraining electron filling and causing weaker H adsorption. On the contrary, a lower $E_{\text{LUS}}$ represents the higher ability of H adsorption. Compared with $\alpha_1$-MoSi$_2$N$_4$, $\alpha_1$-NbSi$_2$N$_4$ with lower $E_{\text{LUS}}$ shows higher activity toward HER.

HER performance of monolayer MA$_2$Z$_4$ family is also examined by the combination of DFT calculations and machine learning algorithms including support vector regression (SVR), kernel ridge regression (KRR), random forest regression (RFR), extreme gradient boosting regression (XGBR), least absolute shrinkage, and selection operator (LASSO)~\cite{Zheng2021-AMI-MoSi2N4-Catalysts}. $\Delta G_{\text{H}^\ast}$ and Gibbs free energy of deuterium ($\Delta G_{\text{D}^\ast}$) can be accurately and rapidly predicted via XGBR by using only simple genetic programming processed elemental features, with a low predictive root-mean-square error of 0.14~eV. $\Delta G_{\text{H}^\ast}$ of group-\uppercase\expandafter{\romannumeral5}B MA$_2$Z$_4$ is closer to zero, indicating the excellent HER capacity. For example, TaSn$_2$P$_4$ (0.07 eV) has a similar absolute value of $\Delta G_{\text{H}^\ast}$ as NbSn$_2$P$_4$ (--0.05 eV), while $\Delta G_{\text{H}^\ast}$ of CrSn$_2$P$_4$ (0.23 eV) is nearly 3.5 times higher than that of TaSn$_2$P$_4$ (0.07 eV). It can be concluded that M element is the crucial factor for the HER performance of MA$_2$Z$_4$ materials. In addition, NbSi$_2$N$_4$ with $\Delta G_{\text{H}^\ast}$~=~--0.041~eV and $\Delta G_{\text{D}^\ast}$~=~--0.102~eV, as well as VSi$_2$N$_4$ with $\Delta G_{\text{H}^\ast}$~=~0.024~eV and $\Delta G_{\text{D}^\ast}$~=~--0.033~eV is screened as the best HER and deuterium evolution reaction (DER) catalysts among the MA$_2$Z$_4$ family. 

As for ORR, the four-electron (4e$^-$) mechanism favors the production of H$_2$O. The order of ORR activity is MGe$_2$As$_4$~$>$~MSi$_2$As$_4$~$>$~MSi$_2$N4~$>$~MSi$_2$2P$_4$~$>$~MGe$_2$P$_4$ $\approx$ MGe$_2$N$_4$~\cite{Chen2021-JPCC-MAZ-ORE}. Among them, VGe$_2$As$_4$, CrGe$_2$As$_4$, VSi$_2$As$_4$ and NbSi$_2$As$_4$ are screened out to be highly promising electrocatalysts with a small overpotential around 0.5--0.6~V. The topmost surface As acts as the active site, and the p-band center of the As atom shows correlation with the adsorption strength of the critical intermediate. Zhang et al.~\cite{Lin2022-AFM-OER} efficiently screened photocatalytic OER catalysts in MA$_2$Z$_4$ family via an automated high-throughput workflow. They found the adsorption ability of O atoms determines the catalytic effect. $\beta_2$-ZrSi$_2$N$_4$ and $\beta_2$-HfSi$_2$N$_4$ are considered as the efficient photocatalytic OER catalysts. 
In particular, CrGe$_2$As$_4$ exhibits outstandingly high ORR activity with ultralow overpotential (0.49~V), which is comparable with the Pt-based catalysts. The metallic conductivity, as well as the moderate adsorption and orbital hybridization between As and O* intermediate, is responsible for the exceptional activity~\cite{Chen2021-JPCC-MAZ-ORE}.

\begin{figure}[!t]
\centering
\includegraphics[width=8.6cm]{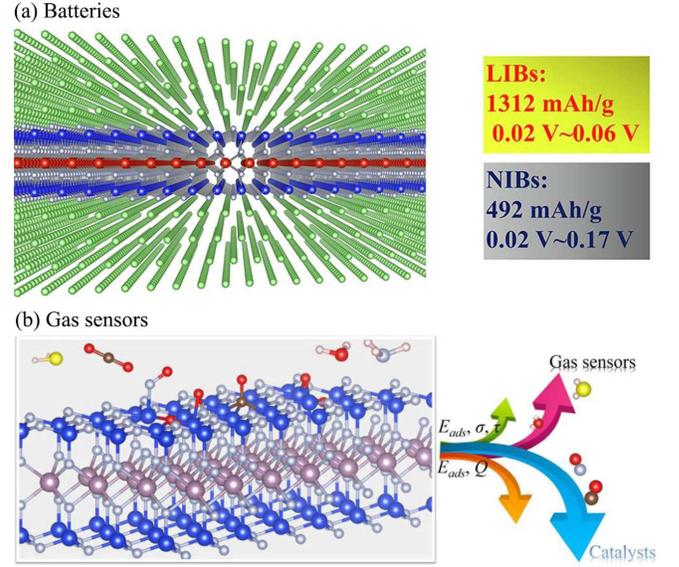}
\caption{Diagram of monolayer MA$_2$Z$_4$ family in applications of (a) Li-Na batteries and (b) gas sensors.~\cite{Wang2022-ASS-VSi2N4-batteries, Xiao2022-ACSomega-MoSi2N4-moleculardope}}
\label{batteries-sensors}
\end{figure}

\subsection{Batteries}
Metal-air batteries have great advantages of high-energy-density metal anodes, active air cathodes, light weight, and simple structure, which are convenient to utilize in portable equipments. Bilayer or multilayer vdW MoSi$_2$N$_4$ shows great potential application as electrodes (both anode and cathode) of Zn-air batteries~\cite{Li2022-PSS-layerMoSi2N4-ZnBattery}. On the anode side, the maximum theoretical capacity of Zn in MoSi$_2$N$_4$ is up to 257~mAh/g. While on the cathode side, O$_2$ reduction reaction on the MoSi$_2$N$_4$ surface is more efficient than the general sluggish four-electron aqueous O$_2$ redox reactions. 
Furthermore, VSi$_2$N$_4$ provides two critical specifications (high specific capacity and full battery open-circuit voltage) for the high-performance secondary Li-ion or Na-ion batteries (LIBs/NIBs) (Fig.~\ref{batteries-sensors}(a))~\cite{Wang2022-ASS-VSi2N4-batteries}. The ionic capacity is up to 1312 mAh/g for Li and 492 mAh/g and Na, while the average open-circuit voltages are 0.02–0.06 V for LIBs, and 0.02–0.17 V for NIBs.

\subsection{Sensors}
Environmental dependence of physical properties of 2D materials promotes the development of potential applications, e.g., sensors~\cite{Munteanu2020-Sensors, Buckley2020-sensors, Tyagi2020-nanoscale-sensors, Yi2021-graphenebased-sensor, Subbanna2022-sensors}. Gas sensor is one of the significant device for detecting gas molecules, especially contaminated and poisonous ones, which can be utilized in the field of industrial harm evaluation, cultivation of agricultural products, assessment of medical drugs, etc. 
The physisorption behaviors of gas molecules of monolayer MoSi$_2$N$_4$ by spin-polarized DFT calculations have been investigated~\cite{Xiao2022-ACSomega-MoSi2N4-moleculardope, Bafekry2021-ASS-MoSi2N4-adsorption}. Due to the weak interaction and small charge transfer, the gas molecules are physically adsorbed on the MoSi$_2$N$_4$ surface (Fig.~\ref{batteries-sensors}(b)). 
The results show that H$_2$, N$_2$, CO, CO$_2$, NO, NO$_2$, H$_2$O, H$_2$S, NH$_3$ and CH$_4$ molecules reduce the bandgap of MoSi$_2$N$_4$ (from 1.73~eV to 1.50~eV)~\cite{Bafekry2021-ASS-MoSi2N4-adsorption}. While the absorption of O$_2$, NO$_2$ and SO$_2$ molecules obviously influences the electronic properties of MoSi$_2$N$_4$ and even induces the spin polarization with magnetic moments (1--2~$\mu_\text{B}$). The magnitude of magnetic moments is sensitive to the concentration of gas molecules, which increases with the increment of concentration of NO$_2$ but decreases with the increment of concentration of SO$_2$. This indicates that MoSi$_2$N$_4$-based gas sensor has a high application potential for O$_2$, NO, NO$_2$ and SO$_2$ detection.
Furthermore, the introduction of N vacancy into MoSi$_2$N$_4$ improves the absorption performance~\cite{Xiao2022-ACSomega-MoSi2N4-moleculardope}, resulting in MoSi$_2$N$_4$ with promising prospects in highly sensitive and reusable gas sensors of H$_2$O and H$_2$S molecules.

\section{Summary and Outlook}
In summary, the experimental achievements have brought novel 2D monolayer MoSi$_2$N$_4$ and WSi$_2$N$_4$, while theoretical predictions have provided much more possibility of the emerging MA$_2$Z$_4$ family due to the structural complexity and component diversity.
In this review, we have summarized the latest progress of this novel 2D MA$_2$Z$_4$ family with a focus on its physical and chemical properties, as well as its promising applications. 
Different from transition metal carbides, nitrides and dichalcogenides, this family exhibits more affluent and intriguing features, such as excellent mechanical properties, interesting electronic properties (from insulator to semiconductor to metal) related to the number of total valence electrons, wide range of thermal conductivity (10$^1$--10$^3$~Wm$^{-1}$K$^{-1}$), high optical absorption of visible and ultraviolet light, spin-valley effect, etc. 
Moreover, the properties of MA$_2$Z$_4$ family are manipulable by external fields, providing more degrees of freedom to realize some specific applications. Strain engineering is demonstrated to shrink the band structure of MoSi$_2$N$_4$ and induce the well-known ‘Mexican hat’, as well as result in an obvious shift in absorption spectra. By the layered strategy, semiconducting MA$_2$Z$_4$-based vdW heterostructures have promising applications in photocatalysts, while metallic ones effectively reduce the Schottky barrier height and are beneficial for the energy efficient and high-performance power devices. Finally, we survey the perspective applications of MA$_2$Z$_4$, from the aspects of transistors, photocatalysts, batteries, and sensors.

The emerging 2D MA$_2$Z$_4$ family with versatile properties and applications opens the mind of low-dimensional structural designs and provides the new possibilities and opportunities to the development of 2D materials. As an outlook, there exist considerable spaces to further understand and exploit the emerging MA$_2$Z$_4$ family.

$\bullet$
Firstly, the experience of synthesizing 2D MA$_2$Z$_4$ could inspire the idea of synthesizing other 2D materials without knowing their bulk counterparts. Experimental attempts following or beyond the synthesis strategy for monolayer MoSi$_2$N$_4$ are intriguing for the advent of new synthetic 2D materials.

$\bullet$
Secondly, since the development of interdisciplinary, the attention to a potential material is no longer limited to some excellent characteristics. The synergistic effect of multiphysics for practical applications should also be the focus. Although the excellent properties of 2D MA$_2$Z$_4$ family have been revealed, synergy of multiphysics coupling would be a crucial method to integrate the versatile properties and maximize their advantages, and even emanate more novel sparks.

$\bullet$
Thirdly, most of the current theoretical predictions on 2D MA$_2$Z$_4$ family are made by first-principles calculations. It is highly recommended that modeling and simulation methodologies across scales be developed to understand 2D MA$_2$Z$_4$ from electronic, atomistic, microstructure, to device level.

$\bullet$
Fourthly, the sandwich structure of 2D MA$_2$Z$_4$ enables more degrees of freedom to elaborate properties and functionalities, calling for more extensive efforts. For instance, the engineering of 2D MA$_2$Z$_4$ by asymmetric design of its two sides could be further explored to achieve more versatile structures and functionalities, such as ferroelectricity, multiferroics, Janus structures, etc.

$\bullet$
Fifthly, multilayer or MA$_2$Z$_4$-based vdW heterostructures show great potential in nanodevices. The practical application of 2D MA$_2$Z$_4$ family requires more reasonable designs and attempts (e.g., assemble and stacking way, interlayer adaptability, in-plane heterostructure, etc.) to ensure the reliable realization of their advantages.

$\bullet$
Last but not least, the successful preparation of semiconducting monolayer MoSi$_2$N$_4$ and WSi$_2$N$_4$ by CVD provides the appropriate growth method for this family. However, most of the reported structures, properties, functionalities, and applications of 2D MA$_2$Z$_4$ are from the theoretical predictions. Continuous experimental efforts are mandatory to verify the existence of other members (e.g., magnetic VSi$_2$N$_4$ and VSi$_2$P$_4$) as well as the theoretically predicted properties, applications, and devices.


\section*{Acknowledgment}
The authors acknowledge the support from the National Natural Science Foundation of China (NSFC 11902150, 12272173), 15th Thousand Youth Talents Program of China, the Research Fund of State Key Laboratory of Mechanics and Control of Mechanical Structures (MCMS-I-0419G01 and MCMS-I-0421K01), the Fundamental Research Funds for the Central Universities (1001-XAC21021), and a project Funded by the Priority Academic Program Development of Jiangsu Higher Education Institutions.

\bibliography{References}

\begin{thebibliography}{200}%
\makeatletter
\providecommand \@ifxundefined [1]{%
 \@ifx{#1\undefined}
}%
\providecommand \@ifnum [1]{%
 \ifnum #1\expandafter \@firstoftwo
 \else \expandafter \@secondoftwo
 \fi
}%
\providecommand \@ifx [1]{%
 \ifx #1\expandafter \@firstoftwo
 \else \expandafter \@secondoftwo
 \fi
}%
\providecommand \natexlab [1]{#1}%
\providecommand \enquote  [1]{``#1''}%
\providecommand \bibnamefont  [1]{#1}%
\providecommand \bibfnamefont [1]{#1}%
\providecommand \citenamefont [1]{#1}%
\providecommand \href@noop [0]{\@secondoftwo}%
\providecommand \href [0]{\begingroup \@sanitize@url \@href}%
\providecommand \@href[1]{\@@startlink{#1}\@@href}%
\providecommand \@@href[1]{\endgroup#1\@@endlink}%
\providecommand \@sanitize@url [0]{\catcode `\\12\catcode `\$12\catcode
  `\&12\catcode `\#12\catcode `\^12\catcode `\_12\catcode `\%12\relax}%
\providecommand \@@startlink[1]{}%
\providecommand \@@endlink[0]{}%
\providecommand \url  [0]{\begingroup\@sanitize@url \@url }%
\providecommand \@url [1]{\endgroup\@href {#1}{\urlprefix }}%
\providecommand \urlprefix  [0]{URL }%
\providecommand \Eprint [0]{\href }%
\providecommand \doibase [0]{https://doi.org/}%
\providecommand \selectlanguage [0]{\@gobble}%
\providecommand \bibinfo  [0]{\@secondoftwo}%
\providecommand \bibfield  [0]{\@secondoftwo}%
\providecommand \translation [1]{[#1]}%
\providecommand \BibitemOpen [0]{}%
\providecommand \bibitemStop [0]{}%
\providecommand \bibitemNoStop [0]{.\EOS\space}%
\providecommand \EOS [0]{\spacefactor3000\relax}%
\providecommand \BibitemShut  [1]{\csname bibitem#1\endcsname}%
\let\auto@bib@innerbib\@empty
\bibitem [{\citenamefont {Akinwande}\ \emph {et~al.}(2017)\citenamefont
  {Akinwande}, \citenamefont {Brennan}, \citenamefont {Bunch}, \citenamefont
  {Egberts}, \citenamefont {Felts}, \citenamefont {Gao}, \citenamefont {Huang},
  \citenamefont {Kim}, \citenamefont {Li}, \citenamefont {Li}, \citenamefont
  {Liechti}, \citenamefont {Lu}, \citenamefont {Park}, \citenamefont {Reed},
  \citenamefont {Wang}, \citenamefont {Yakobson}, \citenamefont {Zhang},
  \citenamefont {Zhang}, \citenamefont {Zhou},\ and\ \citenamefont
  {Zhu}}]{Akinwande2017-EML-Mechanics}%
  \BibitemOpen
  \bibfield  {author} {\bibinfo {author} {\bibfnamefont {D.}~\bibnamefont
  {Akinwande}}, \bibinfo {author} {\bibfnamefont {C.~J.}\ \bibnamefont
  {Brennan}}, \bibinfo {author} {\bibfnamefont {J.~S.}\ \bibnamefont {Bunch}},
  \bibinfo {author} {\bibfnamefont {P.}~\bibnamefont {Egberts}}, \bibinfo
  {author} {\bibfnamefont {J.~R.}\ \bibnamefont {Felts}}, \bibinfo {author}
  {\bibfnamefont {H.}~\bibnamefont {Gao}}, \bibinfo {author} {\bibfnamefont
  {R.}~\bibnamefont {Huang}}, \bibinfo {author} {\bibfnamefont {J.~S.}\
  \bibnamefont {Kim}}, \bibinfo {author} {\bibfnamefont {T.}~\bibnamefont
  {Li}}, \bibinfo {author} {\bibfnamefont {Y.}~\bibnamefont {Li}}, \bibinfo
  {author} {\bibfnamefont {K.~M.}\ \bibnamefont {Liechti}}, \bibinfo {author}
  {\bibfnamefont {N.}~\bibnamefont {Lu}}, \bibinfo {author} {\bibfnamefont
  {H.~S.}\ \bibnamefont {Park}}, \bibinfo {author} {\bibfnamefont {E.~J.}\
  \bibnamefont {Reed}}, \bibinfo {author} {\bibfnamefont {P.}~\bibnamefont
  {Wang}}, \bibinfo {author} {\bibfnamefont {B.~I.}\ \bibnamefont {Yakobson}},
  \bibinfo {author} {\bibfnamefont {T.}~\bibnamefont {Zhang}}, \bibinfo
  {author} {\bibfnamefont {Y.~W.}\ \bibnamefont {Zhang}}, \bibinfo {author}
  {\bibfnamefont {Y.}~\bibnamefont {Zhou}},\ and\ \bibinfo {author}
  {\bibfnamefont {Y.}~\bibnamefont {Zhu}},\ }\bibfield  {title} {\bibinfo
  {title} {{A review on mechanics and mechanical properties of 2D
  materials—Graphene and beyond}},\ }\href
  {https://doi.org/10.1016/j.eml.2017.01.008} {\bibfield  {journal} {\bibinfo
  {journal} {Extreme Mechanics Letters}\ }\textbf {\bibinfo {volume} {13}},\
  \bibinfo {pages} {42} (\bibinfo {year} {2017})}\BibitemShut {NoStop}%
\bibitem [{\citenamefont {Lee}\ \emph {et~al.}(2008)\citenamefont {Lee},
  \citenamefont {Wei}, \citenamefont {Kysar},\ and\ \citenamefont
  {Hone}}]{Lee2008-science-Graphene-mechanics}%
  \BibitemOpen
  \bibfield  {author} {\bibinfo {author} {\bibfnamefont {C.}~\bibnamefont
  {Lee}}, \bibinfo {author} {\bibfnamefont {X.}~\bibnamefont {Wei}}, \bibinfo
  {author} {\bibfnamefont {J.~W.}\ \bibnamefont {Kysar}},\ and\ \bibinfo
  {author} {\bibfnamefont {J.}~\bibnamefont {Hone}},\ }\bibfield  {title}
  {\bibinfo {title} {{Measurement of the elastic properties and intrinsic
  strength of monolayer graphene}},\ }\href
  {https://doi.org/10.1126/science.1157996} {\bibfield  {journal} {\bibinfo
  {journal} {Science}\ }\textbf {\bibinfo {volume} {321}},\ \bibinfo {pages}
  {385} (\bibinfo {year} {2008})}\BibitemShut {NoStop}%
\bibitem [{\citenamefont {Bertolazzi}\ \emph {et~al.}(2011)\citenamefont
  {Bertolazzi}, \citenamefont {Brivio},\ and\ \citenamefont
  {Kis}}]{Bertolazzi2011-ACSNano-MoS2-mechanics}%
  \BibitemOpen
  \bibfield  {author} {\bibinfo {author} {\bibfnamefont {S.}~\bibnamefont
  {Bertolazzi}}, \bibinfo {author} {\bibfnamefont {J.}~\bibnamefont {Brivio}},\
  and\ \bibinfo {author} {\bibfnamefont {A.}~\bibnamefont {Kis}},\ }\bibfield
  {title} {\bibinfo {title} {{Stretching and breaking of ultrathin MoS$_2$}},\
  }\href {https://doi.org/10.1021/nn203879f} {\bibfield  {journal} {\bibinfo
  {journal} {ACS Nano}\ }\textbf {\bibinfo {volume} {5}},\ \bibinfo {pages}
  {9703} (\bibinfo {year} {2011})}\BibitemShut {NoStop}%
\bibitem [{\citenamefont {Balandin}(2011)}]{Balandin2011-NatMater-thermal}%
  \BibitemOpen
  \bibfield  {author} {\bibinfo {author} {\bibfnamefont {A.~A.}\ \bibnamefont
  {Balandin}},\ }\bibfield  {title} {\bibinfo {title} {{Thermal properties of
  graphene and nanostructured carbon materials}},\ }\href
  {https://doi.org/10.1038/nmat3064} {\bibfield  {journal} {\bibinfo  {journal}
  {Nature Materials}\ }\textbf {\bibinfo {volume} {10}},\ \bibinfo {pages}
  {569} (\bibinfo {year} {2011})}\BibitemShut {NoStop}%
\bibitem [{\citenamefont {Li}\ \emph {et~al.}(2020{\natexlab{a}})\citenamefont
  {Li}, \citenamefont {Gao}, \citenamefont {Cheng}, \citenamefont {He},
  \citenamefont {Yin}, \citenamefont {Hu}, \citenamefont {Chen}, \citenamefont
  {Cheng},\ and\ \citenamefont {Zhao}}]{Li2020-AFM-boronthermal}%
  \BibitemOpen
  \bibfield  {author} {\bibinfo {author} {\bibfnamefont {D.}~\bibnamefont
  {Li}}, \bibinfo {author} {\bibfnamefont {J.}~\bibnamefont {Gao}}, \bibinfo
  {author} {\bibfnamefont {P.}~\bibnamefont {Cheng}}, \bibinfo {author}
  {\bibfnamefont {J.}~\bibnamefont {He}}, \bibinfo {author} {\bibfnamefont
  {Y.}~\bibnamefont {Yin}}, \bibinfo {author} {\bibfnamefont {Y.}~\bibnamefont
  {Hu}}, \bibinfo {author} {\bibfnamefont {L.}~\bibnamefont {Chen}}, \bibinfo
  {author} {\bibfnamefont {Y.}~\bibnamefont {Cheng}},\ and\ \bibinfo {author}
  {\bibfnamefont {J.}~\bibnamefont {Zhao}},\ }\bibfield  {title} {\bibinfo
  {title} {{2D boron sheets: Structure, growth, and electronic and thermal
  transport properties}},\ }\href {https://doi.org/10.1002/adfm.201904349}
  {\bibfield  {journal} {\bibinfo  {journal} {Advanced Functional Materials}\
  }\textbf {\bibinfo {volume} {30}},\ \bibinfo {pages} {1904349} (\bibinfo
  {year} {2020}{\natexlab{a}})}\BibitemShut {NoStop}%
\bibitem [{\citenamefont {Qian}\ \emph {et~al.}(2021)\citenamefont {Qian},
  \citenamefont {Zhou},\ and\ \citenamefont
  {Chen}}]{Qian2021-NatMater-thermal}%
  \BibitemOpen
  \bibfield  {author} {\bibinfo {author} {\bibfnamefont {X.}~\bibnamefont
  {Qian}}, \bibinfo {author} {\bibfnamefont {J.}~\bibnamefont {Zhou}},\ and\
  \bibinfo {author} {\bibfnamefont {G.}~\bibnamefont {Chen}},\ }\bibfield
  {title} {\bibinfo {title} {{Phonon-engineered extreme thermal conductivity
  materials}},\ }\href {https://doi.org/10.1038/s41563-021-00918-3} {\bibfield
  {journal} {\bibinfo  {journal} {Nature Materials}\ }\textbf {\bibinfo
  {volume} {20}},\ \bibinfo {pages} {1188} (\bibinfo {year}
  {2021})}\BibitemShut {NoStop}%
\bibitem [{\citenamefont {Wang}\ \emph {et~al.}(2019)\citenamefont {Wang},
  \citenamefont {Lian}, \citenamefont {Guo}, \citenamefont {Mao}, \citenamefont
  {Zhang}, \citenamefont {Zhang}, \citenamefont {Gu}, \citenamefont {Xu},\ and\
  \citenamefont {Duan}}]{Wang2019-PRL-Superconductivity}%
  \BibitemOpen
  \bibfield  {author} {\bibinfo {author} {\bibfnamefont {C.}~\bibnamefont
  {Wang}}, \bibinfo {author} {\bibfnamefont {B.}~\bibnamefont {Lian}}, \bibinfo
  {author} {\bibfnamefont {X.}~\bibnamefont {Guo}}, \bibinfo {author}
  {\bibfnamefont {J.}~\bibnamefont {Mao}}, \bibinfo {author} {\bibfnamefont
  {Z.}~\bibnamefont {Zhang}}, \bibinfo {author} {\bibfnamefont
  {D.}~\bibnamefont {Zhang}}, \bibinfo {author} {\bibfnamefont {B.~L.}\
  \bibnamefont {Gu}}, \bibinfo {author} {\bibfnamefont {Y.}~\bibnamefont
  {Xu}},\ and\ \bibinfo {author} {\bibfnamefont {W.}~\bibnamefont {Duan}},\
  }\bibfield  {title} {\bibinfo {title} {{Type-II Ising superconductivity in
  two-dimensional materials with spin-orbit coupling}},\ }\href
  {https://doi.org/10.1103/PhysRevLett.123.126402} {\bibfield  {journal}
  {\bibinfo  {journal} {Physical Review Letters}\ }\textbf {\bibinfo {volume}
  {123}},\ \bibinfo {pages} {126402} (\bibinfo {year} {2019})}\BibitemShut
  {NoStop}%
\bibitem [{\citenamefont {Bekaert}\ \emph {et~al.}(2019)\citenamefont
  {Bekaert}, \citenamefont {Petrov}, \citenamefont {Aperis}, \citenamefont
  {Oppeneer},\ and\ \citenamefont
  {Milo{\v{s}}evi{\'{c}}}}]{Bekaert2019-PRL-H-MoB2-superconductivity}%
  \BibitemOpen
  \bibfield  {author} {\bibinfo {author} {\bibfnamefont {J.}~\bibnamefont
  {Bekaert}}, \bibinfo {author} {\bibfnamefont {M.}~\bibnamefont {Petrov}},
  \bibinfo {author} {\bibfnamefont {A.}~\bibnamefont {Aperis}}, \bibinfo
  {author} {\bibfnamefont {P.~M.}\ \bibnamefont {Oppeneer}},\ and\ \bibinfo
  {author} {\bibfnamefont {M.~V.}\ \bibnamefont {Milo{\v{s}}evi{\'{c}}}},\
  }\bibfield  {title} {\bibinfo {title} {{Hydrogen-induced high-temperature
  superconductivity in two-dimensional materials: The example of hydrogenated
  monolayer MgB$_2$}},\ }\href {https://doi.org/10.1103/PhysRevLett.123.077001}
  {\bibfield  {journal} {\bibinfo  {journal} {Physical Review Letters}\
  }\textbf {\bibinfo {volume} {123}},\ \bibinfo {pages} {077001} (\bibinfo
  {year} {2019})}\BibitemShut {NoStop}%
\bibitem [{\citenamefont {Li}\ \emph {et~al.}(2021{\natexlab{a}})\citenamefont
  {Li}, \citenamefont {Huang}, \citenamefont {Li}, \citenamefont {Zhao},
  \citenamefont {Lu}, \citenamefont {Han},\ and\ \citenamefont
  {Wang}}]{Li2021-MTP-review-superconductivity}%
  \BibitemOpen
  \bibfield  {author} {\bibinfo {author} {\bibfnamefont {W.}~\bibnamefont
  {Li}}, \bibinfo {author} {\bibfnamefont {J.}~\bibnamefont {Huang}}, \bibinfo
  {author} {\bibfnamefont {X.}~\bibnamefont {Li}}, \bibinfo {author}
  {\bibfnamefont {S.}~\bibnamefont {Zhao}}, \bibinfo {author} {\bibfnamefont
  {J.}~\bibnamefont {Lu}}, \bibinfo {author} {\bibfnamefont {Z.~V.}\
  \bibnamefont {Han}},\ and\ \bibinfo {author} {\bibfnamefont {H.}~\bibnamefont
  {Wang}},\ }\bibfield  {title} {\bibinfo {title} {{Recent progresses in
  two-dimensional Ising superconductivity}},\ }\href
  {https://doi.org/10.1016/j.mtphys.2021.100504} {\bibfield  {journal}
  {\bibinfo  {journal} {Materials Today Physics}\ }\textbf {\bibinfo {volume}
  {21}},\ \bibinfo {pages} {100504} (\bibinfo {year}
  {2021}{\natexlab{a}})}\BibitemShut {NoStop}%
\bibitem [{\citenamefont {Liu}\ \emph {et~al.}(2019)\citenamefont {Liu},
  \citenamefont {Hao}, \citenamefont {Watanabe}, \citenamefont {Taniguchi},
  \citenamefont {Halperin},\ and\ \citenamefont
  {Kim}}]{Liu2019-NatPhys-BiGraphene-QHE}%
  \BibitemOpen
  \bibfield  {author} {\bibinfo {author} {\bibfnamefont {X.}~\bibnamefont
  {Liu}}, \bibinfo {author} {\bibfnamefont {Z.}~\bibnamefont {Hao}}, \bibinfo
  {author} {\bibfnamefont {K.}~\bibnamefont {Watanabe}}, \bibinfo {author}
  {\bibfnamefont {T.}~\bibnamefont {Taniguchi}}, \bibinfo {author}
  {\bibfnamefont {B.~I.}\ \bibnamefont {Halperin}},\ and\ \bibinfo {author}
  {\bibfnamefont {P.}~\bibnamefont {Kim}},\ }\bibfield  {title} {\bibinfo
  {title} {{Interlayer fractional quantum Hall effect in a coupled graphene
  double layer}},\ }\href {https://doi.org/10.1038/s41567-019-0546-0}
  {\bibfield  {journal} {\bibinfo  {journal} {Nature Physics}\ }\textbf
  {\bibinfo {volume} {15}},\ \bibinfo {pages} {893} (\bibinfo {year}
  {2019})}\BibitemShut {NoStop}%
\bibitem [{\citenamefont {Shi}\ \emph {et~al.}(2020)\citenamefont {Shi},
  \citenamefont {Shih}, \citenamefont {Gustafsson}, \citenamefont {Rhodes},
  \citenamefont {Kim}, \citenamefont {Watanabe}, \citenamefont {Taniguchi},
  \citenamefont {Papi{\'{c}}}, \citenamefont {Hone},\ and\ \citenamefont
  {Dean}}]{Shi2020-NatNanotech-WSe2-QHE}%
  \BibitemOpen
  \bibfield  {author} {\bibinfo {author} {\bibfnamefont {Q.}~\bibnamefont
  {Shi}}, \bibinfo {author} {\bibfnamefont {E.~M.}\ \bibnamefont {Shih}},
  \bibinfo {author} {\bibfnamefont {M.~V.}\ \bibnamefont {Gustafsson}},
  \bibinfo {author} {\bibfnamefont {D.~A.}\ \bibnamefont {Rhodes}}, \bibinfo
  {author} {\bibfnamefont {B.}~\bibnamefont {Kim}}, \bibinfo {author}
  {\bibfnamefont {K.}~\bibnamefont {Watanabe}}, \bibinfo {author}
  {\bibfnamefont {T.}~\bibnamefont {Taniguchi}}, \bibinfo {author}
  {\bibfnamefont {Z.}~\bibnamefont {Papi{\'{c}}}}, \bibinfo {author}
  {\bibfnamefont {J.}~\bibnamefont {Hone}},\ and\ \bibinfo {author}
  {\bibfnamefont {C.~R.}\ \bibnamefont {Dean}},\ }\bibfield  {title} {\bibinfo
  {title} {{Odd- and even-denominator fractional quantum Hall states in
  monolayer WSe2}},\ }\href {https://doi.org/10.1038/s41565-020-0685-6}
  {\bibfield  {journal} {\bibinfo  {journal} {Nature Nanotechnology}\ }\textbf
  {\bibinfo {volume} {15}},\ \bibinfo {pages} {569} (\bibinfo {year}
  {2020})}\BibitemShut {NoStop}%
\bibitem [{\citenamefont {{K. S. Novoselov}}\ \emph {et~al.}(2004)\citenamefont
  {{K. S. Novoselov}}, \citenamefont {Geim}, \citenamefont {Jiang},
  \citenamefont {Zhang}, \citenamefont {Dubonos}, \citenamefont {Grigorieva},\
  and\ \citenamefont {Firsov}}]{Geim2004-Science-graphene}%
  \BibitemOpen
  \bibfield  {author} {\bibinfo {author} {\bibnamefont {{K. S. Novoselov}}},
  \bibinfo {author} {\bibfnamefont {A.~K.}\ \bibnamefont {Geim}}, \bibinfo
  {author} {\bibfnamefont {S.~V. M.~D.}\ \bibnamefont {Jiang}}, \bibinfo
  {author} {\bibfnamefont {Y.}~\bibnamefont {Zhang}}, \bibinfo {author}
  {\bibfnamefont {S.~V.}\ \bibnamefont {Dubonos}}, \bibinfo {author}
  {\bibfnamefont {I.~V.}\ \bibnamefont {Grigorieva}},\ and\ \bibinfo {author}
  {\bibfnamefont {A.~A.}\ \bibnamefont {Firsov}},\ }\bibfield  {title}
  {\bibinfo {title} {{Electric field effect in atomically thin carbon films}},\
  }\href {https://doi.org/10.1126/science.1102896} {\bibfield  {journal}
  {\bibinfo  {journal} {Science}\ }\textbf {\bibinfo {volume} {306}},\ \bibinfo
  {pages} {666} (\bibinfo {year} {2004})}\BibitemShut {NoStop}%
\bibitem [{\citenamefont {AK}(2009)}]{Geim2009-Science-graphene}%
  \BibitemOpen
  \bibfield  {author} {\bibinfo {author} {\bibfnamefont {G.}~\bibnamefont
  {AK}},\ }\bibfield  {title} {\bibinfo {title} {{Graphene: Status and
  prospects}},\ }\href {https://doi.org/10.1126/science.1158877} {\bibfield
  {journal} {\bibinfo  {journal} {Science}\ }\textbf {\bibinfo {volume}
  {324}},\ \bibinfo {pages} {1530} (\bibinfo {year} {2009})}\BibitemShut
  {NoStop}%
\bibitem [{\citenamefont {Yi}\ and\ \citenamefont
  {Shen}(2015)}]{Yi2015-JMCA-Graphene-mechanicalExfoliation}%
  \BibitemOpen
  \bibfield  {author} {\bibinfo {author} {\bibfnamefont {M.}~\bibnamefont
  {Yi}}\ and\ \bibinfo {author} {\bibfnamefont {Z.}~\bibnamefont {Shen}},\
  }\bibfield  {title} {\bibinfo {title} {{A review on mechanical exfoliation
  for the scalable production of graphene}},\ }\href
  {https://doi.org/10.1039/c5ta00252d} {\bibfield  {journal} {\bibinfo
  {journal} {Journal of Materials Chemistry A}\ }\textbf {\bibinfo {volume}
  {3}},\ \bibinfo {pages} {11700} (\bibinfo {year} {2015})}\BibitemShut
  {NoStop}%
\bibitem [{\citenamefont {Novoselov}\ \emph {et~al.}(2005)\citenamefont
  {Novoselov}, \citenamefont {Jiang}, \citenamefont {Schedin}, \citenamefont
  {Booth}, \citenamefont {Khotkevich}, \citenamefont {Morozov},\ and\
  \citenamefont {Geim}}]{Novoselov2005-2D-mechanicalexfoliation}%
  \BibitemOpen
  \bibfield  {author} {\bibinfo {author} {\bibfnamefont {K.~S.}\ \bibnamefont
  {Novoselov}}, \bibinfo {author} {\bibfnamefont {D.}~\bibnamefont {Jiang}},
  \bibinfo {author} {\bibfnamefont {F.}~\bibnamefont {Schedin}}, \bibinfo
  {author} {\bibfnamefont {T.~J.}\ \bibnamefont {Booth}}, \bibinfo {author}
  {\bibfnamefont {V.~V.}\ \bibnamefont {Khotkevich}}, \bibinfo {author}
  {\bibfnamefont {S.~V.}\ \bibnamefont {Morozov}},\ and\ \bibinfo {author}
  {\bibfnamefont {A.~K.}\ \bibnamefont {Geim}},\ }\bibfield  {title} {\bibinfo
  {title} {{Two-dimensional atomic crystals}},\ }\href
  {https://doi.org/10.1073/pnas.0502848102} {\bibfield  {journal} {\bibinfo
  {journal} {Proceedings of the National Academy of Sciences of the United
  States of America}\ }\textbf {\bibinfo {volume} {102}},\ \bibinfo {pages}
  {10451} (\bibinfo {year} {2005})}\BibitemShut {NoStop}%
\bibitem [{\citenamefont {Geim}\ and\ \citenamefont
  {Grigorieva}(2013)}]{Geim2013-Nature-vdWs}%
  \BibitemOpen
  \bibfield  {author} {\bibinfo {author} {\bibfnamefont {A.~K.}\ \bibnamefont
  {Geim}}\ and\ \bibinfo {author} {\bibfnamefont {I.~V.}\ \bibnamefont
  {Grigorieva}},\ }\bibfield  {title} {\bibinfo {title} {{Van der Waals
  heterostructures}},\ }\href {https://doi.org/10.1038/nature12385} {\bibfield
  {journal} {\bibinfo  {journal} {Nature}\ }\textbf {\bibinfo {volume} {499}},\
  \bibinfo {pages} {419} (\bibinfo {year} {2013})}\BibitemShut {NoStop}%
\bibitem [{\citenamefont {Otrokov}\ \emph {et~al.}(2019)\citenamefont
  {Otrokov}, \citenamefont {Klimovskikh}, \citenamefont {Bentmann},
  \citenamefont {Estyunin}, \citenamefont {Zeugner}, \citenamefont {Aliev},
  \citenamefont {Ga{\ss}}, \citenamefont {Wolter}, \citenamefont {Koroleva},
  \citenamefont {Shikin}, \citenamefont {Blanco-Rey}, \citenamefont {Hoffmann},
  \citenamefont {Rusinov}, \citenamefont {Vyazovskaya}, \citenamefont
  {Eremeev}, \citenamefont {Koroteev}, \citenamefont {Kuznetsov}, \citenamefont
  {Freyse}, \citenamefont {S{\'{a}}nchez-Barriga}, \citenamefont {Amiraslanov},
  \citenamefont {Babanly}, \citenamefont {Mamedov}, \citenamefont {Abdullayev},
  \citenamefont {Zverev}, \citenamefont {Alfonsov}, \citenamefont {Kataev},
  \citenamefont {B{\"{u}}chner}, \citenamefont {Schwier}, \citenamefont
  {Kumar}, \citenamefont {Kimura}, \citenamefont {Petaccia}, \citenamefont {{Di
  Santo}}, \citenamefont {Vidal}, \citenamefont {Schatz}, \citenamefont
  {Ki{\ss}ner}, \citenamefont {{\"{U}}nzelmann}, \citenamefont {Min},
  \citenamefont {Moser}, \citenamefont {Peixoto}, \citenamefont {Reinert},
  \citenamefont {Ernst}, \citenamefont {Echenique}, \citenamefont {Isaeva},\
  and\ \citenamefont {Chulkov}}]{Otrokov2019-Nature-MnBi2Te4}%
  \BibitemOpen
  \bibfield  {author} {\bibinfo {author} {\bibfnamefont {M.~M.}\ \bibnamefont
  {Otrokov}}, \bibinfo {author} {\bibfnamefont {I.~I.}\ \bibnamefont
  {Klimovskikh}}, \bibinfo {author} {\bibfnamefont {H.}~\bibnamefont
  {Bentmann}}, \bibinfo {author} {\bibfnamefont {D.}~\bibnamefont {Estyunin}},
  \bibinfo {author} {\bibfnamefont {A.}~\bibnamefont {Zeugner}}, \bibinfo
  {author} {\bibfnamefont {Z.~S.}\ \bibnamefont {Aliev}}, \bibinfo {author}
  {\bibfnamefont {S.}~\bibnamefont {Ga{\ss}}}, \bibinfo {author} {\bibfnamefont
  {A.~U.}\ \bibnamefont {Wolter}}, \bibinfo {author} {\bibfnamefont {A.~V.}\
  \bibnamefont {Koroleva}}, \bibinfo {author} {\bibfnamefont {A.~M.}\
  \bibnamefont {Shikin}}, \bibinfo {author} {\bibfnamefont {M.}~\bibnamefont
  {Blanco-Rey}}, \bibinfo {author} {\bibfnamefont {M.}~\bibnamefont
  {Hoffmann}}, \bibinfo {author} {\bibfnamefont {I.~P.}\ \bibnamefont
  {Rusinov}}, \bibinfo {author} {\bibfnamefont {A.~Y.}\ \bibnamefont
  {Vyazovskaya}}, \bibinfo {author} {\bibfnamefont {S.~V.}\ \bibnamefont
  {Eremeev}}, \bibinfo {author} {\bibfnamefont {Y.~M.}\ \bibnamefont
  {Koroteev}}, \bibinfo {author} {\bibfnamefont {V.~M.}\ \bibnamefont
  {Kuznetsov}}, \bibinfo {author} {\bibfnamefont {F.}~\bibnamefont {Freyse}},
  \bibinfo {author} {\bibfnamefont {J.}~\bibnamefont {S{\'{a}}nchez-Barriga}},
  \bibinfo {author} {\bibfnamefont {I.~R.}\ \bibnamefont {Amiraslanov}},
  \bibinfo {author} {\bibfnamefont {M.~B.}\ \bibnamefont {Babanly}}, \bibinfo
  {author} {\bibfnamefont {N.~T.}\ \bibnamefont {Mamedov}}, \bibinfo {author}
  {\bibfnamefont {N.~A.}\ \bibnamefont {Abdullayev}}, \bibinfo {author}
  {\bibfnamefont {V.~N.}\ \bibnamefont {Zverev}}, \bibinfo {author}
  {\bibfnamefont {A.}~\bibnamefont {Alfonsov}}, \bibinfo {author}
  {\bibfnamefont {V.}~\bibnamefont {Kataev}}, \bibinfo {author} {\bibfnamefont
  {B.}~\bibnamefont {B{\"{u}}chner}}, \bibinfo {author} {\bibfnamefont {E.~F.}\
  \bibnamefont {Schwier}}, \bibinfo {author} {\bibfnamefont {S.}~\bibnamefont
  {Kumar}}, \bibinfo {author} {\bibfnamefont {A.}~\bibnamefont {Kimura}},
  \bibinfo {author} {\bibfnamefont {L.}~\bibnamefont {Petaccia}}, \bibinfo
  {author} {\bibfnamefont {G.}~\bibnamefont {{Di Santo}}}, \bibinfo {author}
  {\bibfnamefont {R.~C.}\ \bibnamefont {Vidal}}, \bibinfo {author}
  {\bibfnamefont {S.}~\bibnamefont {Schatz}}, \bibinfo {author} {\bibfnamefont
  {K.}~\bibnamefont {Ki{\ss}ner}}, \bibinfo {author} {\bibfnamefont
  {M.}~\bibnamefont {{\"{U}}nzelmann}}, \bibinfo {author} {\bibfnamefont
  {C.~H.}\ \bibnamefont {Min}}, \bibinfo {author} {\bibfnamefont
  {S.}~\bibnamefont {Moser}}, \bibinfo {author} {\bibfnamefont {T.~R.}\
  \bibnamefont {Peixoto}}, \bibinfo {author} {\bibfnamefont {F.}~\bibnamefont
  {Reinert}}, \bibinfo {author} {\bibfnamefont {A.}~\bibnamefont {Ernst}},
  \bibinfo {author} {\bibfnamefont {P.~M.}\ \bibnamefont {Echenique}}, \bibinfo
  {author} {\bibfnamefont {A.}~\bibnamefont {Isaeva}},\ and\ \bibinfo {author}
  {\bibfnamefont {E.~V.}\ \bibnamefont {Chulkov}},\ }\bibfield  {title}
  {\bibinfo {title} {{Prediction and observation of an antiferromagnetic
  topological insulator}},\ }\href {https://doi.org/10.1038/s41586-019-1840-9}
  {\bibfield  {journal} {\bibinfo  {journal} {Nature}\ }\textbf {\bibinfo
  {volume} {576}},\ \bibinfo {pages} {416} (\bibinfo {year}
  {2019})}\BibitemShut {NoStop}%
\bibitem [{\citenamefont {Dong}\ \emph {et~al.}(2018)\citenamefont {Dong},
  \citenamefont {Zhang},\ and\ \citenamefont
  {Feng}}]{Dong2018-ChemRev-interface-assisted}%
  \BibitemOpen
  \bibfield  {author} {\bibinfo {author} {\bibfnamefont {R.}~\bibnamefont
  {Dong}}, \bibinfo {author} {\bibfnamefont {T.}~\bibnamefont {Zhang}},\ and\
  \bibinfo {author} {\bibfnamefont {X.}~\bibnamefont {Feng}},\ }\bibfield
  {title} {\bibinfo {title} {{Interface-assisted synthesis of 2D materials:
  Trend and challenges}},\ }\href {https://doi.org/10.1021/acs.chemrev.8b00056}
  {\bibfield  {journal} {\bibinfo  {journal} {Chemical Reviews}\ }\textbf
  {\bibinfo {volume} {118}},\ \bibinfo {pages} {6189} (\bibinfo {year}
  {2018})}\BibitemShut {NoStop}%
\bibitem [{\citenamefont {Huang}\ \emph {et~al.}(2020)\citenamefont {Huang},
  \citenamefont {Pan}, \citenamefont {Yang}, \citenamefont {Bao}, \citenamefont
  {Meng}, \citenamefont {Luo}, \citenamefont {Cai}, \citenamefont {Liu},
  \citenamefont {Zhao}, \citenamefont {Zhou}, \citenamefont {Wu}, \citenamefont
  {Zhu}, \citenamefont {Huang}, \citenamefont {Liu}, \citenamefont {Liu},
  \citenamefont {Cheng}, \citenamefont {Wu}, \citenamefont {Tian},
  \citenamefont {Gu}, \citenamefont {Shi}, \citenamefont {Guo}, \citenamefont
  {Cheng}, \citenamefont {Hu}, \citenamefont {Zhao}, \citenamefont {Yang},
  \citenamefont {Sutter}, \citenamefont {Sutter}, \citenamefont {Wang},
  \citenamefont {Ji}, \citenamefont {Zhou},\ and\ \citenamefont
  {Gao}}]{Huang2020-NatCommun-2D-interfaceExfoliation}%
  \BibitemOpen
  \bibfield  {author} {\bibinfo {author} {\bibfnamefont {Y.}~\bibnamefont
  {Huang}}, \bibinfo {author} {\bibfnamefont {Y.~H.}\ \bibnamefont {Pan}},
  \bibinfo {author} {\bibfnamefont {R.}~\bibnamefont {Yang}}, \bibinfo {author}
  {\bibfnamefont {L.~H.}\ \bibnamefont {Bao}}, \bibinfo {author} {\bibfnamefont
  {L.}~\bibnamefont {Meng}}, \bibinfo {author} {\bibfnamefont {H.~L.}\
  \bibnamefont {Luo}}, \bibinfo {author} {\bibfnamefont {Y.~Q.}\ \bibnamefont
  {Cai}}, \bibinfo {author} {\bibfnamefont {G.~D.}\ \bibnamefont {Liu}},
  \bibinfo {author} {\bibfnamefont {W.~J.}\ \bibnamefont {Zhao}}, \bibinfo
  {author} {\bibfnamefont {Z.}~\bibnamefont {Zhou}}, \bibinfo {author}
  {\bibfnamefont {L.~M.}\ \bibnamefont {Wu}}, \bibinfo {author} {\bibfnamefont
  {Z.~L.}\ \bibnamefont {Zhu}}, \bibinfo {author} {\bibfnamefont
  {M.}~\bibnamefont {Huang}}, \bibinfo {author} {\bibfnamefont {L.~W.}\
  \bibnamefont {Liu}}, \bibinfo {author} {\bibfnamefont {L.}~\bibnamefont
  {Liu}}, \bibinfo {author} {\bibfnamefont {P.}~\bibnamefont {Cheng}}, \bibinfo
  {author} {\bibfnamefont {K.~H.}\ \bibnamefont {Wu}}, \bibinfo {author}
  {\bibfnamefont {S.~B.}\ \bibnamefont {Tian}}, \bibinfo {author}
  {\bibfnamefont {C.~Z.}\ \bibnamefont {Gu}}, \bibinfo {author} {\bibfnamefont
  {Y.~G.}\ \bibnamefont {Shi}}, \bibinfo {author} {\bibfnamefont {Y.~F.}\
  \bibnamefont {Guo}}, \bibinfo {author} {\bibfnamefont {Z.~G.}\ \bibnamefont
  {Cheng}}, \bibinfo {author} {\bibfnamefont {J.~P.}\ \bibnamefont {Hu}},
  \bibinfo {author} {\bibfnamefont {L.}~\bibnamefont {Zhao}}, \bibinfo {author}
  {\bibfnamefont {G.~H.}\ \bibnamefont {Yang}}, \bibinfo {author}
  {\bibfnamefont {E.}~\bibnamefont {Sutter}}, \bibinfo {author} {\bibfnamefont
  {P.}~\bibnamefont {Sutter}}, \bibinfo {author} {\bibfnamefont {Y.~L.}\
  \bibnamefont {Wang}}, \bibinfo {author} {\bibfnamefont {W.}~\bibnamefont
  {Ji}}, \bibinfo {author} {\bibfnamefont {X.~J.}\ \bibnamefont {Zhou}},\ and\
  \bibinfo {author} {\bibfnamefont {H.~J.}\ \bibnamefont {Gao}},\ }\bibfield
  {title} {\bibinfo {title} {{Universal mechanical exfoliation of large-area 2D
  crystals}},\ }\href {https://doi.org/10.1038/s41467-020-16266-w} {\bibfield
  {journal} {\bibinfo  {journal} {Nature Communications}\ }\textbf {\bibinfo
  {volume} {11}},\ \bibinfo {pages} {2453} (\bibinfo {year}
  {2020})}\BibitemShut {NoStop}%
\bibitem [{\citenamefont {Yi}\ \emph {et~al.}(2013)\citenamefont {Yi},
  \citenamefont {Shen}, \citenamefont {Zhang}, \citenamefont {Zhu},
  \citenamefont {Liu}, \citenamefont {Liang}, \citenamefont {Zhang},\ and\
  \citenamefont {Ma}}]{Yi2013-BN-hydrodynamics}%
  \BibitemOpen
  \bibfield  {author} {\bibinfo {author} {\bibfnamefont {M.}~\bibnamefont
  {Yi}}, \bibinfo {author} {\bibfnamefont {Z.}~\bibnamefont {Shen}}, \bibinfo
  {author} {\bibfnamefont {W.}~\bibnamefont {Zhang}}, \bibinfo {author}
  {\bibfnamefont {J.}~\bibnamefont {Zhu}}, \bibinfo {author} {\bibfnamefont
  {L.}~\bibnamefont {Liu}}, \bibinfo {author} {\bibfnamefont {S.}~\bibnamefont
  {Liang}}, \bibinfo {author} {\bibfnamefont {X.}~\bibnamefont {Zhang}},\ and\
  \bibinfo {author} {\bibfnamefont {S.}~\bibnamefont {Ma}},\ }\bibfield
  {title} {\bibinfo {title} {{Hydrodynamics-assisted scalable production of
  boron nitride nanosheets and their application in improving oxygen-atom
  erosion resistance of polymeric composites}},\ }\href
  {https://doi.org/10.1039/c3nr03714b} {\bibfield  {journal} {\bibinfo
  {journal} {Nanoscale}\ }\textbf {\bibinfo {volume} {5}},\ \bibinfo {pages}
  {10660} (\bibinfo {year} {2013})}\BibitemShut {NoStop}%
\bibitem [{\citenamefont {Yi}\ and\ \citenamefont
  {Shen}(2014)}]{Yi2014-kitchenBlender}%
  \BibitemOpen
  \bibfield  {author} {\bibinfo {author} {\bibfnamefont {M.}~\bibnamefont
  {Yi}}\ and\ \bibinfo {author} {\bibfnamefont {Z.}~\bibnamefont {Shen}},\
  }\bibfield  {title} {\bibinfo {title} {{Kitchen blender for producing
  high-quality few-layer graphene}},\ }\href
  {https://doi.org/10.1016/j.carbon.2014.07.035} {\bibfield  {journal}
  {\bibinfo  {journal} {Carbon}\ }\textbf {\bibinfo {volume} {78}},\ \bibinfo
  {pages} {622} (\bibinfo {year} {2014})}\BibitemShut {NoStop}%
\bibitem [{\citenamefont {Yi}\ and\ \citenamefont
  {Shen}(2016)}]{Yi2016-fluiddynamics}%
  \BibitemOpen
  \bibfield  {author} {\bibinfo {author} {\bibfnamefont {M.}~\bibnamefont
  {Yi}}\ and\ \bibinfo {author} {\bibfnamefont {Z.}~\bibnamefont {Shen}},\
  }\bibfield  {title} {\bibinfo {title} {{Fluid dynamics: An emerging route for
  the scalable production of graphene in the last five years}},\ }\href
  {https://doi.org/10.1039/c6ra15269d} {\bibfield  {journal} {\bibinfo
  {journal} {RSC Advances}\ }\textbf {\bibinfo {volume} {6}},\ \bibinfo {pages}
  {72525} (\bibinfo {year} {2016})}\BibitemShut {NoStop}%
\bibitem [{\citenamefont {Mannix}\ \emph {et~al.}(2015)\citenamefont {Mannix},
  \citenamefont {Zhou}, \citenamefont {Kiraly}, \citenamefont {Wood},
  \citenamefont {Alducin}, \citenamefont {Myers}, \citenamefont {Liu},
  \citenamefont {Fisher}, \citenamefont {Santiago}, \citenamefont {Guest},
  \citenamefont {Yacaman}, \citenamefont {Ponce}, \citenamefont {Oganov},
  \citenamefont {Hersam},\ and\ \citenamefont
  {Guisinger}}]{Mannix2015-Science-Borophene}%
  \BibitemOpen
  \bibfield  {author} {\bibinfo {author} {\bibfnamefont {A.~J.}\ \bibnamefont
  {Mannix}}, \bibinfo {author} {\bibfnamefont {X.~F.}\ \bibnamefont {Zhou}},
  \bibinfo {author} {\bibfnamefont {B.}~\bibnamefont {Kiraly}}, \bibinfo
  {author} {\bibfnamefont {J.~D.}\ \bibnamefont {Wood}}, \bibinfo {author}
  {\bibfnamefont {D.}~\bibnamefont {Alducin}}, \bibinfo {author} {\bibfnamefont
  {B.~D.}\ \bibnamefont {Myers}}, \bibinfo {author} {\bibfnamefont
  {X.}~\bibnamefont {Liu}}, \bibinfo {author} {\bibfnamefont {B.~L.}\
  \bibnamefont {Fisher}}, \bibinfo {author} {\bibfnamefont {U.}~\bibnamefont
  {Santiago}}, \bibinfo {author} {\bibfnamefont {J.~R.}\ \bibnamefont {Guest}},
  \bibinfo {author} {\bibfnamefont {M.~J.}\ \bibnamefont {Yacaman}}, \bibinfo
  {author} {\bibfnamefont {A.}~\bibnamefont {Ponce}}, \bibinfo {author}
  {\bibfnamefont {A.~R.}\ \bibnamefont {Oganov}}, \bibinfo {author}
  {\bibfnamefont {M.~C.}\ \bibnamefont {Hersam}},\ and\ \bibinfo {author}
  {\bibfnamefont {N.~P.}\ \bibnamefont {Guisinger}},\ }\bibfield  {title}
  {\bibinfo {title} {{Synthesis of borophenes: Anisotropic, two-dimensional
  boron polymorphs}},\ }\href {https://doi.org/10.1126/science.aad1080}
  {\bibfield  {journal} {\bibinfo  {journal} {Science}\ }\textbf {\bibinfo
  {volume} {350}},\ \bibinfo {pages} {1513} (\bibinfo {year}
  {2015})}\BibitemShut {NoStop}%
\bibitem [{\citenamefont {Feng}\ \emph {et~al.}(2016)\citenamefont {Feng},
  \citenamefont {Zhang}, \citenamefont {Zhong}, \citenamefont {Li},
  \citenamefont {Li}, \citenamefont {Li}, \citenamefont {Cheng}, \citenamefont
  {Meng}, \citenamefont {Chen},\ and\ \citenamefont
  {Wu}}]{Feng2016-NatChem-Borophene}%
  \BibitemOpen
  \bibfield  {author} {\bibinfo {author} {\bibfnamefont {B.}~\bibnamefont
  {Feng}}, \bibinfo {author} {\bibfnamefont {J.}~\bibnamefont {Zhang}},
  \bibinfo {author} {\bibfnamefont {Q.}~\bibnamefont {Zhong}}, \bibinfo
  {author} {\bibfnamefont {W.}~\bibnamefont {Li}}, \bibinfo {author}
  {\bibfnamefont {S.}~\bibnamefont {Li}}, \bibinfo {author} {\bibfnamefont
  {H.}~\bibnamefont {Li}}, \bibinfo {author} {\bibfnamefont {P.}~\bibnamefont
  {Cheng}}, \bibinfo {author} {\bibfnamefont {S.}~\bibnamefont {Meng}},
  \bibinfo {author} {\bibfnamefont {L.}~\bibnamefont {Chen}},\ and\ \bibinfo
  {author} {\bibfnamefont {K.}~\bibnamefont {Wu}},\ }\bibfield  {title}
  {\bibinfo {title} {{Experimental realization of two-dimensional boron
  sheets}},\ }\href {https://doi.org/10.1038/nchem.2491} {\bibfield  {journal}
  {\bibinfo  {journal} {Nature Chemistry}\ }\textbf {\bibinfo {volume} {8}},\
  \bibinfo {pages} {563} (\bibinfo {year} {2016})}\BibitemShut {NoStop}%
\bibitem [{\citenamefont {Zhan}\ \emph {et~al.}(2012)\citenamefont {Zhan},
  \citenamefont {Liu}, \citenamefont {Najmaei}, \citenamefont {Ajayan},\ and\
  \citenamefont {Lou}}]{Zhan2012-Small-MoS2-SiO2}%
  \BibitemOpen
  \bibfield  {author} {\bibinfo {author} {\bibfnamefont {Y.}~\bibnamefont
  {Zhan}}, \bibinfo {author} {\bibfnamefont {Z.}~\bibnamefont {Liu}}, \bibinfo
  {author} {\bibfnamefont {S.}~\bibnamefont {Najmaei}}, \bibinfo {author}
  {\bibfnamefont {P.~M.}\ \bibnamefont {Ajayan}},\ and\ \bibinfo {author}
  {\bibfnamefont {J.}~\bibnamefont {Lou}},\ }\bibfield  {title} {\bibinfo
  {title} {{Large-area vapor-phase growth and characterization of MoS$_2$
  atomic layers on a SiO$_2$ substrate}},\ }\href
  {https://doi.org/10.1002/smll.201102654} {\bibfield  {journal} {\bibinfo
  {journal} {Small}\ }\textbf {\bibinfo {volume} {8}},\ \bibinfo {pages} {966}
  (\bibinfo {year} {2012})}\BibitemShut {NoStop}%
\bibitem [{\citenamefont {Zhang}\ \emph {et~al.}(2019)\citenamefont {Zhang},
  \citenamefont {Yao}, \citenamefont {Sendeku}, \citenamefont {Yin},
  \citenamefont {Zhan}, \citenamefont {Wang}, \citenamefont {Wang},\ and\
  \citenamefont {He}}]{Zhang2019-AM-CVD-2DTMDs}%
  \BibitemOpen
  \bibfield  {author} {\bibinfo {author} {\bibfnamefont {Y.}~\bibnamefont
  {Zhang}}, \bibinfo {author} {\bibfnamefont {Y.}~\bibnamefont {Yao}}, \bibinfo
  {author} {\bibfnamefont {M.~G.}\ \bibnamefont {Sendeku}}, \bibinfo {author}
  {\bibfnamefont {L.}~\bibnamefont {Yin}}, \bibinfo {author} {\bibfnamefont
  {X.}~\bibnamefont {Zhan}}, \bibinfo {author} {\bibfnamefont {F.}~\bibnamefont
  {Wang}}, \bibinfo {author} {\bibfnamefont {Z.}~\bibnamefont {Wang}},\ and\
  \bibinfo {author} {\bibfnamefont {J.}~\bibnamefont {He}},\ }\bibfield
  {title} {\bibinfo {title} {{Recent progress in CVD growth of 2D transition
  metal dichalcogenides and related heterostructures}},\ }\href
  {https://doi.org/10.1002/adma.201901694} {\bibfield  {journal} {\bibinfo
  {journal} {Advanced Materials}\ }\textbf {\bibinfo {volume} {31}},\ \bibinfo
  {pages} {1901694} (\bibinfo {year} {2019})}\BibitemShut {NoStop}%
\bibitem [{\citenamefont {Novoselov}\ \emph {et~al.}(2016)\citenamefont
  {Novoselov}, \citenamefont {Mishchenko}, \citenamefont {Carvalho},\ and\
  \citenamefont {{Castro Neto}}}]{Novoselov2016-Science-vdWHetero}%
  \BibitemOpen
  \bibfield  {author} {\bibinfo {author} {\bibfnamefont {K.~S.}\ \bibnamefont
  {Novoselov}}, \bibinfo {author} {\bibfnamefont {A.}~\bibnamefont
  {Mishchenko}}, \bibinfo {author} {\bibfnamefont {A.}~\bibnamefont
  {Carvalho}},\ and\ \bibinfo {author} {\bibfnamefont {A.~H.}\ \bibnamefont
  {{Castro Neto}}},\ }\bibfield  {title} {\bibinfo {title} {{2D materials and
  van der Waals heterostructures}},\ }\href
  {https://doi.org/10.1126/science.aac9439} {\bibfield  {journal} {\bibinfo
  {journal} {Science}\ }\textbf {\bibinfo {volume} {353}},\ \bibinfo {pages}
  {6298} (\bibinfo {year} {2016})}\BibitemShut {NoStop}%
\bibitem [{\citenamefont {Hong}\ \emph {et~al.}(2020)\citenamefont {Hong},
  \citenamefont {Liu}, \citenamefont {Wang}, \citenamefont {Zhou},
  \citenamefont {Ma}, \citenamefont {Xu}, \citenamefont {Feng}, \citenamefont
  {Chen}, \citenamefont {Chen}, \citenamefont {Sun}, \citenamefont {Chen},
  \citenamefont {Cheng},\ and\ \citenamefont {Ren}}]{Hong2020-Science}%
  \BibitemOpen
  \bibfield  {author} {\bibinfo {author} {\bibfnamefont {Y.~L.}\ \bibnamefont
  {Hong}}, \bibinfo {author} {\bibfnamefont {Z.}~\bibnamefont {Liu}}, \bibinfo
  {author} {\bibfnamefont {L.}~\bibnamefont {Wang}}, \bibinfo {author}
  {\bibfnamefont {T.}~\bibnamefont {Zhou}}, \bibinfo {author} {\bibfnamefont
  {W.}~\bibnamefont {Ma}}, \bibinfo {author} {\bibfnamefont {C.}~\bibnamefont
  {Xu}}, \bibinfo {author} {\bibfnamefont {S.}~\bibnamefont {Feng}}, \bibinfo
  {author} {\bibfnamefont {L.}~\bibnamefont {Chen}}, \bibinfo {author}
  {\bibfnamefont {M.~L.}\ \bibnamefont {Chen}}, \bibinfo {author}
  {\bibfnamefont {D.~M.}\ \bibnamefont {Sun}}, \bibinfo {author} {\bibfnamefont
  {X.~Q.}\ \bibnamefont {Chen}}, \bibinfo {author} {\bibfnamefont {H.~M.}\
  \bibnamefont {Cheng}},\ and\ \bibinfo {author} {\bibfnamefont
  {W.}~\bibnamefont {Ren}},\ }\bibfield  {title} {\bibinfo {title} {{Chemical
  vapor deposition of layered two-dimensional MoSi$_2$N$_4$ materials}},\
  }\href {https://doi.org/10.1126/science.abb7023} {\bibfield  {journal}
  {\bibinfo  {journal} {Science}\ }\textbf {\bibinfo {volume} {369}},\ \bibinfo
  {pages} {670} (\bibinfo {year} {2020})}\BibitemShut {NoStop}%
\bibitem [{\citenamefont {Wang}\ \emph
  {et~al.}(2021{\natexlab{a}})\citenamefont {Wang}, \citenamefont {Shi},
  \citenamefont {Liu}, \citenamefont {Zhang}, \citenamefont {Hong},
  \citenamefont {Li}, \citenamefont {Gao}, \citenamefont {Chen}, \citenamefont
  {Ren}, \citenamefont {Cheng}, \citenamefont {Li},\ and\ \citenamefont
  {Chen}}]{Wang2021-NC}%
  \BibitemOpen
  \bibfield  {author} {\bibinfo {author} {\bibfnamefont {L.}~\bibnamefont
  {Wang}}, \bibinfo {author} {\bibfnamefont {Y.}~\bibnamefont {Shi}}, \bibinfo
  {author} {\bibfnamefont {M.}~\bibnamefont {Liu}}, \bibinfo {author}
  {\bibfnamefont {A.}~\bibnamefont {Zhang}}, \bibinfo {author} {\bibfnamefont
  {Y.~L.}\ \bibnamefont {Hong}}, \bibinfo {author} {\bibfnamefont
  {R.}~\bibnamefont {Li}}, \bibinfo {author} {\bibfnamefont {Q.}~\bibnamefont
  {Gao}}, \bibinfo {author} {\bibfnamefont {M.}~\bibnamefont {Chen}}, \bibinfo
  {author} {\bibfnamefont {W.}~\bibnamefont {Ren}}, \bibinfo {author}
  {\bibfnamefont {H.~M.}\ \bibnamefont {Cheng}}, \bibinfo {author}
  {\bibfnamefont {Y.}~\bibnamefont {Li}},\ and\ \bibinfo {author}
  {\bibfnamefont {X.~Q.}\ \bibnamefont {Chen}},\ }\bibfield  {title} {\bibinfo
  {title} {{Intercalated architecture of MA$_2$Z$_4$ family layered van der
  Waals materials with emerging topological, magnetic and superconducting
  properties}},\ }\href {https://doi.org/10.1038/s41467-021-22324-8} {\bibfield
   {journal} {\bibinfo  {journal} {Nature Communications}\ }\textbf {\bibinfo
  {volume} {12}},\ \bibinfo {pages} {2361} (\bibinfo {year}
  {2021}{\natexlab{a}})}\BibitemShut {NoStop}%
\bibitem [{\citenamefont {Yang}\ \emph {et~al.}(2021)\citenamefont {Yang},
  \citenamefont {Zhao}, \citenamefont {Li}, \citenamefont {Liu}, \citenamefont
  {Wang}, \citenamefont {Chen}, \citenamefont {Gao},\ and\ \citenamefont
  {Zhao}}]{Yang2021-Nanoscale-Z-MA2Z4}%
  \BibitemOpen
  \bibfield  {author} {\bibinfo {author} {\bibfnamefont {J.~S.}\ \bibnamefont
  {Yang}}, \bibinfo {author} {\bibfnamefont {L.}~\bibnamefont {Zhao}}, \bibinfo
  {author} {\bibfnamefont {S.~Q.}\ \bibnamefont {Li}}, \bibinfo {author}
  {\bibfnamefont {H.}~\bibnamefont {Liu}}, \bibinfo {author} {\bibfnamefont
  {L.}~\bibnamefont {Wang}}, \bibinfo {author} {\bibfnamefont {M.}~\bibnamefont
  {Chen}}, \bibinfo {author} {\bibfnamefont {J.}~\bibnamefont {Gao}},\ and\
  \bibinfo {author} {\bibfnamefont {J.}~\bibnamefont {Zhao}},\ }\bibfield
  {title} {\bibinfo {title} {{Accurate electronic properties and non-linear
  optical response of two-dimensional MA$_2$Z$_4$}},\ }\href
  {https://doi.org/10.1039/d0nr09146d} {\bibfield  {journal} {\bibinfo
  {journal} {Nanoscale}\ }\textbf {\bibinfo {volume} {13}},\ \bibinfo {pages}
  {5479} (\bibinfo {year} {2021})}\BibitemShut {NoStop}%
\bibitem [{\citenamefont {Kang}\ and\ \citenamefont
  {Lin}(2021)}]{Kang2021-PRB-2ndGeneration}%
  \BibitemOpen
  \bibfield  {author} {\bibinfo {author} {\bibfnamefont {L.}~\bibnamefont
  {Kang}}\ and\ \bibinfo {author} {\bibfnamefont {Z.}~\bibnamefont {Lin}},\
  }\bibfield  {title} {\bibinfo {title} {{Second harmonic generation of
  MoSi$_2$N$_4$-type layers}},\ }\href
  {https://doi.org/10.1103/PhysRevB.103.195404} {\bibfield  {journal} {\bibinfo
   {journal} {Physical Review B}\ }\textbf {\bibinfo {volume} {103}},\ \bibinfo
  {pages} {195404} (\bibinfo {year} {2021})}\BibitemShut {NoStop}%
\bibitem [{\citenamefont {Liang}\ \emph
  {et~al.}(2022{\natexlab{a}})\citenamefont {Liang}, \citenamefont {Xu},
  \citenamefont {Lu},\ and\ \citenamefont {Cai}}]{Liang2022-APS-Exciton}%
  \BibitemOpen
  \bibfield  {author} {\bibinfo {author} {\bibfnamefont {D.}~\bibnamefont
  {Liang}}, \bibinfo {author} {\bibfnamefont {S.}~\bibnamefont {Xu}}, \bibinfo
  {author} {\bibfnamefont {P.}~\bibnamefont {Lu}},\ and\ \bibinfo {author}
  {\bibfnamefont {Y.}~\bibnamefont {Cai}},\ }\bibfield  {title} {\bibinfo
  {title} {{Highly tunable and strongly bound exciton in MoSi$_2$N$_4$ via
  strain engineering}},\ }\href {https://doi.org/10.1103/PhysRevB.105.195302}
  {\bibfield  {journal} {\bibinfo  {journal} {Physical Review B}\ }\textbf
  {\bibinfo {volume} {105}},\ \bibinfo {pages} {195302} (\bibinfo {year}
  {2022}{\natexlab{a}})}\BibitemShut {NoStop}%
\bibitem [{\citenamefont {Huang}\ \emph
  {et~al.}(2022{\natexlab{a}})\citenamefont {Huang}, \citenamefont {Liang},
  \citenamefont {Guo}, \citenamefont {Lu}, \citenamefont {Wang}, \citenamefont
  {Yu},\ and\ \citenamefont {Zhang}}]{Huang2022-AOM-Exciton-Phonon}%
  \BibitemOpen
  \bibfield  {author} {\bibinfo {author} {\bibfnamefont {D.}~\bibnamefont
  {Huang}}, \bibinfo {author} {\bibfnamefont {F.}~\bibnamefont {Liang}},
  \bibinfo {author} {\bibfnamefont {R.}~\bibnamefont {Guo}}, \bibinfo {author}
  {\bibfnamefont {D.}~\bibnamefont {Lu}}, \bibinfo {author} {\bibfnamefont
  {J.}~\bibnamefont {Wang}}, \bibinfo {author} {\bibfnamefont {H.}~\bibnamefont
  {Yu}},\ and\ \bibinfo {author} {\bibfnamefont {H.}~\bibnamefont {Zhang}},\
  }\bibfield  {title} {\bibinfo {title} {{MoSi$_2$N$_4$: A 2D regime with
  strong exciton–phonon coupling}},\ }\href
  {https://doi.org/10.1002/adom.202102612} {\bibfield  {journal} {\bibinfo
  {journal} {Advanced Optical Materials}\ }\textbf {\bibinfo {volume} {10}},\
  \bibinfo {pages} {2102612} (\bibinfo {year}
  {2022}{\natexlab{a}})}\BibitemShut {NoStop}%
\bibitem [{\citenamefont {Wang}\ \emph
  {et~al.}(2021{\natexlab{b}})\citenamefont {Wang}, \citenamefont {Kuang},
  \citenamefont {Yu}, \citenamefont {Zhao}, \citenamefont {Zhong},\ and\
  \citenamefont {Yuan}}]{Wang2021-PRB-Ele-QuasiparticleModel}%
  \BibitemOpen
  \bibfield  {author} {\bibinfo {author} {\bibfnamefont {Z.}~\bibnamefont
  {Wang}}, \bibinfo {author} {\bibfnamefont {X.}~\bibnamefont {Kuang}},
  \bibinfo {author} {\bibfnamefont {G.}~\bibnamefont {Yu}}, \bibinfo {author}
  {\bibfnamefont {P.}~\bibnamefont {Zhao}}, \bibinfo {author} {\bibfnamefont
  {H.}~\bibnamefont {Zhong}},\ and\ \bibinfo {author} {\bibfnamefont
  {S.}~\bibnamefont {Yuan}},\ }\bibfield  {title} {\bibinfo {title}
  {{Electronic properties and quasiparticle model of monolayer}},\ }\href
  {https://doi.org/10.1103/PhysRevB.104.155110} {\bibfield  {journal} {\bibinfo
   {journal} {Physical Review B}\ }\textbf {\bibinfo {volume} {104}},\ \bibinfo
  {pages} {155110} (\bibinfo {year} {2021}{\natexlab{b}})}\BibitemShut
  {NoStop}%
\bibitem [{\citenamefont {Zhao}\ \emph
  {et~al.}(2021{\natexlab{a}})\citenamefont {Zhao}, \citenamefont {Jin},
  \citenamefont {Zeng}, \citenamefont {Yao},\ and\ \citenamefont
  {Yan}}]{Zhao2021-APL-crcl3-hetero-spinvalley}%
  \BibitemOpen
  \bibfield  {author} {\bibinfo {author} {\bibfnamefont {J.}~\bibnamefont
  {Zhao}}, \bibinfo {author} {\bibfnamefont {X.}~\bibnamefont {Jin}}, \bibinfo
  {author} {\bibfnamefont {H.}~\bibnamefont {Zeng}}, \bibinfo {author}
  {\bibfnamefont {C.}~\bibnamefont {Yao}},\ and\ \bibinfo {author}
  {\bibfnamefont {G.}~\bibnamefont {Yan}},\ }\bibfield  {title} {\bibinfo
  {title} {{Spin-valley coupling and valley splitting in the
  MoSi$_2$N$_4$/CrCl$_3$ van der Waals heterostructure}},\ }\href
  {https://doi.org/10.1063/5.0072266} {\bibfield  {journal} {\bibinfo
  {journal} {Applied Physics Letters}\ }\textbf {\bibinfo {volume} {119}},\
  \bibinfo {pages} {213101} (\bibinfo {year} {2021}{\natexlab{a}})}\BibitemShut
  {NoStop}%
\bibitem [{\citenamefont {Yan}\ \emph {et~al.}(2021)\citenamefont {Yan},
  \citenamefont {Wang}, \citenamefont {Huang}, \citenamefont {Li},
  \citenamefont {Xue}, \citenamefont {Zhang}, \citenamefont {Ren},\ and\
  \citenamefont {Zhou}}]{Yan2021-nanoscale-Ta-Nb-superconductivity}%
  \BibitemOpen
  \bibfield  {author} {\bibinfo {author} {\bibfnamefont {L.}~\bibnamefont
  {Yan}}, \bibinfo {author} {\bibfnamefont {B.~T.}\ \bibnamefont {Wang}},
  \bibinfo {author} {\bibfnamefont {X.}~\bibnamefont {Huang}}, \bibinfo
  {author} {\bibfnamefont {Q.}~\bibnamefont {Li}}, \bibinfo {author}
  {\bibfnamefont {K.}~\bibnamefont {Xue}}, \bibinfo {author} {\bibfnamefont
  {J.}~\bibnamefont {Zhang}}, \bibinfo {author} {\bibfnamefont
  {W.}~\bibnamefont {Ren}},\ and\ \bibinfo {author} {\bibfnamefont
  {L.}~\bibnamefont {Zhou}},\ }\bibfield  {title} {\bibinfo {title} {{Surface
  passivation induced a significant enhancement of superconductivity in layered
  two-dimensional MSi$_2$N$_4$(M = Ta and Nb) materials}},\ }\href
  {https://doi.org/10.1039/d1nr05560g} {\bibfield  {journal} {\bibinfo
  {journal} {Nanoscale}\ }\textbf {\bibinfo {volume} {13}},\ \bibinfo {pages}
  {18947} (\bibinfo {year} {2021})}\BibitemShut {NoStop}%
\bibitem [{\citenamefont {Zhou}\ \emph
  {et~al.}(2021{\natexlab{a}})\citenamefont {Zhou}, \citenamefont {Zhang},
  \citenamefont {Zhang}, \citenamefont {Feng}, \citenamefont {Mokrousov},\ and\
  \citenamefont {Yao}}]{Zhou2021-npj-VSi2N4-framework}%
  \BibitemOpen
  \bibfield  {author} {\bibinfo {author} {\bibfnamefont {X.}~\bibnamefont
  {Zhou}}, \bibinfo {author} {\bibfnamefont {R.~W.}\ \bibnamefont {Zhang}},
  \bibinfo {author} {\bibfnamefont {Z.}~\bibnamefont {Zhang}}, \bibinfo
  {author} {\bibfnamefont {W.}~\bibnamefont {Feng}}, \bibinfo {author}
  {\bibfnamefont {Y.}~\bibnamefont {Mokrousov}},\ and\ \bibinfo {author}
  {\bibfnamefont {Y.}~\bibnamefont {Yao}},\ }\bibfield  {title} {\bibinfo
  {title} {{Sign-reversible valley-dependent Berry phase effects in 2D
  valley-half-semiconductors}},\ }\href
  {https://doi.org/10.1038/s41524-021-00632-3} {\bibfield  {journal} {\bibinfo
  {journal} {npj Computational Materials}\ }\textbf {\bibinfo {volume} {7}},\
  \bibinfo {pages} {160} (\bibinfo {year} {2021}{\natexlab{a}})}\BibitemShut
  {NoStop}%
\bibitem [{\citenamefont {Wang}\ \emph
  {et~al.}(2022{\natexlab{a}})\citenamefont {Wang}, \citenamefont {Legut},
  \citenamefont {Liu}, \citenamefont {Li}, \citenamefont {Li}, \citenamefont
  {Sun}, \citenamefont {Zhang},\ and\ \citenamefont
  {Zhang}}]{Wang2022-PRB-XSi2N4-mott}%
  \BibitemOpen
  \bibfield  {author} {\bibinfo {author} {\bibfnamefont {Y.}~\bibnamefont
  {Wang}}, \bibinfo {author} {\bibfnamefont {D.}~\bibnamefont {Legut}},
  \bibinfo {author} {\bibfnamefont {X.}~\bibnamefont {Liu}}, \bibinfo {author}
  {\bibfnamefont {Y.}~\bibnamefont {Li}}, \bibinfo {author} {\bibfnamefont
  {C.}~\bibnamefont {Li}}, \bibinfo {author} {\bibfnamefont {Y.}~\bibnamefont
  {Sun}}, \bibinfo {author} {\bibfnamefont {R.}~\bibnamefont {Zhang}},\ and\
  \bibinfo {author} {\bibfnamefont {Q.}~\bibnamefont {Zhang}},\ }\bibfield
  {title} {\bibinfo {title} {{Mott transition and superexchange mechanism in
  magnetically doped XSi$_2$N$_4$ caused by large 3$d$ orbital onsite Coulomb
  interaction}},\ }\href {https://doi.org/10.1103/PhysRevB.106.104421}
  {\bibfield  {journal} {\bibinfo  {journal} {Physical Review B}\ }\textbf
  {\bibinfo {volume} {106}},\ \bibinfo {pages} {104421} (\bibinfo {year}
  {2022}{\natexlab{a}})}\BibitemShut {NoStop}%
\bibitem [{\citenamefont {Cui}\ \emph {et~al.}(2021{\natexlab{a}})\citenamefont
  {Cui}, \citenamefont {Luo}, \citenamefont {Yu},\ and\ \citenamefont
  {Xu}}]{Cui2021-PhysE-MoSi2N4-moleculardoped}%
  \BibitemOpen
  \bibfield  {author} {\bibinfo {author} {\bibfnamefont {Z.}~\bibnamefont
  {Cui}}, \bibinfo {author} {\bibfnamefont {Y.}~\bibnamefont {Luo}}, \bibinfo
  {author} {\bibfnamefont {J.}~\bibnamefont {Yu}},\ and\ \bibinfo {author}
  {\bibfnamefont {Y.}~\bibnamefont {Xu}},\ }\bibfield  {title} {\bibinfo
  {title} {{Tuning the electronic properties of MoSi$_2$N$_4$ by molecular
  doping: A first principles investigation}},\ }\href
  {https://doi.org/10.1016/j.physe.2021.114873} {\bibfield  {journal} {\bibinfo
   {journal} {Physica E: Low-Dimensional Systems and Nanostructures}\ }\textbf
  {\bibinfo {volume} {134}},\ \bibinfo {pages} {114873} (\bibinfo {year}
  {2021}{\natexlab{a}})}\BibitemShut {NoStop}%
\bibitem [{\citenamefont {Zhou}\ \emph
  {et~al.}(2021{\natexlab{b}})\citenamefont {Zhou}, \citenamefont {Wu},
  \citenamefont {Li}, \citenamefont {Zhang},\ and\ \citenamefont
  {Ouyang}}]{Zhou2021-JPCL-monoWSi2N4-stacking}%
  \BibitemOpen
  \bibfield  {author} {\bibinfo {author} {\bibfnamefont {W.}~\bibnamefont
  {Zhou}}, \bibinfo {author} {\bibfnamefont {L.}~\bibnamefont {Wu}}, \bibinfo
  {author} {\bibfnamefont {A.}~\bibnamefont {Li}}, \bibinfo {author}
  {\bibfnamefont {B.}~\bibnamefont {Zhang}},\ and\ \bibinfo {author}
  {\bibfnamefont {F.}~\bibnamefont {Ouyang}},\ }\bibfield  {title} {\bibinfo
  {title} {{Structural symmetry, spin-Orbit coupling, and valley-related
  properties of monolayer WSi$_2$N$_4$ family}},\ }\href
  {https://doi.org/10.1021/acs.jpclett.1c03197} {\bibfield  {journal} {\bibinfo
   {journal} {Journal of Physical Chemistry Letters}\ }\textbf {\bibinfo
  {volume} {12}},\ \bibinfo {pages} {11622} (\bibinfo {year}
  {2021}{\natexlab{b}})}\BibitemShut {NoStop}%
\bibitem [{\citenamefont {Islam}\ \emph {et~al.}(2021)\citenamefont {Islam},
  \citenamefont {Ghosh}, \citenamefont {Autieri}, \citenamefont {Chowdhury},
  \citenamefont {Bansil}, \citenamefont {Agarwal},\ and\ \citenamefont
  {Singh}}]{Islam2021-PRB-dimensionaleffect-spin}%
  \BibitemOpen
  \bibfield  {author} {\bibinfo {author} {\bibfnamefont {R.}~\bibnamefont
  {Islam}}, \bibinfo {author} {\bibfnamefont {B.}~\bibnamefont {Ghosh}},
  \bibinfo {author} {\bibfnamefont {C.}~\bibnamefont {Autieri}}, \bibinfo
  {author} {\bibfnamefont {S.}~\bibnamefont {Chowdhury}}, \bibinfo {author}
  {\bibfnamefont {A.}~\bibnamefont {Bansil}}, \bibinfo {author} {\bibfnamefont
  {A.}~\bibnamefont {Agarwal}},\ and\ \bibinfo {author} {\bibfnamefont
  {B.}~\bibnamefont {Singh}},\ }\bibfield  {title} {\bibinfo {title} {{Tunable
  spin polarization and electronic structure of bottom-up synthesized
  MoSi$_2$N$_4$ materials}},\ }\href
  {https://doi.org/10.1103/PhysRevB.104.L201112} {\bibfield  {journal}
  {\bibinfo  {journal} {Physical Review B}\ }\textbf {\bibinfo {volume}
  {104}},\ \bibinfo {pages} {L201112} (\bibinfo {year} {2021})}\BibitemShut
  {NoStop}%
\bibitem [{\citenamefont {Wu}\ and\ \citenamefont
  {Ang}(2022)}]{Wu2022-APL-MTJ}%
  \BibitemOpen
  \bibfield  {author} {\bibinfo {author} {\bibfnamefont {Q.}~\bibnamefont
  {Wu}}\ and\ \bibinfo {author} {\bibfnamefont {L.~K.}\ \bibnamefont {Ang}},\
  }\bibfield  {title} {\bibinfo {title} {{Giant tunneling magnetoresistance in
  atomically thin VSi$_2$N$_4$/MoSi$_2$N$_4$/VSi$_2$N$_4$ magnetic tunnel
  junction}},\ }\href {https://doi.org/10.1063/5.0075046} {\bibfield  {journal}
  {\bibinfo  {journal} {Applied Physics Letters}\ }\textbf {\bibinfo {volume}
  {120}},\ \bibinfo {pages} {022401} (\bibinfo {year} {2022})}\BibitemShut
  {NoStop}%
\bibitem [{\citenamefont {Ma}\ \emph {et~al.}(2022)\citenamefont {Ma},
  \citenamefont {Zhao}, \citenamefont {Zhang}, \citenamefont {Liu},
  \citenamefont {Ren}, \citenamefont {Zhu}, \citenamefont {Chi}, \citenamefont
  {Ding},\ and\ \citenamefont {Guo}}]{Ma2022-ASS-MoSi2N4defect}%
  \BibitemOpen
  \bibfield  {author} {\bibinfo {author} {\bibfnamefont {H.}~\bibnamefont
  {Ma}}, \bibinfo {author} {\bibfnamefont {W.}~\bibnamefont {Zhao}}, \bibinfo
  {author} {\bibfnamefont {Q.}~\bibnamefont {Zhang}}, \bibinfo {author}
  {\bibfnamefont {D.}~\bibnamefont {Liu}}, \bibinfo {author} {\bibfnamefont
  {H.}~\bibnamefont {Ren}}, \bibinfo {author} {\bibfnamefont {H.}~\bibnamefont
  {Zhu}}, \bibinfo {author} {\bibfnamefont {Y.}~\bibnamefont {Chi}}, \bibinfo
  {author} {\bibfnamefont {F.}~\bibnamefont {Ding}},\ and\ \bibinfo {author}
  {\bibfnamefont {W.}~\bibnamefont {Guo}},\ }\bibfield  {title} {\bibinfo
  {title} {{Chemical environment dependent stabilities, electronic properties
  and diffusions behaviors of intrinsic point defects in novel two-dimensional
  MoSi$_2$N$_4$ monolayer}},\ }\href
  {https://doi.org/10.1016/j.apsusc.2022.153214} {\bibfield  {journal}
  {\bibinfo  {journal} {Applied Surface Science}\ }\textbf {\bibinfo {volume}
  {592}},\ \bibinfo {pages} {153214} (\bibinfo {year} {2022})}\BibitemShut
  {NoStop}%
\bibitem [{\citenamefont {Zhao}\ \emph
  {et~al.}(2021{\natexlab{b}})\citenamefont {Zhao}, \citenamefont {Yang},
  \citenamefont {Liu}, \citenamefont {Yang}, \citenamefont {Gu}, \citenamefont
  {Wei}, \citenamefont {Xie}, \citenamefont {Zhang}, \citenamefont {Bian},
  \citenamefont {Zhang}, \citenamefont {Huo},\ and\ \citenamefont
  {Lu}}]{Zhao2021-ACSAEM-WGe2N4}%
  \BibitemOpen
  \bibfield  {author} {\bibinfo {author} {\bibfnamefont {H.}~\bibnamefont
  {Zhao}}, \bibinfo {author} {\bibfnamefont {G.}~\bibnamefont {Yang}}, \bibinfo
  {author} {\bibfnamefont {Y.}~\bibnamefont {Liu}}, \bibinfo {author}
  {\bibfnamefont {X.}~\bibnamefont {Yang}}, \bibinfo {author} {\bibfnamefont
  {Y.}~\bibnamefont {Gu}}, \bibinfo {author} {\bibfnamefont {C.}~\bibnamefont
  {Wei}}, \bibinfo {author} {\bibfnamefont {Z.}~\bibnamefont {Xie}}, \bibinfo
  {author} {\bibfnamefont {Q.}~\bibnamefont {Zhang}}, \bibinfo {author}
  {\bibfnamefont {B.}~\bibnamefont {Bian}}, \bibinfo {author} {\bibfnamefont
  {X.}~\bibnamefont {Zhang}}, \bibinfo {author} {\bibfnamefont
  {X.}~\bibnamefont {Huo}},\ and\ \bibinfo {author} {\bibfnamefont
  {N.}~\bibnamefont {Lu}},\ }\bibfield  {title} {\bibinfo {title} {{Quantum
  transport of sub-10 nm monolayer WGe$_2$N$_4$ transistors}},\ }\href
  {https://doi.org/10.1021/acsaelm.1c00829} {\bibfield  {journal} {\bibinfo
  {journal} {ACS Applied Electronic Materials}\ }\textbf {\bibinfo {volume}
  {3}},\ \bibinfo {pages} {5086} (\bibinfo {year}
  {2021}{\natexlab{b}})}\BibitemShut {NoStop}%
\bibitem [{\citenamefont {Xiao}\ \emph {et~al.}(2022)\citenamefont {Xiao},
  \citenamefont {Ma}, \citenamefont {Sa}, \citenamefont {Cui}, \citenamefont
  {Gao}, \citenamefont {Du}, \citenamefont {Sun},\ and\ \citenamefont
  {Li}}]{Xiao2022-ACSomega-MoSi2N4-moleculardope}%
  \BibitemOpen
  \bibfield  {author} {\bibinfo {author} {\bibfnamefont {C.}~\bibnamefont
  {Xiao}}, \bibinfo {author} {\bibfnamefont {Z.}~\bibnamefont {Ma}}, \bibinfo
  {author} {\bibfnamefont {R.}~\bibnamefont {Sa}}, \bibinfo {author}
  {\bibfnamefont {Z.}~\bibnamefont {Cui}}, \bibinfo {author} {\bibfnamefont
  {S.}~\bibnamefont {Gao}}, \bibinfo {author} {\bibfnamefont {W.}~\bibnamefont
  {Du}}, \bibinfo {author} {\bibfnamefont {X.}~\bibnamefont {Sun}},\ and\
  \bibinfo {author} {\bibfnamefont {Q.~H.}\ \bibnamefont {Li}},\ }\bibfield
  {title} {\bibinfo {title} {{Adsorption behavior of environmental gas
  molecules on pristine and defective MoSi$_2$N$_4$: Possible application as
  highly sensitive and reusable gas sensors}},\ }\href
  {https://doi.org/10.1021/acsomega.1c06860} {\bibfield  {journal} {\bibinfo
  {journal} {ACS Omega}\ }\textbf {\bibinfo {volume} {7}},\ \bibinfo {pages}
  {8706} (\bibinfo {year} {2022})}\BibitemShut {NoStop}%
\bibitem [{\citenamefont {Gao}\ \emph {et~al.}(2022)\citenamefont {Gao},
  \citenamefont {Liao}, \citenamefont {Wang}, \citenamefont {Wu}, \citenamefont
  {Li}, \citenamefont {Wang}, \citenamefont {Ma}, \citenamefont {Gong},
  \citenamefont {Wang}, \citenamefont {Dong}, \citenamefont {Jiao},\ and\
  \citenamefont {An}}]{Gao2022-PRAppl-MoSi2P4}%
  \BibitemOpen
  \bibfield  {author} {\bibinfo {author} {\bibfnamefont {Y.}~\bibnamefont
  {Gao}}, \bibinfo {author} {\bibfnamefont {J.}~\bibnamefont {Liao}}, \bibinfo
  {author} {\bibfnamefont {H.}~\bibnamefont {Wang}}, \bibinfo {author}
  {\bibfnamefont {Y.}~\bibnamefont {Wu}}, \bibinfo {author} {\bibfnamefont
  {Y.}~\bibnamefont {Li}}, \bibinfo {author} {\bibfnamefont {K.}~\bibnamefont
  {Wang}}, \bibinfo {author} {\bibfnamefont {C.}~\bibnamefont {Ma}}, \bibinfo
  {author} {\bibfnamefont {S.}~\bibnamefont {Gong}}, \bibinfo {author}
  {\bibfnamefont {T.}~\bibnamefont {Wang}}, \bibinfo {author} {\bibfnamefont
  {X.}~\bibnamefont {Dong}}, \bibinfo {author} {\bibfnamefont {Z.}~\bibnamefont
  {Jiao}},\ and\ \bibinfo {author} {\bibfnamefont {Y.}~\bibnamefont {An}},\
  }\bibfield  {title} {\bibinfo {title} {{Electronic transport properties and
  nanodevice designs for monolayer MoSi$_2$P$_4$}},\ }\href
  {https://doi.org/10.1103/PhysRevApplied.18.034033} {\bibfield  {journal}
  {\bibinfo  {journal} {Physical Review Applied}\ }\textbf {\bibinfo {volume}
  {18}},\ \bibinfo {pages} {034033} (\bibinfo {year} {2022})}\BibitemShut
  {NoStop}%
\bibitem [{\citenamefont {Mortazavi}\ \emph
  {et~al.}(2021{\natexlab{a}})\citenamefont {Mortazavi}, \citenamefont
  {Javvaji}, \citenamefont {Shojaei}, \citenamefont {Rabczuk}, \citenamefont
  {Shapeev},\ and\ \citenamefont {Zhuang}}]{Mortazavi2021-NanoEnergy}%
  \BibitemOpen
  \bibfield  {author} {\bibinfo {author} {\bibfnamefont {B.}~\bibnamefont
  {Mortazavi}}, \bibinfo {author} {\bibfnamefont {B.}~\bibnamefont {Javvaji}},
  \bibinfo {author} {\bibfnamefont {F.}~\bibnamefont {Shojaei}}, \bibinfo
  {author} {\bibfnamefont {T.}~\bibnamefont {Rabczuk}}, \bibinfo {author}
  {\bibfnamefont {A.~V.}\ \bibnamefont {Shapeev}},\ and\ \bibinfo {author}
  {\bibfnamefont {X.}~\bibnamefont {Zhuang}},\ }\bibfield  {title} {\bibinfo
  {title} {{Exceptional piezoelectricity, high thermal conductivity and
  stiffness and promising photocatalysis in two-dimensional MoSi$_2$N$_4$
  family confirmed by first-principles}},\ }\href
  {https://doi.org/10.1016/j.nanoen.2020.105716} {\bibfield  {journal}
  {\bibinfo  {journal} {Nano Energy}\ }\textbf {\bibinfo {volume} {82}},\
  \bibinfo {pages} {105716} (\bibinfo {year} {2021}{\natexlab{a}})}\BibitemShut
  {NoStop}%
\bibitem [{\citenamefont {Liu}\ \emph {et~al.}(2021{\natexlab{a}})\citenamefont
  {Liu}, \citenamefont {Zhang}, \citenamefont {Yang}, \citenamefont {Zhang},
  \citenamefont {Fan},\ and\ \citenamefont {Liu}}]{Liu2021-PLA-MoSi2P4}%
  \BibitemOpen
  \bibfield  {author} {\bibinfo {author} {\bibfnamefont {X.}~\bibnamefont
  {Liu}}, \bibinfo {author} {\bibfnamefont {H.}~\bibnamefont {Zhang}}, \bibinfo
  {author} {\bibfnamefont {Z.}~\bibnamefont {Yang}}, \bibinfo {author}
  {\bibfnamefont {Z.}~\bibnamefont {Zhang}}, \bibinfo {author} {\bibfnamefont
  {X.}~\bibnamefont {Fan}},\ and\ \bibinfo {author} {\bibfnamefont
  {H.}~\bibnamefont {Liu}},\ }\bibfield  {title} {\bibinfo {title} {{Structure
  and electronic properties of MoSi$_2$P$_4$ monolayer}},\ }\href
  {https://doi.org/10.1016/j.physleta.2021.127751} {\bibfield  {journal}
  {\bibinfo  {journal} {Physics Letters A}\ }\textbf {\bibinfo {volume}
  {420}},\ \bibinfo {pages} {127751} (\bibinfo {year}
  {2021}{\natexlab{a}})}\BibitemShut {NoStop}%
\bibitem [{\citenamefont {Yin}\ \emph {et~al.}(2021)\citenamefont {Yin},
  \citenamefont {Yi},\ and\ \citenamefont {Guo}}]{Yin2021-ACSAMI-MSi2Z4}%
  \BibitemOpen
  \bibfield  {author} {\bibinfo {author} {\bibfnamefont {Y.}~\bibnamefont
  {Yin}}, \bibinfo {author} {\bibfnamefont {M.}~\bibnamefont {Yi}},\ and\
  \bibinfo {author} {\bibfnamefont {W.}~\bibnamefont {Guo}},\ }\bibfield
  {title} {\bibinfo {title} {{ High and anomalous thermal conductivity in
  monolayer MSi$_2$Z$_4$ semiconductors }},\ }\href
  {https://doi.org/10.1021/acsami.1c14205} {\bibfield  {journal} {\bibinfo
  {journal} {ACS Applied Materials $\&$ Interfaces}\ }\textbf {\bibinfo
  {volume} {13}},\ \bibinfo {pages} {45907} (\bibinfo {year}
  {2021})}\BibitemShut {NoStop}%
\bibitem [{\citenamefont {Lee}\ \emph {et~al.}(2013)\citenamefont {Lee},
  \citenamefont {Cooper}, \citenamefont {An}, \citenamefont {Lee},
  \citenamefont {{Van Der Zande}}, \citenamefont {Petrone}, \citenamefont
  {Hammerberg}, \citenamefont {Lee}, \citenamefont {Crawford}, \citenamefont
  {Oliver}, \citenamefont {Kysar},\ and\ \citenamefont
  {Hone}}]{Lee2013-Science-graphene-mechanical}%
  \BibitemOpen
  \bibfield  {author} {\bibinfo {author} {\bibfnamefont {G.~H.}\ \bibnamefont
  {Lee}}, \bibinfo {author} {\bibfnamefont {R.~C.}\ \bibnamefont {Cooper}},
  \bibinfo {author} {\bibfnamefont {S.~J.}\ \bibnamefont {An}}, \bibinfo
  {author} {\bibfnamefont {S.}~\bibnamefont {Lee}}, \bibinfo {author}
  {\bibfnamefont {A.}~\bibnamefont {{Van Der Zande}}}, \bibinfo {author}
  {\bibfnamefont {N.}~\bibnamefont {Petrone}}, \bibinfo {author} {\bibfnamefont
  {A.~G.}\ \bibnamefont {Hammerberg}}, \bibinfo {author} {\bibfnamefont
  {C.}~\bibnamefont {Lee}}, \bibinfo {author} {\bibfnamefont {B.}~\bibnamefont
  {Crawford}}, \bibinfo {author} {\bibfnamefont {W.}~\bibnamefont {Oliver}},
  \bibinfo {author} {\bibfnamefont {J.~W.}\ \bibnamefont {Kysar}},\ and\
  \bibinfo {author} {\bibfnamefont {J.}~\bibnamefont {Hone}},\ }\bibfield
  {title} {\bibinfo {title} {{High-strength chemical-vapor-deposited graphene
  and grain boundaries}},\ }\href {https://doi.org/10.1126/science.1235126}
  {\bibfield  {journal} {\bibinfo  {journal} {Science}\ }\textbf {\bibinfo
  {volume} {340}},\ \bibinfo {pages} {1074} (\bibinfo {year}
  {2013})}\BibitemShut {NoStop}%
\bibitem [{\citenamefont {Liu}\ \emph {et~al.}(2014)\citenamefont {Liu},
  \citenamefont {Yan}, \citenamefont {Chen}, \citenamefont {Fan}, \citenamefont
  {Sun}, \citenamefont {Suh}, \citenamefont {Fu}, \citenamefont {Lee},
  \citenamefont {Zhou}, \citenamefont {Tongay}, \citenamefont {Ji},
  \citenamefont {Neaton},\ and\ \citenamefont
  {Wu}}]{Liu2014-NanoLett-MoS2-WS2-mechanics}%
  \BibitemOpen
  \bibfield  {author} {\bibinfo {author} {\bibfnamefont {K.}~\bibnamefont
  {Liu}}, \bibinfo {author} {\bibfnamefont {Q.}~\bibnamefont {Yan}}, \bibinfo
  {author} {\bibfnamefont {M.}~\bibnamefont {Chen}}, \bibinfo {author}
  {\bibfnamefont {W.}~\bibnamefont {Fan}}, \bibinfo {author} {\bibfnamefont
  {Y.}~\bibnamefont {Sun}}, \bibinfo {author} {\bibfnamefont {J.}~\bibnamefont
  {Suh}}, \bibinfo {author} {\bibfnamefont {D.}~\bibnamefont {Fu}}, \bibinfo
  {author} {\bibfnamefont {S.}~\bibnamefont {Lee}}, \bibinfo {author}
  {\bibfnamefont {J.}~\bibnamefont {Zhou}}, \bibinfo {author} {\bibfnamefont
  {S.}~\bibnamefont {Tongay}}, \bibinfo {author} {\bibfnamefont
  {J.}~\bibnamefont {Ji}}, \bibinfo {author} {\bibfnamefont {J.~B.}\
  \bibnamefont {Neaton}},\ and\ \bibinfo {author} {\bibfnamefont
  {J.}~\bibnamefont {Wu}},\ }\bibfield  {title} {\bibinfo {title} {{Elastic
  properties of chemical-vapor-deposited monolayer MoS$_2$, WS$_2$, and their
  bilayer heterostructures}},\ }\href {https://doi.org/10.1021/nl501793a}
  {\bibfield  {journal} {\bibinfo  {journal} {Nano Letters}\ }\textbf {\bibinfo
  {volume} {14}},\ \bibinfo {pages} {5097} (\bibinfo {year}
  {2014})}\BibitemShut {NoStop}%
\bibitem [{\citenamefont {Zhang}\ \emph {et~al.}(2016)\citenamefont {Zhang},
  \citenamefont {Koutsos},\ and\ \citenamefont
  {Cheung}}]{Zhang2016-APL-WSe2-mechanics}%
  \BibitemOpen
  \bibfield  {author} {\bibinfo {author} {\bibfnamefont {R.}~\bibnamefont
  {Zhang}}, \bibinfo {author} {\bibfnamefont {V.}~\bibnamefont {Koutsos}},\
  and\ \bibinfo {author} {\bibfnamefont {R.}~\bibnamefont {Cheung}},\
  }\bibfield  {title} {\bibinfo {title} {{Elastic properties of suspended
  multilayer WSe$_2$}},\ }\href {https://doi.org/10.1063/1.4940982} {\bibfield
  {journal} {\bibinfo  {journal} {Applied Physics Letters}\ }\textbf {\bibinfo
  {volume} {108}},\ \bibinfo {pages} {042104} (\bibinfo {year}
  {2016})}\BibitemShut {NoStop}%
\bibitem [{\citenamefont {Lipatov}\ \emph {et~al.}(2018)\citenamefont
  {Lipatov}, \citenamefont {Lu}, \citenamefont {Alhabeb}, \citenamefont
  {Anasori}, \citenamefont {Gruverman}, \citenamefont {Gogotsi},\ and\
  \citenamefont {Sinitskii}}]{lipatov2018-Ti3C2Tx}%
  \BibitemOpen
  \bibfield  {author} {\bibinfo {author} {\bibfnamefont {A.}~\bibnamefont
  {Lipatov}}, \bibinfo {author} {\bibfnamefont {H.}~\bibnamefont {Lu}},
  \bibinfo {author} {\bibfnamefont {M.}~\bibnamefont {Alhabeb}}, \bibinfo
  {author} {\bibfnamefont {B.}~\bibnamefont {Anasori}}, \bibinfo {author}
  {\bibfnamefont {A.}~\bibnamefont {Gruverman}}, \bibinfo {author}
  {\bibfnamefont {Y.}~\bibnamefont {Gogotsi}},\ and\ \bibinfo {author}
  {\bibfnamefont {A.}~\bibnamefont {Sinitskii}},\ }\bibfield  {title} {\bibinfo
  {title} {{Elastic properties of 2D Ti$_3$C$_2$T$_x$ MXene monolayers and
  bilayers}},\ }\href {https://doi.org/10.1126/sciadv.aat0491} {\bibfield
  {journal} {\bibinfo  {journal} {Science advances}\ }\textbf {\bibinfo
  {volume} {4}},\ \bibinfo {pages} {eaat0491} (\bibinfo {year}
  {2018})}\BibitemShut {NoStop}%
\bibitem [{\citenamefont {Lipatov}\ \emph {et~al.}(2020)\citenamefont
  {Lipatov}, \citenamefont {Alhabeb}, \citenamefont {Lu}, \citenamefont {Zhao},
  \citenamefont {Loes}, \citenamefont {Vorobeva}, \citenamefont {Dall'Agnese},
  \citenamefont {Gao}, \citenamefont {Gruverman}, \citenamefont {Gogotsi},\
  and\ \citenamefont {Sinitskii}}]{Lipatov2020-AEM-Nb4C3Tx}%
  \BibitemOpen
  \bibfield  {author} {\bibinfo {author} {\bibfnamefont {A.}~\bibnamefont
  {Lipatov}}, \bibinfo {author} {\bibfnamefont {M.}~\bibnamefont {Alhabeb}},
  \bibinfo {author} {\bibfnamefont {H.}~\bibnamefont {Lu}}, \bibinfo {author}
  {\bibfnamefont {S.}~\bibnamefont {Zhao}}, \bibinfo {author} {\bibfnamefont
  {M.~J.}\ \bibnamefont {Loes}}, \bibinfo {author} {\bibfnamefont {N.~S.}\
  \bibnamefont {Vorobeva}}, \bibinfo {author} {\bibfnamefont {Y.}~\bibnamefont
  {Dall'Agnese}}, \bibinfo {author} {\bibfnamefont {Y.}~\bibnamefont {Gao}},
  \bibinfo {author} {\bibfnamefont {A.}~\bibnamefont {Gruverman}}, \bibinfo
  {author} {\bibfnamefont {Y.}~\bibnamefont {Gogotsi}},\ and\ \bibinfo {author}
  {\bibfnamefont {A.}~\bibnamefont {Sinitskii}},\ }\bibfield  {title} {\bibinfo
  {title} {{Electrical and elastic properties of individual single-layer
  Nb$_4$C$_3$T$_x$ MXene flakes}},\ }\href
  {https://doi.org/10.1002/aelm.201901382} {\bibfield  {journal} {\bibinfo
  {journal} {Advanced Electronic Materials}\ }\textbf {\bibinfo {volume} {6}},\
  \bibinfo {pages} {1901382} (\bibinfo {year} {2020})}\BibitemShut {NoStop}%
\bibitem [{\citenamefont {Wei}\ and\ \citenamefont
  {Peng}(2014)}]{Wei2014-APL-BlackP-mechanical}%
  \BibitemOpen
  \bibfield  {author} {\bibinfo {author} {\bibfnamefont {Q.}~\bibnamefont
  {Wei}}\ and\ \bibinfo {author} {\bibfnamefont {X.}~\bibnamefont {Peng}},\
  }\bibfield  {title} {\bibinfo {title} {{Superior mechanical flexibility of
  phosphorene and few-layer black phosphorus}},\ }\href
  {https://doi.org/10.1063/1.4885215} {\bibfield  {journal} {\bibinfo
  {journal} {Applied Physics Letters}\ }\textbf {\bibinfo {volume} {104}},\
  \bibinfo {pages} {251915} (\bibinfo {year} {2014})}\BibitemShut {NoStop}%
\bibitem [{\citenamefont {Li}\ \emph {et~al.}(2021{\natexlab{b}})\citenamefont
  {Li}, \citenamefont {Zhou}, \citenamefont {Wan},\ and\ \citenamefont
  {Zhou}}]{Li2021-PhysE-mechanism}%
  \BibitemOpen
  \bibfield  {author} {\bibinfo {author} {\bibfnamefont {Q.}~\bibnamefont
  {Li}}, \bibinfo {author} {\bibfnamefont {W.}~\bibnamefont {Zhou}}, \bibinfo
  {author} {\bibfnamefont {X.}~\bibnamefont {Wan}},\ and\ \bibinfo {author}
  {\bibfnamefont {J.}~\bibnamefont {Zhou}},\ }\bibfield  {title} {\bibinfo
  {title} {{Strain effects on monolayer MoSi$_2$N$_4$: Ideal strength and
  failure mechanism}},\ }\href {https://doi.org/10.1016/j.physe.2021.114753}
  {\bibfield  {journal} {\bibinfo  {journal} {Physica E: Low-Dimensional
  Systems and Nanostructures}\ }\textbf {\bibinfo {volume} {131}},\ \bibinfo
  {pages} {114753} (\bibinfo {year} {2021}{\natexlab{b}})}\BibitemShut
  {NoStop}%
\bibitem [{\citenamefont {Bafekry}\ \emph
  {et~al.}(2021{\natexlab{a}})\citenamefont {Bafekry}, \citenamefont {Faraji},
  \citenamefont {Hoat}, \citenamefont {Shahrokhi6}, \citenamefont {Fadlallah},
  \citenamefont {Shojaei}, \citenamefont {Feghhi}, \citenamefont
  {Ghergherehchi},\ and\ \citenamefont {Gogova}}]{Shojaei2021-JPDAPP-MoSi2N4}%
  \BibitemOpen
  \bibfield  {author} {\bibinfo {author} {\bibfnamefont {A.}~\bibnamefont
  {Bafekry}}, \bibinfo {author} {\bibfnamefont {M.}~\bibnamefont {Faraji}},
  \bibinfo {author} {\bibfnamefont {D.~M.}\ \bibnamefont {Hoat}}, \bibinfo
  {author} {\bibfnamefont {M.}~\bibnamefont {Shahrokhi6}}, \bibinfo {author}
  {\bibfnamefont {M.~M.}\ \bibnamefont {Fadlallah}}, \bibinfo {author}
  {\bibfnamefont {F.}~\bibnamefont {Shojaei}}, \bibinfo {author} {\bibfnamefont
  {S.~A.~H.}\ \bibnamefont {Feghhi}}, \bibinfo {author} {\bibfnamefont
  {M.}~\bibnamefont {Ghergherehchi}},\ and\ \bibinfo {author} {\bibfnamefont
  {D.}~\bibnamefont {Gogova}},\ }\bibfield  {title} {\bibinfo {title}
  {{MoSi$_2$N$_4$ single-layer: A novel two-dimensional material with
  outstanding mechanical, thermal, electronic, optical, and photocatalytic
  properties}},\ }\href {https://doi.org/10.1088/1361-6463/abdb6b} {\bibfield
  {journal} {\bibinfo  {journal} {Journal of Physics D: Applied Physics}\
  }\textbf {\bibinfo {volume} {54}},\ \bibinfo {pages} {155303} (\bibinfo
  {year} {2021}{\natexlab{a}})}\BibitemShut {NoStop}%
\bibitem [{\citenamefont {Marianetti}\ and\ \citenamefont
  {Yevick}(2010)}]{Marianetti2010-PRL-graphene-failure}%
  \BibitemOpen
  \bibfield  {author} {\bibinfo {author} {\bibfnamefont {C.~A.}\ \bibnamefont
  {Marianetti}}\ and\ \bibinfo {author} {\bibfnamefont {H.~G.}\ \bibnamefont
  {Yevick}},\ }\bibfield  {title} {\bibinfo {title} {{Failure mechanisms of
  graphene under tension}},\ }\href
  {https://doi.org/10.1103/PhysRevLett.105.245502} {\bibfield  {journal}
  {\bibinfo  {journal} {Physical Review Letters}\ }\textbf {\bibinfo {volume}
  {105}},\ \bibinfo {pages} {245502} (\bibinfo {year} {2010})}\BibitemShut
  {NoStop}%
\bibitem [{\citenamefont {Li}(2012)}]{Li2012-PRB-MoS2-mechanics}%
  \BibitemOpen
  \bibfield  {author} {\bibinfo {author} {\bibfnamefont {T.}~\bibnamefont
  {Li}},\ }\bibfield  {title} {\bibinfo {title} {{Ideal strength and phonon
  instability in single-layer MoS$_2$}},\ }\href
  {https://doi.org/10.1103/PhysRevB.85.235407} {\bibfield  {journal} {\bibinfo
  {journal} {Physical Review B}\ }\textbf {\bibinfo {volume} {85}},\ \bibinfo
  {pages} {235407} (\bibinfo {year} {2012})}\BibitemShut {NoStop}%
\bibitem [{\citenamefont {Mortazavi}\ \emph
  {et~al.}(2021{\natexlab{b}})\citenamefont {Mortazavi}, \citenamefont
  {Shojaei}, \citenamefont {Javvaji}, \citenamefont {Rabczuk},\ and\
  \citenamefont {Zhuang}}]{Mortazavi2021-MaterTodayEnergy-CrC2N4}%
  \BibitemOpen
  \bibfield  {author} {\bibinfo {author} {\bibfnamefont {B.}~\bibnamefont
  {Mortazavi}}, \bibinfo {author} {\bibfnamefont {F.}~\bibnamefont {Shojaei}},
  \bibinfo {author} {\bibfnamefont {B.}~\bibnamefont {Javvaji}}, \bibinfo
  {author} {\bibfnamefont {T.}~\bibnamefont {Rabczuk}},\ and\ \bibinfo {author}
  {\bibfnamefont {X.}~\bibnamefont {Zhuang}},\ }\bibfield  {title} {\bibinfo
  {title} {{Outstandingly high thermal conductivity, elastic modulus, carrier
  mobility and piezoelectricity in two-dimensional semiconducting CrC$_2$N$_4$:
  A first-principles study}},\ }\href
  {https://doi.org/10.1016/j.mtener.2021.100839} {\bibfield  {journal}
  {\bibinfo  {journal} {Materials Today Energy}\ }\textbf {\bibinfo {volume}
  {22}},\ \bibinfo {pages} {100839} (\bibinfo {year}
  {2021}{\natexlab{b}})}\BibitemShut {NoStop}%
\bibitem [{\citenamefont {Tian}\ \emph {et~al.}(2021)\citenamefont {Tian},
  \citenamefont {Wei}, \citenamefont {Zhang}, \citenamefont {Wang},\ and\
  \citenamefont {Yang}}]{Tian2021-PRB-SnSi2N4}%
  \BibitemOpen
  \bibfield  {author} {\bibinfo {author} {\bibfnamefont {M.}~\bibnamefont
  {Tian}}, \bibinfo {author} {\bibfnamefont {C.}~\bibnamefont {Wei}}, \bibinfo
  {author} {\bibfnamefont {J.}~\bibnamefont {Zhang}}, \bibinfo {author}
  {\bibfnamefont {J.}~\bibnamefont {Wang}},\ and\ \bibinfo {author}
  {\bibfnamefont {R.}~\bibnamefont {Yang}},\ }\bibfield  {title} {\bibinfo
  {title} {{Electronic, optical, and water solubility properties of
  two-dimensional layered SnSi$_2$N$_4$ from first principles}},\ }\href
  {https://doi.org/10.1103/PhysRevB.103.195305} {\bibfield  {journal} {\bibinfo
   {journal} {Physical Review B}\ }\textbf {\bibinfo {volume} {103}},\ \bibinfo
  {pages} {195305} (\bibinfo {year} {2021})}\BibitemShut {NoStop}%
\bibitem [{\citenamefont {Dat}\ and\ \citenamefont
  {Vu}(2022)}]{Dat2022-RSCadv-SnGe2N4}%
  \BibitemOpen
  \bibfield  {author} {\bibinfo {author} {\bibfnamefont {V.~D.}\ \bibnamefont
  {Dat}}\ and\ \bibinfo {author} {\bibfnamefont {T.~V.}\ \bibnamefont {Vu}},\
  }\bibfield  {title} {\bibinfo {title} {{Layered post-transition-metal
  dichalcogenide SnGe$_2$N$_4$ as a promising photoelectric material: A DFT
  study}},\ }\href {https://doi.org/10.1039/d2ra00935h} {\bibfield  {journal}
  {\bibinfo  {journal} {RSC Advances}\ }\textbf {\bibinfo {volume} {12}},\
  \bibinfo {pages} {10249} (\bibinfo {year} {2022})}\BibitemShut {NoStop}%
\bibitem [{\citenamefont {Sibatov}\ \emph {et~al.}(2022)\citenamefont
  {Sibatov}, \citenamefont {Meftakhutdinov},\ and\ \citenamefont
  {Kochaev}}]{Sibatov2022-ASS-XMoSiN2}%
  \BibitemOpen
  \bibfield  {author} {\bibinfo {author} {\bibfnamefont {R.~T.}\ \bibnamefont
  {Sibatov}}, \bibinfo {author} {\bibfnamefont {R.~M.}\ \bibnamefont
  {Meftakhutdinov}},\ and\ \bibinfo {author} {\bibfnamefont {A.~I.}\
  \bibnamefont {Kochaev}},\ }\bibfield  {title} {\bibinfo {title} {{Asymmetric
  XMoSiN$_2$ (X=S, Se, Te) monolayers as novel promising 2D materials for
  nanoelectronics and photovoltaics}},\ }\href
  {https://doi.org/10.1016/j.apsusc.2022.152465} {\bibfield  {journal}
  {\bibinfo  {journal} {Applied Surface Science}\ }\textbf {\bibinfo {volume}
  {585}},\ \bibinfo {pages} {152465} (\bibinfo {year} {2022})}\BibitemShut
  {NoStop}%
\bibitem [{\citenamefont {Guo}\ \emph {et~al.}(2021{\natexlab{a}})\citenamefont
  {Guo}, \citenamefont {Zhu}, \citenamefont {Mu}, \citenamefont {Wang},\ and\
  \citenamefont {Chen}}]{Guo2021-ComputMaterSci-structureEffect}%
  \BibitemOpen
  \bibfield  {author} {\bibinfo {author} {\bibfnamefont {S.~D.}\ \bibnamefont
  {Guo}}, \bibinfo {author} {\bibfnamefont {Y.~T.}\ \bibnamefont {Zhu}},
  \bibinfo {author} {\bibfnamefont {W.~Q.}\ \bibnamefont {Mu}}, \bibinfo
  {author} {\bibfnamefont {L.}~\bibnamefont {Wang}},\ and\ \bibinfo {author}
  {\bibfnamefont {X.~Q.}\ \bibnamefont {Chen}},\ }\bibfield  {title} {\bibinfo
  {title} {{Structure effect on intrinsic piezoelectricity in
  septuple-atomic-layer MSi$_2$N$_4$ (M=Mo and W)}},\ }\href
  {https://doi.org/10.1016/j.commatsci.2020.110223} {\bibfield  {journal}
  {\bibinfo  {journal} {Computational Materials Science}\ }\textbf {\bibinfo
  {volume} {188}},\ \bibinfo {pages} {110223} (\bibinfo {year}
  {2021}{\natexlab{a}})}\BibitemShut {NoStop}%
\bibitem [{\citenamefont {Guo}\ \emph {et~al.}(2020{\natexlab{a}})\citenamefont
  {Guo}, \citenamefont {Zhu}, \citenamefont {Mu},\ and\ \citenamefont
  {Ren}}]{Guo2020-EPL-MSi2N4-piezoelectricity}%
  \BibitemOpen
  \bibfield  {author} {\bibinfo {author} {\bibfnamefont {S.~D.}\ \bibnamefont
  {Guo}}, \bibinfo {author} {\bibfnamefont {Y.~T.}\ \bibnamefont {Zhu}},
  \bibinfo {author} {\bibfnamefont {W.~Q.}\ \bibnamefont {Mu}},\ and\ \bibinfo
  {author} {\bibfnamefont {W.~C.}\ \bibnamefont {Ren}},\ }\bibfield  {title}
  {\bibinfo {title} {{Intrinsic piezoelectricity in monolayer MSi$_2$N$_4$ (M =
  Mo, W, Cr, Ti, Zr and Hf)}},\ }\href
  {https://doi.org/10.1209/0295-5075/132/57002} {\bibfield  {journal} {\bibinfo
   {journal} {Europhysics Letters}\ }\textbf {\bibinfo {volume} {132}},\
  \bibinfo {pages} {57002} (\bibinfo {year} {2020}{\natexlab{a}})}\BibitemShut
  {NoStop}%
\bibitem [{\citenamefont {Duerloo}\ \emph {et~al.}(2012)\citenamefont
  {Duerloo}, \citenamefont {Ong},\ and\ \citenamefont
  {Reed}}]{Duerloo2012-2Dmater-inplane-piezo}%
  \BibitemOpen
  \bibfield  {author} {\bibinfo {author} {\bibfnamefont {K.~A.~N.}\
  \bibnamefont {Duerloo}}, \bibinfo {author} {\bibfnamefont {M.~T.}\
  \bibnamefont {Ong}},\ and\ \bibinfo {author} {\bibfnamefont {E.~J.}\
  \bibnamefont {Reed}},\ }\bibfield  {title} {\bibinfo {title} {{Intrinsic
  piezoelectricity in two-dimensional materials}},\ }\href
  {https://doi.org/10.1021/jz3012436} {\bibfield  {journal} {\bibinfo
  {journal} {Journal of Physical Chemistry Letters}\ }\textbf {\bibinfo
  {volume} {3}},\ \bibinfo {pages} {2871} (\bibinfo {year} {2012})}\BibitemShut
  {NoStop}%
\bibitem [{\citenamefont {Blonsky}\ \emph {et~al.}(2015)\citenamefont
  {Blonsky}, \citenamefont {Zhuang}, \citenamefont {Singh},\ and\ \citenamefont
  {Hennig}}]{Blonsky2015-2Dmater-inplane-piezo}%
  \BibitemOpen
  \bibfield  {author} {\bibinfo {author} {\bibfnamefont {M.~N.}\ \bibnamefont
  {Blonsky}}, \bibinfo {author} {\bibfnamefont {H.~L.}\ \bibnamefont {Zhuang}},
  \bibinfo {author} {\bibfnamefont {A.~K.}\ \bibnamefont {Singh}},\ and\
  \bibinfo {author} {\bibfnamefont {R.~G.}\ \bibnamefont {Hennig}},\ }\bibfield
   {title} {\bibinfo {title} {{Ab initio prediction of piezoelectricity in
  two-dimensional materials}},\ }\href
  {https://doi.org/10.1021/acsnano.5b03394} {\bibfield  {journal} {\bibinfo
  {journal} {ACS Nano}\ }\textbf {\bibinfo {volume} {9}},\ \bibinfo {pages}
  {9885} (\bibinfo {year} {2015})}\BibitemShut {NoStop}%
\bibitem [{\citenamefont {Cui}\ \emph {et~al.}(2018)\citenamefont {Cui},
  \citenamefont {Xue}, \citenamefont {Hu},\ and\ \citenamefont
  {Li}}]{Cui2018-npj2D-piezo}%
  \BibitemOpen
  \bibfield  {author} {\bibinfo {author} {\bibfnamefont {C.}~\bibnamefont
  {Cui}}, \bibinfo {author} {\bibfnamefont {F.}~\bibnamefont {Xue}}, \bibinfo
  {author} {\bibfnamefont {W.~J.}\ \bibnamefont {Hu}},\ and\ \bibinfo {author}
  {\bibfnamefont {L.~J.}\ \bibnamefont {Li}},\ }\bibfield  {title} {\bibinfo
  {title} {{Two-dimensional materials with piezoelectric and ferroelectric
  functionalities}},\ }\href {https://doi.org/10.1038/s41699-018-0063-5}
  {\bibfield  {journal} {\bibinfo  {journal} {npj 2D Materials and
  Applications}\ }\textbf {\bibinfo {volume} {2}},\ \bibinfo {pages} {18}
  (\bibinfo {year} {2018})}\BibitemShut {NoStop}%
\bibitem [{\citenamefont {Lueng}\ \emph {et~al.}(1999)\citenamefont {Lueng},
  \citenamefont {Chan}, \citenamefont {Fong}, \citenamefont {Surya},\ and\
  \citenamefont {Choy}}]{Lueng1999-JAP-BulkAlN-GaN-piezo}%
  \BibitemOpen
  \bibfield  {author} {\bibinfo {author} {\bibfnamefont {C.~M.}\ \bibnamefont
  {Lueng}}, \bibinfo {author} {\bibfnamefont {H.~L.}\ \bibnamefont {Chan}},
  \bibinfo {author} {\bibfnamefont {W.~K.}\ \bibnamefont {Fong}}, \bibinfo
  {author} {\bibfnamefont {C.}~\bibnamefont {Surya}},\ and\ \bibinfo {author}
  {\bibfnamefont {C.~L.}\ \bibnamefont {Choy}},\ }\bibfield  {title} {\bibinfo
  {title} {{Piezoelectric coefficients of aluminum nitride and gallium
  nitride}},\ }\href {https://doi.org/10.1557/proc-572-389} {\bibfield
  {journal} {\bibinfo  {journal} {Materials Research Society Symposium -
  Proceedings}\ }\textbf {\bibinfo {volume} {572}},\ \bibinfo {pages} {389}
  (\bibinfo {year} {1999})}\BibitemShut {NoStop}%
\bibitem [{\citenamefont {Guo}\ \emph {et~al.}(2020{\natexlab{b}})\citenamefont
  {Guo}, \citenamefont {Mu}, \citenamefont {Zhu},\ and\ \citenamefont
  {Chen}}]{Guo2020-PCCP-VSi2P4}%
  \BibitemOpen
  \bibfield  {author} {\bibinfo {author} {\bibfnamefont {S.~D.}\ \bibnamefont
  {Guo}}, \bibinfo {author} {\bibfnamefont {W.~Q.}\ \bibnamefont {Mu}},
  \bibinfo {author} {\bibfnamefont {Y.~T.}\ \bibnamefont {Zhu}},\ and\ \bibinfo
  {author} {\bibfnamefont {X.~Q.}\ \bibnamefont {Chen}},\ }\bibfield  {title}
  {\bibinfo {title} {{Coexistence of intrinsic piezoelectricity and
  ferromagnetism induced by small biaxial strain in septuple-atomic-layer
  VSi$_2$P$_4$}},\ }\href {https://doi.org/10.1039/d0cp05273f} {\bibfield
  {journal} {\bibinfo  {journal} {Physical Chemistry Chemical Physics}\
  }\textbf {\bibinfo {volume} {22}},\ \bibinfo {pages} {28359} (\bibinfo {year}
  {2020}{\natexlab{b}})}\BibitemShut {NoStop}%
\bibitem [{\citenamefont {Guo}\ \emph {et~al.}(2021{\natexlab{b}})\citenamefont
  {Guo}, \citenamefont {Mu}, \citenamefont {Zhu}, \citenamefont {Han},\ and\
  \citenamefont {Ren}}]{Guo2021-JMCC-Janus-MSiGeN4}%
  \BibitemOpen
  \bibfield  {author} {\bibinfo {author} {\bibfnamefont {S.~D.}\ \bibnamefont
  {Guo}}, \bibinfo {author} {\bibfnamefont {W.~Q.}\ \bibnamefont {Mu}},
  \bibinfo {author} {\bibfnamefont {Y.~T.}\ \bibnamefont {Zhu}}, \bibinfo
  {author} {\bibfnamefont {R.~Y.}\ \bibnamefont {Han}},\ and\ \bibinfo {author}
  {\bibfnamefont {W.~C.}\ \bibnamefont {Ren}},\ }\bibfield  {title} {\bibinfo
  {title} {{Predicted septuple-atomic-layer Janus MSiGeN$_4$(M = Mo and W)
  monolayers with Rashba spin splitting and high electron carrier
  mobilities}},\ }\href {https://doi.org/10.1039/d0tc05649a} {\bibfield
  {journal} {\bibinfo  {journal} {Journal of Materials Chemistry C}\ }\textbf
  {\bibinfo {volume} {9}},\ \bibinfo {pages} {2464} (\bibinfo {year}
  {2021}{\natexlab{b}})}\BibitemShut {NoStop}%
\bibitem [{\citenamefont {Guo}\ and\ \citenamefont
  {Guo}(2021{\natexlab{a}})}]{Guo2021-JSemicond-Janus-MSiGeN4}%
  \BibitemOpen
  \bibfield  {author} {\bibinfo {author} {\bibfnamefont {X.}~\bibnamefont
  {Guo}}\ and\ \bibinfo {author} {\bibfnamefont {S.}~\bibnamefont {Guo}},\
  }\bibfield  {title} {\bibinfo {title} {{Janus MSiGeN$_4$ (M = Zr and Hf)
  monolayers derived from centrosymmetric $\beta$-MA$_2$Z$_4$: A
  first-principles study}},\ }\href
  {https://doi.org/10.1088/1674-4926/42/12/122002} {\bibfield  {journal}
  {\bibinfo  {journal} {Journal of Semiconductors}\ }\textbf {\bibinfo {volume}
  {42}},\ \bibinfo {pages} {122002} (\bibinfo {year}
  {2021}{\natexlab{a}})}\BibitemShut {NoStop}%
\bibitem [{\citenamefont {Guo}\ \emph {et~al.}(2021{\natexlab{c}})\citenamefont
  {Guo}, \citenamefont {Zhu}, \citenamefont {Mu},\ and\ \citenamefont
  {Chen}}]{Guo2021-JMCC-SrAlGaSe4}%
  \BibitemOpen
  \bibfield  {author} {\bibinfo {author} {\bibfnamefont {S.~D.}\ \bibnamefont
  {Guo}}, \bibinfo {author} {\bibfnamefont {Y.~T.}\ \bibnamefont {Zhu}},
  \bibinfo {author} {\bibfnamefont {W.~Q.}\ \bibnamefont {Mu}},\ and\ \bibinfo
  {author} {\bibfnamefont {X.~Q.}\ \bibnamefont {Chen}},\ }\bibfield  {title}
  {\bibinfo {title} {{A piezoelectric quantum spin Hall insulator with Rashba
  spin splitting in Janus monolayer SrAlGaSe$_4$}},\ }\href
  {https://doi.org/10.1039/d1tc01165k} {\bibfield  {journal} {\bibinfo
  {journal} {Journal of Materials Chemistry C}\ }\textbf {\bibinfo {volume}
  {9}},\ \bibinfo {pages} {7465} (\bibinfo {year}
  {2021}{\natexlab{c}})}\BibitemShut {NoStop}%
\bibitem [{\citenamefont {Zhong}\ \emph
  {et~al.}(2021{\natexlab{a}})\citenamefont {Zhong}, \citenamefont {Ren},
  \citenamefont {Zhang}, \citenamefont {Gao},\ and\ \citenamefont
  {Wu}}]{Zhong2021-JMCA-SlidingFerroelectricity}%
  \BibitemOpen
  \bibfield  {author} {\bibinfo {author} {\bibfnamefont {T.}~\bibnamefont
  {Zhong}}, \bibinfo {author} {\bibfnamefont {Y.}~\bibnamefont {Ren}}, \bibinfo
  {author} {\bibfnamefont {Z.}~\bibnamefont {Zhang}}, \bibinfo {author}
  {\bibfnamefont {J.}~\bibnamefont {Gao}},\ and\ \bibinfo {author}
  {\bibfnamefont {M.}~\bibnamefont {Wu}},\ }\bibfield  {title} {\bibinfo
  {title} {{Sliding ferroelectricity in two-dimensional MoA$_2$N$_4$(A = Si or
  Ge) bilayers: High polarizations and Moir{\'{e}} potentials}},\ }\href
  {https://doi.org/10.1039/d1ta02645c} {\bibfield  {journal} {\bibinfo
  {journal} {Journal of Materials Chemistry A}\ }\textbf {\bibinfo {volume}
  {9}},\ \bibinfo {pages} {19659} (\bibinfo {year}
  {2021}{\natexlab{a}})}\BibitemShut {NoStop}%
\bibitem [{\citenamefont {Fei}\ \emph {et~al.}(2018{\natexlab{a}})\citenamefont
  {Fei}, \citenamefont {Zhao}, \citenamefont {Palomaki}, \citenamefont {Sun},
  \citenamefont {Miller}, \citenamefont {Zhao}, \citenamefont {Yan},
  \citenamefont {Xu},\ and\ \citenamefont
  {Cobden}}]{Fei2018-Nature-2D-ferroelectric}%
  \BibitemOpen
  \bibfield  {author} {\bibinfo {author} {\bibfnamefont {Z.}~\bibnamefont
  {Fei}}, \bibinfo {author} {\bibfnamefont {W.}~\bibnamefont {Zhao}}, \bibinfo
  {author} {\bibfnamefont {T.~A.}\ \bibnamefont {Palomaki}}, \bibinfo {author}
  {\bibfnamefont {B.}~\bibnamefont {Sun}}, \bibinfo {author} {\bibfnamefont
  {M.~K.}\ \bibnamefont {Miller}}, \bibinfo {author} {\bibfnamefont
  {Z.}~\bibnamefont {Zhao}}, \bibinfo {author} {\bibfnamefont {J.}~\bibnamefont
  {Yan}}, \bibinfo {author} {\bibfnamefont {X.}~\bibnamefont {Xu}},\ and\
  \bibinfo {author} {\bibfnamefont {D.~H.}\ \bibnamefont {Cobden}},\ }\bibfield
   {title} {\bibinfo {title} {{Ferroelectric switching of a two-dimensional
  metal}},\ }\href {https://doi.org/10.1038/s41586-018-0336-3} {\bibfield
  {journal} {\bibinfo  {journal} {Nature}\ }\textbf {\bibinfo {volume} {560}},\
  \bibinfo {pages} {336} (\bibinfo {year} {2018}{\natexlab{a}})}\BibitemShut
  {NoStop}%
\bibitem [{\citenamefont {Li}\ and\ \citenamefont
  {Wu}(2017)}]{Li2017-ACSNano-2D-Verticalpolarization}%
  \BibitemOpen
  \bibfield  {author} {\bibinfo {author} {\bibfnamefont {L.}~\bibnamefont
  {Li}}\ and\ \bibinfo {author} {\bibfnamefont {M.}~\bibnamefont {Wu}},\
  }\bibfield  {title} {\bibinfo {title} {{Binary compound bilayer and
  multilayer with vertical polarizations: Two-dimensional ferroelectrics,
  multiferroics, and nanogenerators}},\ }\href
  {https://doi.org/10.1021/acsnano.7b02756} {\bibfield  {journal} {\bibinfo
  {journal} {ACS Nano}\ }\textbf {\bibinfo {volume} {11}},\ \bibinfo {pages}
  {6382} (\bibinfo {year} {2017})}\BibitemShut {NoStop}%
\bibitem [{\citenamefont {Zhuang}\ \emph {et~al.}(2019)\citenamefont {Zhuang},
  \citenamefont {He}, \citenamefont {Javvaji},\ and\ \citenamefont
  {Park}}]{Zhuang2019-PRB-MoS2-flexo}%
  \BibitemOpen
  \bibfield  {author} {\bibinfo {author} {\bibfnamefont {X.}~\bibnamefont
  {Zhuang}}, \bibinfo {author} {\bibfnamefont {B.}~\bibnamefont {He}}, \bibinfo
  {author} {\bibfnamefont {B.}~\bibnamefont {Javvaji}},\ and\ \bibinfo {author}
  {\bibfnamefont {H.~S.}\ \bibnamefont {Park}},\ }\bibfield  {title} {\bibinfo
  {title} {{Intrinsic bending flexoelectric constants in two-dimensional
  materials}},\ }\href {https://doi.org/10.1103/PhysRevB.99.054105} {\bibfield
  {journal} {\bibinfo  {journal} {Physical Review B}\ }\textbf {\bibinfo
  {volume} {99}},\ \bibinfo {pages} {054105} (\bibinfo {year}
  {2019})}\BibitemShut {NoStop}%
\bibitem [{\citenamefont {Dong}\ \emph {et~al.}(2017)\citenamefont {Dong},
  \citenamefont {Lou},\ and\ \citenamefont
  {Shenoy}}]{Dong2017-ACSNano-JanusMXY-piezo}%
  \BibitemOpen
  \bibfield  {author} {\bibinfo {author} {\bibfnamefont {L.}~\bibnamefont
  {Dong}}, \bibinfo {author} {\bibfnamefont {J.}~\bibnamefont {Lou}},\ and\
  \bibinfo {author} {\bibfnamefont {V.~B.}\ \bibnamefont {Shenoy}},\ }\bibfield
   {title} {\bibinfo {title} {{Large in-plane and vertical piezoelectricity in
  Janus transition metal dichalchogenides}},\ }\href
  {https://doi.org/10.1021/acsnano.7b03313} {\bibfield  {journal} {\bibinfo
  {journal} {ACS Nano}\ }\textbf {\bibinfo {volume} {11}},\ \bibinfo {pages}
  {8242} (\bibinfo {year} {2017})}\BibitemShut {NoStop}%
\bibitem [{\citenamefont {Javvaji}\ \emph {et~al.}(2019)\citenamefont
  {Javvaji}, \citenamefont {He}, \citenamefont {Zhuang},\ and\ \citenamefont
  {Park}}]{Javvaji2019-PRM-JanusTMDs-flexo}%
  \BibitemOpen
  \bibfield  {author} {\bibinfo {author} {\bibfnamefont {B.}~\bibnamefont
  {Javvaji}}, \bibinfo {author} {\bibfnamefont {B.}~\bibnamefont {He}},
  \bibinfo {author} {\bibfnamefont {X.}~\bibnamefont {Zhuang}},\ and\ \bibinfo
  {author} {\bibfnamefont {H.~S.}\ \bibnamefont {Park}},\ }\bibfield  {title}
  {\bibinfo {title} {{High flexoelectric constants in Janus transition-metal
  dichalcogenides}},\ }\href
  {https://doi.org/10.1103/PhysRevMaterials.3.125402} {\bibfield  {journal}
  {\bibinfo  {journal} {Physical Review Materials}\ }\textbf {\bibinfo {volume}
  {3}},\ \bibinfo {pages} {125402} (\bibinfo {year} {2019})}\BibitemShut
  {NoStop}%
\bibitem [{\citenamefont {Yu}\ \emph {et~al.}(2021{\natexlab{a}})\citenamefont
  {Yu}, \citenamefont {Zhou}, \citenamefont {Wan},\ and\ \citenamefont
  {Li}}]{Yu2021-NewJPhys-Thermal}%
  \BibitemOpen
  \bibfield  {author} {\bibinfo {author} {\bibfnamefont {J.}~\bibnamefont
  {Yu}}, \bibinfo {author} {\bibfnamefont {J.}~\bibnamefont {Zhou}}, \bibinfo
  {author} {\bibfnamefont {X.}~\bibnamefont {Wan}},\ and\ \bibinfo {author}
  {\bibfnamefont {Q.}~\bibnamefont {Li}},\ }\bibfield  {title} {\bibinfo
  {title} {{High intrinsic lattice thermal conductivity in monolayer
  MoSi$_2$N$_4$}},\ }\href {https://doi.org/10.1088/1367-2630/abe8f7}
  {\bibfield  {journal} {\bibinfo  {journal} {New Journal of Physics}\ }\textbf
  {\bibinfo {volume} {23}},\ \bibinfo {pages} {033005} (\bibinfo {year}
  {2021}{\natexlab{a}})}\BibitemShut {NoStop}%
\bibitem [{\citenamefont {Shen}\ \emph {et~al.}(2022)\citenamefont {Shen},
  \citenamefont {Wang}, \citenamefont {Wei}, \citenamefont {Zhang},
  \citenamefont {Qin}, \citenamefont {Chen},\ and\ \citenamefont
  {Zhang}}]{Shen2022-PCCP-Thermal}%
  \BibitemOpen
  \bibfield  {author} {\bibinfo {author} {\bibfnamefont {C.}~\bibnamefont
  {Shen}}, \bibinfo {author} {\bibfnamefont {L.}~\bibnamefont {Wang}}, \bibinfo
  {author} {\bibfnamefont {D.}~\bibnamefont {Wei}}, \bibinfo {author}
  {\bibfnamefont {Y.}~\bibnamefont {Zhang}}, \bibinfo {author} {\bibfnamefont
  {G.}~\bibnamefont {Qin}}, \bibinfo {author} {\bibfnamefont {X.~Q.}\
  \bibnamefont {Chen}},\ and\ \bibinfo {author} {\bibfnamefont
  {H.}~\bibnamefont {Zhang}},\ }\bibfield  {title} {\bibinfo {title}
  {{Two-dimensional layered MSi$_2$N$_4$(M = Mo, W) as promising thermal
  management materials: A comparative study}},\ }\href
  {https://doi.org/10.1039/d1cp03941e} {\bibfield  {journal} {\bibinfo
  {journal} {Physical Chemistry Chemical Physics}\ }\textbf {\bibinfo {volume}
  {24}},\ \bibinfo {pages} {3086} (\bibinfo {year} {2022})}\BibitemShut
  {NoStop}%
\bibitem [{\citenamefont {Li}\ \emph {et~al.}(2020{\natexlab{b}})\citenamefont
  {Li}, \citenamefont {Nie},\ and\ \citenamefont
  {Sun}}]{Li2020-NanoRes-BHBFBCl-thermal}%
  \BibitemOpen
  \bibfield  {author} {\bibinfo {author} {\bibfnamefont {T.}~\bibnamefont
  {Li}}, \bibinfo {author} {\bibfnamefont {G.}~\bibnamefont {Nie}},\ and\
  \bibinfo {author} {\bibfnamefont {Q.}~\bibnamefont {Sun}},\ }\bibfield
  {title} {\bibinfo {title} {Highly sensitive tuning of lattice thermal
  conductivity of graphene-like borophene by fluorination and chlorination},\
  }\href {https://doi.org/10.1007/s12274-020-2767-z} {\bibfield  {journal}
  {\bibinfo  {journal} {Nano Research}\ }\textbf {\bibinfo {volume} {13}},\
  \bibinfo {pages} {1171} (\bibinfo {year} {2020}{\natexlab{b}})}\BibitemShut
  {NoStop}%
\bibitem [{\citenamefont {Cai}\ \emph {et~al.}(2014{\natexlab{a}})\citenamefont
  {Cai}, \citenamefont {Lan}, \citenamefont {Zhang},\ and\ \citenamefont
  {Zhang}}]{Cai2014-PRB-MoS2-thermal}%
  \BibitemOpen
  \bibfield  {author} {\bibinfo {author} {\bibfnamefont {Y.}~\bibnamefont
  {Cai}}, \bibinfo {author} {\bibfnamefont {J.}~\bibnamefont {Lan}}, \bibinfo
  {author} {\bibfnamefont {G.}~\bibnamefont {Zhang}},\ and\ \bibinfo {author}
  {\bibfnamefont {Y.}~\bibnamefont {Zhang}},\ }\bibfield  {title} {\bibinfo
  {title} {Lattice vibrational modes and phonon thermal conductivity of
  monolayer {MoS$_2$}},\ }\href {https://doi.org/10.1103/PhysRevB.89.035438}
  {\bibfield  {journal} {\bibinfo  {journal} {Physical Review B}\ }\textbf
  {\bibinfo {volume} {89}},\ \bibinfo {pages} {035438} (\bibinfo {year}
  {2014}{\natexlab{a}})}\BibitemShut {NoStop}%
\bibitem [{\citenamefont {Gu}\ and\ \citenamefont
  {Yang}(2014)}]{Gu2014-APL-TMD}%
  \BibitemOpen
  \bibfield  {author} {\bibinfo {author} {\bibfnamefont {X.}~\bibnamefont
  {Gu}}\ and\ \bibinfo {author} {\bibfnamefont {R.}~\bibnamefont {Yang}},\
  }\bibfield  {title} {\bibinfo {title} {Phonon transport in single-layer
  transition metal dichalcogenides: A first-principles study},\ }\href
  {https://doi.org/10.1063/1.4896685} {\bibfield  {journal} {\bibinfo
  {journal} {Applied Physics Letters}\ }\textbf {\bibinfo {volume} {105}},\
  \bibinfo {pages} {131903} (\bibinfo {year} {2014})}\BibitemShut {NoStop}%
\bibitem [{\citenamefont {Torres}\ \emph {et~al.}(2019)\citenamefont {Torres},
  \citenamefont {Alvarez}, \citenamefont {Cartoix{\`{a} X. }},\ and\
  \citenamefont {Rurali}}]{Torres2019-2DMater-TMD}%
  \BibitemOpen
  \bibfield  {author} {\bibinfo {author} {\bibfnamefont {P.}~\bibnamefont
  {Torres}}, \bibinfo {author} {\bibfnamefont {F.~X.}\ \bibnamefont {Alvarez}},
  \bibinfo {author} {\bibnamefont {Cartoix{\`{a} X. }}},\ and\ \bibinfo
  {author} {\bibfnamefont {R.}~\bibnamefont {Rurali}},\ }\bibfield  {title}
  {\bibinfo {title} {Thermal conductivity and phonon hydrodynamics in
  transition metal dichalcogenides from first-principles},\ }\href
  {https://doi.org/10.1088/2053-1583/ab0c31} {\bibfield  {journal} {\bibinfo
  {journal} {2D Materials}\ }\textbf {\bibinfo {volume} {6}},\ \bibinfo {pages}
  {035002} (\bibinfo {year} {2019})}\BibitemShut {NoStop}%
\bibitem [{\citenamefont {Liu}\ \emph {et~al.}(2018)\citenamefont {Liu},
  \citenamefont {Bo}, \citenamefont {Xu}, \citenamefont {Yin}, \citenamefont
  {Zhang}, \citenamefont {Wang}, \citenamefont {Eriksson},\ and\ \citenamefont
  {Wang}}]{Liu2018-PRB-GroupIVSe-thermal}%
  \BibitemOpen
  \bibfield  {author} {\bibinfo {author} {\bibfnamefont {P.}~\bibnamefont
  {Liu}}, \bibinfo {author} {\bibfnamefont {T.}~\bibnamefont {Bo}}, \bibinfo
  {author} {\bibfnamefont {J.}~\bibnamefont {Xu}}, \bibinfo {author}
  {\bibfnamefont {W.}~\bibnamefont {Yin}}, \bibinfo {author} {\bibfnamefont
  {J.}~\bibnamefont {Zhang}}, \bibinfo {author} {\bibfnamefont
  {F.}~\bibnamefont {Wang}}, \bibinfo {author} {\bibfnamefont {O.}~\bibnamefont
  {Eriksson}},\ and\ \bibinfo {author} {\bibfnamefont {B.~T.}\ \bibnamefont
  {Wang}},\ }\bibfield  {title} {\bibinfo {title} {First-principles
  calculations of the ultralow thermal conductivity in two-dimensional
  group-\uppercase\expandafter{\romannumeral4} selenides},\ }\href
  {https://doi.org/10.1103/PhysRevB.98.235426} {\bibfield  {journal} {\bibinfo
  {journal} {Physical Review B}\ }\textbf {\bibinfo {volume} {98}},\ \bibinfo
  {pages} {235426} (\bibinfo {year} {2018})}\BibitemShut {NoStop}%
\bibitem [{\citenamefont {Sun}\ \emph {et~al.}(2019)\citenamefont {Sun},
  \citenamefont {Shuai},\ and\ \citenamefont
  {Wang}}]{Sun2019-JPCC-SnSe-thermal}%
  \BibitemOpen
  \bibfield  {author} {\bibinfo {author} {\bibfnamefont {Y.}~\bibnamefont
  {Sun}}, \bibinfo {author} {\bibfnamefont {Z.}~\bibnamefont {Shuai}},\ and\
  \bibinfo {author} {\bibfnamefont {D.}~\bibnamefont {Wang}},\ }\bibfield
  {title} {\bibinfo {title} {{Reducing lattice thermal conductivity of the
  thermoelectric SnSe monolayer: Role of phonon–electron coupling}},\ }\href
  {https://doi.org/10.1021/acs.jpcc.9b02344} {\bibfield  {journal} {\bibinfo
  {journal} {Journal of Physical Chemistry C}\ }\textbf {\bibinfo {volume}
  {123}},\ \bibinfo {pages} {12001} (\bibinfo {year} {2019})}\BibitemShut
  {NoStop}%
\bibitem [{\citenamefont {Balandin}\ \emph {et~al.}(2008)\citenamefont
  {Balandin}, \citenamefont {Ghosh}, \citenamefont {Bao}, \citenamefont
  {Calizo},\ and\ \citenamefont
  {Lau}}]{Balandin2008-Nanolett-Graphene-thermal}%
  \BibitemOpen
  \bibfield  {author} {\bibinfo {author} {\bibfnamefont {A.~A.}\ \bibnamefont
  {Balandin}}, \bibinfo {author} {\bibfnamefont {S.}~\bibnamefont {Ghosh}},
  \bibinfo {author} {\bibfnamefont {W.}~\bibnamefont {Bao}}, \bibinfo {author}
  {\bibfnamefont {I.}~\bibnamefont {Calizo}},\ and\ \bibinfo {author}
  {\bibfnamefont {C.}~\bibnamefont {Lau}},\ }\bibfield  {title} {\bibinfo
  {title} {Superior thermal conductivity of single-layer graphene},\ }\href
  {https://doi.org/10.1021/nl0731872} {\bibfield  {journal} {\bibinfo
  {journal} {Nano Letters}\ }\textbf {\bibinfo {volume} {8}},\ \bibinfo {pages}
  {902} (\bibinfo {year} {2008})}\BibitemShut {NoStop}%
\bibitem [{\citenamefont {Ghosh}\ \emph {et~al.}(2008)\citenamefont {Ghosh},
  \citenamefont {Calizo}, \citenamefont {Teweldebrhan}, \citenamefont
  {Pokatilov}, \citenamefont {Nika}, \citenamefont {Balandin}, \citenamefont
  {Bao}, \citenamefont {Miao},\ and\ \citenamefont
  {Lau}}]{Ghosh2008-APL-Graphene3000}%
  \BibitemOpen
  \bibfield  {author} {\bibinfo {author} {\bibfnamefont {S.}~\bibnamefont
  {Ghosh}}, \bibinfo {author} {\bibfnamefont {I.}~\bibnamefont {Calizo}},
  \bibinfo {author} {\bibfnamefont {D.}~\bibnamefont {Teweldebrhan}}, \bibinfo
  {author} {\bibfnamefont {E.~P.}\ \bibnamefont {Pokatilov}}, \bibinfo {author}
  {\bibfnamefont {D.}~\bibnamefont {Nika}}, \bibinfo {author} {\bibfnamefont
  {A.~A.}\ \bibnamefont {Balandin}}, \bibinfo {author} {\bibfnamefont
  {W.}~\bibnamefont {Bao}}, \bibinfo {author} {\bibfnamefont {F.}~\bibnamefont
  {Miao}},\ and\ \bibinfo {author} {\bibfnamefont {C.~N.}\ \bibnamefont
  {Lau}},\ }\bibfield  {title} {\bibinfo {title} {Extremely high thermal
  conductivity of graphene: Prospects for thermal management applications in
  nanoelectronic circuits},\ }\href {https://doi.org/10.1063/1.2907977}
  {\bibfield  {journal} {\bibinfo  {journal} {Applied Physics Letters}\
  }\textbf {\bibinfo {volume} {92}},\ \bibinfo {pages} {151911} (\bibinfo
  {year} {2008})}\BibitemShut {NoStop}%
\bibitem [{\citenamefont {Lindsay}\ and\ \citenamefont
  {Broido}(2011)}]{Lindsay2011-PRB-BN600}%
  \BibitemOpen
  \bibfield  {author} {\bibinfo {author} {\bibfnamefont {L.}~\bibnamefont
  {Lindsay}}\ and\ \bibinfo {author} {\bibfnamefont {D.~A.}\ \bibnamefont
  {Broido}},\ }\bibfield  {title} {\bibinfo {title} {Enhanced thermal
  conductivity and isotope effect in single-layer hexagonal boron nitride},\
  }\href {https://doi.org/10.1103/physrevb.84.155421} {\bibfield  {journal}
  {\bibinfo  {journal} {Physical Review B}\ }\textbf {\bibinfo {volume} {84}},\
  \bibinfo {pages} {155421} (\bibinfo {year} {2011})}\BibitemShut {NoStop}%
\bibitem [{\citenamefont {Slack}(1962)}]{Slack1962-PhysRev}%
  \BibitemOpen
  \bibfield  {author} {\bibinfo {author} {\bibfnamefont {G.~A.}\ \bibnamefont
  {Slack}},\ }\bibfield  {title} {\bibinfo {title} {Anisotropic thermal
  conductivity of pyrolytic graphite},\ }\href
  {https://doi.org/10.1103/PhysRev.127.694} {\bibfield  {journal} {\bibinfo
  {journal} {Physical Reviews}\ }\textbf {\bibinfo {volume} {127}},\ \bibinfo
  {pages} {694} (\bibinfo {year} {1962})}\BibitemShut {NoStop}%
\bibitem [{\citenamefont {Slack}(1973)}]{Slack1973-JPCS}%
  \BibitemOpen
  \bibfield  {author} {\bibinfo {author} {\bibfnamefont {G.~A.}\ \bibnamefont
  {Slack}},\ }\bibfield  {title} {\bibinfo {title} {Nonmetallic crystals with
  high thermal conductivity},\ }\href
  {https://doi.org/10.1016/0022-3697(73)90092-9} {\bibfield  {journal}
  {\bibinfo  {journal} {Journal of Physics and Chemistry of Solids}\ }\textbf
  {\bibinfo {volume} {34}},\ \bibinfo {pages} {321} (\bibinfo {year}
  {1973})}\BibitemShut {NoStop}%
\bibitem [{\citenamefont {Guo}\ and\ \citenamefont
  {Guo}(2021{\natexlab{b}})}]{Guo2021-CPB-TransportCoefficients}%
  \BibitemOpen
  \bibfield  {author} {\bibinfo {author} {\bibfnamefont {X.~S.}\ \bibnamefont
  {Guo}}\ and\ \bibinfo {author} {\bibfnamefont {S.~D.}\ \bibnamefont {Guo}},\
  }\bibfield  {title} {\bibinfo {title} {{Tuning transport coefficients of
  monolayer MoSi$_2$N$_4$ with biaxial strain}},\ }\href
  {https://doi.org/10.1088/1674-1056/abdb22} {\bibfield  {journal} {\bibinfo
  {journal} {Chinese Physics B}\ }\textbf {\bibinfo {volume} {30}},\ \bibinfo
  {pages} {067102} (\bibinfo {year} {2021}{\natexlab{b}})}\BibitemShut
  {NoStop}%
\bibitem [{\citenamefont {Huang}\ \emph
  {et~al.}(2022{\natexlab{b}})\citenamefont {Huang}, \citenamefont {Zhong},
  \citenamefont {Yuan},\ and\ \citenamefont
  {Chen}}]{Huang2022-EurophysLett-MoSi2As4}%
  \BibitemOpen
  \bibfield  {author} {\bibinfo {author} {\bibfnamefont {Y.}~\bibnamefont
  {Huang}}, \bibinfo {author} {\bibfnamefont {X.}~\bibnamefont {Zhong}},
  \bibinfo {author} {\bibfnamefont {H.}~\bibnamefont {Yuan}},\ and\ \bibinfo
  {author} {\bibfnamefont {H.}~\bibnamefont {Chen}},\ }\bibfield  {title}
  {\bibinfo {title} {{ Thermoelectric performance of MoSi$_2$As$_4$ monolayer
  }},\ }\href {https://doi.org/10.1209/0295-5075/ac49d3} {\bibfield  {journal}
  {\bibinfo  {journal} {Europhysics Letters}\ }\textbf {\bibinfo {volume}
  {137}},\ \bibinfo {pages} {16002} (\bibinfo {year}
  {2022}{\natexlab{b}})}\BibitemShut {NoStop}%
\bibitem [{\citenamefont {Zhang}\ \emph {et~al.}(2022)\citenamefont {Zhang},
  \citenamefont {Wei}, \citenamefont {Zhang}, \citenamefont {Chen},
  \citenamefont {Chen}, \citenamefont {Hao},\ and\ \citenamefont
  {Jia}}]{Zhang2022-JSSC-MoSiN-MoGeN-thermoelec}%
  \BibitemOpen
  \bibfield  {author} {\bibinfo {author} {\bibfnamefont {C.}~\bibnamefont
  {Zhang}}, \bibinfo {author} {\bibfnamefont {F.}~\bibnamefont {Wei}}, \bibinfo
  {author} {\bibfnamefont {X.}~\bibnamefont {Zhang}}, \bibinfo {author}
  {\bibfnamefont {W.}~\bibnamefont {Chen}}, \bibinfo {author} {\bibfnamefont
  {C.}~\bibnamefont {Chen}}, \bibinfo {author} {\bibfnamefont {J.}~\bibnamefont
  {Hao}},\ and\ \bibinfo {author} {\bibfnamefont {B.}~\bibnamefont {Jia}},\
  }\bibfield  {title} {\bibinfo {title} {{Thermoelectric properties of
  monolayer MoSi$_2$N$_4$ and MoGe$_2$N$_4$ with large Seebeck coefficient and
  high carrier mobility: A first principles study}},\ }\href
  {https://doi.org/10.1016/j.jssc.2022.123447} {\bibfield  {journal} {\bibinfo
  {journal} {Journal of Solid State Chemistry}\ }\textbf {\bibinfo {volume}
  {315}},\ \bibinfo {pages} {123447} (\bibinfo {year} {2022})}\BibitemShut
  {NoStop}%
\bibitem [{\citenamefont
  {Alavi-rad}(2022)}]{Alavi-rad2022-SST-monoMoSi2N4-strain}%
  \BibitemOpen
  \bibfield  {author} {\bibinfo {author} {\bibfnamefont {H.}~\bibnamefont
  {Alavi-rad}},\ }\bibfield  {title} {\bibinfo {title} {{Strain engineering in
  optoelectronic properties of MoSi$_2$N$_4$ monolayer: Ultrahigh
  tunability}},\ }\href {https://doi.org/10.1088/1361-6641/ac6769} {\bibfield
  {journal} {\bibinfo  {journal} {Semiconductor Science and Technology}\
  }\textbf {\bibinfo {volume} {37}},\ \bibinfo {pages} {065018} (\bibinfo
  {year} {2022})}\BibitemShut {NoStop}%
\bibitem [{\citenamefont {Zhong}\ \emph
  {et~al.}(2021{\natexlab{b}})\citenamefont {Zhong}, \citenamefont {Xiong},
  \citenamefont {Lv}, \citenamefont {Yu},\ and\ \citenamefont
  {Yuan}}]{Zhong2021-PRB-bilayer-verticalstrain}%
  \BibitemOpen
  \bibfield  {author} {\bibinfo {author} {\bibfnamefont {H.}~\bibnamefont
  {Zhong}}, \bibinfo {author} {\bibfnamefont {W.}~\bibnamefont {Xiong}},
  \bibinfo {author} {\bibfnamefont {P.}~\bibnamefont {Lv}}, \bibinfo {author}
  {\bibfnamefont {J.}~\bibnamefont {Yu}},\ and\ \bibinfo {author}
  {\bibfnamefont {S.}~\bibnamefont {Yuan}},\ }\bibfield  {title} {\bibinfo
  {title} {{Strain-induced semiconductor to metal transition in MA$_2$Z$_4$
  bilayers (M= Ti, Cr, Mo; A= Si; Z= N, P)}},\ }\href
  {https://doi.org/10.1103/PhysRevB.103.085124} {\bibfield  {journal} {\bibinfo
   {journal} {Physical Review B}\ }\textbf {\bibinfo {volume} {103}},\ \bibinfo
  {pages} {085124} (\bibinfo {year} {2021}{\natexlab{b}})}\BibitemShut
  {NoStop}%
\bibitem [{\citenamefont {Ding}\ and\ \citenamefont
  {Wang}(2021)}]{Ding2021-JPCC-Y-Cd-Hf-Hg}%
  \BibitemOpen
  \bibfield  {author} {\bibinfo {author} {\bibfnamefont {Y.}~\bibnamefont
  {Ding}}\ and\ \bibinfo {author} {\bibfnamefont {Y.}~\bibnamefont {Wang}},\
  }\bibfield  {title} {\bibinfo {title} {{Computational exploration of stable
  4d/5d transition-metal MSi$_2$N$_4$ (M = Y-Cd and Hf-Hg) nanosheets and their
  versatile electronic and magnetic properties}},\ }\href
  {https://doi.org/10.1021/acs.jpcc.1c06734} {\bibfield  {journal} {\bibinfo
  {journal} {Journal of Physical Chemistry C}\ }\textbf {\bibinfo {volume}
  {125}},\ \bibinfo {pages} {19580} (\bibinfo {year} {2021})}\BibitemShut
  {NoStop}%
\bibitem [{\citenamefont {Wu}\ \emph {et~al.}(2022)\citenamefont {Wu},
  \citenamefont {Tang}, \citenamefont {Xia}, \citenamefont {Gao}, \citenamefont
  {Jia}, \citenamefont {Zhang}, \citenamefont {Zhu}, \citenamefont {Zhang},\
  and\ \citenamefont {Zhang}}]{Wu2022-npjComputMater-robustbandedge}%
  \BibitemOpen
  \bibfield  {author} {\bibinfo {author} {\bibfnamefont {Y.}~\bibnamefont
  {Wu}}, \bibinfo {author} {\bibfnamefont {Z.}~\bibnamefont {Tang}}, \bibinfo
  {author} {\bibfnamefont {W.}~\bibnamefont {Xia}}, \bibinfo {author}
  {\bibfnamefont {W.}~\bibnamefont {Gao}}, \bibinfo {author} {\bibfnamefont
  {F.}~\bibnamefont {Jia}}, \bibinfo {author} {\bibfnamefont {Y.}~\bibnamefont
  {Zhang}}, \bibinfo {author} {\bibfnamefont {W.}~\bibnamefont {Zhu}}, \bibinfo
  {author} {\bibfnamefont {W.}~\bibnamefont {Zhang}},\ and\ \bibinfo {author}
  {\bibfnamefont {P.}~\bibnamefont {Zhang}},\ }\bibfield  {title} {\bibinfo
  {title} {{Prediction of protected band edge states and dielectric tunable
  quasiparticle and excitonic properties of monolayer MoSi$_2$N$_4$}},\ }\href
  {https://doi.org/10.1038/s41524-022-00815-6} {\bibfield  {journal} {\bibinfo
  {journal} {npj Computational Materials}\ }\textbf {\bibinfo {volume} {8}},\
  \bibinfo {pages} {129} (\bibinfo {year} {2022})}\BibitemShut {NoStop}%
\bibitem [{\citenamefont {Cai}\ \emph {et~al.}(2014{\natexlab{b}})\citenamefont
  {Cai}, \citenamefont {Zhang},\ and\ \citenamefont
  {Zhang}}]{Cai2014-JACS-MoS2-elec}%
  \BibitemOpen
  \bibfield  {author} {\bibinfo {author} {\bibfnamefont {Y.}~\bibnamefont
  {Cai}}, \bibinfo {author} {\bibfnamefont {G.}~\bibnamefont {Zhang}},\ and\
  \bibinfo {author} {\bibfnamefont {Y.~W.}\ \bibnamefont {Zhang}},\ }\bibfield
  {title} {\bibinfo {title} {{Polarity-reversed robust carrier mobility in
  monolayer MoS$_2$ nanoribbons}},\ }\href {https://doi.org/10.1021/ja4109787}
  {\bibfield  {journal} {\bibinfo  {journal} {Journal of the American Chemical
  Society}\ }\textbf {\bibinfo {volume} {136}},\ \bibinfo {pages} {6269}
  (\bibinfo {year} {2014}{\natexlab{b}})}\BibitemShut {NoStop}%
\bibitem [{\citenamefont {Kong}\ \emph {et~al.}(2022)\citenamefont {Kong},
  \citenamefont {Murakami},\ and\ \citenamefont
  {Zhang}}]{KONG2022100814Acomprehensive}%
  \BibitemOpen
  \bibfield  {author} {\bibinfo {author} {\bibfnamefont {M.}~\bibnamefont
  {Kong}}, \bibinfo {author} {\bibfnamefont {S.}~\bibnamefont {Murakami}},\
  and\ \bibinfo {author} {\bibfnamefont {T.}~\bibnamefont {Zhang}},\ }\bibfield
   {title} {\bibinfo {title} {A comprehensive study of complex non-adiabatic
  exciton dynamics in mosi2n4},\ }\href
  {https://doi.org/10.1016/j.mtphys.2022.100814} {\bibfield  {journal}
  {\bibinfo  {journal} {Materials Today Physics}\ }\textbf {\bibinfo {volume}
  {27}},\ \bibinfo {pages} {100814} (\bibinfo {year} {2022})}\BibitemShut
  {NoStop}%
\bibitem [{\citenamefont {Yao}\ \emph {et~al.}(2021)\citenamefont {Yao},
  \citenamefont {Zhang}, \citenamefont {Wang}, \citenamefont {Li},
  \citenamefont {Yu}, \citenamefont {Xu}, \citenamefont {Wang},\ and\
  \citenamefont {Wei}}]{Yao2021-Nanomaterials-Bilayer}%
  \BibitemOpen
  \bibfield  {author} {\bibinfo {author} {\bibfnamefont {H.}~\bibnamefont
  {Yao}}, \bibinfo {author} {\bibfnamefont {C.}~\bibnamefont {Zhang}}, \bibinfo
  {author} {\bibfnamefont {Q.}~\bibnamefont {Wang}}, \bibinfo {author}
  {\bibfnamefont {J.}~\bibnamefont {Li}}, \bibinfo {author} {\bibfnamefont
  {Y.}~\bibnamefont {Yu}}, \bibinfo {author} {\bibfnamefont {F.}~\bibnamefont
  {Xu}}, \bibinfo {author} {\bibfnamefont {B.}~\bibnamefont {Wang}},\ and\
  \bibinfo {author} {\bibfnamefont {Y.}~\bibnamefont {Wei}},\ }\bibfield
  {title} {\bibinfo {title} {{Novel two-dimensional layered MoSi$_2$Z$_4$ (Z =
  P, As): New promising optoelectronic materials}},\ }\href
  {https://doi.org/10.3390/nano11030559} {\bibfield  {journal} {\bibinfo
  {journal} {Nanomaterials}\ }\textbf {\bibinfo {volume} {11}},\ \bibinfo
  {pages} {559} (\bibinfo {year} {2021})}\BibitemShut {NoStop}%
\bibitem [{\citenamefont {Wu}\ \emph {et~al.}(2021)\citenamefont {Wu},
  \citenamefont {Cao}, \citenamefont {Ang},\ and\ \citenamefont
  {Ang}}]{Wu2021-APL-biMoSi2N4andWSi2N4}%
  \BibitemOpen
  \bibfield  {author} {\bibinfo {author} {\bibfnamefont {Q.}~\bibnamefont
  {Wu}}, \bibinfo {author} {\bibfnamefont {L.}~\bibnamefont {Cao}}, \bibinfo
  {author} {\bibfnamefont {Y.~S.}\ \bibnamefont {Ang}},\ and\ \bibinfo {author}
  {\bibfnamefont {L.~K.}\ \bibnamefont {Ang}},\ }\bibfield  {title} {\bibinfo
  {title} {{Semiconductor-to-metal transition in bilayer MoSi$_2$N$_4$ and
  WSi$_2$N$_4$ with strain and electric field}},\ }\href
  {https://doi.org/10.1063/5.0044431} {\bibfield  {journal} {\bibinfo
  {journal} {Applied Physics Letters}\ }\textbf {\bibinfo {volume} {118}},\
  \bibinfo {pages} {113102} (\bibinfo {year} {2021})}\BibitemShut {NoStop}%
\bibitem [{\citenamefont {Cai}\ \emph {et~al.}(2021{\natexlab{a}})\citenamefont
  {Cai}, \citenamefont {Zhang}, \citenamefont {Zhu}, \citenamefont {Lin},
  \citenamefont {Yu}, \citenamefont {Wang}, \citenamefont {Yang}, \citenamefont
  {Jia},\ and\ \citenamefont {Jia}}]{Cai2021-JMCC-MoSe2-hetero}%
  \BibitemOpen
  \bibfield  {author} {\bibinfo {author} {\bibfnamefont {X.}~\bibnamefont
  {Cai}}, \bibinfo {author} {\bibfnamefont {Z.}~\bibnamefont {Zhang}}, \bibinfo
  {author} {\bibfnamefont {Y.}~\bibnamefont {Zhu}}, \bibinfo {author}
  {\bibfnamefont {L.}~\bibnamefont {Lin}}, \bibinfo {author} {\bibfnamefont
  {W.}~\bibnamefont {Yu}}, \bibinfo {author} {\bibfnamefont {Q.}~\bibnamefont
  {Wang}}, \bibinfo {author} {\bibfnamefont {X.}~\bibnamefont {Yang}}, \bibinfo
  {author} {\bibfnamefont {X.}~\bibnamefont {Jia}},\ and\ \bibinfo {author}
  {\bibfnamefont {Y.}~\bibnamefont {Jia}},\ }\bibfield  {title} {\bibinfo
  {title} {{A two-dimensional MoSe$_2$/MoSi$_2$N$_4$ van der Waals
  heterostructure with high carrier mobility and diversified regulation of its
  electronic properties}},\ }\href {https://doi.org/10.1039/d1tc01149a}
  {\bibfield  {journal} {\bibinfo  {journal} {Journal of Materials Chemistry
  C}\ }\textbf {\bibinfo {volume} {9}},\ \bibinfo {pages} {10073} (\bibinfo
  {year} {2021}{\natexlab{a}})}\BibitemShut {NoStop}%
\bibitem [{\citenamefont {He}\ \emph {et~al.}(2022)\citenamefont {He},
  \citenamefont {Zhu}, \citenamefont {Zhang}, \citenamefont {Du}, \citenamefont
  {Guo}, \citenamefont {Liu}, \citenamefont {Tian}, \citenamefont {Zhong},
  \citenamefont {Wang},\ and\ \citenamefont
  {Shi}}]{He2022-PCCP-InSe-MoSi2N4-elec-optic-HER}%
  \BibitemOpen
  \bibfield  {author} {\bibinfo {author} {\bibfnamefont {Y.}~\bibnamefont
  {He}}, \bibinfo {author} {\bibfnamefont {Y.~H.}\ \bibnamefont {Zhu}},
  \bibinfo {author} {\bibfnamefont {M.}~\bibnamefont {Zhang}}, \bibinfo
  {author} {\bibfnamefont {J.}~\bibnamefont {Du}}, \bibinfo {author}
  {\bibfnamefont {W.~H.}\ \bibnamefont {Guo}}, \bibinfo {author} {\bibfnamefont
  {S.~M.}\ \bibnamefont {Liu}}, \bibinfo {author} {\bibfnamefont
  {C.}~\bibnamefont {Tian}}, \bibinfo {author} {\bibfnamefont {H.~X.}\
  \bibnamefont {Zhong}}, \bibinfo {author} {\bibfnamefont {X.}~\bibnamefont
  {Wang}},\ and\ \bibinfo {author} {\bibfnamefont {J.~J.}\ \bibnamefont
  {Shi}},\ }\bibfield  {title} {\bibinfo {title} {{High hydrogen production in
  the InSe/MoSi$_2$N$_4$ van der Waals heterostructure for overall water
  splitting}},\ }\href {https://doi.org/10.1039/d1cp04705a} {\bibfield
  {journal} {\bibinfo  {journal} {Physical Chemistry Chemical Physics}\
  }\textbf {\bibinfo {volume} {24}},\ \bibinfo {pages} {2110} (\bibinfo {year}
  {2022})}\BibitemShut {NoStop}%
\bibitem [{\citenamefont {Cai}\ \emph {et~al.}(2021{\natexlab{b}})\citenamefont
  {Cai}, \citenamefont {Zhang}, \citenamefont {Song}, \citenamefont {Chen},
  \citenamefont {Yu}, \citenamefont {Jia}, \citenamefont {Yang}, \citenamefont
  {Liu},\ and\ \citenamefont {Jia}}]{Cai2021-SSRN-WSe-hetero}%
  \BibitemOpen
  \bibfield  {author} {\bibinfo {author} {\bibfnamefont {X.}~\bibnamefont
  {Cai}}, \bibinfo {author} {\bibfnamefont {Z.}~\bibnamefont {Zhang}}, \bibinfo
  {author} {\bibfnamefont {A.}~\bibnamefont {Song}}, \bibinfo {author}
  {\bibfnamefont {G.}~\bibnamefont {Chen}}, \bibinfo {author} {\bibfnamefont
  {W.}~\bibnamefont {Yu}}, \bibinfo {author} {\bibfnamefont {X.}~\bibnamefont
  {Jia}}, \bibinfo {author} {\bibfnamefont {X.}~\bibnamefont {Yang}}, \bibinfo
  {author} {\bibfnamefont {Y.}~\bibnamefont {Liu}},\ and\ \bibinfo {author}
  {\bibfnamefont {Y.}~\bibnamefont {Jia}},\ }\bibfield  {title} {\bibinfo
  {title} {{Indirect to direct bandgap transition and enhanced optoelectronic
  properties in WSe$_2$ monolayer through forming WSe$_2$/MoSi$_2$N$_4$
  bilayer}},\ }\href {http://dx.doi.org/10.2139/ssrn.3968855} {\bibfield
  {journal} {\bibinfo  {journal} {SSRN Electronic Journal}\ } (\bibinfo {year}
  {2021}{\natexlab{b}})}\BibitemShut {NoStop}%
\bibitem [{\citenamefont {Nguyen}\ \emph
  {et~al.}(2022{\natexlab{a}})\citenamefont {Nguyen}, \citenamefont {Ang},
  \citenamefont {Nguyen}, \citenamefont {Hoang}, \citenamefont {Hung},\ and\
  \citenamefont {Nguyen}}]{Nguyen2022-PRB-C3N4-hetero}%
  \BibitemOpen
  \bibfield  {author} {\bibinfo {author} {\bibfnamefont {C.~Q.}\ \bibnamefont
  {Nguyen}}, \bibinfo {author} {\bibfnamefont {Y.~S.}\ \bibnamefont {Ang}},
  \bibinfo {author} {\bibfnamefont {S.~T.}\ \bibnamefont {Nguyen}}, \bibinfo
  {author} {\bibfnamefont {N.~V.}\ \bibnamefont {Hoang}}, \bibinfo {author}
  {\bibfnamefont {N.~M.}\ \bibnamefont {Hung}},\ and\ \bibinfo {author}
  {\bibfnamefont {C.~V.}\ \bibnamefont {Nguyen}},\ }\bibfield  {title}
  {\bibinfo {title} {{Tunable type-II band alignment and electronic structure
  of C$_3$N$_4$/MoSi$_2$N$_4$ heterostructure: Interlayer coupling and electric
  field}},\ }\href {https://doi.org/10.1103/PhysRevB.105.045303} {\bibfield
  {journal} {\bibinfo  {journal} {Physical Review B}\ }\textbf {\bibinfo
  {volume} {105}},\ \bibinfo {pages} {045303} (\bibinfo {year}
  {2022}{\natexlab{a}})}\BibitemShut {NoStop}%
\bibitem [{\citenamefont {Ng}\ \emph {et~al.}(2022)\citenamefont {Ng},
  \citenamefont {Wu}, \citenamefont {Ang},\ and\ \citenamefont
  {Ang}}]{Ng2022-APL-GaN-ZnO-hetero}%
  \BibitemOpen
  \bibfield  {author} {\bibinfo {author} {\bibfnamefont {J.~Q.}\ \bibnamefont
  {Ng}}, \bibinfo {author} {\bibfnamefont {Q.}~\bibnamefont {Wu}}, \bibinfo
  {author} {\bibfnamefont {L.~K.}\ \bibnamefont {Ang}},\ and\ \bibinfo {author}
  {\bibfnamefont {Y.~S.}\ \bibnamefont {Ang}},\ }\bibfield  {title} {\bibinfo
  {title} {{Tunable electronic properties and band alignments of
  MoSi$_2$N$_4$/GaN and MoSi$_2$N$_4$/ZnO van der Waals heterostructures}},\
  }\href {https://doi.org/10.1063/5.0083736} {\bibfield  {journal} {\bibinfo
  {journal} {Applied Physics Letters}\ }\textbf {\bibinfo {volume} {120}},\
  \bibinfo {pages} {103101} (\bibinfo {year} {2022})}\BibitemShut {NoStop}%
\bibitem [{\citenamefont {Liu}\ \emph {et~al.}(2022)\citenamefont {Liu},
  \citenamefont {Wang}, \citenamefont {Xiong}, \citenamefont {Zhong},\ and\
  \citenamefont {Yuan}}]{Liu2022-JAP-Cs3Bi2I9-hetero}%
  \BibitemOpen
  \bibfield  {author} {\bibinfo {author} {\bibfnamefont {C.}~\bibnamefont
  {Liu}}, \bibinfo {author} {\bibfnamefont {Z.}~\bibnamefont {Wang}}, \bibinfo
  {author} {\bibfnamefont {W.}~\bibnamefont {Xiong}}, \bibinfo {author}
  {\bibfnamefont {H.}~\bibnamefont {Zhong}},\ and\ \bibinfo {author}
  {\bibfnamefont {S.}~\bibnamefont {Yuan}},\ }\bibfield  {title} {\bibinfo
  {title} {{Effect of vertical strain and in-plane biaxial strain on type-II
  MoSi$_2$N$_4$/Cs$_3$Bi$_2$I$_9$ van der Waals heterostructure}},\ }\href
  {https://doi.org/10.1063/5.0080224} {\bibfield  {journal} {\bibinfo
  {journal} {Journal of Applied Physics}\ }\textbf {\bibinfo {volume} {131}},\
  \bibinfo {pages} {163102} (\bibinfo {year} {2022})}\BibitemShut {NoStop}%
\bibitem [{\citenamefont {Guo}\ \emph {et~al.}(2022)\citenamefont {Guo},
  \citenamefont {Min}, \citenamefont {Cai}, \citenamefont {Zhang},
  \citenamefont {Liu},\ and\ \citenamefont {Jia}}]{Guo2022-BP-MoSi2P4-hetero}%
  \BibitemOpen
  \bibfield  {author} {\bibinfo {author} {\bibfnamefont {Y.}~\bibnamefont
  {Guo}}, \bibinfo {author} {\bibfnamefont {J.}~\bibnamefont {Min}}, \bibinfo
  {author} {\bibfnamefont {X.}~\bibnamefont {Cai}}, \bibinfo {author}
  {\bibfnamefont {L.}~\bibnamefont {Zhang}}, \bibinfo {author} {\bibfnamefont
  {C.}~\bibnamefont {Liu}},\ and\ \bibinfo {author} {\bibfnamefont
  {Y.}~\bibnamefont {Jia}},\ }\bibfield  {title} {\bibinfo {title}
  {{Two-dimensional type-II BP/MoSi$_2$P$_4$ vdW heterostructures for
  high-performance solar cells}},\ }\href
  {https://doi.org/10.1021/acs.jpcc.1c10476} {\bibfield  {journal} {\bibinfo
  {journal} {Journal of Physical Chemistry C}\ }\textbf {\bibinfo {volume}
  {126}},\ \bibinfo {pages} {4677} (\bibinfo {year} {2022})}\BibitemShut
  {NoStop}%
\bibitem [{\citenamefont {Nguyen}\ \emph {et~al.}(2021)\citenamefont {Nguyen},
  \citenamefont {Hoang}, \citenamefont {Phuc}, \citenamefont {Sin},\ and\
  \citenamefont {Nguyen}}]{Nguyen2021-JPCL-BP-MoGe2N4-hetero}%
  \BibitemOpen
  \bibfield  {author} {\bibinfo {author} {\bibfnamefont {C.}~\bibnamefont
  {Nguyen}}, \bibinfo {author} {\bibfnamefont {N.~V.}\ \bibnamefont {Hoang}},
  \bibinfo {author} {\bibfnamefont {H.~V.}\ \bibnamefont {Phuc}}, \bibinfo
  {author} {\bibfnamefont {A.~Y.}\ \bibnamefont {Sin}},\ and\ \bibinfo {author}
  {\bibfnamefont {C.~V.}\ \bibnamefont {Nguyen}},\ }\bibfield  {title}
  {\bibinfo {title} {{Two-dimensional boron phosphide/MoGe$_2$N$_4$ van der
  Waals heterostructure: A promising tunable optoelectronic material}},\ }\href
  {https://doi.org/10.1021/acs.jpclett.1c01284} {\bibfield  {journal} {\bibinfo
   {journal} {Journal of Physical Chemistry Letters}\ }\textbf {\bibinfo
  {volume} {12}},\ \bibinfo {pages} {5076} (\bibinfo {year}
  {2021})}\BibitemShut {NoStop}%
\bibitem [{\citenamefont {Liang}\ \emph
  {et~al.}(2022{\natexlab{b}})\citenamefont {Liang}, \citenamefont {Luo},
  \citenamefont {Wang}, \citenamefont {Liang},\ and\ \citenamefont
  {Xie}}]{Liang2022-CPB-graphene-hetero}%
  \BibitemOpen
  \bibfield  {author} {\bibinfo {author} {\bibfnamefont {Q.}~\bibnamefont
  {Liang}}, \bibinfo {author} {\bibfnamefont {X.-y.}\ \bibnamefont {Luo}},
  \bibinfo {author} {\bibfnamefont {Y.-x.}\ \bibnamefont {Wang}}, \bibinfo
  {author} {\bibfnamefont {Y.-c.}\ \bibnamefont {Liang}},\ and\ \bibinfo
  {author} {\bibfnamefont {Q.}~\bibnamefont {Xie}},\ }\bibfield  {title}
  {\bibinfo {title} {{Modulation of Schottky barrier in XSi$_2$N$_4$~/~graphene
  (X=Mo and W) heterojunctions by biaxial strain}},\ }\href
  {https://doi.org/10.1088/1674-1056/ac5c3b} {\bibfield  {journal} {\bibinfo
  {journal} {Chinese Physics B}\ }\textbf {\bibinfo {volume} {31}},\ \bibinfo
  {pages} {087101} (\bibinfo {year} {2022}{\natexlab{b}})}\BibitemShut
  {NoStop}%
\bibitem [{\citenamefont {Cao}\ \emph {et~al.}(2021)\citenamefont {Cao},
  \citenamefont {Zhou}, \citenamefont {Wang}, \citenamefont {Ang},\ and\
  \citenamefont {Ang}}]{Cao2021-APL-GrapheneNbS2hetero}%
  \BibitemOpen
  \bibfield  {author} {\bibinfo {author} {\bibfnamefont {L.}~\bibnamefont
  {Cao}}, \bibinfo {author} {\bibfnamefont {G.}~\bibnamefont {Zhou}}, \bibinfo
  {author} {\bibfnamefont {Q.}~\bibnamefont {Wang}}, \bibinfo {author}
  {\bibfnamefont {L.~K.}\ \bibnamefont {Ang}},\ and\ \bibinfo {author}
  {\bibfnamefont {Y.~S.}\ \bibnamefont {Ang}},\ }\bibfield  {title} {\bibinfo
  {title} {{Two-dimensional van der Waals electrical contact to monolayer
  MoSi$_2$N$_4$}},\ }\href {https://doi.org/10.1063/5.0033241} {\bibfield
  {journal} {\bibinfo  {journal} {Applied Physics Letters}\ }\textbf {\bibinfo
  {volume} {118}},\ \bibinfo {pages} {013106} (\bibinfo {year}
  {2021})}\BibitemShut {NoStop}%
\bibitem [{\citenamefont {Pham}\ \emph {et~al.}(2021)\citenamefont {Pham},
  \citenamefont {Nguyen}, \citenamefont {Nguyen}, \citenamefont {Cuong},\ and\
  \citenamefont {Hieu}}]{Pham2021-NJP-GR-schottky}%
  \BibitemOpen
  \bibfield  {author} {\bibinfo {author} {\bibfnamefont {K.~D.}\ \bibnamefont
  {Pham}}, \bibinfo {author} {\bibfnamefont {C.~Q.}\ \bibnamefont {Nguyen}},
  \bibinfo {author} {\bibfnamefont {C.~V.}\ \bibnamefont {Nguyen}}, \bibinfo
  {author} {\bibfnamefont {P.~V.}\ \bibnamefont {Cuong}},\ and\ \bibinfo
  {author} {\bibfnamefont {N.~V.}\ \bibnamefont {Hieu}},\ }\bibfield  {title}
  {\bibinfo {title} {{Two-dimensional van der Waals graphene/transition metal
  nitride heterostructures as promising high-performance nanodevices}},\ }\href
  {https://doi.org/10.1039/d1nj00374g} {\bibfield  {journal} {\bibinfo
  {journal} {New Journal of Chemistry}\ }\textbf {\bibinfo {volume} {45}},\
  \bibinfo {pages} {5509} (\bibinfo {year} {2021})}\BibitemShut {NoStop}%
\bibitem [{\citenamefont {Binh}\ \emph {et~al.}(2021)\citenamefont {Binh},
  \citenamefont {Nguyen}, \citenamefont {Vu},\ and\ \citenamefont
  {Nguyen}}]{Binh2021-JPCL-GR-MoGeSiN4-hetero}%
  \BibitemOpen
  \bibfield  {author} {\bibinfo {author} {\bibfnamefont {N.~T.}\ \bibnamefont
  {Binh}}, \bibinfo {author} {\bibfnamefont {C.~Q.}\ \bibnamefont {Nguyen}},
  \bibinfo {author} {\bibfnamefont {T.~V.}\ \bibnamefont {Vu}},\ and\ \bibinfo
  {author} {\bibfnamefont {C.~V.}\ \bibnamefont {Nguyen}},\ }\bibfield  {title}
  {\bibinfo {title} {{Interfacial electronic properties and tunable contact
  types in Graphene/Janus MoGeSiN$_4$ heterostructures}},\ }\href
  {https://doi.org/10.1021/acs.jpclett.1c00682} {\bibfield  {journal} {\bibinfo
   {journal} {Journal of Physical Chemistry Letters}\ }\textbf {\bibinfo
  {volume} {12}},\ \bibinfo {pages} {3934} (\bibinfo {year}
  {2021})}\BibitemShut {NoStop}%
\bibitem [{\citenamefont {Nguyen}\ \emph
  {et~al.}(2022{\natexlab{b}})\citenamefont {Nguyen}, \citenamefont {Nguyen},
  \citenamefont {Nguyen}, \citenamefont {Ang},\ and\ \citenamefont
  {Hieu}}]{Nguyen2022-JPCL-MoSH-hetero}%
  \BibitemOpen
  \bibfield  {author} {\bibinfo {author} {\bibfnamefont {C.~V.}\ \bibnamefont
  {Nguyen}}, \bibinfo {author} {\bibfnamefont {C.~Q.}\ \bibnamefont {Nguyen}},
  \bibinfo {author} {\bibfnamefont {S.-T.}\ \bibnamefont {Nguyen}}, \bibinfo
  {author} {\bibfnamefont {Y.~S.}\ \bibnamefont {Ang}},\ and\ \bibinfo {author}
  {\bibfnamefont {N.~V.}\ \bibnamefont {Hieu}},\ }\bibfield  {title} {\bibinfo
  {title} {{Two-dimensional metal/semiconductor contact in a Janus
  MoSH/MoSi$_2$N$_4$ van der Waals heterostructure }},\ }\href
  {https://doi.org/10.1021/acs.jpclett.2c00245} {\bibfield  {journal} {\bibinfo
   {journal} {Journal of Physical Chemistry Letters}\ }\textbf {\bibinfo
  {volume} {13}},\ \bibinfo {pages} {2576} (\bibinfo {year}
  {2022}{\natexlab{b}})}\BibitemShut {NoStop}%
\bibitem [{\citenamefont {Pham}(2021)}]{Pham2021-RSCAdv-MoGeN-GR-MoSiN-hetero}%
  \BibitemOpen
  \bibfield  {author} {\bibinfo {author} {\bibfnamefont {D.~K.}\ \bibnamefont
  {Pham}},\ }\bibfield  {title} {\bibinfo {title} {{Electronic properties of a
  two-dimensional van der Waals MoGe$_2$N$_4$/MoSi$_2$N$_4$ heterobilayer:
  Effect of the insertion of a graphene layer and interlayer coupling}},\
  }\href {https://doi.org/10.1039/d1ra04531h} {\bibfield  {journal} {\bibinfo
  {journal} {RSC Advances}\ }\textbf {\bibinfo {volume} {11}},\ \bibinfo
  {pages} {28659} (\bibinfo {year} {2021})}\BibitemShut {NoStop}%
\bibitem [{\citenamefont {Xuefeng}\ \emph {et~al.}(2022)\citenamefont
  {Xuefeng}, \citenamefont {Wenna}, \citenamefont {Minglei}, \citenamefont
  {Fengzhu}, \citenamefont {Chengxiao}, \citenamefont {Qinfen}, \citenamefont
  {Bing},\ and\ \citenamefont {Huabing}}]{Xuefeng2022-JPD-BlueP-hetero}%
  \BibitemOpen
  \bibfield  {author} {\bibinfo {author} {\bibfnamefont {C.}~\bibnamefont
  {Xuefeng}}, \bibinfo {author} {\bibfnamefont {H.}~\bibnamefont {Wenna}},
  \bibinfo {author} {\bibfnamefont {J.}~\bibnamefont {Minglei}}, \bibinfo
  {author} {\bibfnamefont {R.}~\bibnamefont {Fengzhu}}, \bibinfo {author}
  {\bibfnamefont {P.}~\bibnamefont {Chengxiao}}, \bibinfo {author}
  {\bibfnamefont {G.}~\bibnamefont {Qinfen}}, \bibinfo {author} {\bibfnamefont
  {W.}~\bibnamefont {Bing}},\ and\ \bibinfo {author} {\bibfnamefont
  {Y.}~\bibnamefont {Huabing}},\ }\bibfield  {title} {\bibinfo {title} {{A
  direct Z-scheme MoSi$_2$N$_4$/BlueP vdW heterostructure for photocatalytic
  overall water splitting}},\ }\href {https://doi.org/10.1088/1361-6463/ac5662}
  {\bibfield  {journal} {\bibinfo  {journal} {Journal of Physics D: Applied
  Physics}\ }\textbf {\bibinfo {volume} {55}},\ \bibinfo {pages} {215502}
  (\bibinfo {year} {2022})}\BibitemShut {NoStop}%
\bibitem [{\citenamefont {Bafekry}\ \emph
  {et~al.}(2021{\natexlab{b}})\citenamefont {Bafekry}, \citenamefont {Faraji},
  \citenamefont {{Abdollahzadeh Ziabari}}, \citenamefont {Fadlallah},
  \citenamefont {Nguyen}, \citenamefont {Ghergherehchi},\ and\ \citenamefont
  {Feghhi}}]{Bafekry2021-NJC-MoS2-Hetero}%
  \BibitemOpen
  \bibfield  {author} {\bibinfo {author} {\bibfnamefont {A.}~\bibnamefont
  {Bafekry}}, \bibinfo {author} {\bibfnamefont {M.}~\bibnamefont {Faraji}},
  \bibinfo {author} {\bibfnamefont {A.}~\bibnamefont {{Abdollahzadeh
  Ziabari}}}, \bibinfo {author} {\bibfnamefont {M.~M.}\ \bibnamefont
  {Fadlallah}}, \bibinfo {author} {\bibfnamefont {C.~V.}\ \bibnamefont
  {Nguyen}}, \bibinfo {author} {\bibfnamefont {M.}~\bibnamefont
  {Ghergherehchi}},\ and\ \bibinfo {author} {\bibfnamefont {S.~A.}\
  \bibnamefont {Feghhi}},\ }\bibfield  {title} {\bibinfo {title} {{A van der
  Waals heterostructure of MoS$_2$/MoSi$_2$N$_4$: A first-principles study}},\
  }\href {https://doi.org/10.1039/d1nj00344e} {\bibfield  {journal} {\bibinfo
  {journal} {New Journal of Chemistry}\ }\textbf {\bibinfo {volume} {45}},\
  \bibinfo {pages} {8291} (\bibinfo {year} {2021}{\natexlab{b}})}\BibitemShut
  {NoStop}%
\bibitem [{\citenamefont {Wang}\ \emph
  {et~al.}(2021{\natexlab{c}})\citenamefont {Wang}, \citenamefont {Zhao},
  \citenamefont {Hu}, \citenamefont {Ren},\ and\ \citenamefont
  {Yuan}}]{Wang2021-Nanomaterials-hetero-strain}%
  \BibitemOpen
  \bibfield  {author} {\bibinfo {author} {\bibfnamefont {J.}~\bibnamefont
  {Wang}}, \bibinfo {author} {\bibfnamefont {X.}~\bibnamefont {Zhao}}, \bibinfo
  {author} {\bibfnamefont {G.}~\bibnamefont {Hu}}, \bibinfo {author}
  {\bibfnamefont {J.}~\bibnamefont {Ren}},\ and\ \bibinfo {author}
  {\bibfnamefont {X.}~\bibnamefont {Yuan}},\ }\bibfield  {title} {\bibinfo
  {title} {{Manipulable electronic and optical properties of two-dimensional
  MoSTe/MoGe$_2$N$_4$ van der Waals heterostructures}},\ }\href
  {https://doi.org/10.3390/nano11123338} {\bibfield  {journal} {\bibinfo
  {journal} {Nanomaterials}\ }\textbf {\bibinfo {volume} {11}},\ \bibinfo
  {pages} {3338} (\bibinfo {year} {2021}{\natexlab{c}})}\BibitemShut {NoStop}%
\bibitem [{\citenamefont {Yadav}\ \emph {et~al.}(2021)\citenamefont {Yadav},
  \citenamefont {Kangsabanik}, \citenamefont {Singh},\ and\ \citenamefont
  {Alam}}]{Yadav2021-MA2N4-photovoltaic-photocatalytic}%
  \BibitemOpen
  \bibfield  {author} {\bibinfo {author} {\bibfnamefont {A.}~\bibnamefont
  {Yadav}}, \bibinfo {author} {\bibfnamefont {J.}~\bibnamefont {Kangsabanik}},
  \bibinfo {author} {\bibfnamefont {N.}~\bibnamefont {Singh}},\ and\ \bibinfo
  {author} {\bibfnamefont {A.}~\bibnamefont {Alam}},\ }\bibfield  {title}
  {\bibinfo {title} {{Novel two-dimensional MA$_2$N$_4$ materials for
  photovoltaic and spintronic applications}},\ }\href
  {https://doi.org/10.1021/acs.jpclett.1c02650} {\bibfield  {journal} {\bibinfo
   {journal} {Journal of Physical Chemistry Letters}\ }\textbf {\bibinfo
  {volume} {12}},\ \bibinfo {pages} {10120} (\bibinfo {year}
  {2021})}\BibitemShut {NoStop}%
\bibitem [{\citenamefont {{Norouzi Azizabad}}\ and\ \citenamefont
  {Alavi-Rad}(2021)}]{Norouzi2021-PhysScrip-WSi2N4-G0W0}%
  \BibitemOpen
  \bibfield  {author} {\bibinfo {author} {\bibfnamefont {M.}~\bibnamefont
  {{Norouzi Azizabad}}}\ and\ \bibinfo {author} {\bibfnamefont
  {H.}~\bibnamefont {Alavi-Rad}},\ }\bibfield  {title} {\bibinfo {title}
  {{Quasiparticle and excitonic effects in WSi$_2$N$_4$ monolayer}},\ }\href
  {https://doi.org/10.1088/1402-4896/ac2858} {\bibfield  {journal} {\bibinfo
  {journal} {Physica Scripta}\ }\textbf {\bibinfo {volume} {96}},\ \bibinfo
  {pages} {125826} (\bibinfo {year} {2021})}\BibitemShut {NoStop}%
\bibitem [{\citenamefont {Yu}\ \emph {et~al.}(2021{\natexlab{b}})\citenamefont
  {Yu}, \citenamefont {Zhou}, \citenamefont {Guo},\ and\ \citenamefont
  {Sun}}]{Yu2021-AMI-Janus-HER-OER}%
  \BibitemOpen
  \bibfield  {author} {\bibinfo {author} {\bibfnamefont {Y.}~\bibnamefont
  {Yu}}, \bibinfo {author} {\bibfnamefont {J.}~\bibnamefont {Zhou}}, \bibinfo
  {author} {\bibfnamefont {Z.}~\bibnamefont {Guo}},\ and\ \bibinfo {author}
  {\bibfnamefont {Z.}~\bibnamefont {Sun}},\ }\bibfield  {title} {\bibinfo
  {title} {{Novel two-dimensional Janus MoSiGeN$_4$ and WSiGeN$_4$ as highly
  efficient photocatalysts for spontaneous overall water splitting}},\ }\href
  {https://doi.org/10.1021/acsami.1c04138} {\bibfield  {journal} {\bibinfo
  {journal} {ACS Applied Materials and Interfaces}\ }\textbf {\bibinfo {volume}
  {13}},\ \bibinfo {pages} {28090} (\bibinfo {year}
  {2021}{\natexlab{b}})}\BibitemShut {NoStop}%
\bibitem [{\citenamefont {Jian}\ \emph {et~al.}(2021)\citenamefont {Jian},
  \citenamefont {Ma}, \citenamefont {Zhang},\ and\ \citenamefont
  {Yong}}]{Jian2021-JPCC-monoMoSi2N4-strain}%
  \BibitemOpen
  \bibfield  {author} {\bibinfo {author} {\bibfnamefont {C.-c.}\ \bibnamefont
  {Jian}}, \bibinfo {author} {\bibfnamefont {X.}~\bibnamefont {Ma}}, \bibinfo
  {author} {\bibfnamefont {J.}~\bibnamefont {Zhang}},\ and\ \bibinfo {author}
  {\bibfnamefont {X.}~\bibnamefont {Yong}},\ }\bibfield  {title} {\bibinfo
  {title} {{Strained MoSi$_2$N$_4$ monolayers with excellent solar energy
  absorption and carrier transport properties}},\ }\href
  {https://doi.org/10.1021/acs.jpcc.1c03585} {\bibfield  {journal} {\bibinfo
  {journal} {Journal of Physical Chemistry C}\ }\textbf {\bibinfo {volume}
  {125}},\ \bibinfo {pages} {15185} (\bibinfo {year} {2021})}\BibitemShut
  {NoStop}%
\bibitem [{\citenamefont {Lv}\ \emph {et~al.}(2022)\citenamefont {Lv},
  \citenamefont {Xu}, \citenamefont {Mao}, \citenamefont {Liu}, \citenamefont
  {Zhao},\ and\ \citenamefont {Yang}}]{Lv2022-PhysE-monolayer}%
  \BibitemOpen
  \bibfield  {author} {\bibinfo {author} {\bibfnamefont {X.}~\bibnamefont
  {Lv}}, \bibinfo {author} {\bibfnamefont {Y.}~\bibnamefont {Xu}}, \bibinfo
  {author} {\bibfnamefont {B.}~\bibnamefont {Mao}}, \bibinfo {author}
  {\bibfnamefont {G.}~\bibnamefont {Liu}}, \bibinfo {author} {\bibfnamefont
  {G.}~\bibnamefont {Zhao}},\ and\ \bibinfo {author} {\bibfnamefont
  {J.}~\bibnamefont {Yang}},\ }\bibfield  {title} {\bibinfo {title} {{Strain
  modulation of electronic and optical properties of monolayer
  MoSi$_2$N$_4$}},\ }\href {https://doi.org/10.1016/j.physe.2021.114964}
  {\bibfield  {journal} {\bibinfo  {journal} {Physica E: Low-Dimensional
  Systems and Nanostructures}\ }\textbf {\bibinfo {volume} {135}},\ \bibinfo
  {pages} {114964} (\bibinfo {year} {2022})}\BibitemShut {NoStop}%
\bibitem [{\citenamefont {Mwankemwa}\ \emph {et~al.}(2022)\citenamefont
  {Mwankemwa}, \citenamefont {Wang}, \citenamefont {Zhu}, \citenamefont {Fan},
  \citenamefont {Zhang},\ and\ \citenamefont
  {Zhang}}]{Mwankemwa2022-ResPhys-MoSi2N4-MoSiGeN4}%
  \BibitemOpen
  \bibfield  {author} {\bibinfo {author} {\bibfnamefont {N.}~\bibnamefont
  {Mwankemwa}}, \bibinfo {author} {\bibfnamefont {H.-e.}\ \bibnamefont {Wang}},
  \bibinfo {author} {\bibfnamefont {T.}~\bibnamefont {Zhu}}, \bibinfo {author}
  {\bibfnamefont {Q.}~\bibnamefont {Fan}}, \bibinfo {author} {\bibfnamefont
  {F.}~\bibnamefont {Zhang}},\ and\ \bibinfo {author} {\bibfnamefont
  {W.}~\bibnamefont {Zhang}},\ }\bibfield  {title} {\bibinfo {title} {{First
  principles calculations investigation of optoelectronic properties and
  photocatalytic CO$_2$ reduction of (MoSi$_2$N$_4$)$_{5-n}$/(MoSiGeN$_4$)$_n$
  in-plane heterostructures}},\ }\href
  {https://doi.org/10.1016/j.rinp.2022.105549} {\bibfield  {journal} {\bibinfo
  {journal} {Results in Physics}\ }\textbf {\bibinfo {volume} {37}},\ \bibinfo
  {pages} {105549} (\bibinfo {year} {2022})}\BibitemShut {NoStop}%
\bibitem [{\citenamefont {Bafekry}\ \emph
  {et~al.}(2021{\natexlab{c}})\citenamefont {Bafekry}, \citenamefont {Stampfl},
  \citenamefont {Naseri}, \citenamefont {Fadlallah}, \citenamefont {Faraji},
  \citenamefont {Ghergherehchi}, \citenamefont {Gogova},\ and\ \citenamefont
  {Feghhi}}]{Bafekry2021-JAP-biMoSi2N4-ElecField-strain}%
  \BibitemOpen
  \bibfield  {author} {\bibinfo {author} {\bibfnamefont {A.}~\bibnamefont
  {Bafekry}}, \bibinfo {author} {\bibfnamefont {C.}~\bibnamefont {Stampfl}},
  \bibinfo {author} {\bibfnamefont {M.}~\bibnamefont {Naseri}}, \bibinfo
  {author} {\bibfnamefont {M.~M.}\ \bibnamefont {Fadlallah}}, \bibinfo {author}
  {\bibfnamefont {M.}~\bibnamefont {Faraji}}, \bibinfo {author} {\bibfnamefont
  {M.}~\bibnamefont {Ghergherehchi}}, \bibinfo {author} {\bibfnamefont
  {D.}~\bibnamefont {Gogova}},\ and\ \bibinfo {author} {\bibfnamefont {S.~A.}\
  \bibnamefont {Feghhi}},\ }\bibfield  {title} {\bibinfo {title} {{Effect of
  electric field and vertical strain on the electro-optical properties of the
  MoSi$_2$N$_4$ bilayer: A first-principles calculation}},\ }\href
  {https://doi.org/10.1063/5.0044976} {\bibfield  {journal} {\bibinfo
  {journal} {Journal of Applied Physics}\ }\textbf {\bibinfo {volume} {129}},\
  \bibinfo {pages} {155103} (\bibinfo {year} {2021}{\natexlab{c}})}\BibitemShut
  {NoStop}%
\bibitem [{\citenamefont {Xu}\ \emph {et~al.}(2022)\citenamefont {Xu},
  \citenamefont {Wu}, \citenamefont {Sun}, \citenamefont {Mwankemwa},
  \citenamefont {bin Zhang},\ and\ \citenamefont {xing
  Yang}}]{Xu2022-JPCS-AuAdsorption}%
  \BibitemOpen
  \bibfield  {author} {\bibinfo {author} {\bibfnamefont {J.}~\bibnamefont
  {Xu}}, \bibinfo {author} {\bibfnamefont {Q.}~\bibnamefont {Wu}}, \bibinfo
  {author} {\bibfnamefont {Z.}~\bibnamefont {Sun}}, \bibinfo {author}
  {\bibfnamefont {N.}~\bibnamefont {Mwankemwa}}, \bibinfo {author}
  {\bibfnamefont {W.}~\bibnamefont {bin Zhang}},\ and\ \bibinfo {author}
  {\bibfnamefont {W.}~\bibnamefont {xing Yang}},\ }\bibfield  {title} {\bibinfo
  {title} {{First-principles investigations of electronic, optical, and
  photocatalytic properties of Au-adsorbed MoSi$_2$N$_4$ monolayer}},\ }\href
  {https://doi.org/10.1016/j.jpcs.2021.110494} {\bibfield  {journal} {\bibinfo
  {journal} {Journal of Physics and Chemistry of Solids}\ }\textbf {\bibinfo
  {volume} {162}},\ \bibinfo {pages} {110494} (\bibinfo {year}
  {2022})}\BibitemShut {NoStop}%
\bibitem [{\citenamefont {Sun}\ \emph {et~al.}(2022)\citenamefont {Sun},
  \citenamefont {Xu}, \citenamefont {Mwankemwa}, \citenamefont {Yang},
  \citenamefont {Wu}, \citenamefont {Yi}, \citenamefont {Chen},\ and\
  \citenamefont {Zhang}}]{Sun2022-CommuniTheoerticPhys-Li-Na-K}%
  \BibitemOpen
  \bibfield  {author} {\bibinfo {author} {\bibfnamefont {Z.}~\bibnamefont
  {Sun}}, \bibinfo {author} {\bibfnamefont {J.}~\bibnamefont {Xu}}, \bibinfo
  {author} {\bibfnamefont {N.}~\bibnamefont {Mwankemwa}}, \bibinfo {author}
  {\bibfnamefont {W.}~\bibnamefont {Yang}}, \bibinfo {author} {\bibfnamefont
  {X.}~\bibnamefont {Wu}}, \bibinfo {author} {\bibfnamefont {Z.}~\bibnamefont
  {Yi}}, \bibinfo {author} {\bibfnamefont {S.}~\bibnamefont {Chen}},\ and\
  \bibinfo {author} {\bibfnamefont {W.}~\bibnamefont {Zhang}},\ }\bibfield
  {title} {\bibinfo {title} {{Alkali-metal (Li, Na, and K)-adsorbed
  MoSi$_2$N$_4$ monolayer: An investigation of its outstanding electronic,
  optical, and photocatalytic properties}},\ }\href
  {https://doi.org/10.1088/1572-9494/ac3ada} {\bibfield  {journal} {\bibinfo
  {journal} {Communications in Theoretical Physics}\ }\textbf {\bibinfo
  {volume} {74}},\ \bibinfo {pages} {015503} (\bibinfo {year}
  {2022})}\BibitemShut {NoStop}%
\bibitem [{\citenamefont {Chen}\ \emph
  {et~al.}(2021{\natexlab{a}})\citenamefont {Chen}, \citenamefont {Chen},\ and\
  \citenamefont {Zhang}}]{Chen2021-ResPhys-F-MoSi2N4}%
  \BibitemOpen
  \bibfield  {author} {\bibinfo {author} {\bibfnamefont {R.}~\bibnamefont
  {Chen}}, \bibinfo {author} {\bibfnamefont {D.}~\bibnamefont {Chen}},\ and\
  \bibinfo {author} {\bibfnamefont {W.}~\bibnamefont {Zhang}},\ }\bibfield
  {title} {\bibinfo {title} {{First-principles calculations to investigate
  stability, electronic and optical properties of fluorinated MoSi$_2$N$_4$
  monolayer}},\ }\href {https://doi.org/10.1016/j.rinp.2021.104864} {\bibfield
  {journal} {\bibinfo  {journal} {Results in Physics}\ }\textbf {\bibinfo
  {volume} {30}},\ \bibinfo {pages} {104864} (\bibinfo {year}
  {2021}{\natexlab{a}})}\BibitemShut {NoStop}%
\bibitem [{\citenamefont {Zeng}\ \emph {et~al.}(2021)\citenamefont {Zeng},
  \citenamefont {Xu}, \citenamefont {Yang}, \citenamefont {Luo}, \citenamefont
  {Li}, \citenamefont {Xiong},\ and\ \citenamefont
  {Wang}}]{Zeng2021-PCCP-optical-C2N-herto}%
  \BibitemOpen
  \bibfield  {author} {\bibinfo {author} {\bibfnamefont {J.}~\bibnamefont
  {Zeng}}, \bibinfo {author} {\bibfnamefont {L.}~\bibnamefont {Xu}}, \bibinfo
  {author} {\bibfnamefont {Y.}~\bibnamefont {Yang}}, \bibinfo {author}
  {\bibfnamefont {X.}~\bibnamefont {Luo}}, \bibinfo {author} {\bibfnamefont
  {H.~J.}\ \bibnamefont {Li}}, \bibinfo {author} {\bibfnamefont {S.~X.}\
  \bibnamefont {Xiong}},\ and\ \bibinfo {author} {\bibfnamefont {L.~L.}\
  \bibnamefont {Wang}},\ }\bibfield  {title} {\bibinfo {title} {{Boosting the
  photocatalytic hydrogen evolution performance of monolayer C2N coupled with
  MoSi2N4: Density-functional theory calculations}},\ }\href
  {https://doi.org/10.1039/d1cp00364j} {\bibfield  {journal} {\bibinfo
  {journal} {Physical Chemistry Chemical Physics}\ }\textbf {\bibinfo {volume}
  {23}},\ \bibinfo {pages} {8318} (\bibinfo {year} {2021})}\BibitemShut
  {NoStop}%
\bibitem [{\citenamefont {Chen}\ and\ \citenamefont
  {Tang}(2021)}]{Chen2021-CAEJ-MA2Z4-magnetic}%
  \BibitemOpen
  \bibfield  {author} {\bibinfo {author} {\bibfnamefont {J.}~\bibnamefont
  {Chen}}\ and\ \bibinfo {author} {\bibfnamefont {Q.}~\bibnamefont {Tang}},\
  }\bibfield  {title} {\bibinfo {title} {{The versatile electronic, magnetic
  and photo-electro catalytic activity of a new 2D MA$_2$Z$_4$ Family}},\
  }\href {https://doi.org/10.1002/chem.202100851} {\bibfield  {journal}
  {\bibinfo  {journal} {Chemistry - A European Journal}\ }\textbf {\bibinfo
  {volume} {27}},\ \bibinfo {pages} {9925} (\bibinfo {year}
  {2021})}\BibitemShut {NoStop}%
\bibitem [{\citenamefont {Formed}\ \emph {et~al.}(2022)\citenamefont {Formed},
  \citenamefont {Yoon}, \citenamefont {Lee}, \citenamefont {Touski},
  \citenamefont {Xu}, \citenamefont {Mao},\ and\ \citenamefont
  {Xie}}]{Formed2022-JPDAP-MoSi2N4-NRs}%
  \BibitemOpen
  \bibfield  {author} {\bibinfo {author} {\bibfnamefont {C.}~\bibnamefont
  {Formed}}, \bibinfo {author} {\bibfnamefont {J.-k.}\ \bibnamefont {Yoon}},
  \bibinfo {author} {\bibfnamefont {K.-h.}\ \bibnamefont {Lee}}, \bibinfo
  {author} {\bibfnamefont {S.~B.}\ \bibnamefont {Touski}}, \bibinfo {author}
  {\bibfnamefont {J.}~\bibnamefont {Xu}}, \bibinfo {author} {\bibfnamefont
  {X.}~\bibnamefont {Mao}},\ and\ \bibinfo {author} {\bibfnamefont {Z.-h.}\
  \bibnamefont {Xie}},\ }\bibfield  {title} {\bibinfo {title} {{Band-gap
  engineering, magnetic behavior and Dirac-semimetal character in the
  MoSi$_2$N$_4$ nanoribbon with armchair and zigzag edges}},\ }\href
  {https://doi.org/10.1088/1361-6463/ac2cab} {\bibfield  {journal} {\bibinfo
  {journal} {Journal of Physics D: Applied Physics}\ }\textbf {\bibinfo
  {volume} {55}},\ \bibinfo {pages} {035301} (\bibinfo {year}
  {2022})}\BibitemShut {NoStop}%
\bibitem [{\citenamefont {Ding}\ and\ \citenamefont
  {Wang}(2022)}]{Ding2022-ASS-MoN2X2Y2}%
  \BibitemOpen
  \bibfield  {author} {\bibinfo {author} {\bibfnamefont {Y.}~\bibnamefont
  {Ding}}\ and\ \bibinfo {author} {\bibfnamefont {Y.}~\bibnamefont {Wang}},\
  }\bibfield  {title} {\bibinfo {title} {{First-principles study of
  two-dimensional MoN$_2$X$_2$Y$_2$ (X=B--In, Y=N--Te) nanosheets: The III-VI
  analogues of MoSi$_2$N$_4$ with peculiar electronic and magnetic
  properties}},\ }\href {https://doi.org/10.1016/j.apsusc.2022.153317}
  {\bibfield  {journal} {\bibinfo  {journal} {Applied Surface Science}\
  }\textbf {\bibinfo {volume} {593}},\ \bibinfo {pages} {153317} (\bibinfo
  {year} {2022})}\BibitemShut {NoStop}%
\bibitem [{\citenamefont {Feng}\ \emph {et~al.}(2022)\citenamefont {Feng},
  \citenamefont {Wang}, \citenamefont {Zuo},\ and\ \citenamefont
  {Gao}}]{Feng2022-APL-VSi2X4-mag}%
  \BibitemOpen
  \bibfield  {author} {\bibinfo {author} {\bibfnamefont {Y.}~\bibnamefont
  {Feng}}, \bibinfo {author} {\bibfnamefont {Z.}~\bibnamefont {Wang}}, \bibinfo
  {author} {\bibfnamefont {X.}~\bibnamefont {Zuo}},\ and\ \bibinfo {author}
  {\bibfnamefont {G.}~\bibnamefont {Gao}},\ }\bibfield  {title} {\bibinfo
  {title} {{Electronic phase transition, spin filtering effect, and spin
  Seebeck effect in 2D high-spin-polarized VSi$_2$X$_4$ (X = N, P, As)}},\
  }\href {https://doi.org/10.1063/5.0086990} {\bibfield  {journal} {\bibinfo
  {journal} {Applied Physics Letters}\ }\textbf {\bibinfo {volume} {120}},\
  \bibinfo {pages} {092405} (\bibinfo {year} {2022})}\BibitemShut {NoStop}%
\bibitem [{\citenamefont {Dey}\ \emph {et~al.}(2022)\citenamefont {Dey},
  \citenamefont {Ray},\ and\ \citenamefont
  {Yu}}]{Dey2022-arxiv-MAZ-VSiXN4-Janus-mag}%
  \BibitemOpen
  \bibfield  {author} {\bibinfo {author} {\bibfnamefont {D.}~\bibnamefont
  {Dey}}, \bibinfo {author} {\bibfnamefont {A.}~\bibnamefont {Ray}},\ and\
  \bibinfo {author} {\bibfnamefont {L.}~\bibnamefont {Yu}},\ }\bibfield
  {title} {\bibinfo {title} {{Intrinsic ferromagnetism and restrictive
  thermodynamic stability in MA$_2$N$_4$ and Janus VSiGeN$_4$ monolayers}},\
  }\href {http://arxiv.org/abs/2203.11605} {\bibfield  {journal} {\bibinfo
  {journal} {arXiv:2203.11605}\ } (\bibinfo {year} {2022})}\BibitemShut
  {NoStop}%
\bibitem [{\citenamefont {Wang}(2017)}]{Wang2017-NSR-SpinGaplessSemiconductor}%
  \BibitemOpen
  \bibfield  {author} {\bibinfo {author} {\bibfnamefont {X.~L.}\ \bibnamefont
  {Wang}},\ }\bibfield  {title} {\bibinfo {title} {{Dirac spin-gapless
  semiconductors: Promising platforms for massless and dissipationless
  spintronics and new (quantum) anomalous spin Hall effects}},\ }\href
  {https://doi.org/10.1093/nsr/nww069} {\bibfield  {journal} {\bibinfo
  {journal} {National Science Review}\ }\textbf {\bibinfo {volume} {4}},\
  \bibinfo {pages} {252} (\bibinfo {year} {2017})}\BibitemShut {NoStop}%
\bibitem [{\citenamefont {Wang}\ \emph {et~al.}(2018)\citenamefont {Wang},
  \citenamefont {Li}, \citenamefont {Cheng}, \citenamefont {Wang},\ and\
  \citenamefont {Chen}}]{Wang2018-APL-SpinGaplessSemiconductor}%
  \BibitemOpen
  \bibfield  {author} {\bibinfo {author} {\bibfnamefont {X.}~\bibnamefont
  {Wang}}, \bibinfo {author} {\bibfnamefont {T.}~\bibnamefont {Li}}, \bibinfo
  {author} {\bibfnamefont {Z.}~\bibnamefont {Cheng}}, \bibinfo {author}
  {\bibfnamefont {X.~L.}\ \bibnamefont {Wang}},\ and\ \bibinfo {author}
  {\bibfnamefont {H.}~\bibnamefont {Chen}},\ }\bibfield  {title} {\bibinfo
  {title} {{Recent advances in Dirac spin-gapless semiconductors}},\ }\href
  {https://doi.org/10.1063/1.5042604} {\bibfield  {journal} {\bibinfo
  {journal} {Applied Physics Reviews}\ }\textbf {\bibinfo {volume} {5}},\
  \bibinfo {pages} {041103} (\bibinfo {year} {2018})}\BibitemShut {NoStop}%
\bibitem [{\citenamefont {Cui}\ \emph {et~al.}(2021{\natexlab{b}})\citenamefont
  {Cui}, \citenamefont {Zhu}, \citenamefont {Liang}, \citenamefont {Cui},\ and\
  \citenamefont {Yang}}]{Cui2021-PRB-VSi2N4-mag-spinvalley}%
  \BibitemOpen
  \bibfield  {author} {\bibinfo {author} {\bibfnamefont {Q.}~\bibnamefont
  {Cui}}, \bibinfo {author} {\bibfnamefont {Y.}~\bibnamefont {Zhu}}, \bibinfo
  {author} {\bibfnamefont {J.}~\bibnamefont {Liang}}, \bibinfo {author}
  {\bibfnamefont {P.}~\bibnamefont {Cui}},\ and\ \bibinfo {author}
  {\bibfnamefont {H.}~\bibnamefont {Yang}},\ }\bibfield  {title} {\bibinfo
  {title} {{Spin-valley coupling in a two-dimensional VSi$_2$N$_4$
  monolayer}},\ }\href {https://doi.org/10.1103/PhysRevB.103.085421} {\bibfield
   {journal} {\bibinfo  {journal} {Physical Review B}\ }\textbf {\bibinfo
  {volume} {103}},\ \bibinfo {pages} {085421} (\bibinfo {year}
  {2021}{\natexlab{b}})}\BibitemShut {NoStop}%
\bibitem [{\citenamefont {Akanda}\ and\ \citenamefont
  {Lake}(2021)}]{Akanda2021-APL-Nb-VSi2N4-VSi2P4-mag}%
  \BibitemOpen
  \bibfield  {author} {\bibinfo {author} {\bibfnamefont {M.~R.~K.}\
  \bibnamefont {Akanda}}\ and\ \bibinfo {author} {\bibfnamefont {R.~K.}\
  \bibnamefont {Lake}},\ }\bibfield  {title} {\bibinfo {title} {{Magnetic
  properties of NbSi$_2$N$_4$, VSi$_2$N$_4$, and VSi$_2$P$_4$ monolayers}},\
  }\href {https://doi.org/10.1063/5.0055878} {\bibfield  {journal} {\bibinfo
  {journal} {Applied Physics Letters}\ }\textbf {\bibinfo {volume} {119}},\
  \bibinfo {pages} {052402} (\bibinfo {year} {2021})}\BibitemShut {NoStop}%
\bibitem [{\citenamefont {Xue}\ \emph {et~al.}(2022)\citenamefont {Xue},
  \citenamefont {He}, \citenamefont {Gong}, \citenamefont {Yi},\ and\
  \citenamefont {Guo}}]{XUE20221Nonlinear}%
  \BibitemOpen
  \bibfield  {author} {\bibinfo {author} {\bibfnamefont {M.}~\bibnamefont
  {Xue}}, \bibinfo {author} {\bibfnamefont {W.}~\bibnamefont {He}}, \bibinfo
  {author} {\bibfnamefont {Q.}~\bibnamefont {Gong}}, \bibinfo {author}
  {\bibfnamefont {M.}~\bibnamefont {Yi}},\ and\ \bibinfo {author}
  {\bibfnamefont {W.}~\bibnamefont {Guo}},\ }\bibfield  {title} {\bibinfo
  {title} {Nonlinear elasticity and strain-tunable magnetocalorics of
  antiferromagnetic monolayer mnps3},\ }\href
  {https://doi.org/10.1016/j.eml.2022.101900} {\bibfield  {journal} {\bibinfo
  {journal} {Extreme Mechanics Letters}\ ,\ \bibinfo {pages} {101900}}
  (\bibinfo {year} {2022})}\BibitemShut {NoStop}%
\bibitem [{\citenamefont {Webster}\ and\ \citenamefont
  {Yan}(2018)}]{Webster2018-PRB-CrX3-Mag}%
  \BibitemOpen
  \bibfield  {author} {\bibinfo {author} {\bibfnamefont {L.}~\bibnamefont
  {Webster}}\ and\ \bibinfo {author} {\bibfnamefont {J.~A.}\ \bibnamefont
  {Yan}},\ }\bibfield  {title} {\bibinfo {title} {{Strain-tunable magnetic
  anisotropy in monolayer CrCl$_3$, CrBr$_3$, and CrI$_3$}},\ }\href
  {https://doi.org/10.1103/PhysRevB.98.144411} {\bibfield  {journal} {\bibinfo
  {journal} {Physical Review B}\ }\textbf {\bibinfo {volume} {98}},\ \bibinfo
  {pages} {144411} (\bibinfo {year} {2018})}\BibitemShut {NoStop}%
\bibitem [{\citenamefont {Zheng}\ \emph {et~al.}(2019)\citenamefont {Zheng},
  \citenamefont {Huang}, \citenamefont {Yu}, \citenamefont {Xu}, \citenamefont
  {Zhang}, \citenamefont {Xu}, \citenamefont {Liu}, \citenamefont {Kan},
  \citenamefont {Wang},\ and\ \citenamefont {Yang}}]{Zheng2019-JPCL-Fe3P-Mag}%
  \BibitemOpen
  \bibfield  {author} {\bibinfo {author} {\bibfnamefont {S.}~\bibnamefont
  {Zheng}}, \bibinfo {author} {\bibfnamefont {C.}~\bibnamefont {Huang}},
  \bibinfo {author} {\bibfnamefont {T.}~\bibnamefont {Yu}}, \bibinfo {author}
  {\bibfnamefont {M.}~\bibnamefont {Xu}}, \bibinfo {author} {\bibfnamefont
  {S.}~\bibnamefont {Zhang}}, \bibinfo {author} {\bibfnamefont
  {H.}~\bibnamefont {Xu}}, \bibinfo {author} {\bibfnamefont {Y.}~\bibnamefont
  {Liu}}, \bibinfo {author} {\bibfnamefont {E.}~\bibnamefont {Kan}}, \bibinfo
  {author} {\bibfnamefont {Y.}~\bibnamefont {Wang}},\ and\ \bibinfo {author}
  {\bibfnamefont {G.}~\bibnamefont {Yang}},\ }\bibfield  {title} {\bibinfo
  {title} {{High-temperature ferromagnetism in an Fe$_3$P monolayer with a
  large magnetic anisotropy}},\ }\href
  {https://doi.org/10.1021/acs.jpclett.9b00970} {\bibfield  {journal} {\bibinfo
   {journal} {Journal of Physical Chemistry Letters}\ }\textbf {\bibinfo
  {volume} {10}},\ \bibinfo {pages} {2733} (\bibinfo {year}
  {2019})}\BibitemShut {NoStop}%
\bibitem [{\citenamefont {Zhuang}\ \emph {et~al.}(2016)\citenamefont {Zhuang},
  \citenamefont {Kent},\ and\ \citenamefont
  {Hennig}}]{Zhuang2016-PRB-FeGeTe-Mag}%
  \BibitemOpen
  \bibfield  {author} {\bibinfo {author} {\bibfnamefont {H.~L.}\ \bibnamefont
  {Zhuang}}, \bibinfo {author} {\bibfnamefont {P.~R.}\ \bibnamefont {Kent}},\
  and\ \bibinfo {author} {\bibfnamefont {R.~G.}\ \bibnamefont {Hennig}},\
  }\bibfield  {title} {\bibinfo {title} {{Strong anisotropy and
  magnetostriction in the two-dimensional Stoner ferromagnet Fe$_3$GeTe$_2$}},\
  }\href {https://doi.org/10.1103/PhysRevB.93.134407} {\bibfield  {journal}
  {\bibinfo  {journal} {Physical Review B}\ }\textbf {\bibinfo {volume} {93}},\
  \bibinfo {pages} {134407} (\bibinfo {year} {2016})}\BibitemShut {NoStop}%
\bibitem [{\citenamefont {Han}\ \emph {et~al.}(2020)\citenamefont {Han},
  \citenamefont {Zheng}, \citenamefont {Wang},\ and\ \citenamefont
  {Yan}}]{Han2020-PCCP-NiI2-magnetic}%
  \BibitemOpen
  \bibfield  {author} {\bibinfo {author} {\bibfnamefont {H.}~\bibnamefont
  {Han}}, \bibinfo {author} {\bibfnamefont {H.}~\bibnamefont {Zheng}}, \bibinfo
  {author} {\bibfnamefont {Q.}~\bibnamefont {Wang}},\ and\ \bibinfo {author}
  {\bibfnamefont {Y.}~\bibnamefont {Yan}},\ }\bibfield  {title} {\bibinfo
  {title} {{Enhanced magnetic anisotropy and Curie temperature of the NiI$_2$
  monolayer by applying strain: A first-principles study}},\ }\href
  {https://doi.org/10.1039/d0cp03803b} {\bibfield  {journal} {\bibinfo
  {journal} {Physical Chemistry Chemical Physics}\ }\textbf {\bibinfo {volume}
  {22}},\ \bibinfo {pages} {26917} (\bibinfo {year} {2020})}\BibitemShut
  {NoStop}%
\bibitem [{\citenamefont {Torun}\ \emph {et~al.}(2015)\citenamefont {Torun},
  \citenamefont {Sahin}, \citenamefont {Bacaksiz}, \citenamefont {Senger},\
  and\ \citenamefont {Peeters}}]{Torun2015-PRB-FeCl2-Mag}%
  \BibitemOpen
  \bibfield  {author} {\bibinfo {author} {\bibfnamefont {E.}~\bibnamefont
  {Torun}}, \bibinfo {author} {\bibfnamefont {H.}~\bibnamefont {Sahin}},
  \bibinfo {author} {\bibfnamefont {C.}~\bibnamefont {Bacaksiz}}, \bibinfo
  {author} {\bibfnamefont {R.~T.}\ \bibnamefont {Senger}},\ and\ \bibinfo
  {author} {\bibfnamefont {F.~M.}\ \bibnamefont {Peeters}},\ }\bibfield
  {title} {\bibinfo {title} {{Tuning the magnetic anisotropy in single-layer
  crystal structures}},\ }\href {https://doi.org/10.1103/PhysRevB.92.104407}
  {\bibfield  {journal} {\bibinfo  {journal} {Physical Review B}\ }\textbf
  {\bibinfo {volume} {92}},\ \bibinfo {pages} {104407} (\bibinfo {year}
  {2015})}\BibitemShut {NoStop}%
\bibitem [{\citenamefont {Huang}\ \emph {et~al.}(2017)\citenamefont {Huang},
  \citenamefont {Clark}, \citenamefont {Navarro-Moratalla}, \citenamefont
  {Klein}, \citenamefont {Cheng}, \citenamefont {Seyler}, \citenamefont
  {Zhong}, \citenamefont {Schmidgall}, \citenamefont {McGuire}, \citenamefont
  {Cobden}, \citenamefont {Yao}, \citenamefont {Xiao}, \citenamefont
  {Jarillo-Herrero},\ and\ \citenamefont
  {Xu}}]{Huang2017-Nature-CrI3-ferromagnetism}%
  \BibitemOpen
  \bibfield  {author} {\bibinfo {author} {\bibfnamefont {B.}~\bibnamefont
  {Huang}}, \bibinfo {author} {\bibfnamefont {G.}~\bibnamefont {Clark}},
  \bibinfo {author} {\bibfnamefont {E.}~\bibnamefont {Navarro-Moratalla}},
  \bibinfo {author} {\bibfnamefont {D.~R.}\ \bibnamefont {Klein}}, \bibinfo
  {author} {\bibfnamefont {R.}~\bibnamefont {Cheng}}, \bibinfo {author}
  {\bibfnamefont {K.~L.}\ \bibnamefont {Seyler}}, \bibinfo {author}
  {\bibfnamefont {D.}~\bibnamefont {Zhong}}, \bibinfo {author} {\bibfnamefont
  {E.}~\bibnamefont {Schmidgall}}, \bibinfo {author} {\bibfnamefont {M.~A.}\
  \bibnamefont {McGuire}}, \bibinfo {author} {\bibfnamefont {D.~H.}\
  \bibnamefont {Cobden}}, \bibinfo {author} {\bibfnamefont {W.}~\bibnamefont
  {Yao}}, \bibinfo {author} {\bibfnamefont {D.}~\bibnamefont {Xiao}}, \bibinfo
  {author} {\bibfnamefont {P.}~\bibnamefont {Jarillo-Herrero}},\ and\ \bibinfo
  {author} {\bibfnamefont {X.}~\bibnamefont {Xu}},\ }\bibfield  {title}
  {\bibinfo {title} {{Layer-dependent ferromagnetism in a van der Waals crystal
  down to the monolayer limit}},\ }\href {https://doi.org/10.1038/nature22391}
  {\bibfield  {journal} {\bibinfo  {journal} {Nature}\ }\textbf {\bibinfo
  {volume} {546}},\ \bibinfo {pages} {270} (\bibinfo {year}
  {2017})}\BibitemShut {NoStop}%
\bibitem [{\citenamefont {Fei}\ \emph {et~al.}(2018{\natexlab{b}})\citenamefont
  {Fei}, \citenamefont {Huang}, \citenamefont {Malinowski}, \citenamefont
  {Wang}, \citenamefont {Song}, \citenamefont {Sanchez}, \citenamefont {Yao},
  \citenamefont {Xiao}, \citenamefont {Zhu}, \citenamefont {May}, \citenamefont
  {Wu}, \citenamefont {Cobden}, \citenamefont {Chu},\ and\ \citenamefont
  {Xu}}]{Fei2018-NatMater-Fe3GeTe2-ferromagnetism}%
  \BibitemOpen
  \bibfield  {author} {\bibinfo {author} {\bibfnamefont {Z.}~\bibnamefont
  {Fei}}, \bibinfo {author} {\bibfnamefont {B.}~\bibnamefont {Huang}}, \bibinfo
  {author} {\bibfnamefont {P.}~\bibnamefont {Malinowski}}, \bibinfo {author}
  {\bibfnamefont {W.}~\bibnamefont {Wang}}, \bibinfo {author} {\bibfnamefont
  {T.}~\bibnamefont {Song}}, \bibinfo {author} {\bibfnamefont {J.}~\bibnamefont
  {Sanchez}}, \bibinfo {author} {\bibfnamefont {W.}~\bibnamefont {Yao}},
  \bibinfo {author} {\bibfnamefont {D.}~\bibnamefont {Xiao}}, \bibinfo {author}
  {\bibfnamefont {X.}~\bibnamefont {Zhu}}, \bibinfo {author} {\bibfnamefont
  {A.~F.}\ \bibnamefont {May}}, \bibinfo {author} {\bibfnamefont
  {W.}~\bibnamefont {Wu}}, \bibinfo {author} {\bibfnamefont {D.~H.}\
  \bibnamefont {Cobden}}, \bibinfo {author} {\bibfnamefont {J.~H.}\
  \bibnamefont {Chu}},\ and\ \bibinfo {author} {\bibfnamefont {X.}~\bibnamefont
  {Xu}},\ }\bibfield  {title} {\bibinfo {title} {{Two-dimensional itinerant
  ferromagnetism in atomically thin Fe$_3$GeTe$_2$}},\ }\href
  {https://doi.org/10.1038/s41563-018-0149-7} {\bibfield  {journal} {\bibinfo
  {journal} {Nature Materials}\ }\textbf {\bibinfo {volume} {17}},\ \bibinfo
  {pages} {778} (\bibinfo {year} {2018}{\natexlab{b}})}\BibitemShut {NoStop}%
\bibitem [{\citenamefont {You}\ \emph {et~al.}(2020)\citenamefont {You},
  \citenamefont {Zhang}, \citenamefont {Dong}, \citenamefont {Gu},\ and\
  \citenamefont {Su}}]{You2020-PRRes-2D-Tc}%
  \BibitemOpen
  \bibfield  {author} {\bibinfo {author} {\bibfnamefont {J.~Y.}\ \bibnamefont
  {You}}, \bibinfo {author} {\bibfnamefont {Z.}~\bibnamefont {Zhang}}, \bibinfo
  {author} {\bibfnamefont {X.~J.}\ \bibnamefont {Dong}}, \bibinfo {author}
  {\bibfnamefont {B.}~\bibnamefont {Gu}},\ and\ \bibinfo {author}
  {\bibfnamefont {G.}~\bibnamefont {Su}},\ }\bibfield  {title} {\bibinfo
  {title} {{Two-dimensional magnetic semiconductors with room Curie
  temperatures}},\ }\href {https://doi.org/10.1103/PhysRevResearch.2.013002}
  {\bibfield  {journal} {\bibinfo  {journal} {Physical Review Research}\
  }\textbf {\bibinfo {volume} {2}},\ \bibinfo {pages} {013002} (\bibinfo {year}
  {2020})}\BibitemShut {NoStop}%
\bibitem [{\citenamefont {Jiang}\ \emph {et~al.}(2018)\citenamefont {Jiang},
  \citenamefont {Wang}, \citenamefont {Xing}, \citenamefont {Jiang},\ and\
  \citenamefont {Zhao}}]{Jiang2018-ACSAMI-2D-Tc}%
  \BibitemOpen
  \bibfield  {author} {\bibinfo {author} {\bibfnamefont {Z.}~\bibnamefont
  {Jiang}}, \bibinfo {author} {\bibfnamefont {P.}~\bibnamefont {Wang}},
  \bibinfo {author} {\bibfnamefont {J.}~\bibnamefont {Xing}}, \bibinfo {author}
  {\bibfnamefont {X.}~\bibnamefont {Jiang}},\ and\ \bibinfo {author}
  {\bibfnamefont {J.}~\bibnamefont {Zhao}},\ }\bibfield  {title} {\bibinfo
  {title} {{Screening and design of novel 2D ferromagnetic materials with high
  Curie temperature above room temperature}},\ }\href
  {https://doi.org/10.1021/acsami.8b14037} {\bibfield  {journal} {\bibinfo
  {journal} {ACS Applied Materials and Interfaces}\ }\textbf {\bibinfo {volume}
  {10}},\ \bibinfo {pages} {39032} (\bibinfo {year} {2018})}\BibitemShut
  {NoStop}%
\bibitem [{\citenamefont {Li}\ \emph {et~al.}(2021{\natexlab{c}})\citenamefont
  {Li}, \citenamefont {Wang}, \citenamefont {Yang},\ and\ \citenamefont
  {Liu}}]{Li2021-AnnalenderPhysik-VSi2N4-CrSi2N4-straininducedMagnetic}%
  \BibitemOpen
  \bibfield  {author} {\bibinfo {author} {\bibfnamefont {Y.}~\bibnamefont
  {Li}}, \bibinfo {author} {\bibfnamefont {J.}~\bibnamefont {Wang}}, \bibinfo
  {author} {\bibfnamefont {G.}~\bibnamefont {Yang}},\ and\ \bibinfo {author}
  {\bibfnamefont {Y.}~\bibnamefont {Liu}},\ }\bibfield  {title} {\bibinfo
  {title} {{Strain-induced magnetism in MSi$_2$N$_4$ (M = V, Cr): A
  first-principles study}},\ }\href {https://doi.org/10.1002/andp.202100273}
  {\bibfield  {journal} {\bibinfo  {journal} {Annalen der Physik}\ }\textbf
  {\bibinfo {volume} {533}},\ \bibinfo {pages} {2100273} (\bibinfo {year}
  {2021}{\natexlab{c}})}\BibitemShut {NoStop}%
\bibitem [{\citenamefont {Ray}\ \emph {et~al.}(2021)\citenamefont {Ray},
  \citenamefont {Tyagi}, \citenamefont {Singh},\ and\ \citenamefont
  {Schwingenschl{\"{o}}gl}}]{Ray2021-ACSomega-MoSi2N4-halfmetal}%
  \BibitemOpen
  \bibfield  {author} {\bibinfo {author} {\bibfnamefont {A.}~\bibnamefont
  {Ray}}, \bibinfo {author} {\bibfnamefont {S.}~\bibnamefont {Tyagi}}, \bibinfo
  {author} {\bibfnamefont {N.}~\bibnamefont {Singh}},\ and\ \bibinfo {author}
  {\bibfnamefont {U.}~\bibnamefont {Schwingenschl{\"{o}}gl}},\ }\bibfield
  {title} {\bibinfo {title} {{Inducing half-metallicity in monolayer
  MoSi$_2$N$_4$}},\ }\href {https://doi.org/10.1021/acsomega.1c03444}
  {\bibfield  {journal} {\bibinfo  {journal} {ACS Omega}\ }\textbf {\bibinfo
  {volume} {6}},\ \bibinfo {pages} {30371} (\bibinfo {year}
  {2021})}\BibitemShut {NoStop}%
\bibitem [{\citenamefont {Abdelati}\ \emph {et~al.}(2022)\citenamefont
  {Abdelati}, \citenamefont {Maarouf},\ and\ \citenamefont
  {Fadlallah}}]{Abdelati2022-PCCP-MetalDoping}%
  \BibitemOpen
  \bibfield  {author} {\bibinfo {author} {\bibfnamefont {M.~A.}\ \bibnamefont
  {Abdelati}}, \bibinfo {author} {\bibfnamefont {A.~A.}\ \bibnamefont
  {Maarouf}},\ and\ \bibinfo {author} {\bibfnamefont {M.~M.}\ \bibnamefont
  {Fadlallah}},\ }\bibfield  {title} {\bibinfo {title} {{Substitutional
  transition metal doping in MoSi$_2$N$_4$ monolayer: Structural, electronic
  and magnetic properties}},\ }\href {https://doi.org/10.1039/d1cp04191f}
  {\bibfield  {journal} {\bibinfo  {journal} {Physical Chemistry Chemical
  Physics}\ }\textbf {\bibinfo {volume} {24}},\ \bibinfo {pages} {3035}
  (\bibinfo {year} {2022})}\BibitemShut {NoStop}%
\bibitem [{\citenamefont {Yi}\ \emph {et~al.}(2019)\citenamefont {Yi},
  \citenamefont {Xu}, \citenamefont {M{\"{u}}ller},\ and\ \citenamefont
  {Gross}}]{Yi2019strain}%
  \BibitemOpen
  \bibfield  {author} {\bibinfo {author} {\bibfnamefont {M.}~\bibnamefont
  {Yi}}, \bibinfo {author} {\bibfnamefont {B.~X.}\ \bibnamefont {Xu}}, \bibinfo
  {author} {\bibfnamefont {R.}~\bibnamefont {M{\"{u}}ller}},\ and\ \bibinfo
  {author} {\bibfnamefont {D.}~\bibnamefont {Gross}},\ }\bibfield  {title}
  {\bibinfo {title} {{Strain-mediated magnetoelectric effect for the
  electric-field control of magnetic states in nanomagnets}},\ }\href
  {https://doi.org/10.1007/s00707-017-2029-7} {\bibfield  {journal} {\bibinfo
  {journal} {Acta Mechanica}\ }\textbf {\bibinfo {volume} {230}},\ \bibinfo
  {pages} {1247} (\bibinfo {year} {2019})}\BibitemShut {NoStop}%
\bibitem [{\citenamefont {{D Sander}}(1999)}]{Sander1999the}%
  \BibitemOpen
  \bibfield  {author} {\bibinfo {author} {\bibnamefont {{D Sander}}},\
  }\bibfield  {title} {\bibinfo {title} {{The correlation between mechanical
  stress and magnetic anisotropy in ultrathin films}},\ }\href
  {https://doi.org/10.1088/0034-4885/62/5/204} {\bibfield  {journal} {\bibinfo
  {journal} {Reports on Progress in Physics}\ }\textbf {\bibinfo {volume}
  {62}},\ \bibinfo {pages} {809} (\bibinfo {year} {1999})}\BibitemShut
  {NoStop}%
\bibitem [{\citenamefont {Gong}\ \emph {et~al.}(2020)\citenamefont {Gong},
  \citenamefont {Yi},\ and\ \citenamefont {Xu}}]{Gong2020Electric}%
  \BibitemOpen
  \bibfield  {author} {\bibinfo {author} {\bibfnamefont {Q.}~\bibnamefont
  {Gong}}, \bibinfo {author} {\bibfnamefont {M.}~\bibnamefont {Yi}},\ and\
  \bibinfo {author} {\bibfnamefont {B.~X.}\ \bibnamefont {Xu}},\ }\bibfield
  {title} {\bibinfo {title} {{Electric field induced magnetization reversal in
  magnet/insulator nanoheterostructure}},\ }\href
  {https://doi.org/10.1080/19475411.2020.1815132} {\bibfield  {journal}
  {\bibinfo  {journal} {International Journal of Smart and Nano Materials}\
  }\textbf {\bibinfo {volume} {11}},\ \bibinfo {pages} {298} (\bibinfo {year}
  {2020})}\BibitemShut {NoStop}%
\bibitem [{\citenamefont {Park}\ \emph {et~al.}(2020)\citenamefont {Park},
  \citenamefont {Kwon},\ and\ \citenamefont {Lake}}]{Park2020-PRB-CrSb-AFM}%
  \BibitemOpen
  \bibfield  {author} {\bibinfo {author} {\bibfnamefont {I.~J.}\ \bibnamefont
  {Park}}, \bibinfo {author} {\bibfnamefont {S.}~\bibnamefont {Kwon}},\ and\
  \bibinfo {author} {\bibfnamefont {R.~K.}\ \bibnamefont {Lake}},\ }\bibfield
  {title} {\bibinfo {title} {{Effects of filling, strain, and electric field on
  the N{\'{e}}el vector in antiferromagnetic CrSb}},\ }\href
  {https://doi.org/10.1103/PhysRevB.102.224426} {\bibfield  {journal} {\bibinfo
   {journal} {Physical Review B}\ }\textbf {\bibinfo {volume} {102}},\ \bibinfo
  {pages} {224426} (\bibinfo {year} {2020})}\BibitemShut {NoStop}%
\bibitem [{\citenamefont {Yuan}\ \emph {et~al.}(2022)\citenamefont {Yuan},
  \citenamefont {Wei}, \citenamefont {Sun}, \citenamefont {Yan}, \citenamefont
  {Cai}, \citenamefont {Shen},\ and\ \citenamefont
  {Schwingenschl{\"{o}}gl}}]{Yuan2022-PRB-MA2Z4-valley}%
  \BibitemOpen
  \bibfield  {author} {\bibinfo {author} {\bibfnamefont {J.}~\bibnamefont
  {Yuan}}, \bibinfo {author} {\bibfnamefont {Q.}~\bibnamefont {Wei}}, \bibinfo
  {author} {\bibfnamefont {M.}~\bibnamefont {Sun}}, \bibinfo {author}
  {\bibfnamefont {X.}~\bibnamefont {Yan}}, \bibinfo {author} {\bibfnamefont
  {Y.}~\bibnamefont {Cai}}, \bibinfo {author} {\bibfnamefont {L.}~\bibnamefont
  {Shen}},\ and\ \bibinfo {author} {\bibfnamefont {U.}~\bibnamefont
  {Schwingenschl{\"{o}}gl}},\ }\bibfield  {title} {\bibinfo {title} {{Protected
  valley states and generation of valley- and spin-polarized current in
  monolayer MA$_2$Z$_4$}},\ }\href
  {https://doi.org/10.1103/physrevb.105.195151} {\bibfield  {journal} {\bibinfo
   {journal} {Physical Review B}\ }\textbf {\bibinfo {volume} {105}},\ \bibinfo
  {pages} {195151} (\bibinfo {year} {2022})}\BibitemShut {NoStop}%
\bibitem [{\citenamefont {Cui}\ \emph {et~al.}(2022)\citenamefont {Cui},
  \citenamefont {Yang}, \citenamefont {Ren}, \citenamefont {Zhang},\ and\
  \citenamefont {Wang}}]{Cui2022-metalAtomAdsorption}%
  \BibitemOpen
  \bibfield  {author} {\bibinfo {author} {\bibfnamefont {Z.}~\bibnamefont
  {Cui}}, \bibinfo {author} {\bibfnamefont {K.}~\bibnamefont {Yang}}, \bibinfo
  {author} {\bibfnamefont {K.}~\bibnamefont {Ren}}, \bibinfo {author}
  {\bibfnamefont {S.}~\bibnamefont {Zhang}},\ and\ \bibinfo {author}
  {\bibfnamefont {L.}~\bibnamefont {Wang}},\ }\bibfield  {title} {\bibinfo
  {title} {{Adsorption of metal atoms on MoSi$_2$N$_4$ monolayer: A first
  principles study}},\ }\href {https://doi.org/10.1016/j.mssp.2022.107072}
  {\bibfield  {journal} {\bibinfo  {journal} {Materials Science in
  Semiconductor Processing}\ }\textbf {\bibinfo {volume} {152}},\ \bibinfo
  {pages} {107072} (\bibinfo {year} {2022})}\BibitemShut {NoStop}%
\bibitem [{\citenamefont {Li}\ \emph {et~al.}(2021{\natexlab{d}})\citenamefont
  {Li}, \citenamefont {Geng}, \citenamefont {Ai}, \citenamefont {Kong},
  \citenamefont {Bai}, \citenamefont {Lo}, \citenamefont {Ng}, \citenamefont
  {Kawazoe},\ and\ \citenamefont {Pan}}]{Li2021-Nanoscale-MSi2CxN4-x}%
  \BibitemOpen
  \bibfield  {author} {\bibinfo {author} {\bibfnamefont {B.}~\bibnamefont
  {Li}}, \bibinfo {author} {\bibfnamefont {J.}~\bibnamefont {Geng}}, \bibinfo
  {author} {\bibfnamefont {H.}~\bibnamefont {Ai}}, \bibinfo {author}
  {\bibfnamefont {Y.}~\bibnamefont {Kong}}, \bibinfo {author} {\bibfnamefont
  {H.}~\bibnamefont {Bai}}, \bibinfo {author} {\bibfnamefont {K.~H.}\
  \bibnamefont {Lo}}, \bibinfo {author} {\bibfnamefont {K.~W.}\ \bibnamefont
  {Ng}}, \bibinfo {author} {\bibfnamefont {Y.}~\bibnamefont {Kawazoe}},\ and\
  \bibinfo {author} {\bibfnamefont {H.}~\bibnamefont {Pan}},\ }\bibfield
  {title} {\bibinfo {title} {{Design of 2D materials-MSi$_2$C:XN$_{4-x}$ (M =
  Cr, Mo, and W; X=1 and 2)-with tunable electronic and magnetic properties}},\
  }\href {https://doi.org/10.1039/d1nr00461a} {\bibfield  {journal} {\bibinfo
  {journal} {Nanoscale}\ }\textbf {\bibinfo {volume} {13}},\ \bibinfo {pages}
  {8038} (\bibinfo {year} {2021}{\natexlab{d}})}\BibitemShut {NoStop}%
\bibitem [{\citenamefont {Li}\ \emph {et~al.}(2020{\natexlab{c}})\citenamefont
  {Li}, \citenamefont {Wu}, \citenamefont {Feng}, \citenamefont {Guan},
  \citenamefont {Feng}, \citenamefont {Yao},\ and\ \citenamefont
  {Yang}}]{Li2020-PRB-MoSi2N4-WSi2N4-MoSi2As4-spinvalley}%
  \BibitemOpen
  \bibfield  {author} {\bibinfo {author} {\bibfnamefont {S.}~\bibnamefont
  {Li}}, \bibinfo {author} {\bibfnamefont {W.}~\bibnamefont {Wu}}, \bibinfo
  {author} {\bibfnamefont {X.}~\bibnamefont {Feng}}, \bibinfo {author}
  {\bibfnamefont {S.}~\bibnamefont {Guan}}, \bibinfo {author} {\bibfnamefont
  {W.}~\bibnamefont {Feng}}, \bibinfo {author} {\bibfnamefont {Y.}~\bibnamefont
  {Yao}},\ and\ \bibinfo {author} {\bibfnamefont {S.~A.}\ \bibnamefont
  {Yang}},\ }\bibfield  {title} {\bibinfo {title} {{Valley-dependent properties
  of monolayer MoSi$_2$N$_4$,WSi$_2$N$_4$, and MoSi$_2$As$_4$}},\ }\href
  {https://doi.org/10.1103/PhysRevB.102.235435} {\bibfield  {journal} {\bibinfo
   {journal} {Physical Review B}\ }\textbf {\bibinfo {volume} {102}},\ \bibinfo
  {pages} {235435} (\bibinfo {year} {2020}{\natexlab{c}})}\BibitemShut
  {NoStop}%
\bibitem [{\citenamefont {Ai}\ \emph {et~al.}(2021)\citenamefont {Ai},
  \citenamefont {Liu}, \citenamefont {Geng}, \citenamefont {Wang},
  \citenamefont {Lo},\ and\ \citenamefont
  {Pan}}]{Ai2021-PCCP-MoSi2X4-spinvalley}%
  \BibitemOpen
  \bibfield  {author} {\bibinfo {author} {\bibfnamefont {H.}~\bibnamefont
  {Ai}}, \bibinfo {author} {\bibfnamefont {D.}~\bibnamefont {Liu}}, \bibinfo
  {author} {\bibfnamefont {J.}~\bibnamefont {Geng}}, \bibinfo {author}
  {\bibfnamefont {S.}~\bibnamefont {Wang}}, \bibinfo {author} {\bibfnamefont
  {K.~H.}\ \bibnamefont {Lo}},\ and\ \bibinfo {author} {\bibfnamefont
  {H.}~\bibnamefont {Pan}},\ }\bibfield  {title} {\bibinfo {title}
  {{Theoretical evidence of the spin-valley coupling and valley polarization in
  two-dimensional MoSi$_2$X$_4$(X = N, P, and As)}},\ }\href
  {https://doi.org/10.1039/d0cp05926a} {\bibfield  {journal} {\bibinfo
  {journal} {Physical Chemistry Chemical Physics}\ }\textbf {\bibinfo {volume}
  {23}},\ \bibinfo {pages} {3144} (\bibinfo {year} {2021})}\BibitemShut
  {NoStop}%
\bibitem [{\citenamefont {Liu}\ \emph {et~al.}(2021{\natexlab{b}})\citenamefont
  {Liu}, \citenamefont {Zhang}, \citenamefont {Dou}, \citenamefont {Du},
  \citenamefont {Peng}, \citenamefont {Dai}, \citenamefont {Huang},\ and\
  \citenamefont {Ma}}]{Liu2021-JPCL-CrSiN-CrSiP}%
  \BibitemOpen
  \bibfield  {author} {\bibinfo {author} {\bibfnamefont {Y.}~\bibnamefont
  {Liu}}, \bibinfo {author} {\bibfnamefont {T.}~\bibnamefont {Zhang}}, \bibinfo
  {author} {\bibfnamefont {K.}~\bibnamefont {Dou}}, \bibinfo {author}
  {\bibfnamefont {W.}~\bibnamefont {Du}}, \bibinfo {author} {\bibfnamefont
  {R.}~\bibnamefont {Peng}}, \bibinfo {author} {\bibfnamefont {Y.}~\bibnamefont
  {Dai}}, \bibinfo {author} {\bibfnamefont {B.}~\bibnamefont {Huang}},\ and\
  \bibinfo {author} {\bibfnamefont {Y.}~\bibnamefont {Ma}},\ }\bibfield
  {title} {\bibinfo {title} {{Valley-contrasting physics in single-Layer
  CrSi$_2$N$_4$ and CrSi$_2$P$_4$}},\ }\href
  {https://doi.org/10.1021/acs.jpclett.1c02069} {\bibfield  {journal} {\bibinfo
   {journal} {Journal of Physical Chemistry Letters}\ }\textbf {\bibinfo
  {volume} {12}},\ \bibinfo {pages} {8341} (\bibinfo {year}
  {2021}{\natexlab{b}})}\BibitemShut {NoStop}%
\bibitem [{\citenamefont {Hussain}\ \emph
  {et~al.}(2022{\natexlab{a}})\citenamefont {Hussain}, \citenamefont {Samad},
  \citenamefont {{Ur Rehman}}, \citenamefont {Cuono},\ and\ \citenamefont
  {Autieri}}]{HUSSAIN2022169897Emergence}%
  \BibitemOpen
  \bibfield  {author} {\bibinfo {author} {\bibfnamefont {G.}~\bibnamefont
  {Hussain}}, \bibinfo {author} {\bibfnamefont {A.}~\bibnamefont {Samad}},
  \bibinfo {author} {\bibfnamefont {M.}~\bibnamefont {{Ur Rehman}}}, \bibinfo
  {author} {\bibfnamefont {G.}~\bibnamefont {Cuono}},\ and\ \bibinfo {author}
  {\bibfnamefont {C.}~\bibnamefont {Autieri}},\ }\bibfield  {title} {\bibinfo
  {title} {Emergence of rashba splitting and spin-valley properties in janus
  mogesip$_2$as$_2$ and wgesip$_2$as$_2$ monolayers},\ }\href
  {https://doi.org/10.1016/j.jmmm.2022.169897} {\bibfield  {journal} {\bibinfo
  {journal} {Journal of Magnetism and Magnetic Materials}\ }\textbf {\bibinfo
  {volume} {563}},\ \bibinfo {pages} {169897} (\bibinfo {year}
  {2022}{\natexlab{a}})}\BibitemShut {NoStop}%
\bibitem [{\citenamefont {Sheoran}\ \emph {et~al.}(2022)\citenamefont
  {Sheoran}, \citenamefont {Gill}, \citenamefont {Phutela},\ and\ \citenamefont
  {Bhattacharya}}]{Sheoran2022-arxiv-WSi2N4-spinvally}%
  \BibitemOpen
  \bibfield  {author} {\bibinfo {author} {\bibfnamefont {S.}~\bibnamefont
  {Sheoran}}, \bibinfo {author} {\bibfnamefont {D.}~\bibnamefont {Gill}},
  \bibinfo {author} {\bibfnamefont {A.}~\bibnamefont {Phutela}},\ and\ \bibinfo
  {author} {\bibfnamefont {S.}~\bibnamefont {Bhattacharya}},\ }\bibfield
  {title} {\bibinfo {title} {{Coupled spin-valley, Rashba effect and hidden
  persistent spin polarization in WSi$_2$N$_4$ family}},\ }\href
  {http://arxiv.org/abs/2208.00127} {\bibfield  {journal} {\bibinfo  {journal}
  {arXiv:2208.00127}\ } (\bibinfo {year} {2022})}\BibitemShut {NoStop}%
\bibitem [{\citenamefont {Yang}\ \emph {et~al.}(2020)\citenamefont {Yang},
  \citenamefont {Song}, \citenamefont {Sun}, \citenamefont {Lu}, \citenamefont
  {Berkeley},\ and\ \citenamefont
  {Materials}}]{Yang2020-PRB-MoSi2N4-MoSi2As4-spinvalley}%
  \BibitemOpen
  \bibfield  {author} {\bibinfo {author} {\bibfnamefont {C.}~\bibnamefont
  {Yang}}, \bibinfo {author} {\bibfnamefont {Z.}~\bibnamefont {Song}}, \bibinfo
  {author} {\bibfnamefont {X.}~\bibnamefont {Sun}}, \bibinfo {author}
  {\bibfnamefont {J.}~\bibnamefont {Lu}}, \bibinfo {author} {\bibfnamefont
  {L.}~\bibnamefont {Berkeley}},\ and\ \bibinfo {author} {\bibfnamefont
  {O.~P.}\ \bibnamefont {Materials}},\ }\bibfield  {title} {\bibinfo {title}
  {{Valley pseudospin in monolayer MoSi$_2$N$_4$ and MoSi$_2$As$_4$}},\ }\href
  {https://doi.org/10.1103/PhysRevB.103.035308} {\bibfield  {journal} {\bibinfo
   {journal} {Physical Review B}\ }\textbf {\bibinfo {volume} {103}},\ \bibinfo
  {pages} {035308} (\bibinfo {year} {2020})}\BibitemShut {NoStop}%
\bibitem [{\citenamefont {Feng}\ \emph {et~al.}(2021)\citenamefont {Feng},
  \citenamefont {Xu}, \citenamefont {He}, \citenamefont {Peng}, \citenamefont
  {Dai}, \citenamefont {Huang},\ and\ \citenamefont
  {Ma}}]{Feng2021-PRB-VSi2P4-spinvalley}%
  \BibitemOpen
  \bibfield  {author} {\bibinfo {author} {\bibfnamefont {X.}~\bibnamefont
  {Feng}}, \bibinfo {author} {\bibfnamefont {X.}~\bibnamefont {Xu}}, \bibinfo
  {author} {\bibfnamefont {Z.}~\bibnamefont {He}}, \bibinfo {author}
  {\bibfnamefont {R.}~\bibnamefont {Peng}}, \bibinfo {author} {\bibfnamefont
  {Y.}~\bibnamefont {Dai}}, \bibinfo {author} {\bibfnamefont {B.}~\bibnamefont
  {Huang}},\ and\ \bibinfo {author} {\bibfnamefont {Y.}~\bibnamefont {Ma}},\
  }\bibfield  {title} {\bibinfo {title} {{Valley-related multiple Hall effect
  in monolayer VSi$_2$P$_4$}},\ }\href
  {https://doi.org/10.1103/PhysRevB.104.075421} {\bibfield  {journal} {\bibinfo
   {journal} {Physical Review B}\ }\textbf {\bibinfo {volume} {104}},\ \bibinfo
  {pages} {075421} (\bibinfo {year} {2021})}\BibitemShut {NoStop}%
\bibitem [{\citenamefont {Li}\ \emph {et~al.}(2021{\natexlab{e}})\citenamefont
  {Li}, \citenamefont {Wang}, \citenamefont {Zhang}, \citenamefont {Guo},\ and\
  \citenamefont {Yang}}]{Li2021-PRB-VSi2P4-QAH}%
  \BibitemOpen
  \bibfield  {author} {\bibinfo {author} {\bibfnamefont {S.}~\bibnamefont
  {Li}}, \bibinfo {author} {\bibfnamefont {Q.}~\bibnamefont {Wang}}, \bibinfo
  {author} {\bibfnamefont {C.}~\bibnamefont {Zhang}}, \bibinfo {author}
  {\bibfnamefont {P.}~\bibnamefont {Guo}},\ and\ \bibinfo {author}
  {\bibfnamefont {S.~A.}\ \bibnamefont {Yang}},\ }\bibfield  {title} {\bibinfo
  {title} {{Correlation-driven topological and valley states in monolayer
  VSi$_2$P$_4$}},\ }\href {https://doi.org/10.1103/PhysRevB.104.085149}
  {\bibfield  {journal} {\bibinfo  {journal} {Physical Review B}\ }\textbf
  {\bibinfo {volume} {104}},\ \bibinfo {pages} {085149} (\bibinfo {year}
  {2021}{\natexlab{e}})}\BibitemShut {NoStop}%
\bibitem [{\citenamefont {Wang}\ and\ \citenamefont
  {Ding}(2021)}]{Wang2021-APL-VN2X2Y2-mag}%
  \BibitemOpen
  \bibfield  {author} {\bibinfo {author} {\bibfnamefont {Y.}~\bibnamefont
  {Wang}}\ and\ \bibinfo {author} {\bibfnamefont {Y.}~\bibnamefont {Ding}},\
  }\bibfield  {title} {\bibinfo {title} {{Switchable valley polarization and
  quantum anomalous Hall state in the VN$_2$X$_2$Y$_2$ nanosheets (X =
  group-III and Y = group-VI elements)}},\ }\href
  {https://doi.org/10.1063/5.0072220} {\bibfield  {journal} {\bibinfo
  {journal} {Applied Physics Letters}\ }\textbf {\bibinfo {volume} {119}},\
  \bibinfo {pages} {193101} (\bibinfo {year} {2021})}\BibitemShut {NoStop}%
\bibitem [{\citenamefont {Islam}\ \emph {et~al.}(2022)\citenamefont {Islam},
  \citenamefont {Verma}, \citenamefont {Ghosh}, \citenamefont {Muhammad},
  \citenamefont {Bansil}, \citenamefont {Autieri},\ and\ \citenamefont
  {Singh}}]{Islam2022Switchable}%
  \BibitemOpen
  \bibfield  {author} {\bibinfo {author} {\bibfnamefont {R.}~\bibnamefont
  {Islam}}, \bibinfo {author} {\bibfnamefont {R.}~\bibnamefont {Verma}},
  \bibinfo {author} {\bibfnamefont {B.}~\bibnamefont {Ghosh}}, \bibinfo
  {author} {\bibfnamefont {Z.}~\bibnamefont {Muhammad}}, \bibinfo {author}
  {\bibfnamefont {A.}~\bibnamefont {Bansil}}, \bibinfo {author} {\bibfnamefont
  {C.}~\bibnamefont {Autieri}},\ and\ \bibinfo {author} {\bibfnamefont
  {B.}~\bibnamefont {Singh}},\ }\bibfield  {title} {\bibinfo {title}
  {{Switchable large-gap quantum spin Hall state in two-dimensional
  MSi$_2$Z$_4$ materials class}},\ }\href {https://arxiv.org/abs/2207.08407}
  {\bibfield  {journal} {\bibinfo  {journal} {arXiv:2207.08407}\ } (\bibinfo
  {year} {2022})}\BibitemShut {NoStop}%
\bibitem [{\citenamefont {Sun}\ \emph {et~al.}(2021)\citenamefont {Sun},
  \citenamefont {Song}, \citenamefont {Huo}, \citenamefont {Liu}, \citenamefont
  {Yang}, \citenamefont {Yang}, \citenamefont {Wang},\ and\ \citenamefont
  {Lu}}]{Sun2021-JMCC-MoSi2N4-transistor}%
  \BibitemOpen
  \bibfield  {author} {\bibinfo {author} {\bibfnamefont {X.}~\bibnamefont
  {Sun}}, \bibinfo {author} {\bibfnamefont {Z.}~\bibnamefont {Song}}, \bibinfo
  {author} {\bibfnamefont {N.}~\bibnamefont {Huo}}, \bibinfo {author}
  {\bibfnamefont {S.}~\bibnamefont {Liu}}, \bibinfo {author} {\bibfnamefont
  {C.}~\bibnamefont {Yang}}, \bibinfo {author} {\bibfnamefont {J.}~\bibnamefont
  {Yang}}, \bibinfo {author} {\bibfnamefont {W.}~\bibnamefont {Wang}},\ and\
  \bibinfo {author} {\bibfnamefont {J.}~\bibnamefont {Lu}},\ }\bibfield
  {title} {\bibinfo {title} {{Performance limit of monolayer MoSi$_2$N$_4$
  transistors}},\ }\href {https://doi.org/10.1039/d1tc02937a} {\bibfield
  {journal} {\bibinfo  {journal} {Journal of Materials Chemistry C}\ }\textbf
  {\bibinfo {volume} {9}},\ \bibinfo {pages} {14683} (\bibinfo {year}
  {2021})}\BibitemShut {NoStop}%
\bibitem [{\citenamefont {Huang}\ \emph {et~al.}(2021)\citenamefont {Huang},
  \citenamefont {Li}, \citenamefont {Ren},\ and\ \citenamefont
  {Guo}}]{Huang2021-PRAppli-MoSi2N4-transistor5nm}%
  \BibitemOpen
  \bibfield  {author} {\bibinfo {author} {\bibfnamefont {J.}~\bibnamefont
  {Huang}}, \bibinfo {author} {\bibfnamefont {P.}~\bibnamefont {Li}}, \bibinfo
  {author} {\bibfnamefont {X.}~\bibnamefont {Ren}},\ and\ \bibinfo {author}
  {\bibfnamefont {Z.~X.}\ \bibnamefont {Guo}},\ }\bibfield  {title} {\bibinfo
  {title} {{Promising properties of a sub-5-nm monolayer MoSi$_2$N$_4$
  transistor}},\ }\href {https://doi.org/10.1103/PhysRevApplied.16.044022}
  {\bibfield  {journal} {\bibinfo  {journal} {Physical Review Applied}\
  }\textbf {\bibinfo {volume} {16}},\ \bibinfo {pages} {044022} (\bibinfo
  {year} {2021})}\BibitemShut {NoStop}%
\bibitem [{\citenamefont {Ye}\ \emph {et~al.}(2022)\citenamefont {Ye},
  \citenamefont {Jiang}, \citenamefont {Gu}, \citenamefont {Yang},
  \citenamefont {Liu}, \citenamefont {Zhao}, \citenamefont {Yang},
  \citenamefont {Wei}, \citenamefont {Zhang},\ and\ \citenamefont
  {Lu}}]{Ye2022-PCCP-MoSi2N4-transistor3nm}%
  \BibitemOpen
  \bibfield  {author} {\bibinfo {author} {\bibfnamefont {B.}~\bibnamefont
  {Ye}}, \bibinfo {author} {\bibfnamefont {X.}~\bibnamefont {Jiang}}, \bibinfo
  {author} {\bibfnamefont {Y.}~\bibnamefont {Gu}}, \bibinfo {author}
  {\bibfnamefont {G.}~\bibnamefont {Yang}}, \bibinfo {author} {\bibfnamefont
  {Y.}~\bibnamefont {Liu}}, \bibinfo {author} {\bibfnamefont {H.}~\bibnamefont
  {Zhao}}, \bibinfo {author} {\bibfnamefont {X.}~\bibnamefont {Yang}}, \bibinfo
  {author} {\bibfnamefont {C.}~\bibnamefont {Wei}}, \bibinfo {author}
  {\bibfnamefont {X.}~\bibnamefont {Zhang}},\ and\ \bibinfo {author}
  {\bibfnamefont {N.}~\bibnamefont {Lu}},\ }\bibfield  {title} {\bibinfo
  {title} {{Quantum transport of short-gate MOSFETs based on monolayer
  MoSi$_2$N$_4$ }},\ }\href {https://doi.org/10.1039/d2cp00086e} {\bibfield
  {journal} {\bibinfo  {journal} {Physical Chemistry Chemical Physics}\
  }\textbf {\bibinfo {volume} {24}},\ \bibinfo {pages} {6616} (\bibinfo {year}
  {2022})}\BibitemShut {NoStop}%
\bibitem [{\citenamefont {Ghobadi}\ \emph {et~al.}(2022)\citenamefont
  {Ghobadi}, \citenamefont {Hosseini},\ and\ \citenamefont
  {Touski}}]{Ghobadi2022-IEEE-MoSi2N4-Transistor}%
  \BibitemOpen
  \bibfield  {author} {\bibinfo {author} {\bibfnamefont {N.}~\bibnamefont
  {Ghobadi}}, \bibinfo {author} {\bibfnamefont {M.}~\bibnamefont {Hosseini}},\
  and\ \bibinfo {author} {\bibfnamefont {S.~B.}\ \bibnamefont {Touski}},\
  }\bibfield  {title} {\bibinfo {title} {{Field-effect transistor based on
  MoSi$_2$N$_4$ and WSi$_2$N$_4$ monolayers under biaxial strain : A
  computational study of the electronic properties}},\ }\href
  {http://doi.org/10.1109/TED.2021.3138377} {\bibfield  {journal} {\bibinfo
  {journal} {IEEE Transactions on Electron Devices}\ }\textbf {\bibinfo
  {volume} {69}},\ \bibinfo {pages} {863} (\bibinfo {year} {2022})}\BibitemShut
  {NoStop}%
\bibitem [{\citenamefont {Nandan}\ \emph {et~al.}(2022)\citenamefont {Nandan},
  \citenamefont {Member}, \citenamefont {Ghosh}, \citenamefont {Agarwal},\ and\
  \citenamefont {Bhowmick}}]{Nandan2022-IEEE-MoSi2N4-Transistor}%
  \BibitemOpen
  \bibfield  {author} {\bibinfo {author} {\bibfnamefont {K.}~\bibnamefont
  {Nandan}}, \bibinfo {author} {\bibfnamefont {G.~S.}\ \bibnamefont {Member}},
  \bibinfo {author} {\bibfnamefont {B.}~\bibnamefont {Ghosh}}, \bibinfo
  {author} {\bibfnamefont {A.}~\bibnamefont {Agarwal}},\ and\ \bibinfo {author}
  {\bibfnamefont {S.}~\bibnamefont {Bhowmick}},\ }\bibfield  {title} {\bibinfo
  {title} {{Two-dimensional MoSi$_2$N$_4$: An excellent 2-D semiconductor for
  field-effect tsransistors}},\ }\href
  {http://doi.org/10.1109/TED.2021.3130834} {\bibfield  {journal} {\bibinfo
  {journal} {IEEE Transactions on Electron Devices}\ }\textbf {\bibinfo
  {volume} {69}},\ \bibinfo {pages} {406} (\bibinfo {year} {2022})}\BibitemShut
  {NoStop}%
\bibitem [{\citenamefont {Zhang}\ \emph {et~al.}(2021)\citenamefont {Zhang},
  \citenamefont {Shi}, \citenamefont {Xu}, \citenamefont {Yan}, \citenamefont
  {Zhao}, \citenamefont {Zhang}, \citenamefont {Zhang},\ and\ \citenamefont
  {Lu}}]{Zhang2021-ACSAEM-MoS2-DGMOSFET}%
  \BibitemOpen
  \bibfield  {author} {\bibinfo {author} {\bibfnamefont {H.}~\bibnamefont
  {Zhang}}, \bibinfo {author} {\bibfnamefont {B.}~\bibnamefont {Shi}}, \bibinfo
  {author} {\bibfnamefont {L.}~\bibnamefont {Xu}}, \bibinfo {author}
  {\bibfnamefont {J.}~\bibnamefont {Yan}}, \bibinfo {author} {\bibfnamefont
  {W.}~\bibnamefont {Zhao}}, \bibinfo {author} {\bibfnamefont {Z.}~\bibnamefont
  {Zhang}}, \bibinfo {author} {\bibfnamefont {Z.}~\bibnamefont {Zhang}},\ and\
  \bibinfo {author} {\bibfnamefont {J.}~\bibnamefont {Lu}},\ }\bibfield
  {title} {\bibinfo {title} {{Sub-5 nm monolayer MoS$_2$ transistors toward
  low-power devices}},\ }\href {https://doi.org/10.1021/acsaelm.0c00840}
  {\bibfield  {journal} {\bibinfo  {journal} {ACS Applied Electronic
  Materials}\ }\textbf {\bibinfo {volume} {3}},\ \bibinfo {pages} {1560}
  (\bibinfo {year} {2021})}\BibitemShut {NoStop}%
\bibitem [{\citenamefont {Zhu}\ and\ \citenamefont
  {Wang}(2017)}]{Zhu2017-AdvEnergyMater-Photocatlysis}%
  \BibitemOpen
  \bibfield  {author} {\bibinfo {author} {\bibfnamefont {S.}~\bibnamefont
  {Zhu}}\ and\ \bibinfo {author} {\bibfnamefont {D.}~\bibnamefont {Wang}},\
  }\bibfield  {title} {\bibinfo {title} {{Photocatalysis: Basic principles,
  diverse forms of implementations and emerging scientific opportunities}},\
  }\href {https://doi.org/10.1002/aenm.201700841} {\bibfield  {journal}
  {\bibinfo  {journal} {Advanced Energy Materials}\ }\textbf {\bibinfo {volume}
  {7}},\ \bibinfo {pages} {1700841} (\bibinfo {year} {2017})}\BibitemShut
  {NoStop}%
\bibitem [{\citenamefont {Li}\ \emph {et~al.}(2019)\citenamefont {Li},
  \citenamefont {Gao}, \citenamefont {Long},\ and\ \citenamefont
  {Xiong}}]{Li2019-MaterTodayChem-2DPhotocatalysts}%
  \BibitemOpen
  \bibfield  {author} {\bibinfo {author} {\bibfnamefont {Y.}~\bibnamefont
  {Li}}, \bibinfo {author} {\bibfnamefont {C.}~\bibnamefont {Gao}}, \bibinfo
  {author} {\bibfnamefont {R.}~\bibnamefont {Long}},\ and\ \bibinfo {author}
  {\bibfnamefont {Y.}~\bibnamefont {Xiong}},\ }\bibfield  {title} {\bibinfo
  {title} {{Photocatalyst design based on two-dimensional materials}},\ }\href
  {https://doi.org/10.1016/j.mtchem.2018.11.002} {\bibfield  {journal}
  {\bibinfo  {journal} {Materials Today Chemistry}\ }\textbf {\bibinfo {volume}
  {11}},\ \bibinfo {pages} {197} (\bibinfo {year} {2019})}\BibitemShut
  {NoStop}%
\bibitem [{\citenamefont {Shi}\ \emph {et~al.}(2022)\citenamefont {Shi},
  \citenamefont {Yin}, \citenamefont {Yu}, \citenamefont {Hu},\ and\
  \citenamefont {Wang}}]{Shi2022-JMS-MoSi2N4-HER-adatoms}%
  \BibitemOpen
  \bibfield  {author} {\bibinfo {author} {\bibfnamefont {W.}~\bibnamefont
  {Shi}}, \bibinfo {author} {\bibfnamefont {G.}~\bibnamefont {Yin}}, \bibinfo
  {author} {\bibfnamefont {S.}~\bibnamefont {Yu}}, \bibinfo {author}
  {\bibfnamefont {T.}~\bibnamefont {Hu}},\ and\ \bibinfo {author}
  {\bibfnamefont {X.}~\bibnamefont {Wang}},\ }\bibfield  {title} {\bibinfo
  {title} {{Atomic precision tailoring of two-dimensional MoSi$_2$N$_4$ as
  electrocatalyst for hydrogen evolution reaction}},\ }\href
  {https://doi.org/10.1007/s10853-022-07755-y} {\bibfield  {journal} {\bibinfo
  {journal} {Journal of Materials Science}\ } (\bibinfo {year}
  {2022})}\BibitemShut {NoStop}%
\bibitem [{\citenamefont {Zhao}\ \emph
  {et~al.}(2021{\natexlab{c}})\citenamefont {Zhao}, \citenamefont {Zhao},
  \citenamefont {He}, \citenamefont {Zhou}, \citenamefont {Liang},\ and\
  \citenamefont {Frauenheim}}]{Zhao2021-JPCL-stackingEngineering}%
  \BibitemOpen
  \bibfield  {author} {\bibinfo {author} {\bibfnamefont {J.}~\bibnamefont
  {Zhao}}, \bibinfo {author} {\bibfnamefont {Y.}~\bibnamefont {Zhao}}, \bibinfo
  {author} {\bibfnamefont {H.}~\bibnamefont {He}}, \bibinfo {author}
  {\bibfnamefont {P.}~\bibnamefont {Zhou}}, \bibinfo {author} {\bibfnamefont
  {Y.}~\bibnamefont {Liang}},\ and\ \bibinfo {author} {\bibfnamefont
  {T.}~\bibnamefont {Frauenheim}},\ }\bibfield  {title} {\bibinfo {title}
  {{Stacking engineering: A boosting strategy for 2D photocatalysts}},\ }\href
  {https://doi.org/10.1021/acs.jpclett.1c03089} {\bibfield  {journal} {\bibinfo
   {journal} {Journal of Physical Chemistry Letters}\ }\textbf {\bibinfo
  {volume} {12}},\ \bibinfo {pages} {10190} (\bibinfo {year}
  {2021}{\natexlab{c}})}\BibitemShut {NoStop}%
\bibitem [{\citenamefont {Hussain}\ \emph
  {et~al.}(2022{\natexlab{b}})\citenamefont {Hussain}, \citenamefont {Manzoor},
  \citenamefont {Iqbal}, \citenamefont {Muhammad}, \citenamefont {Bafekry},
  \citenamefont {Ullah},\ and\ \citenamefont
  {Autieri}}]{Hussain2022-PhysE-MoSi2N4-XSi2N4-hetero}%
  \BibitemOpen
  \bibfield  {author} {\bibinfo {author} {\bibfnamefont {G.}~\bibnamefont
  {Hussain}}, \bibinfo {author} {\bibfnamefont {M.}~\bibnamefont {Manzoor}},
  \bibinfo {author} {\bibfnamefont {M.~W.}\ \bibnamefont {Iqbal}}, \bibinfo
  {author} {\bibfnamefont {I.}~\bibnamefont {Muhammad}}, \bibinfo {author}
  {\bibfnamefont {A.}~\bibnamefont {Bafekry}}, \bibinfo {author} {\bibfnamefont
  {H.}~\bibnamefont {Ullah}},\ and\ \bibinfo {author} {\bibfnamefont
  {C.}~\bibnamefont {Autieri}},\ }\bibfield  {title} {\bibinfo {title} {{Strain
  modulated electronic and optical properties of laterally stitched
  MoSi$_2$N$_4$/XSi$_2$N$_4$ (X=W, Ti) 2D heterostructures}},\ }\href
  {https://doi.org/10.1016/j.physe.2022.115471} {\bibfield  {journal} {\bibinfo
   {journal} {Physica E: Low-dimensional Systems and Nanostructures}\ }\textbf
  {\bibinfo {volume} {144}},\ \bibinfo {pages} {115471} (\bibinfo {year}
  {2022}{\natexlab{b}})}\BibitemShut {NoStop}%
\bibitem [{\citenamefont {Ren}\ \emph {et~al.}(2022)\citenamefont {Ren},
  \citenamefont {Hu}, \citenamefont {Chen}, \citenamefont {Hu}, \citenamefont
  {Wang}, \citenamefont {Gong}, \citenamefont {Zhang}, \citenamefont {Huang},\
  and\ \citenamefont {Shi}}]{Ren2022-PRM-MSi2N4-bandAlignments}%
  \BibitemOpen
  \bibfield  {author} {\bibinfo {author} {\bibfnamefont {Y.~T.}\ \bibnamefont
  {Ren}}, \bibinfo {author} {\bibfnamefont {L.}~\bibnamefont {Hu}}, \bibinfo
  {author} {\bibfnamefont {Y.~T.}\ \bibnamefont {Chen}}, \bibinfo {author}
  {\bibfnamefont {Y.~J.}\ \bibnamefont {Hu}}, \bibinfo {author} {\bibfnamefont
  {J.~L.}\ \bibnamefont {Wang}}, \bibinfo {author} {\bibfnamefont {P.~L.}\
  \bibnamefont {Gong}}, \bibinfo {author} {\bibfnamefont {H.}~\bibnamefont
  {Zhang}}, \bibinfo {author} {\bibfnamefont {L.}~\bibnamefont {Huang}},\ and\
  \bibinfo {author} {\bibfnamefont {X.~Q.}\ \bibnamefont {Shi}},\ }\bibfield
  {title} {\bibinfo {title} {{Two-dimensional MSi$_2$N$_4$ monolayers and van
  der Waals heterostructures: Promising spintronic properties and band
  alignments}},\ }\href {https://doi.org/10.1103/PhysRevMaterials.6.064006}
  {\bibfield  {journal} {\bibinfo  {journal} {Physical Review Materials}\
  }\textbf {\bibinfo {volume} {6}},\ \bibinfo {pages} {064006} (\bibinfo {year}
  {2022})}\BibitemShut {NoStop}%
\bibitem [{\citenamefont {Zang}\ \emph {et~al.}(2021)\citenamefont {Zang},
  \citenamefont {Wu}, \citenamefont {Du}, \citenamefont {Dai}, \citenamefont
  {Huang},\ and\ \citenamefont {Ma}}]{Zang2021-PRM-MoSi2N4-WSi2N4-HER}%
  \BibitemOpen
  \bibfield  {author} {\bibinfo {author} {\bibfnamefont {Y.}~\bibnamefont
  {Zang}}, \bibinfo {author} {\bibfnamefont {Q.}~\bibnamefont {Wu}}, \bibinfo
  {author} {\bibfnamefont {W.}~\bibnamefont {Du}}, \bibinfo {author}
  {\bibfnamefont {Y.}~\bibnamefont {Dai}}, \bibinfo {author} {\bibfnamefont
  {B.}~\bibnamefont {Huang}},\ and\ \bibinfo {author} {\bibfnamefont
  {Y.}~\bibnamefont {Ma}},\ }\bibfield  {title} {\bibinfo {title} {{Activating
  electrocatalytic hydrogen evolution performance of two-dimensional
  MSi$_2$N$_4$(M=Mo, W): A theoretical prediction}},\ }\href
  {https://doi.org/10.1103/PhysRevMaterials.5.045801} {\bibfield  {journal}
  {\bibinfo  {journal} {Physical Review Materials}\ }\textbf {\bibinfo {volume}
  {5}},\ \bibinfo {pages} {045801} (\bibinfo {year} {2021})}\BibitemShut
  {NoStop}%
\bibitem [{\citenamefont {Qian}\ \emph {et~al.}(2022)\citenamefont {Qian},
  \citenamefont {Chen}, \citenamefont {Zhang},\ and\ \citenamefont
  {Yin}}]{Qian2022-JMST-MoSi2N4-x-HER}%
  \BibitemOpen
  \bibfield  {author} {\bibinfo {author} {\bibfnamefont {W.}~\bibnamefont
  {Qian}}, \bibinfo {author} {\bibfnamefont {Z.}~\bibnamefont {Chen}}, \bibinfo
  {author} {\bibfnamefont {J.}~\bibnamefont {Zhang}},\ and\ \bibinfo {author}
  {\bibfnamefont {L.}~\bibnamefont {Yin}},\ }\bibfield  {title} {\bibinfo
  {title} {{Monolayer MoSi$_2$N$_{4-x}$ as promising electrocatalyst for
  hydrogen evolution reaction: A DFT prediction}},\ }\href
  {https://doi.org/10.1016/j.jmst.2021.06.004} {\bibfield  {journal} {\bibinfo
  {journal} {Journal of Materials Science and Technology}\ }\textbf {\bibinfo
  {volume} {99}},\ \bibinfo {pages} {215} (\bibinfo {year} {2022})}\BibitemShut
  {NoStop}%
\bibitem [{\citenamefont {Xiao}\ \emph {et~al.}(2021)\citenamefont {Xiao},
  \citenamefont {Sa}, \citenamefont {Cui}, \citenamefont {Gao}, \citenamefont
  {Du}, \citenamefont {Sun}, \citenamefont {Zhang}, \citenamefont {Li},\ and\
  \citenamefont {Ma}}]{Xiao2021-ASS-defectiveMoSi2N4-HER}%
  \BibitemOpen
  \bibfield  {author} {\bibinfo {author} {\bibfnamefont {C.}~\bibnamefont
  {Xiao}}, \bibinfo {author} {\bibfnamefont {R.}~\bibnamefont {Sa}}, \bibinfo
  {author} {\bibfnamefont {Z.}~\bibnamefont {Cui}}, \bibinfo {author}
  {\bibfnamefont {S.}~\bibnamefont {Gao}}, \bibinfo {author} {\bibfnamefont
  {W.}~\bibnamefont {Du}}, \bibinfo {author} {\bibfnamefont {X.}~\bibnamefont
  {Sun}}, \bibinfo {author} {\bibfnamefont {X.}~\bibnamefont {Zhang}}, \bibinfo
  {author} {\bibfnamefont {Q.}~\bibnamefont {Li}},\ and\ \bibinfo {author}
  {\bibfnamefont {Z.}~\bibnamefont {Ma}},\ }\bibfield  {title} {\bibinfo
  {title} {{Enhancing the hydrogen evolution reaction by non-precious
  transition metal (Non-metal) atom doping in defective MoSi$_2$N$_4$
  monolayer}},\ }\href {https://doi.org/10.1016/j.apsusc.2021.150388}
  {\bibfield  {journal} {\bibinfo  {journal} {Applied Surface Science}\
  }\textbf {\bibinfo {volume} {563}},\ \bibinfo {pages} {150388} (\bibinfo
  {year} {2021})}\BibitemShut {NoStop}%
\bibitem [{\citenamefont {Luo}\ \emph {et~al.}(2021)\citenamefont {Luo},
  \citenamefont {Li}, \citenamefont {Dai}, \citenamefont {Zhang}, \citenamefont
  {Zhao}, \citenamefont {Jiang}, \citenamefont {Ling},\ and\ \citenamefont
  {Huang}}]{Luo2021-JMCA-MSi2N4-adsorption}%
  \BibitemOpen
  \bibfield  {author} {\bibinfo {author} {\bibfnamefont {Y.}~\bibnamefont
  {Luo}}, \bibinfo {author} {\bibfnamefont {M.}~\bibnamefont {Li}}, \bibinfo
  {author} {\bibfnamefont {Y.}~\bibnamefont {Dai}}, \bibinfo {author}
  {\bibfnamefont {X.}~\bibnamefont {Zhang}}, \bibinfo {author} {\bibfnamefont
  {R.}~\bibnamefont {Zhao}}, \bibinfo {author} {\bibfnamefont {F.}~\bibnamefont
  {Jiang}}, \bibinfo {author} {\bibfnamefont {C.}~\bibnamefont {Ling}},\ and\
  \bibinfo {author} {\bibfnamefont {Y.}~\bibnamefont {Huang}},\ }\bibfield
  {title} {\bibinfo {title} {{Screening of effective NRR electrocatalysts among
  the Si-based MSi$_2$N$_4$(M = Ti, Zr, Hf, V, Nb, Ta, Cr, Mo, and W)
  monolayers}},\ }\href {https://doi.org/10.1039/d1ta02998c} {\bibfield
  {journal} {\bibinfo  {journal} {Journal of Materials Chemistry A}\ }\textbf
  {\bibinfo {volume} {9}},\ \bibinfo {pages} {15217} (\bibinfo {year}
  {2021})}\BibitemShut {NoStop}%
\bibitem [{\citenamefont {Lu}\ \emph {et~al.}(2022)\citenamefont {Lu},
  \citenamefont {Zhang}, \citenamefont {Lou}, \citenamefont {Guo},\ and\
  \citenamefont {Yu}}]{Lu2022-MetalEmbeddedMoSi2N4ASS-OER-ORR}%
  \BibitemOpen
  \bibfield  {author} {\bibinfo {author} {\bibfnamefont {S.}~\bibnamefont
  {Lu}}, \bibinfo {author} {\bibfnamefont {Y.}~\bibnamefont {Zhang}}, \bibinfo
  {author} {\bibfnamefont {F.}~\bibnamefont {Lou}}, \bibinfo {author}
  {\bibfnamefont {K.}~\bibnamefont {Guo}},\ and\ \bibinfo {author}
  {\bibfnamefont {Z.}~\bibnamefont {Yu}},\ }\bibfield  {title} {\bibinfo
  {title} {{Non-precious metal activated MoSi$_2$N$_4$ monolayers for
  high-performance OER and ORR electrocatalysts: A first-principles study}},\
  }\href {https://doi.org/10.1016/j.apsusc.2021.152234} {\bibfield  {journal}
  {\bibinfo  {journal} {Applied Surface Science}\ }\textbf {\bibinfo {volume}
  {579}},\ \bibinfo {pages} {152234} (\bibinfo {year} {2022})}\BibitemShut
  {NoStop}%
\bibitem [{\citenamefont {Sahoo}\ \emph {et~al.}(2022)\citenamefont {Sahoo},
  \citenamefont {Ray},\ and\ \citenamefont
  {Singh}}]{Sahoo2022-ACSomega-VGe2N4-NbGe2N4-strain}%
  \BibitemOpen
  \bibfield  {author} {\bibinfo {author} {\bibfnamefont {M.~R.}\ \bibnamefont
  {Sahoo}}, \bibinfo {author} {\bibfnamefont {A.}~\bibnamefont {Ray}},\ and\
  \bibinfo {author} {\bibfnamefont {N.}~\bibnamefont {Singh}},\ }\bibfield
  {title} {\bibinfo {title} {{Theoretical insights into the hydrogen evolution
  reaction on VGe$_2$N$_4$ and NbGe$_2$N$_4$ monolayers}},\ }\href
  {https://doi.org/10.1021/acsomega.1c06730} {\bibfield  {journal} {\bibinfo
  {journal} {ACS Omega}\ }\textbf {\bibinfo {volume} {7}},\ \bibinfo {pages}
  {7837} (\bibinfo {year} {2022})}\BibitemShut {NoStop}%
\bibitem [{\citenamefont {Liu}\ \emph {et~al.}(2021{\natexlab{c}})\citenamefont
  {Liu}, \citenamefont {Ji},\ and\ \citenamefont {Li}}]{Liu2021-JPCC-MAZ-HER}%
  \BibitemOpen
  \bibfield  {author} {\bibinfo {author} {\bibfnamefont {Y.}~\bibnamefont
  {Liu}}, \bibinfo {author} {\bibfnamefont {Y.}~\bibnamefont {Ji}},\ and\
  \bibinfo {author} {\bibfnamefont {Y.}~\bibnamefont {Li}},\ }\bibfield
  {title} {\bibinfo {title} {{Multilevel theoretical screening of novel
  two-dimensional MA$_2$Z$_4$ family for hydrogen evolution}},\ }\href
  {https://doi.org/10.1021/acs.jpclett.1c02487} {\bibfield  {journal} {\bibinfo
   {journal} {Journal of Physical Chemistry Letters}\ }\textbf {\bibinfo
  {volume} {12}},\ \bibinfo {pages} {9149} (\bibinfo {year}
  {2021}{\natexlab{c}})}\BibitemShut {NoStop}%
\bibitem [{\citenamefont {Zheng}\ \emph {et~al.}(2021)\citenamefont {Zheng},
  \citenamefont {Sun}, \citenamefont {Hu}, \citenamefont {Wang}, \citenamefont
  {Yao}, \citenamefont {Deng}, \citenamefont {Pan}, \citenamefont {Pan},\ and\
  \citenamefont {Wang}}]{Zheng2021-AMI-MoSi2N4-Catalysts}%
  \BibitemOpen
  \bibfield  {author} {\bibinfo {author} {\bibfnamefont {J.}~\bibnamefont
  {Zheng}}, \bibinfo {author} {\bibfnamefont {X.}~\bibnamefont {Sun}}, \bibinfo
  {author} {\bibfnamefont {J.}~\bibnamefont {Hu}}, \bibinfo {author}
  {\bibfnamefont {S.}~\bibnamefont {Wang}}, \bibinfo {author} {\bibfnamefont
  {Z.}~\bibnamefont {Yao}}, \bibinfo {author} {\bibfnamefont {S.}~\bibnamefont
  {Deng}}, \bibinfo {author} {\bibfnamefont {X.}~\bibnamefont {Pan}}, \bibinfo
  {author} {\bibfnamefont {Z.}~\bibnamefont {Pan}},\ and\ \bibinfo {author}
  {\bibfnamefont {J.}~\bibnamefont {Wang}},\ }\bibfield  {title} {\bibinfo
  {title} {{Symbolic transformer accelerating machine learning screening of
  hydrogen and deuterium evolution reaction catalysts in MA$_2$Z$_4$
  materials}},\ }\href {https://doi.org/10.1021/acsami.1c13236} {\bibfield
  {journal} {\bibinfo  {journal} {ACS Applied Materials and Interfaces}\
  }\textbf {\bibinfo {volume} {13}},\ \bibinfo {pages} {50878} (\bibinfo {year}
  {2021})}\BibitemShut {NoStop}%
\bibitem [{\citenamefont {Chen}\ \emph
  {et~al.}(2021{\natexlab{b}})\citenamefont {Chen}, \citenamefont {Tian},\ and\
  \citenamefont {Tang}}]{Chen2021-JPCC-MAZ-ORE}%
  \BibitemOpen
  \bibfield  {author} {\bibinfo {author} {\bibfnamefont {Y.}~\bibnamefont
  {Chen}}, \bibinfo {author} {\bibfnamefont {S.}~\bibnamefont {Tian}},\ and\
  \bibinfo {author} {\bibfnamefont {Q.}~\bibnamefont {Tang}},\ }\bibfield
  {title} {\bibinfo {title} {{First-principles studies on electrocatalytic
  activity of novel two-dimensional MA$_2$Z$_4$ monolayers toward oxygen
  reduction reaction}},\ }\href {https://doi.org/10.1021/acs.jpcc.1c07044}
  {\bibfield  {journal} {\bibinfo  {journal} {Journal of Physical Chemistry C}\
  }\textbf {\bibinfo {volume} {125}},\ \bibinfo {pages} {22581} (\bibinfo
  {year} {2021}{\natexlab{b}})}\BibitemShut {NoStop}%
\bibitem [{\citenamefont {Lin}\ \emph {et~al.}(2022)\citenamefont {Lin},
  \citenamefont {Feng}, \citenamefont {Legut}, \citenamefont {Liu},
  \citenamefont {Seh}, \citenamefont {Zhang},\ and\ \citenamefont
  {Zhang}}]{Lin2022-AFM-OER}%
  \BibitemOpen
  \bibfield  {author} {\bibinfo {author} {\bibfnamefont {C.}~\bibnamefont
  {Lin}}, \bibinfo {author} {\bibfnamefont {X.}~\bibnamefont {Feng}}, \bibinfo
  {author} {\bibfnamefont {D.}~\bibnamefont {Legut}}, \bibinfo {author}
  {\bibfnamefont {X.}~\bibnamefont {Liu}}, \bibinfo {author} {\bibfnamefont
  {Z.~W.}\ \bibnamefont {Seh}}, \bibinfo {author} {\bibfnamefont
  {R.}~\bibnamefont {Zhang}},\ and\ \bibinfo {author} {\bibfnamefont
  {Q.}~\bibnamefont {Zhang}},\ }\bibfield  {title} {\bibinfo {title}
  {{Discovery of efficient visible-light driven oxygen evolution
  photocatalysts: Automated high-throughput computational screening of
  MA$_2$Z$_4$}},\ }\href {https://doi.org/10.1002/adfm.202207415} {\bibfield
  {journal} {\bibinfo  {journal} {Advanced Functional Materials}\ ,\ \bibinfo
  {pages} {2207415}} (\bibinfo {year} {2022})}\BibitemShut {NoStop}%
\bibitem [{\citenamefont {Wang}\ \emph
  {et~al.}(2022{\natexlab{b}})\citenamefont {Wang}, \citenamefont {Zhang},
  \citenamefont {Wang}, \citenamefont {Huang}, \citenamefont {Liu},
  \citenamefont {Ouayng},\ and\ \citenamefont
  {Hu}}]{Wang2022-ASS-VSi2N4-batteries}%
  \BibitemOpen
  \bibfield  {author} {\bibinfo {author} {\bibfnamefont {Z.}~\bibnamefont
  {Wang}}, \bibinfo {author} {\bibfnamefont {G.}~\bibnamefont {Zhang}},
  \bibinfo {author} {\bibfnamefont {Y.}~\bibnamefont {Wang}}, \bibinfo {author}
  {\bibfnamefont {C.}~\bibnamefont {Huang}}, \bibinfo {author} {\bibfnamefont
  {Y.}~\bibnamefont {Liu}}, \bibinfo {author} {\bibfnamefont {C.}~\bibnamefont
  {Ouayng}},\ and\ \bibinfo {author} {\bibfnamefont {J.}~\bibnamefont {Hu}},\
  }\bibfield  {title} {\bibinfo {title} {{Heavy 2D VSi$_2$N$_4$: High capacity
  and full battery open-circuit voltage as Li/Na-ion batteries anode}},\ }\href
  {https://doi.org/10.1016/j.apsusc.2022.153354} {\bibfield  {journal}
  {\bibinfo  {journal} {Applied Surface Science}\ }\textbf {\bibinfo {volume}
  {593}},\ \bibinfo {pages} {153354} (\bibinfo {year}
  {2022}{\natexlab{b}})}\BibitemShut {NoStop}%
\bibitem [{\citenamefont {Li}\ \emph {et~al.}(2022)\citenamefont {Li},
  \citenamefont {Lin}, \citenamefont {Cheng},\ and\ \citenamefont
  {Chen}}]{Li2022-PSS-layerMoSi2N4-ZnBattery}%
  \BibitemOpen
  \bibfield  {author} {\bibinfo {author} {\bibfnamefont {X.~M.}\ \bibnamefont
  {Li}}, \bibinfo {author} {\bibfnamefont {Z.~Z.}\ \bibnamefont {Lin}},
  \bibinfo {author} {\bibfnamefont {L.~R.}\ \bibnamefont {Cheng}},\ and\
  \bibinfo {author} {\bibfnamefont {X.}~\bibnamefont {Chen}},\ }\bibfield
  {title} {\bibinfo {title} {{Layered MoSi$_2$N$_4$ as electrode material of
  Zn–Air battery}},\ }\href {https://doi.org/10.1002/pssr.202200007}
  {\bibfield  {journal} {\bibinfo  {journal} {Physica Status Solidi - Rapid
  Research Letters}\ }\textbf {\bibinfo {volume} {16}},\ \bibinfo {pages}
  {2200007} (\bibinfo {year} {2022})}\BibitemShut {NoStop}%
\bibitem [{\citenamefont {Munteanu}\ \emph {et~al.}(2020)\citenamefont
  {Munteanu}, \citenamefont {Moreno}, \citenamefont {Bramini},\ and\
  \citenamefont {G{\'{a}}sp{\'{a}}r}}]{Munteanu2020-Sensors}%
  \BibitemOpen
  \bibfield  {author} {\bibinfo {author} {\bibfnamefont {R.~E.}\ \bibnamefont
  {Munteanu}}, \bibinfo {author} {\bibfnamefont {P.~S.}\ \bibnamefont
  {Moreno}}, \bibinfo {author} {\bibfnamefont {M.}~\bibnamefont {Bramini}},\
  and\ \bibinfo {author} {\bibfnamefont {S.}~\bibnamefont
  {G{\'{a}}sp{\'{a}}r}},\ }\bibfield  {title} {\bibinfo {title} {{2D materials
  in electrochemical sensors for in vitro or in vivo use}},\ }\href
  {https://doi.org/10.1007/s00216-020-02831-1} {\bibfield  {journal} {\bibinfo
  {journal} {Analytical and Bioanalytical Chemistry}\ }\textbf {\bibinfo
  {volume} {413}},\ \bibinfo {pages} {701} (\bibinfo {year}
  {2020})}\BibitemShut {NoStop}%
\bibitem [{\citenamefont {Buckley}\ \emph {et~al.}(2020)\citenamefont
  {Buckley}, \citenamefont {Black}, \citenamefont {Castanon}, \citenamefont
  {Melios}, \citenamefont {Hardman},\ and\ \citenamefont
  {Kazakova}}]{Buckley2020-sensors}%
  \BibitemOpen
  \bibfield  {author} {\bibinfo {author} {\bibfnamefont {D.~J.}\ \bibnamefont
  {Buckley}}, \bibinfo {author} {\bibfnamefont {N.~C.}\ \bibnamefont {Black}},
  \bibinfo {author} {\bibfnamefont {E.~G.}\ \bibnamefont {Castanon}}, \bibinfo
  {author} {\bibfnamefont {C.}~\bibnamefont {Melios}}, \bibinfo {author}
  {\bibfnamefont {M.}~\bibnamefont {Hardman}},\ and\ \bibinfo {author}
  {\bibfnamefont {O.}~\bibnamefont {Kazakova}},\ }\bibfield  {title} {\bibinfo
  {title} {{Frontiers of graphene and 2D material-based gas sensors for
  environmental monitoring}},\ }\href
  {https://doi.org/10.1088/2053-1583/ab7bc5} {\bibfield  {journal} {\bibinfo
  {journal} {2D Materials}\ }\textbf {\bibinfo {volume} {7}},\ \bibinfo {pages}
  {032002} (\bibinfo {year} {2020})}\BibitemShut {NoStop}%
\bibitem [{\citenamefont {Tyagi}\ \emph {et~al.}(2020)\citenamefont {Tyagi},
  \citenamefont {Wang}, \citenamefont {Huang}, \citenamefont {Hu},
  \citenamefont {Tang}, \citenamefont {Guo}, \citenamefont {Ouyang},\ and\
  \citenamefont {Zhang}}]{Tyagi2020-nanoscale-sensors}%
  \BibitemOpen
  \bibfield  {author} {\bibinfo {author} {\bibfnamefont {D.}~\bibnamefont
  {Tyagi}}, \bibinfo {author} {\bibfnamefont {H.}~\bibnamefont {Wang}},
  \bibinfo {author} {\bibfnamefont {W.}~\bibnamefont {Huang}}, \bibinfo
  {author} {\bibfnamefont {L.}~\bibnamefont {Hu}}, \bibinfo {author}
  {\bibfnamefont {Y.}~\bibnamefont {Tang}}, \bibinfo {author} {\bibfnamefont
  {Z.}~\bibnamefont {Guo}}, \bibinfo {author} {\bibfnamefont {Z.}~\bibnamefont
  {Ouyang}},\ and\ \bibinfo {author} {\bibfnamefont {H.}~\bibnamefont
  {Zhang}},\ }\bibfield  {title} {\bibinfo {title} {{Recent advances in
  two-dimensional-material-based sensing technology toward health and
  environmental monitoring applications}},\ }\href
  {https://doi.org/10.1039/C9NR10178K} {\bibfield  {journal} {\bibinfo
  {journal} {Nanoscale}\ }\textbf {\bibinfo {volume} {12}},\ \bibinfo {pages}
  {3535} (\bibinfo {year} {2020})}\BibitemShut {NoStop}%
\bibitem [{\citenamefont {Li}\ \emph {et~al.}(2021{\natexlab{f}})\citenamefont
  {Li}, \citenamefont {Yang}, \citenamefont {Yi},\ and\ \citenamefont
  {Shen}}]{Yi2021-graphenebased-sensor}%
  \BibitemOpen
  \bibfield  {author} {\bibinfo {author} {\bibfnamefont {K.}~\bibnamefont
  {Li}}, \bibinfo {author} {\bibfnamefont {W.}~\bibnamefont {Yang}}, \bibinfo
  {author} {\bibfnamefont {M.}~\bibnamefont {Yi}},\ and\ \bibinfo {author}
  {\bibfnamefont {Z.}~\bibnamefont {Shen}},\ }\bibfield  {title} {\bibinfo
  {title} {{Graphene-based pressure sensor and strain sensor for detecting
  human activities}},\ }\href {https://doi.org/10.1088/1361-665X/ac0d8b}
  {\bibfield  {journal} {\bibinfo  {journal} {Smart Materials and Structures}\
  }\textbf {\bibinfo {volume} {30}},\ \bibinfo {pages} {085027} (\bibinfo
  {year} {2021}{\natexlab{f}})}\BibitemShut {NoStop}%
\bibitem [{\citenamefont {Subbanna}\ \emph {et~al.}(2022)\citenamefont
  {Subbanna}, \citenamefont {Choudhary}, \citenamefont {Singh},\ and\
  \citenamefont {Kumar}}]{Subbanna2022-sensors}%
  \BibitemOpen
  \bibfield  {author} {\bibinfo {author} {\bibfnamefont {B.~B.}\ \bibnamefont
  {Subbanna}}, \bibinfo {author} {\bibfnamefont {K.}~\bibnamefont {Choudhary}},
  \bibinfo {author} {\bibfnamefont {S.}~\bibnamefont {Singh}},\ and\ \bibinfo
  {author} {\bibfnamefont {S.}~\bibnamefont {Kumar}},\ }\bibfield  {title}
  {\bibinfo {title} {{2D material-based optical sensors: a review}},\ }\href
  {https://doi.org/10.1007/s41683-021-00083-4} {\bibfield  {journal} {\bibinfo
  {journal} {ISSS Journal of Micro and Smart Systems}\ }\textbf {\bibinfo
  {volume} {11}},\ \bibinfo {pages} {169} (\bibinfo {year} {2022})}\BibitemShut
  {NoStop}%
\bibitem [{\citenamefont {Bafekry}\ \emph
  {et~al.}(2021{\natexlab{d}})\citenamefont {Bafekry}, \citenamefont {Faraji},
  \citenamefont {Fadlallah}, \citenamefont {{Abdolahzadeh Ziabari}},
  \citenamefont {{Bagheri Khatibani}}, \citenamefont {Feghhi}, \citenamefont
  {Ghergherehchi},\ and\ \citenamefont
  {Gogova}}]{Bafekry2021-ASS-MoSi2N4-adsorption}%
  \BibitemOpen
  \bibfield  {author} {\bibinfo {author} {\bibfnamefont {A.}~\bibnamefont
  {Bafekry}}, \bibinfo {author} {\bibfnamefont {M.}~\bibnamefont {Faraji}},
  \bibinfo {author} {\bibfnamefont {M.~M.}\ \bibnamefont {Fadlallah}}, \bibinfo
  {author} {\bibfnamefont {A.}~\bibnamefont {{Abdolahzadeh Ziabari}}}, \bibinfo
  {author} {\bibfnamefont {A.}~\bibnamefont {{Bagheri Khatibani}}}, \bibinfo
  {author} {\bibfnamefont {S.~A.}\ \bibnamefont {Feghhi}}, \bibinfo {author}
  {\bibfnamefont {M.}~\bibnamefont {Ghergherehchi}},\ and\ \bibinfo {author}
  {\bibfnamefont {D.}~\bibnamefont {Gogova}},\ }\bibfield  {title} {\bibinfo
  {title} {{Adsorption of habitat and industry-relevant molecules on the
  MoSi$_2$N$_4$ monolayer}},\ }\href
  {https://doi.org/10.1016/j.apsusc.2021.150326} {\bibfield  {journal}
  {\bibinfo  {journal} {Applied Surface Science}\ }\textbf {\bibinfo {volume}
  {564}},\ \bibinfo {pages} {150326} (\bibinfo {year}
  {2021}{\natexlab{d}})}\BibitemShut {NoStop}%
\end{thebibliography}%

\end{document}